\documentclass[11pt, twoside]{Thesis} % The default font size and one-sided printing (no margin offsets)

\usepackage[square, numbers, comma, sort&compress]{natbib} % Use the natbib reference package - read up on this to edit the reference style; if you want text (e.g. Smith et al., 2012) for the in-text references (instead of numbers), remove 'numbers' 
\title{\ttitle} % Defines the thesis title - don't touch this
\usepackage[english]{babel}
\usepackage[utf8]{inputenc}
\usepackage{amsfonts,euscript,amssymb,stmaryrd,braket}
\usepackage{graphics,tikz}
\usetikzlibrary{arrows,decorations.markings,patterns}
\usepackage{slashed}
\usepackage{multirow}
\usepackage{epsfig}
\usepackage{graphicx}
\usepackage{tabularx}
\usepackage{longtable}
\usepackage{bbold}
\usepackage{amsthm}

\newcommand{\txpf}{\texorpdfstring}
\newcommand*\cc[1]{
\hspace{2pt}\begin{tikzpicture}[baseline=-3.5pt]
\node[draw,circle,inner sep=1pt] at (0,0) {\scriptsize $#1$};
\node at (0,-0.25) {\scriptsize};
\end{tikzpicture}\hspace{2pt} }
\newcommand*\cct[2]{
\hspace{-1pt}\begin{tikzpicture}[baseline=-3.5pt]
\node[draw,circle,inner sep=1pt] at (0,0) {\scriptsize $#1$};
\node at (0,-0.25) {\scriptsize $#2$};
\end{tikzpicture}\hspace{-1pt} }

\newcommand{\ointc}{\oint}
\newcommand{\pa}{\partial}
\newcommand{\la}{\leftarrow}
\newcommand{\ra}{\rightarrow}
\newcommand{\ua}{\uparrow}

\newcommand{\spm}{{\pm}}

\newcommand{\Li}{\operatorname{Li}_2}
\newcommand{\Tr}{\operatorname{Tr}}
\newcommand{\Str}{\operatorname{Str}}
\newcommand{\as}{\operatorname{arcsinh}}
\newcommand{\qh}{\sqrt{1-q^2}}
\newcommand{\qp}{\sqrt{1+q}}
\newcommand{\qm}{\sqrt{1-q}}
\newcommand{\BES}{\text{BES}}
\newcommand{\AFS}{\text{AFS}}
\newcommand{\HL}{\text{HL}}

\newcommand{\Jc}{\mathcal{J}}
\newcommand{\cl}{\ell}
\newcommand{\IndA}{A}
\newcommand{\IndB}{B}
\newcommand{\IndC}{C}
\newcommand{\IndD}{D}
\newcommand{\IndE}{E}
\newcommand{\IndF}{F}
\newcommand{\ms}{1}

\newcommand{\hh}{\text{h}}
\newcommand{\rmp}{\text{p}}
\newcommand{\rmq}{\text{q}}
\newcommand{\rml}{\text{l}}
\newcommand{\rmP}{\text{P}}
\newcommand{\rmk}{\text{k}}
\renewcommand{\a}{\alpha}

\renewcommand{\b}{\beta}

\renewcommand{\c}{\gamma}
\newcommand{\g}{\gamma}
\renewcommand{\G}{\Gamma}
\renewcommand{\d}{\delta}
\newcommand{\D}{\Delta}
\newcommand{\e}{\epsilon}
\newcommand{\z}{\zeta}

\newcommand{\q}{\theta}

\renewcommand{\k}{\kappa}

\renewcommand{\l}{\lambda}
\renewcommand{\L}{\Lambda}
\newcommand{\m}{\mu}

\newcommand{\n}{\nu}

\newcommand{\p}{\pi}
\newcommand{\vp}{\varpi}

\renewcommand{\r}{\rho}

\newcommand{\s}{\sigma}
\renewcommand{\S}{\Sigma}
\renewcommand{\t}{\tau}

\newcommand{\vf}{\varphi}

\renewcommand{\o}{\omega}
\renewcommand{\O}{\Omega}
\newcommand{\ve}{\varepsilon}
\newcommand{\ppp}{\text{p}}
\newcommand{\qqq}{\text{q}}
\newcommand{\rrr}{\text{r}}

\newcommand{\cp}{\mathbb{CP}}

\newcommand{\adscp}{$AdS_4\times \mathbb{CP}^3$}
 \newcommand{\abjm}{{\rm ABJM}}
  \newcommand{\YM}{{\rm YM}}
\newcommand\bp{\Pi}
\def\[{\begin{equation}}
\def\]{\end{equation}}
\newcommand{\comm}[2]{[#1,#2]}
\newcommand{\acomm}[2]{\{#1,#2\}}
\newcommand{\co}{\zeta}
\newcommand{\arsinh}{\operatorname{arsinh}}

\begin{document}

\frontmatter % Use roman page numbering style (i, ii, iii, iv...) for the pre-content pages

\setstretch{1.3} % Line spacing of 1.3

% Define the page headers using the FancyHdr package and set up for one-sided printing
\fancyhead{} % Clears all page headers and footers
\rhead{\thepage} % Sets the right side header to show the page number
\lhead{} % Clears the left side page header

\pagestyle{fancy} % Finally, use the "fancy" page style to implement the FancyHdr headers

\newcommand{\HRule}{\rule{\linewidth}{0.5mm}} % New command to make the lines in the title page

% PDF meta-data
%\hypersetup{pdftitle={\ttitle}}
%\hypersetup{pdfsubject=\subjectname}
%\hypersetup{pdfauthor=\authornames}
%\hypersetup{pdfkeywords=\keywordnames}

%----------------------------------------------------------------------------------------
%	TITLE PAGE
%----------------------------------------------------------------------------------------

%\begin{titlepage}
%\begin{center}

%\textsc{\LARGE \univname}\\[1.5cm] % University name
%\textsc{\Large Doctoral Thesis}\\[0.5cm] % Thesis type

%\HRule \\[0.4cm] % Horizontal line
%{\huge \bfseries \ttitle}\\[0.4cm] % Thesis title
%\HRule \\[1.5cm] % Horizontal line
 
%\begin{minipage}{0.4\textwidth}
%\begin{flushleft} \large
%\emph{Author:}\\
%\href{http://qft.physik.hu-berlin.de/research/emmy-noether-research-group/lorenzo-bianchi/}{\authornames} % Author name - remove the \href bracket to remove the link
%\end{flushleft}
%\end{minipage}
%\begin{minipage}{0.4\textwidth}
%\begin{flushright} \large
%\emph{Supervisor:} \\
%\href{http://qft.physik.hu-berlin.de/research/emmy-noether-research-group/valentina-forini/}{\supname} % Supervisor name - remove the \href bracket to remove the link  
%\end{flushright}
%\end{minipage}\\[3cm]
 
%\large \textit{A thesis submitted in fulfilment of the requirements\\ for the degree of \degreename}\\[0.3cm] % University requirement text
%\textit{in the}\\[0.4cm]
%\groupname\\\deptname\\[2cm] % Research group name and department name
 
%{\large \today}\\[4cm] % Date
%\includegraphics{Logo} % University/department logo - uncomment to place it
 
%\vfill
%\end{center}

%\end{titlepage}
\pagestyle{empty}
\begingroup

                                             \centering
\textsf
{\huge{\textbf{Perturbation theory for string sigma models}}}\\[2\baselineskip]

                                \sffamily   \Large      DISSERTATION \\[1\baselineskip]
            \large               zur Erlangung des akademischen Grades\\

                           Doctor           rerum         naturalium 
                                             (Dr. rer. nat.)
                                   im Fach Physik\\
Spezialisierung: Theoretische Physik\\ [1\baselineskip]

                                           eingereicht an der\\

                      Mathematisch-Naturwissenschaftlichen Fakult\"at\\

                              der Humboldt-Universit\"at zu Berlin\\[1\baselineskip]

                                                     von\\

                   {\textbf{Lorenzo Bianchi}}

                   \bigskip                     

\textit{ Institut f\"ur Physik,
Humboldt-Universit\"at
zu Berlin\\ 
Zum Gro\ss en Windkanal 6, 12489 Berlin, Germany}

\bigskip 

\textit{II. Institut f\"ur Theoretische Physik,
Universit\"at Hamburg\\
Luruper Chaussee 149,
22761 Hamburg, Germany}
% \begin{table}[h]
% \begin{flushleft}
% \begin{tabular}{ll}
% %\textbf{\textsf{\large Eingereicht am:}}             & \textsf{\large xx.xx.2015} \\
% \textbf{\textsf{\large Tag der mündlichen Prüfung:}} & 
% \end{tabular}
% \end{flushleft}
% \end{table}

% \textbf{\textsf{Eingereicht am:}} 31.01.2015 \\
% \textbf{\textsf{Tag der mündlichen Prüfung:}} 31.03.2015

\endgroup

\clearpage % Start a new page

%----------------------------------------------------------------------------------------
%	QUOTATION PAGE
%----------------------------------------------------------------------------------------

\pagestyle{empty} % No headers or footers for the following pages

\null\vfill % Add some space to move the quote down the page a bit

\textit{Distress not yourself if you cannot at first understand the deeper mysteries of Spaceland. By degrees they will dawn upon you.}

\begin{flushright}
Edwin A. Abbott, ``Flatland: A Romance of Many Dimensions''.
\end{flushright}

\vfill\vfill\vfill\vfill\vfill\vfill\null % Add some space at the bottom to position the quote just right

\clearpage % Start a new page

%----------------------------------------------------------------------------------------
%	ABSTRACT PAGE
%----------------------------------------------------------------------------------------

\addtotoc{Abstract} % Add the "Abstract" page entry to the Contents

\abstract{\addtocontents{toc}{\vspace{1em}} % Add a gap in the Contents, for aesthetics
In this thesis we investigate quantum aspects of the Green-Schwarz superstring in various $AdS$ %(Anti-de-Sitter) 
backgrounds relevant for the $AdS/CFT$ correspondence, providing several examples of perturbative computations in the corresponding integrable sigma-models.\\
We start by reviewing in details the construction of the type IIB superstring action in $AdS_5\times S^5$ background defined as a supercoset sigma model, pointing out the limits of this procedure for backgrounds -- such as \adscp and $AdS_3\times S^3\times M^4$ -- interesting in lower-dimensional examples of the gauge/gravity duality. For the \adscp case we give a thorough derivation of an alternative action, based on the double-dimensional reduction of eleven-dimensional super-membranes in $AdS_4\times S^7$.\\
We then consider the light-cone gauge fixed $AdS_5\times S^5$ and $AdS_3\times S^3\times M^4$ Lagrangians in an expansion about the BMN vacuum. In this setup a particularly interesting object is the S-matrix for the scattering of worldsheet excitations in the decompactification limit. To evaluate its elements efficiently, inspired by the four-dimensional case we develop a unitarity-based method for general (relativistic and not) massive two-dimensional field theories. The outcome is a very compact formula yielding the cut-constructible part of any one-loop two-dimensional S-matrix in terms of the tree-level one. We apply the method to the perturbative calculation of worldsheet S-matrices in $AdS_5\times S^5$ and (via a partially off-shell extension of the method) in $AdS_3\times S^3\times M^4$.\\
We also analyze the AdS light-cone gauge fixed string in \adscp expanded around a “null cusp” vacuum. The free energy of this model is related to the cusp anomalous dimension of $\mathcal{N} = 6$ Chern-Simons-Matter (ABJM) theory and, indirectly, to a non-trivial effective coupling $h(\l)$ entering all integrability-based calculations in $AdS_4/CFT_3$. We calculate corrections to the  superstring partition function of the model, thus deriving the cusp anomalous dimension of ABJM theory at strong coupling up to two-loop order and giving support to a recent conjecture for the exact form of $h(\l)$. Finally, we calculate at one-loop the dispersion relation of excitations about the GKP vacuum. Results are in general agreement with the predictions from integrability, up to some expected discrepancies on which we comment.\\
Our successful application of unitarity-cut techniques on several examples supports the conjecture that S-matrices of two-dimensional integrable field theories are cut-constructible.
%, suggesting unitarity methods as a viable way to evaluate further interesting worldsheet observables as form factors. 
Furthermore, our results provide valuable data in support of the quantum consistency of the string actions - often debated due to possible issues with cancellation of UV divergences and the lack of manifest power-counting renormalizability - and furnish  non-trivial stringent tests for the quantum integrability of the analyzed models.
}

\clearpage % Start a new page
\addtotoc{Kurzfassung}

\begin{center}
 \addtocontents{toc}{\vspace{1em}}
 \huge{\textbf{Kurzfassung}}
\end{center}
In dieser Arbeit untersuchen wir Quanten-Aspekte des Green-Schwarz Superstrings in verschiedenen $AdS$-Hintergr{\"u}nden, die f{\"u}r die $AdS/CFT$ Korrespondenz von Bedeutung sind, und geben einige Beispiele f{\"u}r perturbative Rechnungen in den entsprechenden integrablen Sigma-Modellen. Wir beginnen mit einer detaillierten Darstellung der Konstruktion der Wirkung des Typ-IIB-Superstrings auf dem $AdS_5 \times S^5$-Hintergrund, die durch eine Supercoset-Sigma-Modell definiert wird, und zeigen die Grenzen dieser Herangehensweise f{\"u}r Hintergr{\"u}nde -- wie zum Beispiel $AdS_4 \times \mathbb{CP}^3$ und $AdS_3\times S^3 \times M^4$ -- auf, die in niedrig-dimensionalen Beispielen der Eich/Gravitations-Dualit{\"a}t von Interesse sind. Im Falle des $AdS_4 \times \mathbb{CP}^3$-Hintergrunds geben wir eine sorgf{\"a}ltige Herleitung einer alternativen Wirkung an, welche auf einer doppelten dimensionalen Reduktion von elfdimensionalen Super-Membranen auf $AdS_4 \times S^7$ beruht. \\
Daraufhin betrachten wir die Lichtkegel-eichfixierten Lagrangedichten auf $AdS_5 \times S^5$ und $AdS_3 \times S^3 \times M^4$ in einer Entwicklung um das BMN-Vakuum. In diesem Zusammenhang ist die S-Matrix f{\"u}r die Streuung von Weltfl{\"a}chen-Anregungen im Dekompaktifizierungs-Limes von besonderem Interesse. Um ihre Elemente effizient auszuwerten, entwickeln wir -- inspiriert durch den vierdimensionalen Fall -- eine auf Unitarit{\"a}t basierende Methode f{\"u}r allgemeine, d.h. sowohl relativistische als auch nicht-relativistische, massive, zweidimensionale Feldtheorien. Das Ergebnis ist eine sehr kompakte Formel, die den cut-konstruierbaren Anteil jeder zweidimensionalen S-Matrix auf Einschleifen-Ebene durch ihren Wert auf Baumgraphen-Niveau ausdr{\"u}ckt. Wir wenden diese Methode auf die perturbative Berechnung von Weltfl{\"a}chen-S-Matrizen in $AdS_5 \times S^5$ und (vermittels einer teilweisen Fortsetzung der Methode ins ``off-shell''-Regime) in $AdS_3 \times S^3 \times M^4$ an.\\
Weiterhin betrachten wir den $AdS$-Lichtkegel eichfixierten String in $AdS_4 \times \mathbb{CP}^3$ in einer Entwicklung um das ``null-cusp''-Vakuum. Die freie Energie dieses Modells h{\"a}ngt zusammen mit der anomalen Cusp-Dimension der $\mathcal{N}=6$ Chern-Simons-Materie (ABJM) Theorie und indirekt auch mit einer nicht-trivialen effektiven Kopplung $h(\lambda)$, die in allen auf Integrabilit{\"a}t basierenden Rechnungen in $AdS_4/CFT_3$ auftritt. Wir berechnen Korrekturen zur Zustandssumme des Superstring-Modells und leiten somit die anomale Cusp-Dimension der ABJM-Theorie bei starker Kopplung bis zur Zweischleifen-Ordnung her, wobei wir eine k{\"u}rzlich vorgebrachte Vermutung {\"u}ber die exakte Form von $h(\lambda)$ belegen. Schlie{\ss}lich berechnen wir auf Einschleifen-Ebene die Dispersionsrelation von Anregungen um das GKP-Vakuum. Unsere Ergebnisse stimmen abgesehen von einigen erwarteten Abweichungen, auf die wir eingehen, mit den Vorhersagen aus der Anwendung der Integrabilit{\"a}t {\"u}berein. \\
Unsere erfolgreiche Anwendung von auf Unitarit{\"a}t basierenden Cut-Tech\-ni\-ken auf verschiedene Beispiele st{\"u}tzt die Vermutung, dass die S-Matrizen zweidimensionaler, integrabler Feldtheorien cut-konstruierbar sind. Weiterhin liefern unsere Ergebnisse wertvolle Daten, die die Konsistenz der String-Wirkung auf Quanten-Niveau belegen -- diese ist aufgrund m{\"o}glicher Probleme bez{\"u}glich der gegenseitigen Aufhebung von UV-Divergenzen und des Fehlens eines auf Dimensionsanalyse basierenden Arguments f{\"u}r die Renormierbarkeit oft Gegenstand von Diskussionen -- und stellen nicht-triviale stringente Tests der Quanten-Integrabilit{\"a}t der untersuchten Modelle dar.

\clearpage % Start a new page
%----------------------------------------------------------------------------------------
%	ACKNOWLEDGEMENTS
%----------------------------------------------------------------------------------------

\setstretch{1.3} % Reset the line-spacing to 1.3 for body text (if it has changed)

\acknowledgements{\addtocontents{toc}{\vspace{1em}} % Add a gap in the Contents, for aesthetics
First of all, it is a great pleasure to thank my supervisor Valentina Forini for introducing me to the world of scientific research with passion, knowledge and enthusiasm. I will hardly forget the exciting discussions along these three years, and I am grateful to her for suggesting challenging and promising research topics. 

Most of the work of this thesis was done with the invaluable help of Marco Stefano Bianchi and Ben Hoare. Their expertise, which was kindly and patiently transmitted to me, was extremely precious for the success of our investigations and their company, which was no less important, made my lifetime in Berlin extremely enjoyable. 

It was a pleasure to collaborate and share the office with Edoardo Vescovi. His immeasurable aid in the final stage of this thesis will never be rewarded enough. I would also like to thank Alexis Bres and Anatoly V. Kotikov for their fruitful collaboration.

I am extremely grateful to Gabriele Travaglini for the numerous and fruitful discussions, for the ongoing collaboration and for agreeing to be my referee in various successful occasions. His presence in Berlin constituted a personal and scientific enrichment for me.

I am thankful to Arkady Tseytlin for agreeing to be the external referee in my doctoral committee and for the valuable comments he always provided on the drafts of our papers. I would also like to thank Radu Roiban for supporting me in the past round of applications and for many useful discussions.

The nice working atmosphere that I could enjoy during the last three years is mostly due to the daily effort made by Jan Plefka and Matthias Staudacher for creating such a stimulating and pleasant environment. It was a honour and a pleasure to be part of this large and excellent research group.

A special acknowledgement goes to Marco Meineri for always sharing with me his wide and deep knowledge and most importantly for the long-lasting friendship. 

The continuation of my career in science was possible thanks to the great opportunity offered to me by Volker Schomerus and Sven-Olaf Moch. I am looking forward to starting this new experience and I am grateful to them for giving me this chance.

I am indebted to Johannes Henn for the many useful suggestions and discussions, for the enjoyable time spent during various conferences around the world and especially for proposing me to be part of his research group. 

I am grateful to Lorenzo Magnea, my Master thesis supervisor, because his advices are always very helpful and his wide knowledge makes any conversation about physics extremely rich and pleasant. 

During the past three years, I enjoyed many interesting and useful conversations with Fernando Alday, Benjamin Basso, Agnese Bissi, Mariaelena Boglione, Diego Bombardelli, Riccardo Borsato, Joao Caetano, Alessandra Cagnazzo, Andrea Cavagli\`a, Vittorio Del Duca, Daniele Dorigoni, Burkhard Eden, Oluf Tang Engelund, Livia Ferro, Davide Fioravanti, Ilmar Gahramanov, Simone Giombi, Luca Griguolo, Nikolay Gromov, Omer Can Gurdogan, Martin Heinze, George Jorjadze, Zac Kenton, Thomas Klose, Gregory Korchemsky, Shota Komatsu, Martyna Kostacinska, Laura Koster, Bj\"orn Leder, Matias Leoni, Tomasz Lukowski, Andrea Mauri, Dalimil Mazac,  Tristan McLoughlin, Carlo Meneghelli, Vladimir Mitev, Emanuele Moscato, Dennis Müller, Hagen Münkler, Rob Myers, Stefano Negro, Fabrizio Nieri, Sara Pasquetti, Brenda Penante, Silvia Penati, Simone Piscaglia, Antonio Pittelli, Jonas Pollock, Elli Pomoni, Michelangelo Preti, Marco Rossi, Matteo Rosso, Domenico Seminara, Fidel I. Schaposnik, Christoph Sieg, Alessandro Sfondrini, Michael 
Smolkin, Roberto Tateo, Jonathan Toledo, Alessandro Torrielli, Stijn Van Tongeren, Pedro Vieira, Gang Yang, Linus Wulff, Kostantin Zarembo, Stefan Zieme.

A special thank to my friends Francesco Biglieri, Marta Campigotto, Andrea Celon, Mattia Bruno, Riccardo Longo, Amedeo Primo and Michele Re Fiorentin, with whom the discussions about physics filled just a little part of the enjoyable time spent together. 

Finally, the greatest acknowledgement goes to Martina, who offered me, beyond her useful and competent suggestions on this and other pieces of work, continuous support, especially in the last few weeks. Thanks for loving me the way I am.
}
\clearpage % Start a new page

%----------------------------------------------------------------------------------------
%	LIST OF CONTENTS/FIGURES/TABLES PAGES
%----------------------------------------------------------------------------------------

\pagestyle{fancy} % The page style headers have been "empty" all this time, now use the "fancy" headers as defined before to bring them back

\lhead{\emph{Contents}} % Set the left side page header to "Contents"
\tableofcontents % Write out the Table of Contents

\setstretch{1.3} % Return the line spacing back to 1.3

\pagestyle{empty} % Page style needs to be empty for this page

\dedicatory{In memory of Claudia} % Dedication text

\addtocontents{toc}{\vspace{2em}} % Add a gap in the Contents, for aesthetics

%----------------------------------------------------------------------------------------
%	THESIS CONTENT - CHAPTERS
%----------------------------------------------------------------------------------------

\mainmatter % Begin numeric (1,2,3...) page numbering

\pagestyle{fancy} % Return the page headers back to the "fancy" style

% Include the chapters of the thesis as separate files from the Chapters folder
% Uncomment the lines as you write the chapters

% Chapter 1

\chapter{Introduction} % Main chapter title

\label{introduction} % For referencing the chapter elsewhere, use \ref{Chapter1} 

\lhead{Chapter 1. \emph{Introduction}} % This is for the header on each page - perhaps a shortened title

Soon after the proposal of the AdS/CFT correspondence~\cite{Maldacena:1997re,Witten:1998qj} the construction of superstring theory for various $AdS$ backgrounds became an urgent and challenging question. For the prototypical instance of the duality, stating the equivalence of type II B superstring theory in $AdS_5 \times S^5$ background and four-dimensional $\mathcal{N}=4$ Super Yang Mills (SYM) theory, this problem was solved in~\cite{Metsaev:1998it}, where the authors -- inspired by the flat space case~\cite{Henneaux:1984mh} -- exploited the maximal supersymmetry of the background to construct the superstring action as a supercoset non-linear sigma model. The convenience of this formulation became even more manifest when it turned out to encode the classical integrability of the model~\cite{Bena:2003wd}. Hints of the presence of an integrable structure were first observed on the gauge theory side looking at the structure of the one-loop dilatation operator in the planar limit~\cite{Minahan:2002ve}. 
The latter is realized for a $SU(N)$ gauge theory taking $N\to \infty$ with the 't Hooft coupling $\l=g^2 N$ kept fixed~\cite{'tHooft:1973jz}. The AdS/CFT dictionary provides a map between the two gauge theories parameters $\{\l,N\}$ and the string theory ones $\{T,g_s\}$, where $T$ is the string tension, appearing as an overall factor in the Polyakov action and $g_s$  is the string coupling, entering the genus expansion for target space string interaction. The planar limit is then translated into the $g_s\to 0$ limit and we are left with one single parameter, the 't Hooft coupling $\l$ or equivalently the string tension $T$. 
The precise relation between $\l$ and $T$ depends on the specific example of AdS/CFT one is to consider, however as a general rule the two coupling are related by  a monotonic function which maps large values of $\l$ to large values of the string tension, or equivalently small values of $\a'$. This statement can be reformulated saying that AdS/CFT is a strong/weak duality, i.e. the natural superstring perturbative expansion ($\a'\to 0$, $\l\to \infty$) explores a regime which is not accessible to the standard perturbative gauge theory ($\l\to 0$). The discovery of an integrable structure and the assumption of its all-loop validity have therefore offered a valuable tool for testing the conjecture and solving the model exactly.

\section{Integrability in AdS/CFT}

The first hints of an integrable structure in the context of AdS/CFT emerged in the study of the one-loop anomalous dimension of scalar operators in $\mathcal{N}=4$ SYM~\cite{Minahan:2002ve}. In that case Minahan and Zarembo observed that the planar one-loop dilatation operator in the $SO(6)$ sector is isomorphic to the Hamiltonian of a $SO(6)$ integrable spin chain and it can be diagonalized using the (coordinate~\cite{Bethe:1931hc} or algebraic~\cite{Faddeev:1996iy}) Bethe Ansatz technique. This observation was then extended to the full one-loop dilatation operator and to higher loops for some sectors~\cite{Beisert:2003tq,Beisert:2003jj,Beisert:2003yb,Beisert:2004hm}. At the same time the worldsheet sigma model on $AdS_5\times S^5$ background supported by a self-dual Ramond-Ramond (RR) five-form flux was observed to be classically integrable by constructing explicitly a Lax pair~\cite{Bena:2003wd}. 

Motivated by encouraging indications coming from both sides of the duality~\cite{Kazakov:2004nh,Arutyunov:2004vx}, integrability was assumed to be preserved at the quantum level, allowing to formulate an all-loop  Asymptotic Bethe Ansatz (ABA)~\cite{Beisert:2005tm,Beisert:2005fw,Beisert:2006ez}, whose solution would provide the exact anomalous dimension of any long single-trace local gauge-invariant operator in $\mathcal{N}=4$ SYM. Equivalently, on the string theory side an exact S-matrix for the worldsheet excitations of the superstring in light-cone gauge was extrapolated using the off-shell symmetry algebra~\cite{Arutyunov:2006ak}, and shown to be equivalent to the gauge theory one~\cite{Arutyunov:2006yd}. 

The ABA solves the spectral problem for the case of very long operators (or very long strings); the non-trivial generalization of this setting to include finite-size corrections has been achieved by the introduction of the Thermodynamic Bethe Ansatz (TBA) (or equivalently the ``Y-system'')~\cite{Arutyunov:2009zu,Gromov:2009tv,Bombardelli:2009ns,Gromov:2009bc,Arutyunov:2009ur,Cavaglia:2010nm}, which then has evolved into the successful technique of the quantum spectral curve~\cite{Gromov:2013pga}.

Significant progresses in our understanding of the correspondence were also achieved by the discovery of integrable structures for other examples of AdS/CFT. Integrability for the $AdS_4/CFT_3$ model relating type
II A superstring theory in $AdS_4\times \mathbb{CP}^3$ \footnote{Supported by RR four-form flux through $AdS_4$ and RR two-form flux  through a $\cp^1$ in $\cp^3$.} with the three-dimensional $\mathcal{N}=6$ super Chern-Simons theory proposed by Aharony, Bergman, Jafferis and Maldacena (ABJM)~\cite{Aharony:2008gk} was pointed out soon after the original ABJM paper~\cite{Minahan:2008hf} and already in~\cite{Gromov:2008qe} Gromov and Vieira proposed an all-loop ABA. Further indications of the validity of such a Bethe Ansatz came from the analysis of the string sigma model in the supercoset description~\cite{Stefanski:2008ik,Arutyunov:2008if}(on which we comment further in the following) and from the exact S-matrix obtained in~\cite{Ahn:2008aa} postulating the off-shell symmetry later derived by~\cite{Bykov:2009jy}. The surprising result of all this analysis is that the integrable structure underlying the $AdS_4/CFT_3$ system is basically the same as the one describing $AdS_5/CFT_4$, and the difference resides 
in an interpolating function of the coupling $h(\l)$ which we will extensively study in the following. Using 
this 
similarity between the two models significant progresses were made towards the solution of the spectral problem~\cite{Bombardelli:2009xz,Gromov:2009at,Cavaglia:2014exa,Gromov:2014bva}.

The use of integrability techniques for the $AdS_3/CFT_2$ system is more recent (for a review see~\cite{Sfondrini:2014via}) and was initiated in~\cite{Babichenko:2009dk} by studying the classical integrability of the superstring action in $AdS_3\times S^3\times M^4$ backgrounds supported by RR flux (see section \ref{sec:stringAdS3} for a detailed discussion). Further indications were then collected in~\cite{Sundin:2012gc,Sundin:2013uca}. Interestingly, integrability turned out to be present also when the background is supported by a mix of Ramond-Ramond (RR) and Neveu-Schwarz-Neveu-Schwarz (NSNS) fluxes~\cite{Cagnazzo:2012se}.
All these elements came from an analysis of the string theory side of the duality since in this case, despite some recent progresses~\cite{Pakman:2009mi,Sax:2014mea}, it is not clear how integrability plays a role on the gauge theory side.
The main peculiarity of these models, compared to the higher dimensional relatives, is the presence in the string spectrum of massless modes, whose treatment in two dimensions can be rather tricky. For this reason the first works on the subject focused on the massive subsector of the $AdS_3\times S^3\times T^4$ and $AdS_3\times S^3\times S^3 \times S^1$, both supported by pure RR flux~\cite{OhlssonSax:2011ms,Borsato:2012ud,Borsato:2012ss,Borsato:2013qpa,Borsato:2013hoa} and mixed flux~\cite{Hoare:2013ida,Hoare:2013lja}. Massless modes were then included in the integrable description~\cite{Sax:2012jv,Lloyd:2013wza,Borsato:2014exa,Lloyd:2014bsa}, although the perturbative interpretation for large string tension remains, to our knowledge, an open problem.

We conclude this section with a remark. Along the way that leads to an exact solution of the $AdS/CFT$ system via integrability, one has to make a series of assumptions, whose correctness can be tested only by internal consistency and comparison with perturbative results (or, when available, finite coupling predictions obtained by different techniques). That is why the development of new techniques to improve our computational efficiency at the perturbative level on both sides of the correspondence is highly desirable. This thesis  is devoted to this kind of investigation for large values of the 't Hooft coupling. In this regime quantum string  corrections are 
in general non-trivial to calculate, in connection with issues of potential UV divergences and the lack of 
manifest power-counting renormalizability of the string action when expanded around a particular 
background, but have the additional important role 
 of establishing the quantum consistency of the proposed string actions.

\section{Superstring theory for \txpf{$AdS$}{AdS} backgrounds}

String theory can be seen as a non-linear sigma model mapping the two-dimensional worldsheet to an arbitrary target space. The dimension of the latter is arbitrary as long as one considers the classical theory, but it's fixed by consistency after quantization. For superstring theory the cancellation of quantum anomalies fixes the spacetime dimension to $d=10$ (for a review we refer to classical textbooks~\cite{Green:1987sp,Green:1987mn,Polchinski:1998rq,Polchinski:1998rr}). Since in general $AdS$ backgrounds are supported by RR fluxes, the Neveu-Schwarz Ramond (NSR) approach~\cite{Neveu:1971rx,Ramond:1971gb} is not applicable in a straightforward way. On the other hand, the backgrounds we are mostly interested in are chosen in order to preserve a certain amount of spacetime supersymmetries and therefore the Green Schwarz (GS) approach~\cite{Green:1983wt}, which automatically ensures supersymmetry in target space, seems to be more adequate in this context.

As mentioned above, the construction of the GS superstring for the $AdS_5\times S^5$ background was carried out in~\cite{Metsaev:1998it} using a sigma model on a supercoset target space. This construction is tied to the high (super)symmetry of the background. Indeed, $AdS_5\times S^5$ supported by RR 5-form flux, together with its two limits $pp$-wave and flat space preserves all the 32 supercharges of type IIB supergravity~\cite{Schwarz:1983qr}. In the coset construction, this is translated into a superstring action with 32 fermionic degrees of freedom and all the necessary physical properties one is to expect from a GS action. Among those, a relevant one is $\k$-symmetry, a local fermionic symmetry which constitutes a distinguishing feature of the GS superstring and allows to halve the fermionic degrees of freedom obtaining the expected 16 real fermions.

When trying to apply the same procedure to the case of $AdS_4\times \mathbb{CP}^3$~\cite{Arutyunov:2008if,Stefanski:2008ik} one realizes that the output is noticeably different, in that the number of fermionic degrees of freedom in the coset construction equals the number of preserved supercharges of the background, in this case 24~\cite{Nilsson:1984bj}. The puzzle is solved by noticing that the resulting action can be interpreted as a partially gauge fixed GS action where the residual $\k$-symmetry freedom allows to eliminate only 8 fermionic degrees of freedom. A further complication comes into the game if one is to consider string configurations lying only in the $AdS_4$ part of the space~\cite{Arutyunov:2008if}. For these ``singular'' configurations the coset approach has to be discarded and one has to rely on the full superstring action derived as a double dimensional reduction of a supermembrane action  in eleven dimensions~\cite{deWit:1998yu,Gomis:2008jt,Grassi:2009yj}. In section \ref{sec:stringAdS4} 
we analyze this issue 
in some details.  

A similar situation is encountered when studying GS superstring in $AdS_3\times S^3\times T^4$ background. In that case the coset approach~\cite{Babichenko:2009dk} yields an action with fully fixed $\k$-symmetry gauge and, unfortunately, that gauge turns out not to be compatible with any of the possible bosonic light-cone gauges one may fix. Additionally, unlike the higher dimensional case there is no known way to write down the full GS action for such backgrounds and the only possible strategy is to expand the general expression for GS string on curved backgrounds in higher powers of the fields~\cite{Sundin:2012gc,Rughoonauth:2012qd,Sundin:2013ypa,Wulff:2013kga,Wulff:2014kja}.

\section{The BMN vacuum}\label{sec:BMNvacuum}

The AdS/CFT duality establishes a correspondence between the anomalous dimension of local gauge-invariant operators in a CFT and the energy of string states in $AdS$ backgrounds. The simplest operators one can choose in the gauge theory are chains of scalar operators of the form
\begin{equation}\label{eq:BMNop}
 \mathcal{O}=\Tr\{\overbrace{ZZZ...ZZZ}^{L \text{ times}}\}\, ,
\end{equation}
where $Z$ is one of the three complex scalars in $\mathcal{N}=4$ SYM. This operator has various nice features (see also the very nice review~\cite{Minahan:2010js}). First of all it is a superconformal chiral primary, as one can argue noticing that the dimension $\D$ equals the R-charge $J$ 
\begin{equation}\label{eq:BPScond}
 \D=J=L\, .
\end{equation}
Moreover the same condition \eqref{eq:BPScond} implies that such operator is annihilated by half of the supercharges of the superconformal algebra, i.e. it is a BPS operator. The most relevant consequence of this fact is that the BPS requirement imposes the condition \eqref{eq:BPScond} for any value of the coupling, implying that the dimension $\D$ is protected from quantum corrections. Due to these particular properties this state seems to be very convenient to be considered as a vacuum state.

\begin{figure}[htbp]
\begin{center}
\includegraphics[width=0.3 \linewidth]{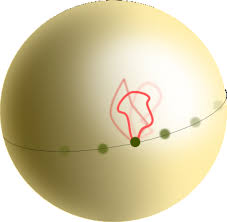}
\caption{The classical solution associated to the BMN vacuum is simply a point-like string rotating on a circle in $S^5$. Courtesy of~\cite{McLoughlin:2010jw}}.
\label{fig:BMN}
\end{center}
 \end{figure}

The name of BMN vacuum goes back to the paper by Berenstein, Maldacena and Nastase (BMN)~\cite{Berenstein:2002jq}, where a precise connection was established between a class of operators in the gauge theory (BMN operators) and the spectrum of superstring theory on a $pp$-wave background~\cite{Blau:2001ne,Blau:2002dy,Metsaev:2001bj,Metsaev:2002re}. 
The operator \eqref{eq:BMNop} is the simplest possible BMN operator and it is associated to the vacuum state in the string theory spectrum. Even without restricting to the $pp$-wave limit, it is interesting to understand which classical string configuration is associated to the operator \eqref{eq:BMNop}. 
In particular, the AdS/CFT dictionary translates equation \eqref{eq:BPScond} to the requirement $E=J$, where in this case $E$ is the target space energy (conjugated to the time variable $t$ in $AdS$) and $J$ is an angular momentum in $S^5$. Therefore the simplest classical solution we can think about is a point-like string rotating on a circle in $S^5$ (see figure \ref{fig:BMN})
\begin{equation}\label{eq:BMNtraj}
t=\frac{\t}{2}\, , \qquad \phi=\frac{\t}{2} \, ,
\end{equation}
where $\phi$ is an angle coordinate in $S^5$ and the factor $\frac12$ is introduced for future convenience. Equation \eqref{eq:BMNtraj} clearly implies $x^+=t+\phi=\t$ and suggests the perturbative quantization of the string in light-cone gauge~\cite{Frolov:2006cc}. It turns out there is a precise connection between the light-cone gauge excitations of the string and the possible impurities one can insert in the operator \eqref{eq:BMNop}.

\section{Spin-chain vs. worldsheet excitations}

The main observation of~\cite{Minahan:2002ve} was that operators like \eqref{eq:BMNop} can be paralleled to the vacuum state of a spin chain whose excited states are constructed by insertion of other fundamental fields of the theory inside the operator \eqref{eq:BMNop}. This assumption was motivated by the crucial observation that the one-loop dilatation operator has the structure of the Hamiltonian of an integrable spin chain. In the $SU(2)$ sector, i.e. the one featuring only the complex excitations $Z$ and $X$, the picture is rather clear since it maps to the familiar one-dimensional $SU(2)$ spin-chain
\begin{equation}\label{eq:BMNexc}
\mathcal{O}=\Tr\{ZZZZXZZZ\}\qquad  \longleftrightarrow \qquad 
 \begin{tikzpicture}[baseline=0]
 \draw [dotted] (0,0) circle (1cm);
\node (a1) at (0,1) {$\downarrow$};
\node (a2) at (0.707,0.707) {$\uparrow$};
\node (a3) at (-0.707,0.707) {$\uparrow$};
\node (a4) at (0.707,-0.707) {$\uparrow$};
\node (a5) at (-0.707,-0.707) {$\uparrow$};
\node (a6) at (0,-1) {$\uparrow$};
\node (a6) at (1,0) {$\uparrow$};
\node (a6) at (-1,0) {$\uparrow$};
 \end{tikzpicture}
\end{equation}
and the dilatation operator has simply the structure of the Hamiltonian for an XXX Heisenberg spin-chain. 

The problem is then reduced to the diagonalization of such a Hamiltonian, a task which becomes increasingly difficult when adding additional excitations and considering higher orders in the perturbative expansion of the dilatation operator. Nevertheless Hans Bethe, back in 1931, developed a powerful technique whose range of application is much wider than he would probably have imagined~\cite{Bethe:1931hc} (see also the review~\cite{Faddeev:1996iy}). The idea is to consider the perturbations of the spin chain vacuum (insertion of fundamental fields in the gauge theory picture) as fundamental excitations (called magnons) with their own wave-functions and dispersion relations. When a magnon moves along the chain it sequentially scatters with its neighbour and for every scattering one has a scattering phase which modifies the magnon wave function. Assuming periodic boundary conditions, the consistency of the whole construction imposes that a magnon moving around the whole chain and scattering through all the 
other 
excitations should reproduce the initial state when coming back to the initial site. This consistency condition imposes a set of algebraic equations commonly known as Bethe equations, whose main ingredient is the S-matrix for the scattering of magnons.

Considering the full field content of $\mathcal{N}=4$ SYM, the vacuum \eqref{eq:BMNop} in the $L\to \infty$ limit\footnote{While the one-loop dilatation operator is isomorphic to a Hamiltonian with only nearest-neighbour interaction, higher order corrections involve long-range interactions which, in the finite $L$ case, can wrap around the chain~\cite{Rej:2010ju}.} (so called asymptotic region) can be excited by 8 bosonic and 8 fermionic fundamental magnons\footnote{A very convenient way to represent generic single trace operators of $\mathcal{N}=4$ SYM is in terms of excitations of different oscillators, using bosonic and fermionic magnon-creation operators~\cite{Gunaydin,Gunaydin:1998sw,Gunaydin:1998jc} (see also~\cite{Beisert:2004ry,Minahan:2010js})} which transform under a centrally extended $\mathfrak{psu}(2|2)\oplus \mathfrak{psu}(2|2)$ algebra~\cite{Beisert:2004ry}. This symmetry suffices to fix completely the form of the magnons dispersion relation~\cite{Beisert:2004hm,Beisert:2005tm}
\begin{equation}\label{eq:disprel}
 \o(p)=\sqrt{1+4h(\l)^2 \sin^2 \frac{p}{2}}
\end{equation}
and of the S-matrix up to an overall factor (see section \ref{sec:intSmat}). In \eqref{eq:disprel} $h(\l)$ is a function of the 't Hooft coupling which acts as an effective coupling for every integrability-based calculation and in general is not fixed by symmetries. In the case of $\mathcal{N}=4$ SYM the simple relation $h(\l)=\frac{\sqrt{\l}}{2\pi}$ has been first suggested by various weak and strong coupling arguments~\cite{Gross:2002su,Santambrogio:2002sb,Hofman:2006xt,Minahan:2006bd,Papathanasiou:2007gd,Klose:2007wq,Sieg:2010tz}(see also~\cite{Berenstein:2009qd}) and then proven by comparing two computations of the Bremsstrahlung function by TBA and by supersymmetric localization~\cite{Correa:2012at,Correa:2012hh,Gromov:2012eu}. The same is not true for lower dimensional examples of $AdS/CFT$. For the $AdS_4/CFT_3$ system, for instance, the computation of $h(\l)$ at finite coupling is an open and challenging problem and a conjecture has been recently proposed 
in~\cite{Gromov:2014eha},
supported by various weak and strong coupling perturbative 
results~\cite{Gaiotto:2008cg,Grignani:2008is,Nishioka:2008gz,Minahan:2009aq,
Minahan:2009wg,Leoni:2010tb,McLoughlin:2008ms,Alday:2008ut,Krishnan:2008zs,McLoughlin:2008he,Gromov:2008fy,
 Astolfi:2008ji,Bandres:2009kw, Abbott:2010yb,Abbott:2011xp,Astolfi:2011ju,Astolfi:2011bg,LopezArcos:2012gb,Forini:2012bb,Bianchi:2014ada}, the latest of which is reviewed in section \ref{sec:cuspAdS4}.
 
 A symmetry pattern similar to the one we described for the spin chain excitations over the BMN vacuum has been found in the study of the light-cone gauge fixed superstring action~\cite{Frolov:2006cc}. In this case, the counterparts of the $8+8$ fundamental magnons are the $8$ bosonic and $8$ fermionic worldsheet excitations characterizing a general light-cone gauge fixed string. They all have the same mass and transform under $\mathfrak{psu}(2|2)\oplus \mathfrak{psu}(2|2)$. As any worldsheet action for closed string, the light-cone gauge fixed action is defined on a cylinder and this prevents the definition of asymptotic states. 
To be able to define the scattering among worldsheet excitations and compare them with the spin-chain picture, one decompactifies the worldsheet and relaxes the level matching condition~\cite{Ambjorn:2005wa,Klose:2006dd,Arutyunov:2006iu}. 
As a consequence, in this unphysical setup the algebra gets centrally extended and the parallel with the spin chain picture works perfectly~\cite{Arutyunov:2006ak} as one can check, for instance, via a perturbative (or exact) study of the worldsheet S-matrix~\cite{Klose:2006zd,Klose:2007wq,Klose:2007rz,Bianchi:2013nra,Engelund:2013fja,Arutyunov:2006ak}.
 
 \section{Exact worldsheet S-matrices}\label{sec:intSmat}
 
The light-cone gauge fixed sigma model in the decompactification limit is classically integrable and enjoys a centrally extended $\mathfrak{psu}(2|2)\oplus \mathfrak{psu}(2|2)$ algebra. The presence of integrability at the quantum level is much harder to establish. The complicated non-polynomial form of the Hamiltonian prevents a canonical quantization, and in the process of perturbative quantization the definition of the quantum model seems to be related to the possibility of finding a symmetry-preserving regulator for the UV and IR divergences arising in higher order computations. Therefore one can follow two different approaches. Either one assumes that integrability is preserved at the quantum level and extracts finite coupling results, or one sticks to the perturbative quantization and performs some checks of quantum integrability pushing the calculation to higher orders in perturbation theory. In this thesis we follow the latter option (see section \ref{sec:Smat}), however let 
us briefly describe the successes of the former.

The exact S-matrix for the $\mathcal{N}=4$ SYM spin chain, up to an overall phase, has been determined in~\cite{Beisert:2005tm} using the residual global symmetry algebra of an infinitely long spin-chain. In~\cite{Arutyunov:2006yd} a parallel analysis was carried out using the Zamolodchikov-Faddeev algebra for the worldsheet excitations. The result was an exact S-matrix physically equivalent to that of~\cite{Beisert:2005tm}, although related by a non-local transformation of the basis state \footnote{An important implication of this fact is that, while the S-matrix of~\cite{Arutyunov:2006yd} satisfies the standard YB equation, that of~\cite{Beisert:2005tm} satisfies a twisted version of it.}. We refer to the S-matrix in~\cite{Beisert:2005tm} as the spin-chain frame and to the one in~\cite{Arutyunov:2006yd} as the string frame.

The undetermined overall factor, often called dressing factor, has been object of a long debate (for a review see~\cite{Vieira:2010kb}). The idea of exploiting a non-relativistic generalization of the crossing symmetry was put forward in~\cite{Janik:2006dc}. The strong coupling leading order of the phase appeared in~\cite{Arutyunov:2004vx}, while a method for determining the next-to-leading order was proposed in~\cite{Beisert:2005cw} and then applied in~\cite{Hernandez:2006tk,Freyhult:2006vr}. A final all-order proposal was made in~\cite{Beisert:2006ez} and it passed all the tests performed so far. In~\cite{Volin:2009uv} the same expression was shown to constitute a minimal solution to the crossing functional equation of~\cite{Janik:2006dc}. It is worthwhile mentioning that in all this process 
of derivation of the dressing phase, perturbative data both 
from the string and gauge theory sides have been crucially 
important~\cite{Frolov:2002av,Frolov:2003qc,Frolov:2003tu,Frolov:2003xy,Arutyunov:2003uj,Frolov:2004bh,Park:2005ji,Beisert:2006ib,Beisert:2006ez}. 

The determination of the dressing phase is even subtler in the case of $AdS_3\times S^3 \times M^4$ backgrounds, where $M^4=T^4$ or $M^4=S^3\times S^1$. Also in this case the worldsheet S-matrix can be fixed by symmetries and integrability~\cite{Borsato:2012ud,Borsato:2012ss,Borsato:2013qpa,Borsato:2013hoa,Hoare:2013ida,Hoare:2013lja}, however much less is known about the corresponding dressing phases. Thus far there is only an all-loop conjecture (supported
by semiclassical one-loop computations in~\cite{Abbott:2012dd,Beccaria:2012kb,Bianchi:2014rfa}\footnote{The semiclassical one-loop %v2+
computation~\cite{Beccaria:2012kb} is not in complete
agreement with the others. While the logarithmic terms match, the rational terms in~\cite{Beccaria:2012kb} and~\cite{Abbott:2012dd,Bianchi:2014rfa} are different and the latter agree with the expansion of the exact result proposed in~\cite{Borsato:2013hoa}. The
precise reason for the disagreement is currently unclear.\label{clarification}}) for the phases in the $AdS_3
\times S^3 \times T^4$ case supported by RR flux~\cite{Borsato:2013hoa}.  There
is also a semiclassical one-loop computation of the phases in the $AdS_3 \times
S^3 \times S^3 \times S^1$ case in~\cite{Abbott:2013ixa} and one by unitarity methods~\cite{Bianchi:2014rfa} which we will review in section \ref{sec:AdS3}. As far as the mixed flux case is concerned the only available results come from the simultaneous and independent calculations of~\cite{Babichenko:2014yaa,Bianchi:2014rfa} then confirmed by~\cite{Sundin:2014ema}. For the latter two cases however an all-order proposal is still lacking.

\section{Perturbative scattering and unitarity methods}
The perturbative study of the two-body S-matrix for the world-sheet sigma-model  (for a review, see~\cite{Arutyunov:2009ga,McLoughlin:2010jw}) was initiated in~\cite{Klose:2006zd}~\footnote{Earlier work on related models  with truncated field content appeared in~\cite{Klose:2006dd,Roiban:2006yc}.}, where the full tree-level result was first derived. As for the one-loop~\cite{Klose:2007wq} and two-loop~\cite{Klose:2007rz} scattering, computations have been carried out firstly in the simpler near-flat-space limit~\cite{Maldacena:2006rv}, where interactions are at most quartic in the fields. These studies have also explicitly shown some consequences of the integrability of the model, such as the factorization of the many-body S-matrix  and the absence of particle production in the scattering processes~\cite{Puletti:2007hq}.

The first one-loop result for the full sigma model was obtained by unitarity methods~\cite{Bern:1994zx} in~\cite{Bianchi:2013nra} shortly followed by one- and two-loop calculations for the logarithmic part of the S-matrix~\cite{Engelund:2013fja}. Finally, the standard Feynman diagram computation appeared in~\cite{Roiban:2014cia}. Perturbative results on the worldsheet S-matrix for strings in $AdS_4\times \mathbb{CP}^3$ and $AdS_3\times S^3\times M^4$ backgrounds are available in~\cite{Zarembo:2009au,Sundin:2013ypa,Engelund:2013fja,Bianchi:2014rfa,Roiban:2014cia,Sundin:2014ema}. One of the aims of this thesis is to review the power and the limits of the application of unitarity methods to such processes.

\begin{figure}[htbp]
 \begin{center}
  \begin{tikzpicture}[line width=0.5pt,scale=0.8]
  \draw[-] (1,1) -- (2,0);
\draw[-] (1,-1) -- (2,0);
\draw[-] (4,0) -- (5,1);
\draw[-] (4,0) -- (5,-1);
\draw    (3,0) circle (1cm);
\draw    [draw=black!100,fill=black!5,opacity=.95] (2,0) circle (0.6cm);
\draw    [draw=black!100,fill=black!5,opacity=.95] (4,0) circle (0.6cm);
\draw[|-|,dashed,blue!70,line width=0.5pt] (3,1.5) -- (3,-1.5);
\node at (1.3,1.2) {$\scriptstyle{p_1}$};
\node at (1.3,-1.2) {$\scriptstyle{p_2}$};
\node at (4.7,1.2) {$\scriptstyle{p_3}$};
\node at (4.7,-1.2) {$\scriptstyle{p_4}$};
\node at (2,0) {$\scriptstyle{\mathcal{A}^{(0)}}$};
\node at (4,0) {$\scriptstyle{\mathcal{A}^{(0)}}$};
\node at (-1,0) {$\mathcal{A}^{(1)}|_{\textup{s-cut}}=$};
%%%TWO-LOOP%%%
\node at (-1,-4) {$\mathcal{A}^{(2)}|_{\textup{s-cut}}=$};
\draw[-] (1,-3) -- (2,-4);
\draw[-] (1,-5) -- (2,-4);
\draw[-] (4,-4) -- (5,-3);
\draw[-] (4,-4) -- (5,-5);
\draw    (3,-4) circle (1cm);
\draw    [draw=black!100,fill=black!5,opacity=.95] (2,-4) circle (0.6cm);
\draw    [draw=black!100,fill=black!5,opacity=.95] (4,-4) circle (0.6cm);
\draw[|-|,dashed,blue!70,line width=0.5pt] (3,-2.5) -- (3,-5.5);
\node at (1.3,-2.8) {$\scriptstyle{p_1}$};
\node at (1.3,-5.2) {$\scriptstyle{p_2}$};
\node at (4.7,-2.8) {$\scriptstyle{p_3}$};
\node at (4.7,-5.2) {$\scriptstyle{p_4}$};
\node at (2,-4) {$\scriptstyle{\mathcal{A}^{(1)}}$};
\node at (4,-4) {$\scriptstyle{\mathcal{A}^{(0)}}$};
\node at (6,-4) {+};
\draw[-] (7,-3) -- (8,-4);
\draw[-] (7,-5) -- (8,-4);
\draw[-] (10,-4) -- (11,-3);
\draw[-] (10,-4) -- (11,-5);
\draw[-] (8,-4) -- (10,-4);
\draw    (9,-4) circle (1cm);
\draw    [draw=black!100,fill=black!5,opacity=.95] (8,-4) circle (0.6cm);
\draw    [draw=black!100,fill=black!5,opacity=.95] (10,-4) circle (0.6cm);
\draw[|-|,dashed,blue!70,line width=0.5pt] (9,-2.5) -- (9,-5.5);
\node at (7.3,-2.8) {$\scriptstyle{p_1}$};
\node at (7.3,-5.2) {$\scriptstyle{p_2}$};
\node at (10.7,-2.8) {$\scriptstyle{p_3}$};
\node at (10.7,-5.2) {$\scriptstyle{p_4}$};
\node at (8,-4) {$\scriptstyle{\mathcal{A}^{(0)}}$};
\node at (10,-4) {$\scriptstyle{\mathcal{A}^{(0)}}$};
\node at (12,-4) {+ $\cdots$};
\end{tikzpicture}
 \end{center}
 \caption{The s-channel cut for a one- and two-loop example}	\label{fig:cut1loop}
\end{figure}
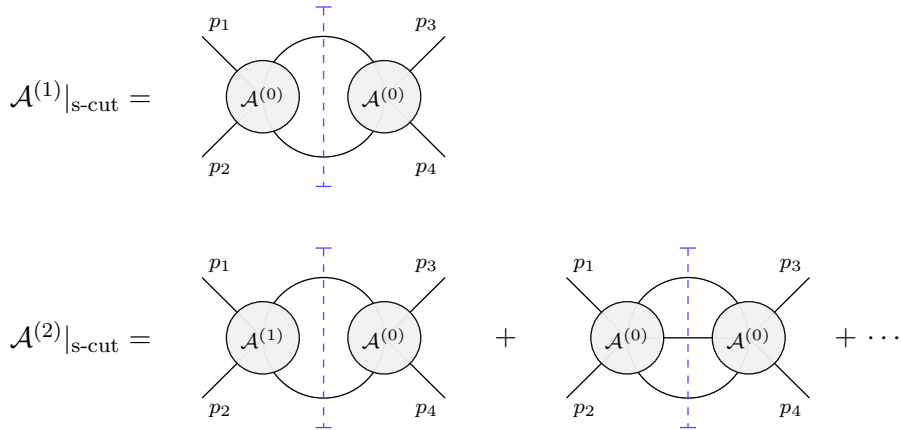

Unitarity techniques have been successfully applied to the computation of scattering amplitudes in four-dimensional gauge theories up to very high orders in perturbation theory~\cite{Bern:1994zx,Bern:1994cg,Roib_Review,Elvang:2013cua,Henn:2014yza,Bern:2012uc}\footnote{See also~\cite{Chen:2011vv,Brandhuber:2013gda,Bianchi:2013pfa} for some three-dimensional applications.}. The textbook strategy to compute loop-level scattering amplitudes consists in writing down all the possible Feynman diagrams and perform the tensor integral reduction on every single integral. This leads to express the result as a linear combination of scalar master integrals\footnote{A basis of master integrals valid for any possible process is known, at present, only at one loop~\cite{Passarino:1978jh}.}. This process turns out to be quite laborious and it soon exceeds the computational power even of our modern best computers. The unitarity methods provide a short-cut for this reduction procedure. The basic idea is that most of the 
structure of the $l$-loop amplitude can be recovered by the knowledge of the $(l-1)$-loop one. This is done in some particular kinematical channel by studying the discontinuities of the amplitude, which are known to be related by the Cutkosky rules~\cite{Cutkosky:1960sp} to some product of lower-order amplitudes. For a one- and two-loop example this is shown pictorially in figure \ref{fig:cut1loop}. The discontinuity in a particular channel is related to an imaginary part in the amplitude, given in general by some multi-valued function, such as logarithms, polylogarithms or generalized polylogs,  
of the kinematical variables. In the following, we will refer to all these possible dependences collectively as the ``logarithmic part of the amplitude''. The latter constitutes that part of the amplitude which can be unambiguously reconstructed by the unitarity methods and it is also known as the cut-constructible part.

In~\cite{Bianchi:2013nra,Engelund:2013fja,Bianchi:2014rfa} unitarity techniques were applied to the worldsheet scattering in two dimensions. On the one hand, since the one-loop basis of scalar integrals in two dimensions consists only of bubbles and tadpoles, the computations are simpler than in higher dimensions. Indeed, for a two-particle cut the loop momenta are frozen to specific values and due to the constraining two-dimensional kinematics the integral  degenerates to a sum over discrete solutions of the on-shell conditions. 
This is reminiscent of the  framework  of generalized unitarity in the four-dimensional case when quadruple  cuts (maximal cuts~\cite{Britto:2004nc}) are used. There, the quadruple-cut integral is completely localized by the four delta-functions of the cut propagators, and it reduces to a product of four tree-level amplitudes. 
On the other hand, the two-dimensional constrained kinematics yields as a drawback the presence of some ill-defined cut, whose interpretation is quite subtle and involves an order of limits problem. 

One of the intriguing consequences of this analysis is
the observation that this approach is particularly powerful when applied to
integrable field theories. In~\cite{Bianchi:2013nra,Bianchi:2014rfa} it was
observed that the full (including rational terms) one-loop S-matrices for a number
of integrable theories (including worldsheet scattering in $AdS_5 \times S^5$ and $AdS_3\times S^3 \times M^4$) are
completely cut-constructible (up to possible finite shifts in the coupling). Furthermore, as the
unitarity construction reduces the one-loop computation to scalar bubble
integrals, which are finite in two dimensions, issues of regularization are
bypassed. All these issues are widely discussed in section \ref{sec:unitarity}.

\section{GKP vacuum}

The BMN vacuum discussed in section \ref{sec:BMNvacuum} is certainly not the only possible choice. In particular we recall that the non-compact group $PSU(2,2|4)$ is rank six and therefore any operator will have a sextuplet of charges usually chosen as $(\D,S_1,S_2;J_1,J_2,J_3)$, where $\D$ is the scaling dimension, $S_1$ and $S_2$ the two $SO(1,3)$ Lorentz spins and $J_a$ are the $SO(6)$ R-charges. From the string theory point of view, the first three charges are associated to $AdS_5$ and the last three are angular momenta on $S^5$. The BMN vacuum was chosen to have large R-charge $J$ (one can choose any of the three) and large dimension $\D$. The Gubser-Klebanov-Polyakov (GKP)~\cite{Gubser:2002tv} vacuum can be seen as the $SO(1,3)$ analogue of the BMN vacuum, i.e. one considers a twist-two operator with large spin $S$ and large dimension $\D$ of the kind~\cite{Gubser:2002tv}~\footnote{The operator twist, defined as the bare scaling dimension minus the Lorentz spin, is the number of complex scalars $Z$ in \eqref{eq:GKPvacuum}}
\begin{equation}\label{eq:GKPvacuum}
 \mathcal{O_S}=\Tr\{Z \overbrace{D_+...D_+}^{S\text{ times}} Z\}+... \, ,
\end{equation}
where $D_{+}$ is the covariant light-cone derivative carrying one unit of dimension and one unit of spin, and the dots indicate that the form of
the operator is renormalized.

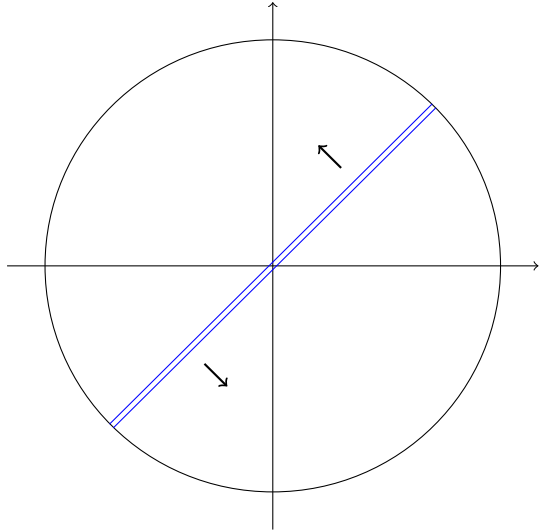
\begin{figure}[htbp]
\begin{center}
\begin{tikzpicture}
  \draw (0,0) circle (3 cm);
  \draw [->] (0,-3.5)--(0,3.5);
  \draw [->] (-3.5,0)--(3.5,0);
  \draw [-,blue] (2.10,2.15)--(-2.15,-2.10); 
  \draw [-,blue] (2.15,2.10)--(-2.10,-2.15); 
  \draw [->,thick] (0.90,1.30)--(0.60,1.60);
  \draw [->,thick] (-0.90,-1.30)--(-0.60,-1.60);
  \end{tikzpicture}
  \end{center}
  \caption{The spinning folded string in the large spin limit. $AdS_3$ is represented as a filled cylinder and the black circle is the boundary. The time direction is orthogonal to the paper. In the large spin limit the string stretches up to the boundary of $AdS_3$.}\label{fig:GKP}
\end{figure}

At strong coupling the GKP vacuum is described by the classical solution parametrizing a folded string rotating
around its center of mass in $AdS_3\subset AdS_5$~\cite{Frolov:2002av,Gubser:2002tv}. At generic values of the spin, it corresponds to a complicated
solution to the classical string equations~\cite{Frolov:2002av} and thus represents itself an intricate background for the
semiclassical expansion of the string sigma model. In the large spin limit, the string gets long with a proper length $2 \log S$ and it stretches up to the boundary for $S\to\infty$ (see figure \ref{fig:GKP}). The energy is then uniformly distributed and leads to the logarithmic scaling~\cite{Gubser:2002tv,Belitsky:2006en}
\begin{equation}
 E-S\sim f(\l) \log S \, ,
\end{equation}
where the function $f(\l)$ assumes the immediate interpretation of the energy per unit of length.

\begin{figure}[htbp]
\begin{minipage}{0.32 \linewidth}
$$(a)$$
\begin{center}
 \includegraphics[height=3 cm]{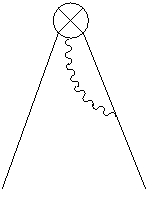}
 \end{center}
\end{minipage}
\begin{minipage}{0.32 \linewidth}
 $$(b)$$
\begin{center}
\includegraphics[height=3 cm]{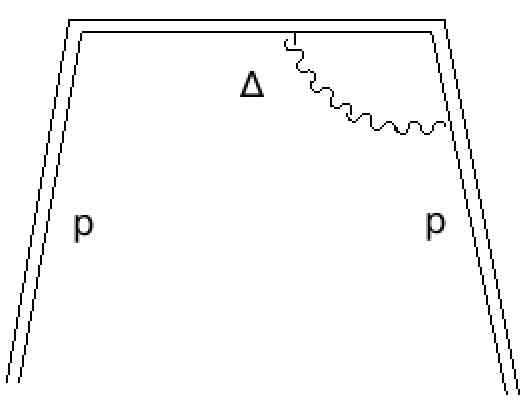}
 \end{center}
\end{minipage}
\begin{minipage}{0.32 \linewidth}
 $$(c)$$
\begin{center}
\includegraphics[height=3 cm]{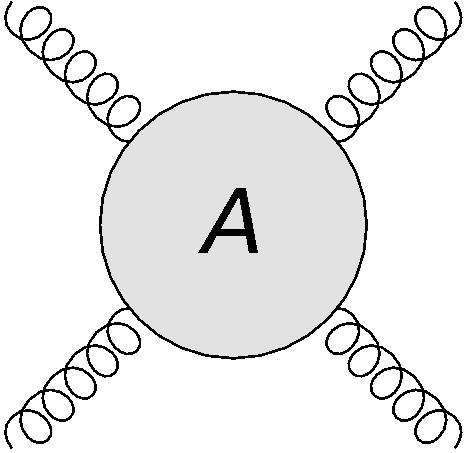}
\end{center}
\end{minipage}
\begin{minipage}{0.32 \linewidth}
 $$\Delta-S \sim 2 + f(\l)\log S$$
\end{minipage}
\begin{minipage}{0.32 \linewidth}
 
 $$\braket{W} \sim \left(\frac{L}{\e}\right)^{ f(\l) \log(p\cdot \D)}$$
\end{minipage}
\begin{minipage}{0.32 \linewidth}
$$
\log \frac{A}{A^{(0)}}=\frac{2\, f(\l)}{\e^2}+\mathcal{O}(\frac{1}{\e}) 
$$
\end{minipage}
\caption{Three possible situations where the cusp anomalous plays a distinctive role: (a) the anomalous dimension of twist-two operators at large spin; (b) the UV divergence generated a Wilson line with sharp angles; (c) the IR divergence of gluon scattering amplitudes. } \label{fig:cusp}
\end{figure}

The same logarithmic scaling emerges naturally at all loops in the gauge theory picture~\cite{Korchemsky:1988si,Korchemsky:1992xv} where the function $f(\l)$ is identified with twice the cusp anomalous dimension~\cite{Korchemsky:1985xj,Korchemsky:1987wg,Polyakov:1980ca}. The latter is an ubiquitous function in gauge theories and it emerges in many different contexts, three of which are particularly relevant and are summarized in figure \ref{fig:cusp}. Remarkably, in the context of $\mathcal{N}=4$ SYM the integrability-based BES equation~\cite{Beisert:2006ez} allows in principle to compute $f(\l)$ to any desired order in both regimes.

At strong coupling, the relation between the scaling dimension of twist-two operators and the expectation value of cusped Wilson lines is translated into the equivalence of the correspondent classical solutions~\cite{Kruczenski:2007cy}, i.e. the folded spinning string and a minimal surface ending on a null cusp respectively~\cite{Kruczenski:2002fb}. Perturbative computations about these two vacua allowed to compute the cusp anomalous dimension at strong coupling up to two loops~\cite{Frolov:2002av,Frolov:2006qe,Kruczenski:2007cy,Roiban:2007dq,Roiban:2007jf,Roiban:2007ju,Giombi:2009gd}.

The $AdS_4/CFT_3$ correspondence offers another setting where to study analogous problems~\cite{Basso:2013pxa}.  The main difference with respect to $\mathcal{N}=4$ SYM resides in the absence of a closed subsector with derivatives and scalar fields only. The simplest set of operators dual to the spinning string solution is  
%\begin{equation}
%\mathcal{O}_S=\Tr(D_+...D_+Y^1D_+...D_+ \psi_{4+}^\dagger D_+\psi^1_+D_+Y^\dagger_4)\, ,
%\end{equation}
built out of bifundamental matter fields $(Y^1,\psi_+^1), (Y_4^\dagger,\psi_{4+}^\dagger)$ and light-cone covariant derivatives $D_+$. To identify the GKP vacuum one has to look for the state with the lowest possible twist. In this case it is provided by a twist-one~\footnote{Notice that both scalar fields and fermions in three dimensions have twist 1/2.} operator containing two bifundamental matter fields and a large number $S$ of covariant derivatives.

The corresponding spinning string solution has been extensively studied up to one loop in sigma-model perturbation theory~\cite{Alday:2008ut,McLoughlin:2008ms,Krishnan:2008zs,McLoughlin:2008he,Forini:2012bb}.
This corresponds to the computation of the strong coupling cusp anomaly of ABJM theory, predicted from integrability~\cite{Gromov:2008qe} to be the same as for ${\cal N}=4$ SYM up to the presence of the effective coupling $h(\lambda)$. Therefore the comparison of the two results yields the strong coupling expansion of the interpolating function. The two-loop correction to $h(\l)$ was first computed in~\cite{Bianchi:2014ada} and will be extensively reviewed in section \ref{sec:cuspAdS4}.

\section{GKP excitations}

Paralleling the case of BMN, one may wonder how to excite the GKP vacuum. An immediate generalization of \eqref{eq:BMNexc} would be to construct a one-particle state of the form
\begin{equation}
 \Tr\{ZD_{+}...D_{+}\Phi D_+...D_+Z\} \, ,
\end{equation}
where $\Phi$ is a generic local operator of $\mathcal{N}=4$ SYM. Under renormalization this operator mixes with similar operators differing for the relative number of covariant derivatives on the left and on the right side of $\Phi$. This is interpreted as the operator $\Phi$ propagating in the background of covariant derivatives, and carrying some momentum $p$ quantized by the condition of the operator having definite scaling dimension $\D$. The main difference with respect to the BMN case is that, since the vacuum here has a complicated structure with non-trivial mixing, it would be difficult to deal with it outside the realm of integrability.

Pushing further the analogy with the BMN picture one may wonder what are the lightest elementary excitations. For the BMN case, they are those with the smallest BMN energy $\Delta-J$. For the GKP operators, the energy\footnote{One should be careful here to avoid confusion between the target space energy $E$ of the string  and the energy $\o$ of the states. The former is the variable conjugated to the time direction in $AdS$ and is mapped to the dimension of the operator in the gauge theory, the latter is an eigenvalue of the worldsheet hamiltonian and is mapped to the energy of the magnons in the spin chain picture.} is proportional to $\Delta-S$ and therefore, by definition, the elementary excitations are those with the minimal twist.
  The latter are known as light-cone operators~\cite{Belitsky:2003sh,Belitsky:2004yg,Belitsky:2005gr} and they are the building blocks of the quasi-partonic operators~\cite{Bukhvostov:1985rn}. The elementary spectrum of excitations is then given by 6 (real) scalar fields in the $\bf{6}$ of $\mathfrak{su}(4)\cong \mathfrak{so}(6)$, $4/4$ twist-one components of left/right Weyl spinors in the $\bf{4}/\bar{\bf{4}}$, and 2 twist-one components of the gluon field strength tensor in the $\bf{1}$. The leading order energy of all these excitations at weak coupling is simply given by their mass, i.e. their twist, and equals 1. 

  The mapping with the strong coupling side is not completely straightforward, since the semiclassical analysis of the GKP string shows that the elementary worldsheet excitations are 5 massless bosons for rotations in $\rm S^{5}$, 2 mass-$\sqrt{2}$ bosons for rotations of $\rm AdS_{3}$ in $\rm AdS_{5}$, 1 mass-2 boson for the transverse fluctuation in $\rm AdS_{3}$, and finally 8 mass-$1$ fermions. These states are relativistic for $\l\to \infty$, but their dispersion relation receives quantum corrections leading to highly non-trivial dispersion relations at finite coupling. The latter were derived for all the excitations in~\cite{Basso:2010in} using integrability, and a precise, though subtle, interpolation between strong and weak coupling became possible. We review the details of these relations in chapter \ref{GKP}.
 
 A similar, though somehow complementary, picture emerges in the study of the elementary excitations about the GKP vacuum in the $AdS_4/CFT_3$ case. At weak coupling the lowest lying excitations are the twist-1/2 matter fields which transform in the $\mathbf{4}$ and $\mathbf{\bar{4}}$ representations of $\mathfrak{su}(4)$. They are accompanied by twist-one fermions in the $\mathbf{6}$ and a twist-one 
excitation, neutral under $\mathfrak{su}(4)$, corresponding to the transverse component of the gauge field~\cite{Basso:2013pxa}. 

On the string theory side one finds a bosonic spectrum composed of 3 complex massless bosons for rotations in $\cp^{3}$, 1 mass-$\sqrt{2}$ boson for the direction in $AdS_{4}$ outside $ AdS_{3}$ and 1 mass-2 boson for the transverse fluctuation in $ AdS_{3}$. The 8 fermionic degrees of freedom appear as 6 mass-$1$ and 2 massless fermions. The exact dispersion relations for these excitations was found in~\cite{Basso:2013pxa} and despite the qualitative difference with respect to $\mathcal{N}=4$ SYM, the similarity of the two integrable models predicts closely related dispersion relations for the excitations in the two theories. The precise connection is investigated further in chapter \ref{GKP}.

\section*{Plan of the thesis}
In chapter \ref{Chapter2} we review the construction of superstring theory for $AdS_5\times S^5$, \adscp and $AdS_3\times S^3 \times M^4$ pointing out the advantages and the limits of the coset approach. %We also explore different possible $\k$-symmetry gauges which will be useful for the following applications.

In chapter \ref{BMN} we discuss the near-BMN expansion of the light-cone gauge fixed sigma model in $AdS_5\times S^5$ and compute the worldsheet S-matrix perturbatively up to the one-loop approximation. To perform the one-loop computation we introduce the unitarity methods, which we then apply to the study of worldsheet scattering in $AdS_3\times S^3\times M^4$ theories.

In chapter \ref{GKP} we perform two perturbative computations in the context of \adscp. First we compute the two-loop correction to the cusp anomalous dimension providing support for a recent conjecture for the interpolating function $h(\l)$, secondly we compute the quantum dispersion relation for excitations on top of the GKP vacuum, finding agreement with the  the Bethe Ansatz predictions up to some known discrepancies on which we comment.

Finally in chapter \ref{conclusion} we summarize our results and propose some future related directions. We collect in four appendices some technical details of the derivations.

% Chapter 1

\chapter{The supercoset sigma model} % Main chapter title

\label{Chapter2} % For referencing the chapter elsewhere, use \ref{Chapter1} 

\lhead{Chapter 2. \emph{The supercoset sigma model}} % This is for the header on each page - perhaps a shortened title

%----------------------------------------------------------------------------------------

This chapter is devoted to the construction of a superstring action in various $AdS$ backgrounds.  The main example is surely $AdS_5\times S^5$, which is a maximally symmetric space and can be described 
as the coset $\frac{SO(2,4)\times SO(6)}{SO(1,4)\times SO(5)}$. It was realized in~\cite{Schwarz:1983qr} that, together with flat space, the $AdS_5\times S^5$ background supported by RR flux preserves all the supersymmetries of type IIB supergravity. Therefore, it is a maximally supersymmetric background and the introduction of fermionic degrees of freedom in string theory can be achieved through the replacement of the bosonic group $SO(2,4)\times SO(6)$ with its supersymmetric extension $SU(2,2|4)$. Other backgrounds that are particularly interesting for their integrable properties are the ones relevant for lower dimensional examples of $AdS/CFT$. Here we will be mostly concerned with $AdS_4\times \mathbb{CP}^3$ and $AdS_3\times S^3 \times M^4$. In these cases interpreting the corresponding GS type II action as a coset sigma-model is not completely straightforward. We will then discuss various subtleties arising in this approach and the way to overcome them.

\section{\txpf{$\mathbb{Z}_4$}{Z4} grading and supercoset action}\label{sec:action}
 Consider a homogeneous space which can be expressed as a coset $G/H$, where $G$ is the group of isometries of the space and $H$ is the stabilizer subgroup. One can formulate GS superstring theory considering the supersymmetric extension $\tilde{G}$ of the group $G$ and taking the supercoset $\tilde{G}/H$ as the target space for the sigma model. This was first realized for flat space in~\cite{Henneaux:1984mh} and then applied to $AdS_5\times S^5$ in~\cite{Metsaev:1998it}. In the following we will specify this general construction to various examples of $AdS$ backgrounds. However, let us first restrict to a particular class of supergroups. Consider the superalgebra $\mathfrak{G}$ associated to the supergroup $\tilde G$ and an automorphism $\Omega$ such that the superalgebra $\mathfrak{G}$, as a vector space, can be decomposed into a direct sum of graded subspaces
 \begin{equation}
  \mathfrak{G}=\mathfrak{G}^{(0)}\oplus \mathfrak{G}^{(1)} \oplus \mathfrak{G}^{(2)} \oplus \mathfrak{G}^{(3)}\, ,
 \end{equation}
with
 \begin{equation}\label{eq:omega}
  \mathfrak{G}^{(k)}=\left\{ A\in \mathfrak{G},\  \O(A)=i^k A\right\}	\, . 
 \end{equation}
If such an automorphism exists, the superalgebra inherits a $\mathbb{Z}_4$ grading which turns out to be a crucial property of the model. By definition of grading, it is clear that $[\mathfrak{G}^{(k)},\mathfrak{G}^{(l)}]\subset \mathfrak{G}^{(l+k)} $, which implies that $\mathfrak{G}^{(0)}$ is a subalgebra. We will explicitly see that in all the examples of interest here the subalgebra $\mathfrak{G}^{(0)}$ will coincide with the subalgebra $\mathfrak{H}$ associated to the subgroup $H$. Additionally, since $\mathfrak{G}$ is a superalgebra, it contains already a $\mathbb{Z}_2$ grading separating bosonic from fermionic variables. Under this grading $\mathfrak{G}^{(0)}$ and $\mathfrak{G}^{(2)}$ are bosonic, whereas $\mathfrak{G}^{(1)}$ and $\mathfrak{G}^{(3)}$ are fermionic.

Given this fairly general structure, we want to find an action for the two-dimensional sigma-model on $\tilde G/H$. This is most conveniently expressed in terms of the left-invariant Cartan form 
\begin{equation}\label{eq:Adef}
 A= g^{-1} d g \in \mathfrak{G}\, ,
\end{equation}
where $g(\s_\a)\in \tilde{G}$ is a coset representative, function of the worldsheet coordinates $\s_\a, \a=1,2$. The current $A$ has the following property
\begin{itemize}
 \item $\mathbb{Z}_4$ decomposition
 \begin{equation}\label{eq:Adec}
 A=A^{(0)}+A^{(1)}+A^{(2)}+A^{(3)}.
\end{equation}
 \item Invariance under global left transformation $g\to hg$ with $h\in \tilde{G}$.
 \item Definite variation under local right transformation $g(\s_\a)\to g(\s_\a) h(\s_\a)$ with $h(\s_\a)\in H$ (and therefore $h^{-1}dh\in \mathfrak{G}^{(0)}$)
\begin{equation}
 A^{(1,2,3)}\to h^{-1}A^{(1,2,3)}h\, , \qquad A^{(0)}\to h^{-1}A^{(0)}h-h^{-1}dh\,.
\end{equation}
\item Vanishing curvature: $dA-A\wedge A=0$\, .
\end{itemize}
In this notation the action of the supercoset sigma model with $\mathbb{Z}_4$ grading reads
\begin{equation}\label{eq:action}
 S=-\frac{T}{2} \int d^2\s\, \mathcal{L}\, , \qquad \qquad \mathcal{L}=\left[\g^{\a\b} \Str\left(A^{(2)}_\a A^{(2)}_\b\right)+\k\, \e^{\a\b} \Str\left(A^{(1)}_\a A^{(3)}_\b\right)\right]\, ,
\end{equation}
where $T$ is the string tension, $\g^{\a\b}=\sqrt{-g}g^{\a\b}$ is the Weyl invariant combination of the worldsheet metric with $\det \g=-1$ and $\e^{\a\b}$ is defined with $\e^{01}=1$. The structure is, as usual, a sum of a ``kinetic'' term\footnote{Here the quotes are a reminder of the fact that we call kinetic the term which comes from $\g^{\a\b}A^{(2)}_\a A^{(2)}_\b$ and contains the kinetic terms, but also many interactions for the presence of fermions and of a non-trivial target space metric.} and a WZ term whose coefficient $\k$ will be the subject of a further discussion in section \ref{sec:ksymm}. It may not be obvious that the second term is actually a WZ term when comparing it, for instance, to the sigma-model action on a group manifold, where it appears in the usual non-local fashion ~\cite{Witten:1983ar}. Indeed one can think of that term as coming from the integration over a three-cycle of the closed three-form $\Theta_3=\Str\left(A^{(2)}\wedge A^{(3)} \wedge A^{(3)}-A^{(2)}\wedge A^{(1)} \wedge 
A^{(1)}\right)$. Nevertheless, the flatness condition of $A$ ensures that actually $\Theta_3$ is not only closed, but also exact $\Theta_3=\frac12 d\left(A^{(1)}\wedge A^{(3)}\right)$.

A few comments about the action \eqref{eq:action} are in order. First of all, one can show that the action is fixed uniquely by some well-motivated physical constraints like reproducing the Polyakov action for the $G/H$ background when the fermions are switched off, reducing to Green-Schwarz string in the flat space limit and having global $\tilde G$ invariance. This last constraint is guaranteed by the fact that the action depends only on $A$, which is invariant under the (left) action of the group $\tilde G$. Notice also that, despite the action depends on the group element $g$, being a function of $A^{(1)}$, $A^{(2)}$ and $A^{(3)}$ only, it is invariant under right multiplication by an element of $H$. As a consequence, the action actually depends only on a coset element in $\tilde G/H$ rather than a group element in $\tilde G$. The last necessary requirement to fix the form of the action is the presence of a local fermionic symmetry known as $\k$-symmetry. Since this feature is crucial for the integrable 
properties of the theory, we discuss it in some detail in the next section. 

\section{\txpf{$\k$}{k}-symmetry and integrability}\label{sec:ksymm}

The Green-Schwarz superstring in flat space enjoys a local fermionic symmetry which goes under the name of $\k$-symmetry. In this section we will discuss the presence of this symmetry in the supercoset sigma model \eqref{eq:action} and its relation with the parameter $\k$ there. We will also show that the presence of $\k$-symmetry and $\mathbb{Z}_4$ grading~\footnote{The relevance of a $\mathbb{Z}_4$-automorphism of $G$ in the construction of the coset sigma-model in this context was first understood in~\cite{Berkovits:1999zq}.} constitute a sufficient condition for the classical integrability of the model.

Let us consider the action of a group element $e^{\ve}\in\tilde G$ with $\ve \in \mathfrak{G}$ and let us assume that $\ve$ is a fermionic variable, i.e. $\ve=\ve^{(1)}+\ve^{(3)}$. The infinitesimal variation of the four components of the current $A$ read
\begin{align}
 \d_{\ve}A^{(1)}&=-d\ve^{(1)}+[A^{(0)},\ve^{(1)}]+[A^{(2)},\ve^{(3)}]\, ,\nonumber\\
\d_{\ve}A^{(3)}&=-d\ve^{(3)}+[A^{(2)},\ve^{(1)}]+[A^{(0)},\ve^{(3)}]\, ,\nonumber\\
\d_{\ve}A^{(2)}&=[A^{(1)},\ve^{(1)}]+[A^{(3)},\ve^{(3)}]\, ,\nonumber\\
\d_{\ve}A^{(0)}&=[A^{(3)},\ve^{(1)}]+[A^{(1)},\ve^{(3)}]\, . \label{eq:transfA}
\end{align}
Using these expressions and the flatness condition for $A$, one can easily extract the variation of the Lagrangian density
\begin{equation}\label{eq:lagvar}
 \d_\ve \mathcal{L}=\d \g^{\a\b} \Str\left(A^{(2)}_\a A^{(2)}_\b\right)-4\Str \left(P^{\a\b}_{+}\left[A^{(1)}_\b,A^{(2)}_\a\right]\ve^{(1)}+P^{\a\b}_- \left[A^{(3)}_\b,A^{(2)}_\a\right]\ve^{(3)}\right)\, ,
\end{equation}
where we introduced the notation
\begin{equation}
 P^{\a\b}_{\pm}=\frac12(\g^{\a\b}\pm \k\, \e^{\a\b})\, .
\end{equation}
The crucial point here is that for $\k=\pm 1$ the tensors $P^{\a\b}_{\pm}$ become orthogonal projectors 
\begin{equation}
P^{\a\b}_{\pm}{P_{\pm}}_\b^\g=P^{\a\g}_{\pm}\, , \qquad  P^{\a\b}_{\pm}{P_{\mp}}_\b^\g=0\,  ,
\end{equation}
and this turns out to be a necessary requirement for the invariance under $\k$-symmetry. Notice also that the relation $P_\pm^{\a\b} A^{(2)}_{\mp, \b}=0$ implies 
\begin{equation}\label{eq:Aprop}
 A_{\pm,\t}=-\frac{\g^{\t\s}\mp \k}{\g^{\t\t}} A_{\pm, \s}\, .
\end{equation}
In equation \eqref{eq:lagvar} we left the variation of the worldsheet metric undetermined so that we can fix it to our convenience once we manage to factor out a $\Str\left(A^{(2)}_\a A^{(2)}_\b\right)$ in the second term. In order to do this, one can change the parametrization for the $\k$-symmetry transformations such that the second term in \eqref{eq:lagvar} contains $\Str\left(A^{(2)}_\a A^{(2)}_\b\right)$. Since this change of parametrization is different for different supergroups, we discuss it on a case by case basis in the following sections. Here we anticipate that all the backgrounds we are concerned with in this review enjoy $k$-symmetry, provided we set $\k=\pm 1$ in the Lagrangian \eqref{eq:action}.

In the rest of the section we will briefly comment on the importance of this additional fermionic symmetry for the classical integrability of the theory. The literature on integrable two-dimensional quantum field theories is extremely vast and we refer the reader to the books~\cite{9780511535024,trove.nla.gov.au/work/28149925}. Here we only state some facts about the classical integrability of a two-dimensional model, and show that a supercoset sigma model with $\mathbb{Z}_4$ grading meets the conditions for being classically integrable. Quantum field theories have an infinite number of degrees of freedom and therefore, to solve a model exactly, one would need an infinite tower of conserved charges. It turns out that in some two-dimensional quantum field theories this can be achieved. In particular, one can show that the existence of a one-parameter family of connections $L_\a(\s_\a,z)$ with vanishing curvature is equivalent to the presence of an infinite tower of conserved charges. 
The parameter $z$ is called spectral parameter, and the connection $L_\a$ usually goes under the name of Lax connection (or Lax pair). The zero curvature condition
\begin{equation}
 \pa_\a L_\b -\pa_\b L_\a-[L_\a,L_\b]=0
\end{equation}
should be fulfilled for any value of the spectral parameter $z$. Of course the statement is true only at the classical level, since there is no general property preventing the quantum corrections from breaking some of the infinite symmetries of the problem. The quantum integrability of the string sigma models will be discussed in full details in the following chapters, since testing it is one of the main purposes of this work.

The upshot of the previous discussion is that an explicit expression for the Lax connection would constitute a sufficient condition for the  classical integrability of a physical system. For the supercoset sigma model, given the $\mathbb{Z}_4$ decomposition \eqref{eq:Adec}, the Lax connection is given by
\begin{equation}
 L_\a=A^{(0)}_\a+\frac12\left(z^2+\frac{1}{z^2}\right) A^{(2)}_\a-\frac{1}{2\k}\left(z^2-\frac{1}{z^2}\right)\g_{\a\b}\e^{\b\g}A^{(2)}_\g+z A^{(1)}_\a+\frac{1}{z}A^{(3)}_\a.
\end{equation}
The zero curvature condition would impose $\k=\pm 1$. In order to prove that the curvature is actually vanishing, one has to compute the curvature of $L_\a$, separate the four $\mathbb{Z}_4$ components and check that they vanish separately once the equations of motion are imposed. In other terms, one can say that the equations of motion of the supercoset sigma model are reformulated as zero curvature conditions for the Lax connection, implying their integrability. We now move to the specific analysis of some $AdS$ backgrounds.

\section{Superstring theory in \txpf{$AdS_5\times S^5$}{AdS5 x S5}}\label{sec:stringAdS5}
The action for the superstring theory in $AdS_5\times S^5$ was first written down in~\cite{Metsaev:1998it} using the aforementioned coset approach with target space $\frac{SU(2,2|4)}{SO(1,4)\times SO(5)}$\footnote{\label{fn:u1} Here we follow the original work~\cite{Metsaev:1998it} where the group of superisometries of $AdS_5\times S^5$ was taken to be $SU(2,2|4)$ and not $PSU(2,2|4)$. The difference resides in the fact that the identity is actually a matrix of the algebra $\mathfrak{su}(2,2|4)$ and the bosonic subalgebra of $\mathfrak{su}(2,2|4)$ is effectively $\mathfrak{su}(2,2)\oplus \mathfrak{su}(4)\oplus \mathfrak{u}(1)$. This additional $\mathfrak{u}(1)$ can be seen as a gauge freedom which we use to set the $A^{(2)}$ part of the decomposition \eqref{eq:Adec} to be traceless~\cite{Arutyunov:2009ga}.}. However, the action \eqref{eq:action} is still very abstract and its physical properties are not apparent. In order to bring the action to a more familiar form, let us add some more information about the 
supergroup we 
are dealing with. The superalgebra $\mathfrak{su}(2,2|4)$ can be represented by $8 \times 8$ supermatrices 
\begin{equation}\label{eq:supermatrix}
 M=\left(\begin{array}{cc} m & \theta \\
         \eta & n
        \end{array}\right)\, ,
\end{equation}
where $m$ and $n$ are bosonic $4\times 4$ matrices, whereas $\theta$ and $\eta$ are fermionic. The matrix $M$ has to satisfy
\begin{equation}
 \Str M =0\, , \qquad M=M^{\star}\, ,
\end{equation}
where the supertrace is defined in the usual way as $\Str M=\Tr m-\Tr n$ and $M^{\star}$ is given by $M^{\star}=-HM^\dagger H^{-1}$. The matrix $H$ is a diagonal matrix \begin{equation}  \label{eq:HandSigma}
H=\left(\begin{array}{cc} \S & 0 \\ 0 & \mathbb{1}_4\end{array}\right)\, , \qquad \S=\left(\begin{array}{cc} \mathbb{1}_2 & 0 \\ 0 & -\mathbb{1}_2\end{array}\right)\, ,
\end{equation}
which carries information about the signature of the target space. The automorphism $\O$ introduced in \eqref{eq:omega} in this specific case reads
\begin{equation}\label{eq:omegaAdS5}
 \O(M)=-\mathcal{K}M^{st} \mathcal{K}^{-1}, 
 \qquad M^{st}=\left(\begin{array}{cc} m^{t} & -\eta^t \\
					 \theta^t & n^t
        \end{array}\right)\, ,
 \qquad \mathcal{K}=\left(\begin{array}{cc} K & 0 \\
         0 & K
        \end{array}\right)\, ,
\end{equation}
with the matrix K given in terms of $2\times 2$ blocks as
\begin{equation}\label{eq:Kdef}
 K=\left(\begin{array}{cc} \e & 0  \\
			    0 & \e
        \end{array}\right) \qquad \e=\left(\begin{array}{cc} 0 & 1  \\
			    -1 & 0
        \end{array}\right)
\end{equation}
Looking at the projection $M^{(0)}$ in the $\mathbb{Z}_4$ decomposition 
\begin{equation}
 M^{(0)}=\frac12\left(\begin{array}{cc} m-K m^t K^{-1} & 0 \\
         0 & n-K n^t K^{-1}
        \end{array}\right)\, ,
\end{equation}
one finds that it is an element of the subalgebra $\mathfrak{so}(4,1)\oplus \mathfrak{so}(5)$, as we anticipated in section \ref{sec:action}. We can also introduce the block matrix $\mathcal{\tilde K}=\textup{diag}(K,-K)$ and express in a compact form the $\mathbb{Z}_4$ projections of an arbitrary matrix $M$. Given the separation in terms of 
\begin{equation}
 M=M_{even}+M_{odd},\qquad M_{even}=\left(\begin{array}{cc} m & 0 \\
         0 & n
        \end{array}\right), \qquad M_{odd}=\left(\begin{array}{cc} 0 & \theta \\
         \eta & 0
        \end{array}\right) ,
\end{equation}
we can project the supermatrix using
\begin{align}
 M^{(0)}&=\frac12 (M_{even}-\mathcal{K} M^t_{even} \mathcal{K}), &  M^{(2)}&=\frac12 (M_{even}+\mathcal{K} M^t_{even} \mathcal{K}), \label{eq:bospro}\\
 M^{(1)}&=\frac12 (M_{odd}-i\mathcal{\tilde K} M^t_{odd} \mathcal{K}), &  M^{(3)}&=\frac12 (M_{odd}+i\mathcal{\tilde K} M^t_{odd} \mathcal{K}). \label{eq:ferpro}
\end{align}

To give an explicit expression for the Lagrangian \eqref{eq:action} in terms of the coset degrees of freedom, it is necessary to choose an embedding of the coset element in the supergroup $SU(2,2|4)$. There are of course many basis of generators one can use to describe the algebra $\mathfrak{su}(2,2|4)$, and consequently many different coset representative one can choose to represent the group element $g$ in \eqref{eq:Adef}. They are all related by non-linear field redefinitions and the convenience of the choice is linked to the quantization approach one is to follow. In this review we focus on two possible choices that are important for our discussion. 

\subsection{Two possible light-cone gauge fixings}
Unlike flat space, the $AdS_5\times S^5$ background admits two inequivalent sets of null geodesics. Either the geodesic wraps a big circle of $S^5$ or it lies entirely in $AdS_5$. Both possibilities, as far as the bosonic coordinates are concerned, are particular instances of the general GGRT formulation~\cite{Goddard:1973qh}, based on writing the Nambu action in the first order form and fixing
the diffeomorphisms by the two conditions – on one coordinate and on one canonical
momentum
\begin{equation}
 x^+ =\tau\, , \qquad p^+=\text{const.}
\end{equation}
The whole difference resides in the choice of the coordinates defining $x^+$. 

In the former case~\cite{Kruczenski:2004cn,Kruczenski:2004kw,Arutyunov:2004yx,Arutyunov:2005hd,Arutyunov:2006gs,Frolov:2006cc,Arutyunov:2006ak}, which we label as uniform light-cone gauge\footnote{Let us stress that, strictly speaking, both gauges are uniform, in that the momentum is distributed uniformly along the string. Nevertheless, although this terminology may be quite misleading, it is now widespread in the literature and we will stick to it hereafter. }, we introduce the coordinate $\phi$ parameterizing a circle on $S^5$ and consequently 
\begin{equation}\label{eq:xpBMN}
 x^+_{S^5}=t+\phi
\end{equation}
where $t$ is the time coordinate in $AdS_5$. In fact, in chapter \ref{BMN} we will consider a generalization of \eqref{eq:xpBMN}, where a residual gauge freedom, parametrized by a parameter $a$, is left unfixed. Nevertheless no significant conceptual difference is introduced by such a modification.

The latter case~\cite{Metsaev:2000yf,Metsaev:2000yu}, usually referred to as $AdS$ light-cone gauge, is better described in Poincar\'e parametrization
\begin{equation}\label{eq:AdS5metric}
 ds_{AdS_5}^2=\frac{dx^{\mu}dx_{\mu}+d z^2}{z^2}\, ,
\end{equation}
with $\mu=0,...,3$, so that 
\begin{equation}
 x^{+}_{AdS_5}=\frac{x^3+x^0}{\sqrt2}\, .
\end{equation}
In section \ref{sec:adslccoset} we analyze a suitable coset representative for this gauge choice, which will then be exploited in chapter \ref{GKP}.

\subsection{Coset representative for uniform light-cone gauge}\label{sec:unilccoset}

A convenient coset representative for the uniform light-cone gauge needs to have nice transformation properties under translations along the $t$ and $\phi$ directions parameterizing the time in $AdS$ and the big circle on the sphere. 

In section \ref{sec:action} we learned that local $PSU(2,2|4)$ transformations act through left multiplication on the group element $g$. Here we would like to find a coset representative such that fermions are neutral under the action of translations along $t$ and $\phi$. Therefore, let us consider the following coset element
\begin{equation}\label{eq:ulccosetrep}
 g(t,\phi,y_i,z_i,\chi)=\L(t,\phi)g(\chi)g(y_i,z_i)\, ,
\end{equation}
with $y_i$ and $z_i$ parameterizing the remaining 8 coordinates of $AdS^5$ and $S^5$ respectively. The fermions are incorporated in the element $g(\chi)$, where $\chi$ is a generic Grassmann odd algebra element whose parametrization is presented in \eqref{eq:ulcfermions}. The bosonic group elements $\L(t,\phi)$ and $g(y_i,z_i)$ are given naturally in terms of exponentials of linear combinations of generators. We consider the bosonic subalgebra $\mathfrak{su}(2,2)\oplus \mathfrak{su(4)}$ generated by $\{\G^0,\G^i,\G^{i0},\G^{ij}\}\oplus\{\tilde\G^{A},\tilde\G^{AB}\}$, with $i,j=1,...,4$ and $A,B=1,...,5$ (commutation relations and supermatrix representations are provided in appendix \ref{notation}). In this context, $\G^0$ generates translations along $t$ and $\tilde\G^5$ along $\phi$. The matrix $\L(t,\phi)$ is then given by
\begin{equation}\label{eq:Lambda}
 \L(t,\phi)=e^{t\G^0+\phi \tilde\G^5}\, .
\end{equation}
The explicit supermatrix representation of the two generators given in \eqref{eq:Gsupermat} shows that $\G^0$ and $\tilde\G^5$ are chosen to be diagonal, so that $\L(t_1+t_2,\phi_1+\phi_2)=\L(t_1,\phi_1)\L(t_2,\phi_2)$. As a consequence, the action of a translation $t\to t+a$ and $\phi\to \phi+b$ can be identified with a left multiplication by $\L(a,b)$
\begin{equation}
 \L(a,b)\L(t,\phi)g(\chi)g(y_i,z_i)=\L(t+a,\phi+b)g(\chi)g(y_i,z_i)\, ,
\end{equation}
and clearly both $g(\chi)$ and $g(y_i,z_i)$ are unaffected by this transformation. Therefore, we achieved our goal of having a coset representative with neutral fermions under $t$ and $\phi$ translation and we get, as a bonus, that also the other bosonic degrees of freedom are neutral.

Let us now describe in more details the structure of the coset representatives $g(\chi)$ and $g(y_i,z_i)$. The latter is expressed naturally using the generators introduced above
\begin{equation}\label{eq:gyz}
 g(y_i,z_i)=e^{z_i \G^i+y_i \tilde \G^i}\equiv e^X\, \qquad X\equiv z_i \G^i+y_i \tilde \G^i=\left(\begin{array}{cc} \frac12 z_i\g^i &0 \\ 0& \frac{i}{2}y_i\g^i \end{array}\right)\, .
\end{equation}
In principle one could follow the same procedure for fermions, taking the generic odd element of the superalgebra \eqref{eq:ulcfermions} and exponentiating it. However, the choice 
\begin{equation}\label{eq:gchi}
 g(\chi)=\chi +\sqrt{1+\chi^2}
\end{equation}
turns out to be more convenient.
One may wonder whether \eqref{eq:gchi} is obviously an element of $PSU(2,2|4)$. To see this one should note that $(\chi+\sqrt{1+\chi^2})^\star=\chi-\sqrt{1+\chi^2}$, which implies that $g(\chi)$ is pseudounitary. Note also that the standard exponential form is achieved with the change of variable $\chi\to\sinh \chi$.

\subsubsection{\txpf{$\k$}{k}-symmetry}

Before using the coset representative \eqref{eq:ulccosetrep} to build the Cartan form, let us constrain its form further using $\k$ symmetry. We start from equation \eqref{eq:lagvar} and perform the following change of variables
\begin{align}\label{eq:ksymmredef}
 \ve^{(1)}&= A^{(2),\a}_{-} \k_\a^{(1)}+\k_\a^{(1)} A^{(2),\a}_{-}\, ,\\
 \ve^{(2)}&= A^{(2),\a}_{+} \k^{(3)}_\a+\k^{(3)}_{\a} A^{(2),\a}_{+}\, ,
\end{align}
where $\k^{(1,3)}_\a$ are new independent parameters of the $\k$-symmetry transformation and $A^{(2),\a}_{\pm}$ stands for $P_{\pm}^{\a\b}A^{(2)}_\b$. After some algebra, whose details can be found in~\cite{Arutyunov:2009ga}, one finds that the necessary variation of the worldsheet metric in equation \eqref{eq:lagvar} is 
\begin{equation}
 \d \g^{\a\b}=\frac12\Tr \left(\left[\k^{(1),\a},A^{(1),\b}_+\right]+\left[\k^{(3),\a},A^{(3),\b}_-\right]\right)\, . \label{eq:transfg}
\end{equation}
Therefore, we showed that the supercoset action in $AdS_5\times S^5$ enjoys a non-trivial local fermionic symmetry, provided the parameter $\k$ in the action \eqref{eq:action} is set to $\pm 1$. The next question one would like to answer is how many degrees of freedom can be gauged away using this symmetry. We will show that the 32 real degrees of freedom one starts from (in the matrix \eqref{eq:ulcfermions} there are sixteen complex fermions $\theta_{ij}$) can be reduced to 16. Of course this has to be done in a way which is compatible with the imposed bosonic gauge. To see how this works for the uniform light-cone gauge, let us consider, without loss of generality, a Cartan form given only by\footnote{Of course in general the Cartan form is a combination of all the $\mathfrak{psu}(2,2|4)$ generators, but for the purpose of finding a fermion structure compatible with the light-cone gauge fixing it is enough to consider this simplified combination.}
\begin{equation}
 A^{(2)}=A_t \G^0+A_\phi \tilde \G^5\, .
\end{equation}
The Virasoro constraint $\Str(A^{(2)}_\a A^{(2)}_\b)=0$ in this case imposes $A_t=\pm A_\phi$ and 
using equations \eqref{eq:ksymmredef} and \eqref{eq:Aprop} we can write the parameter $\ve^{(1)}$ as 
\begin{equation}
 \ve^{(1)}=A^{(2),\t}_- \varkappa + \varkappa A^{(2),\t}_-, \qquad \varkappa=\k^{(1)}_\t-\frac{\g^{\t\t}}{\g^{\t\s}\mp \k} \k^{(1)}_\s\, .
\end{equation}
Picking the solution $A_t=A_\phi$ we find the structure
\begin{equation}
 \ve^{(1)}=2i A_t \left(\begin{array}{cc} 0 & \ve \\ -\ve^\dagger \Sigma & 0 \end{array}\right)\, ,                                                              
\end{equation}
where the matrix $\ve$ is given in terms of the entries of $\varkappa$ 
\begin{equation}\label{eq:epsmatrix}
 \ve=\left(\begin{array}{cccc}
     \varkappa_{11} & \varkappa_{12} & 0 & 0 \\
     \varkappa_{21} & \varkappa_{22} & 0 & 0 \\
     0 & 0 & -\varkappa_{33} & -\varkappa_{34}  \\
     0 & 0 & -\varkappa_{43} & -\varkappa_{44} 
    \end{array}\right)\, .
\end{equation}
This equation shows that $\ve^{(1)}$ depends on 8 independent complex bosonic coordinates. However, the fact that it belongs to the homogeneous component $\mathfrak{G}^{(1)}$ reduces such coordinates by a half, leading to 8 independent real fermionic parameters. A similar analysis shows that also $\ve^{(3)}$ depends on 8 free parameters, yielding a total of 16 fermionic degrees of freedom that can be gauged away fixing $\k$-symmetry. The matrix structure of \eqref{eq:epsmatrix} also shows that a generic odd matrix of $\mathfrak{psu}(2,2|4)$, like \eqref{eq:ulcfermions}, can be brought to the form
\begin{equation}\label{eq:chifix}
 \chi=\left(\begin{array}{cc}
             0 & \Theta\\
             -\Theta^\dagger \S & 0
            \end{array}\right)\, ,
\qquad
\Theta =\left(\begin{array}{cccc}
         0 & 0 & \theta_{13} & \theta_{14}\\
         0 & 0 & \theta_{23} & \theta_{24}\\
         \theta_{31} & \theta_{32} & 0 & 0\\
         \theta_{41} & \theta_{42} & 0 & 0\\
         \end{array}\right)\, .
\end{equation}

\subsubsection{Cartan form}
Given all the ingredients to build the coset representative \eqref{eq:ulccosetrep}, it is straightforward to compute the current \eqref{eq:Adef}. Let us start from the bosonic part. Setting $\chi$ to zero, we expect to recover the standard Polyakov lagrangian for some parametrization of $AdS_5\times S^5$. In this case, using \eqref{eq:ulccosetrep}, \eqref{eq:Lambda} and \eqref{eq:gyz} we find
\begin{equation}\label{eq:ulcLbos}
 \mathcal{L}_{\textup{bos}}=\g^{\a\b}\left(-G_{tt}\, \pa_\a t\, \pa_\b t+ G_{\phi\phi}\, \pa_\a \phi\, \pa_\b \phi+ G_{zz}\, \pa_\a z^i\, \pa_\b z_i +G_{yy}\, \pa_\a y^i\, \pa_\b y_i\right) \, ,
\end{equation}
with
\begin{equation}
 G_{tt}=\left(\frac{1+\frac{z^2}{4}}{1-\frac{z^2}{4}}\right)^2\, , \quad  G_{\phi\phi}=\left(\frac{1-\frac{y^2}{4}}{1+\frac{y^2}{4}}\right)^2\, , \quad G_{zz}=\frac{1}{\big(1-\frac{z^2}{4}\big)^2}\, , \quad G_{yy}=\frac{1}{\big(1-\frac{y^2}{4}\big)^2}\, .
\end{equation}
For future convenience, let us also introduce the ``light-cone'' coordinates 
\begin{equation}
 x^+=a\, \phi+(1-a)\, t\, , \qquad x^-=\phi-t \, ,
\end{equation}
where $a$ is a parameter whose meaning will become clearer in the uniform light-cone gauge discussion in chapter \ref{BMN}. For the moment let us point out that $a=\frac12$ corresponds to the standard light-cone gauge parametrization. In this new system of coordinates the $\k$-symmetry fixed current \eqref{eq:Adef} reads
\begin{align}
 A&=A_{even}+A_{odd}\, \\
 A_{even}%= A^{(0)}+A^{(2)}
 &=-g^{-1}(y_i,z_i)\left[\frac{i}2 \left(dx^+ +\left(\frac12-a\right)dx^-\right)\S_+ (1+2\chi^2)+\frac{i}{4} dx_- \S_-\right]g(y_i,z_i)\nonumber\\
 &\phantom{=}-g^{-1}(y_i,z_i)\left[\sqrt{1+\chi^2} d \sqrt{1+\chi^2}-\chi d \chi\right]g(y_i,z_i)-g^{-1}(y_i,z_i)dg(y_i,z_i)\, , \label{eq:ulcAeven}\\
 A_{odd}%=A^{(1)}+A^{(3)}
 &=-g^{-1}(y_i,z_i)\left[i\, \left(dx^+ +\left(\frac12-a\right)dx^-\right)\S_+ \chi \sqrt{1+\chi^2}\right]g(y_i,z_i)\nonumber\\
 &\phantom{=}+g^{-1}(y_i,z_i)\left[ \sqrt{1+\chi^2}d\chi-\chi d\sqrt{1+\chi^2}\right]g(y_i,z_i)\, ,\label{eq:ulcAodd}
\end{align}
where the $8\times8$ matrices $\S_+$ and $\S_-$ are defined in terms of the $\S$ matrix \eqref{eq:HandSigma}
\begin{equation}\label{eq:sigmapm}
 \S_+=\left(\begin{array}{cc}
            \S & 0 \\
            0 & \S
           \end{array}\right)\, ,\qquad
 \S_-=\left(\begin{array}{cc}
            -\S & 0 \\
            0 & \S
           \end{array}\right)\, .       
\end{equation}
Equations \eqref{eq:ulcAeven} and \eqref{eq:ulcAodd} clearly show that the expression of the current drastically simplifies in the limit $a=\frac12$. Indeed, in this case the odd part of the Cartan form does not depend on the light-cone coordinate $x^-$ and this constitutes a dramatic simplification in the gauge fixing procedure, as we will see in chapter \ref{BMN}. Let us now analyze in some details which bosonic symmetries are still linearly realized after the choice \eqref{eq:ulccosetrep}.

\subsubsection{\txpf{$SU(2)^4$}{SU(2)} parametrization}\label{sec:SU24}
Although very convenient for the light-cone gauge fixing, the parametrization \eqref{eq:ulccosetrep} does not allow the linear realization of the whole bosonic subgroup of $PSU(2,2|4)$. In this section we will derive the maximal bosonic subgroup which acts linearly on the dynamical fields $y_i, z_i$ and $\chi$. This subgroup will coincide with the manifest bosonic symmetry of the light-cone gauge fixed string Lagrangian. 

A group theory analysis of the $\mathfrak{su}(2,2)\oplus \mathfrak{su}(4)$ algebra (see commutation relations \eqref{eq:ulccommrel} and comments below) shows that the centralizer of the $\mathfrak{u}(1)$ isometries, associated to shifts of $t$ and $\phi$, coincides with
\begin{equation}\label{eq:foursu2}
 \mathfrak{so}(4)\oplus \mathfrak{so}(4) =\mathfrak{su}(2)\oplus\mathfrak{su}(2)\oplus\mathfrak{su}(2)\oplus\mathfrak{su}(2)\, ,
\end{equation}
where the first $\mathfrak{so}(4)\subset \mathfrak{so}(1,4) \subset \mathfrak{so}(2,4)$, whereas the second $\mathfrak{so}(4)\subset \mathfrak{so}(5)\subset\mathfrak{so}(6)$. As a consequence, if $G$ is an element of the subgroup associated to the algebra \eqref{eq:foursu2}, we have $G^{-1}\L(t,\phi)G=\L(t,\phi)$ and consequently
\begin{equation}
G\, g(t,\phi,y_i,z_i,\chi)=\L(t,\phi) (G g(\chi)G^{-1}) (G g(y_i,z_i)G^{-1}) G\, .
\end{equation}
In this formula we recognize the last $G$ as a compensating element of the coset denominator $SO(1,4)\times SO(5)$. Therefore, under the action of $G$ both the bosons and the fermions undergo a linear transformation 
\begin{equation}
 \chi\to G\chi G^{-1}\, , \qquad X\to G X G^{-1} \, ,
\end{equation}
and it is natural to ask whether we can introduce a parametrization of the physical degrees of freedom such that this $SU(2)^4$ invariance becomes manifest. Let us first use the supermatrix representation \eqref{eq:Gsupermat} to see explicitly that the elements $X$ and $\chi$ can be represented in terms of $2\times2$ matrices as follows
\begin{equation}
X=\left(\begin{array}{cccc} 
                               0	& Z	& 0	&0\\
                               Z^\dagger&0	&0	&0\\
                               0	&0	&0	&iY\\
                               0	&0	&iY^\dagger&0
                              \end{array} \right)\, , \qquad
\chi=\left(\begin{array}{cccc} 
                               0			&0			&0			&\Theta_1\\
                               0			&0			&\Theta_2^\dagger		&0\\
                               0			&\Theta_2		&0			&0\\
                               -\Theta_1^\dagger	&0			&0			&0
                              \end{array} \right)\, ,
\end{equation}
where the second equation is just another way to express \eqref{eq:chifix} and the matrices $Z$ and $Y$ are
\begin{equation}
 Z=\frac12\left(\begin{array}{cc}
          z_3-iz_4	&-z_1+iz_2\\
          z_1+iz_2	&z_3+iz_4
         \end{array}\right)\, ,
\qquad         
Y=\frac12\left(\begin{array}{cc}
          y_3-iy_4	&-y_1+iy_2\\
          y_1+iy_2	&y_3+iy_4
         \end{array}\right)\, .
\end{equation}
These matrices satisfy the following reality conditions
\begin{equation}\label{eq:realYZ}
 Z^{\dagger}=\e Z^t \e^{-1},\qquad  Y^{\dagger}=\e Y^t \e^{-1}\, ,
\end{equation}
where $\e$ is defined in \eqref{eq:Kdef}. 
To find the action of $G$ on the components $Y,Z,\Theta_1$ and $\Theta_2$, one should realize that the two $SO(4)$ factors in \eqref{eq:foursu2} are generated by $\G^{ij}$ and $\tilde \G^{ij}$ (see \eqref{eq:Gsupermat}) with $i,j=1,...,4$. Therefore the supermatrix representation of $G$ assumes the form
\begin{equation}
 G=\left(\begin{array}{cccc}
 g_1	&0	&0	&0\\
0	&g_2	&0	&0\\
0	&0	&g_3	&0\\
0	&0	&0	&g_4\\
\end{array}\right)\, ,
\end{equation}
with $g_i\in SU(2)$. Simple matrix multiplication yields
\begin{align}
 GXG^{-1}&=\left(\begin{array}{cccc} 
                               0			&g_1 Zg_2^{-1}	&0			&0\\
                               g_2 Z^\dagger g_1^{-1}	&0		&0			&0\\
                               0			&0		&0			&ig_3Yg_4^{-1}\\
                               0			&0		&ig_4Y^\dagger g_3^{-1}	&0
                              \end{array} \right)\, , \\
G\chi G^{-1}&=\left(\begin{array}{cccc} 
                               0				&0				&0				&g_1\Theta_1g_4^{-1}\\
                               0				&0				&g_2\Theta_2^\dagger g_3^{-1}	&0\\
                               0				&g_3\Theta_2 g_2^{-1}		&0				&0\\
                               -g_4\Theta_1^\dagger g_1^{-1}	&0				&0				&0
                              \end{array} \right)\, . \label{eq:chitransf}
\end{align}
If we now consider, e.g., the matrix $Y$ and we multiply it by $\e$ on the right we find that 
\begin{equation}\label{eq:Yeptr}
 Y\e\to g_3 Y g_4^{-1}\e=g_3 Y\e g_4^t\, ,
\end{equation}
where we used the equality $g_4^{-1}=\e g_4^t \e^{-1}$, which provides the equivalence of an irrep of $SU(2)$ and its complex conjugate (stated differently there is no antifundamental representation for $SU(2)$). Therefore, the matrix $Y\e$ transforms in the bifundamental representation of the third and the fourth $SU(2)$ in \eqref{eq:foursu2}. Associating an index $a=1,2$ to the fundamental representation of $g_3$ and $\dot{a}=\dot{1},\dot{2}$ to the fundamental representation of $g_4$, we can rewrite equation \eqref{eq:Yeptr} as
\begin{equation}
 {Y'}^{a \dot a}={{g_3}^a}_b \ {{g_4}^{\dot a}}_{\dot b}\  Y^{b\dot b}\, ,
\end{equation}
where $Y^{a \dot a}$ are the entries of the matrix $Y\e$. This implies that the matrix $Y$ in these new variables assumes the form
\begin{equation}
 Y=\left(\begin{array}{cc} 
Y^{1\dot{2}} &-Y^{1 \dot{1}}\\
Y^{2 \dot{2}} & -Y^{2 \dot{1}}
   \end{array}\right)\, .
\end{equation}
A parallel argument can be applied to the matrix $Z$ introducing an index $\a=3,4$ and $\dot\a=\dot 3,\dot 4$ for the first two copies of $SU(2)$ in \eqref{eq:foursu2}. Finally, using the reality condition \eqref{eq:realYZ}, the supermatrix $X$ in terms of these new degrees of freedom reads
\begin{equation}\label{eq:X}
X=\left(\begin{array}{cccc|cccc} 0 & 0 & Z^{3\dot{4}}
& -Z^{3\dot{3}} & 0 & 0 & 0 & 0 \\
0 & 0 & Z^{4\dot{4}} & -Z^{4\dot{3}} & 0 & 0 & 0 & 0 \\
-Z^{4\dot{3}} & Z^{3\dot{3}} & 0 & 0 & 0 & 0 & 0 & 0 \\
-Z^{4\dot{4}} & Z^{3\dot{4}} & 0 & 0 & 0 & 0 & 0 & 0 \\
\hline
0 & 0 & 0 & 0 & 0 & 0 & iY^{1\dot{2}} & -iY^{1\dot{1}}\\
0 & 0 & 0 & 0 & 0 & 0 & iY^{2\dot{2}} & -iY^{2\dot{1}} \\
0 & 0 & 0 & 0 & -iY^{2\dot{1}} & iY^{1\dot{1}} & 0 & 0 \\
0 & 0 & 0 & 0 & -iY^{2\dot{2}} & iY^{1\dot{2}} & 0 & 0
\end{array}\right)\, .
\end{equation}

It is not difficult to carry out the same argument for the fermions, in light of the fact that $\Theta_1$ in \eqref{eq:chitransf} is related to the bifundamental of $g_1$ and $g_4$, whereas $\Theta_2$ is transformed by $g_2$ and $g_3$. It is then natural to parametrize $\Theta_1$ by the entries $\eta^{\a \dot a}$ and $\Theta_2$ by $\theta^{a \dot{\a}}$. The new parametrization for $\chi$ is therefore
\begin{equation}\label{eq:chi}
 \chi={ \left(\begin{array}{cccc|cccc} 0 & 0 & 0 & 0 & 0 & 0
&
\eta^{3\dot{2}} & -\eta^{3\dot{1}}\\
 0 & 0 & 0 & 0 & 0 & 0 & \eta^{4\dot{2}}
& -\eta^{4\dot{1}} \\
 0 & 0 & 0 & 0 & ~~\theta^\dagger_{1\dot{4}} & ~~\theta^\dagger_{2\dot{4}} & 0
& 0 \\
 0 & 0 & 0 & 0 & -\theta^\dagger_{1\dot{3}} & -\theta^\dagger_{2\dot{3}} & 0
& 0 \\ \hline
 0 & 0 & \theta^{1\dot{4}} & -\theta^{1\dot{3}} & 0 & 0 & 0 & 0
\\
 0 & 0 & \theta^{2\dot{4}}
& -\theta^{2\dot{3}} & 0 & 0 & 0 & 0\\
-\eta^\dagger_{3\dot{2}} & -\eta^\dagger_{4\dot{2}} & 0 & 0 & 0 &
0 & 0 & 0
\\
~~ \eta^\dagger_{3\dot{1}} &~~ \eta^\dagger_{4\dot{1}} & 0 & 0 & 0
& 0 & 0 & 0
\end{array}\right) }\, ,
\end{equation}
where, by definition,  $\theta_{a\dot \a}^\dagger$ and
$\eta_{\a\dot a}^\dagger$ are understood as complex conjugate of
$\theta^{a\dot \a}$ and $\eta^{\a\dot a}$, respectively, 
\begin{equation}
(\theta^{a\dot \a})^*\equiv\theta_{a\dot \a}^{\dagger}\, ,~~~~~~ (\eta^{\a
\dot a})^*\equiv \eta_{\a \dot a}^\dagger\,  . 
\end{equation}

To sum up, we have shown that, after choosing a coset representative particularly suitable for the uniform light-cone gauge fixing, the bosonic subgroup which acts linearly on the physical degrees of freedom is constituted by four different copies of $SU(2)$. Hence, we parametrize those degrees of freedom by a convenient double index notation
\begin{equation}
Z^{\a\dot{\a}}\, , ~~~~Y^{a\dot{a}}\, ,~~~~\theta^{a\dot{\a}}\, ,
~~~~~\eta^{a\dot{\a}}\, ,
\end{equation}
pointing out the bifundamental transformation properties under $SU(2)$. This will be our starting point in chapter \ref{BMN} for deriving a superstring action in uniform light-cone gauge. 

\subsection{Coset representative for \txpf{$AdS$}{AdS} light-cone gauge}\label{sec:adslccoset}
In order to describe a useful choice of coset representative for the $AdS$ light-cone gauge, we use the Poincar\'e parametrization \eqref{eq:AdS5metric}
and we introduce the light-cone coordinates\footnote{The unnatural choice of coordinates is chosen to facilitate the comparison with the existing literature (see for instance~\cite{Alday:2005gi}.)}
\begin{equation}\label{eq:lccoord}
 x^{\pm}=\frac{x^3\pm x^0}{\sqrt2}\,, \qquad x=\frac{-x^2+i\, x^1}{\sqrt2}\, , \qquad \bar x=\frac{-x^2-i\, x^1}{\sqrt2}\, .
\end{equation}
In this context, the set of bosonic generators which is more appropriate is such that the bosonic subalgebra $\mathfrak{so}(4,2)\sim \mathfrak{su}(2,2)$ is interpreted as the conformal group in four spacetime dimensions. Therefore, we introduce the set of bosonic generators $\{J^{\mu\nu},P^\mu,K^{\mu},D\}$, where $P^\mu$ and $J^{\mu\nu}$ describe the four-dimensional Poincar\'e group, $K^\mu$ denote conformal boosts and $D$ is the dilatation generator. In the light-cone coordinates \eqref{eq:lccoord} the set of $\mathfrak{su}(2,2)$ generators can be rewritten as
\begin{equation}\label{eq:lcbosgen}
 \{J^{+-},J^{+x},J^{+\bar x}J^{x \bar x}, P^\pm, P, \bar P, K^\pm, K, \bar K, D\}\, .
\end{equation}

In order to complete the list of generators of the full $\mathfrak{psu}(2,2|4)$ superalgebra , we introduce also fifteen $\mathfrak{su}(4)$ generators ${J^i}_j$ and a set of 32 supercharges $\{Q^{\pm i},Q^{\pm}_i,S^{\pm i},S^{\pm}_i\}$, which are diagonal under the action of $D$, $J^{+-}$ and $J^{x\bar x}$ (see \eqref{eq:dilQS}, \eqref{eq:JpmQS},\eqref{eq:JxxQS}). The full set of commutation relations is given in appendix \ref{notation}, where we also provide a supermatrix representation of this basis. 

In order to choose a convenient coset representative for $\frac{PSU(2,2|4)}{SO(1,4)\times SO(5)}$, we can exploit the fact that in Poincar\'e coordinates the relation between the isometries of $AdS_5$ and the conformal group in four dimensions is apparent. Indeed $P^\mu$ are the generators associated to translations of the $x^\m$ coordinates and $D$ generates translations of $\phi$, with $z=e^\phi$. Therefore, a natural choice for the coset representative of $\frac{SO(2,4)}{SO(1,4)}$ is
\begin{equation}
 g_{\frac{SO(2,4)}{SO(1,4)}}=g(x)g(\phi)=e^{x^\mu P_\mu} e^{\phi D}\, .
\end{equation}
In a similar way we can project the $SU(4)$ generators using $SO(5)$ matrices \eqref{eq:gammamatrices} and write down a coset representative of $\frac{SO(6)}{SO(5)}$ as
\begin{equation}\label{eq:ycoset}
 g_{\frac{SO(6)}{SO(5)}}=g(y)=e^{{y^i}_j {J^j}_i}\, , \qquad {y^i}_j=\frac{i}{2} y_A {(\g^A)^i}_j\, ,
\end{equation}
with $A=1,...,5$.
The inclusion of the fermions is implemented by adding a generic element of the odd part of the algebra, as represented by \eqref{eq:lcfermions}. 

We can finally write down the expression of the full $\frac{PSU(2,2|4)}{SO(1,4)\times SO(5)}$ coset representative
\begin{equation}\label{eq:cosetrep}
 g= g(x,\theta)\,  g(\eta)\,  g(y)\,  g(\phi)\, ,  \qquad g(x,\theta)=e^{x^\mu P_\mu+\theta \cdot Q}\,,  \qquad g(\eta)=e^{\eta\cdot S}\, ,
\end{equation}
with $\theta \cdot Q=\theta^-_i Q^{+\, i}+\theta^{-\, i}Q^+_i+ (+\leftrightarrow -)$ and $\eta \cdot S=\eta^-_i S^{+\, i}+\eta^{-\, i}S^+_i+ (+\leftrightarrow -)$. Given this coset element, it is a straightforward (though long) exercise to derive the left-invariant Cartan form \eqref{eq:Adef}. However, as we did for the uniform light-cone gauge, we fix the $\k$-symmetry gauge before deriving the Cartan form. The standard prescription to fix $\k$-symmetry in light-cone gauge is to impose $\Gamma^+ \theta_I=0$. One can show that, in the light-cone basis we introduced in \eqref{eq:lcbosgen}, this is equivalent to setting to zero all the fermions which has positive charges under $J^{+-}$. Therefore, from \eqref{eq:JpmQS} we conclude that the $\k$-symmetry gauge fixing simply amounts to 
\begin{equation}
 \theta^+_i=\theta^{+\, i}=0\, ,\qquad \eta^+_i=\eta^{+\, i}=0\, .
\end{equation}
To simplify the notation we also set $\theta^-_i=\theta_i$, $\eta^-_i=\eta_i$ and similarly for upper indices. Let us point out that fermions here are assumed to be complex and the procedure of raising and lowering indices is equivalent to take the complex conjugate, therefore
\begin{equation}\label{eq:ccferm}
 \eta^i=\eta_i^\dagger\, ,\qquad \theta^i=\theta_i^\dagger\, .
\end{equation}
Fermions with (upper)lower indices change in the (anti-)fundamental representation of $SU(4)$.

The $\k$-symmetry gauge fixing simplifies the expression of a generic odd element of the algebra to
 \begin{equation}
 \theta_i Q^{+\, i}+\theta^{i}Q^+_i+\eta_i S^{+\, i}+\eta^{i}S^+_i=2^{\frac14}
 \left(\begin{array}{cccc|cccc}
        0 & 0 & 0 & 0 & 0 & 0 & 0 & 0\\
        0 & 0 & 0 & 0 & \eta^{1} & \eta^{2} & \eta^{3} & \eta^{4}\\
        0 & 0 & 0 & 0 & \theta^{1} & \theta^{2} & \theta^{3} & \theta^{4}\\
        0 & 0 & 0 & 0 & 0 & 0 & 0 & 0\\ \hline
        \theta_1 & 0 & 0 & \eta_1 & 0 & 0 & 0 & 0\\
        \theta_2 & 0 & 0 & \eta_2 & 0 & 0 & 0 & 0\\
        \theta_3 & 0 & 0 & \eta_3 & 0 & 0 & 0 & 0\\
        \theta_4 & 0 & 0 & \eta_4 & 0 & 0 & 0 & 0\\
       \end{array}\right)\, .
\end{equation}

The $\k$-symmetry fixed Cartan form assumes then the general form
\begin{align}
A&= g^{-1}d g=A^{(0)}+A^{(2)}+A^{(1)}+A^{(3)}\, ,\\
A^{(0)}+A^{(2)}&=A_P^\mu P_\m+ A_K^\mu K_\m+A_D D+\frac12 A^{\mu\nu}_J J_{\mu\nu}+{A^j_i}{J^i}_j\, ,\nonumber\\
A^{(1)}+A^{(3)}&= {A_Q}^+_i Q^{-\, i}+ {A_Q}^-_i Q^{+\, i}+{A_Q}^{+\, i} Q^{-}_i +{A_Q}^{-\, i} Q^{+}_i\, \nonumber\\
&+ {A_S}^+_i S^{-\, i}+ {A_S}^-_i S^{+\, i}+{A_S}^{+\, i} S^{-}_i +{A_S}^{-\, i} S^{+}_i\, .\nonumber
\end{align}
The coefficients of this linear combination were first derived in~\cite{Metsaev:2000yf} and here we provide explicit expressions only for the ones that are relevant for the construction of the Lagrangian (see equations \eqref{eq:A2A2} and \eqref{eq:A1A3}). The bosonic contributions are
\begin{align}
 A_P^+&=e^{\phi} dx^+   & A_P^-&=e^{\phi}\left[ dx^- +\frac12 i\, \theta_i d\theta^i-\frac12 i\, \theta^i d\theta_i\right] &  A_P&=e^{\phi} d x & \bar A_P&= e^{\phi} d\bar x\, ,\\
A_K^+&=0 & A_{K}^-&=\frac12 e^{-\phi} \left[-(\tilde \eta^2)^2 dx^+ +i\, \tilde \eta^i d \tilde \eta_i +i\, \tilde \eta_i d \tilde \eta^i\right]&A_K&=0 & \bar A_K&=0\, , \\
A_D&=d\phi & A^i_j&=(dUU^{-1})^i_j +2\, i\, \tilde \eta^i\tilde\eta_j  dx^+\, . \label{eq:current3}
 \end{align}
The non-vanishing supercharge coefficients read
\begin{align}
  {A_Q}^-_i&=e^{\frac{\phi}{2}} (\tilde d\theta_i+ \sqrt2\, \tilde \eta_i d\bar x)\, ,& {A_Q}^{-\, i}&=e^{\frac{\phi}{2}} (\tilde d\theta^i+\sqrt2\, \tilde \eta^i dx)\, , \\
  {A_Q}^+_i&=i\,\sqrt2\, e^{\frac{\phi}{2}} \tilde \eta_i dx^+\, , & {A_Q}^{+\, i}&=-i\,\sqrt2\, e^{\frac{\phi}{2}} \tilde \eta^i dx^+ \, , \\
   {A_S}^-_i&=e^{-\frac{\phi}{2}} (\tilde d\eta_i+ i\, \tilde \eta^2 \tilde \eta_i dx^+)\, , & {A_S}^{-\, i}&=e^{-\frac{\phi}{2}} (\tilde d\eta^i-i\, \tilde\eta^2 \tilde \eta^i d x^+)\, , %&{A_S}^+_i&=0 & {A_S}^{+\, i}&=0 
   \end{align}
where we introduced the notation
 \begin{equation}\label{eq:tildedef}
 \tilde \eta^i={U^i}_j \theta^j\, , \qquad \tilde \eta_i = \eta_j {(U^\dagger)^j}_i\, \, ,
\end{equation}
and similarly for $\theta$. The tilde on the differential sign indicates that the rotation is performed after the derivative ($\tilde d\theta^i=U^i_j d\theta^j$). Notice that $\tilde\eta^2=\tilde\eta^i\tilde\eta_i=\eta^2$ and $\tilde\eta^i\tilde d\eta_i=\eta^i d\eta_i$. To better understand the meaning of this rotation, let us consider one of the various terms that can appear in the construction of the Cartan current \eqref{eq:Adef}
\begin{align}
 g^{-1}(y) \eta^i S^+_i g(y)&=\eta^i e^{-{y^k}_j [{J^j}_k,\bullet]} S^+_i\\
 &=\eta^i (S^+_i-{y^k}_j [{J^j}_k,S^+_i]+\frac12 {y^k}_j {y^m}_n [{J^n}_m,[{J^j}_k,S^+_i]]+...)\, .
\end{align}
Since $S^+_i$ is an eigenvector under the adjoint action of ${J^k}_l$ (see equation \eqref{eq:QSSU4}), this expression can be recast into the form
\begin{equation}
 g^{-1}(y) \eta^i S^+_i g(y)=\eta^i {(e^{y})^j}_i S^+_j\equiv {U^j}_i \eta^i S^+_j\, ,
\end{equation}
which provides a definition for the matrix $U$ appearing in \eqref{eq:current3} and \eqref{eq:tildedef}. Using the definition of ${y^i}_j$ in \eqref{eq:ycoset}, the matrix $U$ can be also expressed as
\begin{equation}\label{eq:Uexpl}
 {U^i}_j=\cos\frac{|y|}{2}\d^i_j+i\, (\g^A)^i_j n_A \sin \frac{|y|}{2}\, ,
\end{equation}
with $n_A=\frac{y_A}{|y|}$ and $|y|=\sqrt{y^Ay_A}$.

Using equations \eqref{eq:bospro} and \eqref{eq:ferpro} one can find the supermatrix representations of the $\mathbb{Z}_4$ projections of $A$. Taking products and supertraces one obtains
\begin{align}
 \Str (A^{(2)} A^{(2)} )&=A_{D} A_{D} +2\, (A_K+A_P) (\bar A_{K}+\bar A_{P})+2\, (A^-_{K}+A^-_{P}) (A^+_{K}+A^+_{P})+A^A {A_A}, \label{eq:A2A2}\\
 \Str (A^{(1)} A^{(3)} )&=i\,\sqrt2\, C_{ij}({A_Q}^{+\, i}{A_Q}^{-\, j}+{A_S}^{+\, i}{A_S}^{-\, j})+i\,\sqrt2\, C^{ij}({A_Q}^+_i{A_Q}^-_j + {A_S}^+_i {A_S}^-_j) . \label{eq:A1A3}
\end{align}
Here $C_{ij}$ is a charge conjugation matrix and its explicit expression is provided by the equality $C_{ij}=\rho^6_{ij}$ with the matrix $\rho^6$ given in \eqref{eq:rhomat}. 
The matrix $A^A$ in \eqref{eq:A1A3} is determined by the decomposition
\begin{equation}
 A^i_j=\frac{i}{2} A_A (\g^A)^i_j+\frac14 A_{AB} (\g^{AB})^i_j\, , \qquad \g^{AB}=\frac12 [\g^A,\g^B]\, ,
\end{equation}
separating the $SO(5)$ contribution from the $\frac{SO(6)}{SO(5)}$ one. In this context $A^A$ assumes the geometrical interpretation of the supervielbein of $S^5$, i.e. the standard  bosonic $S^5$ vielbein suitably covariantized due to the presence of fermions. Using the expression given in \eqref{eq:current3} for ${A^i}_j$ and projecting with a gamma matrix, one finds
\begin{equation}\label{eq:S5vielbein}
 A^A_\a=e^A_\a - \pa_\a x^+ \tilde\eta_i (\g^A)^i_j \tilde\eta^j\, , \qquad  e^A_\a=-\frac{i}{2}\Tr(\g^A\pa_\a U U^{-1})\, .
\end{equation}
Here $e^A_\a$ is the bosonic $S^5$ vielbein
\begin{equation}\label{eq:S5metric}
e^A_\a e_{A \b}=G_{AB}\pa_\a y^A \pa_\b y^B\, , \qquad  G_{AB}=\frac{\sin |y|}{|y|} (\d_{AB}-n_An_B) + n_A n_B\, .
\end{equation}

We finally have at our disposal all the ingredients to build the Lagrangian \eqref{eq:action}. We consider the ``kinetic'' and the Wess-Zumino part separately. The former reads
\begin{align}
 \mathcal{L}_{kin}&= \g^{\a\b} \Str\left(A^{(2)}_\a A^{(2)}_\b\right)=\g^{\a\b}\Big(\pa_\a \phi \pa_\b \phi+ 2\, e^{2\phi}\pa_\a x \pa_\b \bar x +2\, e^{2\phi} \pa_\a x^+ \pa_\b  x^- + A^A_\a A_{A\b} \nonumber \\
 &-i\, \pa_\a x^+(\tilde \eta^i\tilde\pa_\b \eta_i +\tilde\eta_i\tilde\pa_\b \eta^i+e^{2\phi}\tilde \theta^i \tilde\pa_\b \theta_i +e^{2\phi}\tilde\theta_i\tilde\pa_\b \theta^i)- \pa_\a x^+ \pa_\b x^+ (\tilde\eta^2)^2\Big) \, .
 \end{align}
In the first line we clearly recognize the Polyakov action for a bosonic string in $AdS_5\times S^5$, with $AdS_5$ in Poincar\'e coordinates \eqref{eq:AdS5metric} and $S^5$ parametrized by \eqref{eq:S5metric}. An appealing feature of this Lagrangian is that it is quartic in fermions. This is a consequence of this particular $\k$ symmetry gauge fixing, and in chapter \ref{GKP} we will see that this property simplifies higher-order computations in pertubation theory. The Wess Zumino term is even simpler
\begin{align}
 \mathcal{L}_{WZ}= \e^{\a\b} \Str\left(A^{(1)}_\a A^{(3)}_\b\right)&=2\, \e^{\a\b}\pa_\a x^+ e^{\phi}\tilde\eta_i C^{ij} (\tilde\pa_\b \theta_j+i\,\sqrt2 \tilde\eta_j\pa_\b x)+h.c.\, .
\end{align}
Another nice property of this Lagrangian is that $x^-$ appears only in the kinetic term $\pa_\a x^+ \pa_\b  x^-$. In chapter \ref{GKP} we will see how this drastically simplifies the gauge fixing procedure. In order to have a convenient sign in front of the fermionic kinetic terms and to deal with fermions with the same scaling dimensions, we apply the transformations
\begin{equation}
 x^a\to-x^a\, , \qquad \tilde \eta^i\to e^{\phi}\tilde \eta^i\, , \qquad \tilde \eta_i\to e^{\phi}\tilde \eta_i\, .
\end{equation}
The new Lagrangian is
\begin{align}\label{eq:Lkin}
 \mathcal{L}_{kin}&=\g^{\a\b}\Big(\pa_\a \phi \pa_\b \phi+ 2\, e^{2\phi}\pa_\a x \pa_\b \bar x +2\, e^{2\phi} \pa_\a x^+ \pa_\b  x^- + A^A_\a A_{A\b} \\
 &+i\, e^{2\phi} \pa_\a x^+(\tilde \eta^i\tilde\pa_\b \eta_i +\tilde\eta_i\tilde\pa_\b \eta^i+\tilde \theta^i \tilde\pa_\b \theta_i +\tilde\theta_i\tilde\pa_\b \theta^i)- \pa_\a x^+ \pa_\b x^+ e^{4\phi}(\tilde\eta^2)^2\Big)\, , \nonumber \\
 \label{eq:LWZ}
 \mathcal{L}_{WZ}&=-2\, \e^{\a\b}\pa_\a x^+ e^{2\phi}\tilde\eta_i C^{ij} (\tilde\pa_\b \theta_j-i\, e^{\phi} \sqrt2 \tilde\eta_j\pa_\b x)+h.c. \, .
 \end{align}
 
Arguably, the one feature of this Lagrangian that is slightly tedious is the matrix $U$ rotating all the fermionic degrees of freedom. It turns out there are different ways to reabsorb that rotation. Here we focus on two strategies, both developed in~\cite{Metsaev:2000yf}. The first one consists in eliminating the rotation introducing a covariant derivative for fermions and we call it the Wess-Zumino (WZ) parametrization. The second one is a nice change of variables after which the Lagrangian looks even simpler and, for a reason that will be clear in the following, we call it the ``4+6'' parametrization. The reason why we analyze also the former, whose Lagrangian is slightly more involved, is that in the lower dimensional example of superstring theory in $AdS_4\times \mathbb{
CP}^3$ (see section \eqref{sec:stringAdS4}) the second strategy does not seem to be applicable and therefore it will be useful to make a comparison with the first one.

\subsubsection{WZ parametrization}\label{sec:WZpar}
The aim of this section is to perform a transformation on the fermions appearing in \eqref{eq:Lkin} and \eqref{eq:LWZ}, such that the rotation by the $U$ matrix are reabsorbed in covariant derivatives. This is easily achieved by the following transformation
\begin{equation}\label{eq:rot}
 \eta_i\to \eta_j {U^j}_i\, , \qquad \eta^i\to {(U^{\dagger})^i}_j \eta^j\, , \qquad \theta_i\to \theta_j {U^j}_i\, , \qquad \theta^i\to {(U^{\dagger})^i}_j \theta^j\, .
\end{equation}
The rotation \eqref{eq:rot} clearly eliminates the matrix $U$ from all the terms where the fermions are not derived. Terms involving derivatives of fermions have the following transformation property
\begin{equation}
 \tilde d \eta^i\equiv {U^i}_j d \eta^j \to d \eta^i-{(dUU^{-1})^i}_j \eta^j\equiv D \eta^i \, ,
\end{equation}
and similarly for fermions with lower indices.
Introducing the notation ${\Omega^i}_j={(dUU^{-1})^i}_j$ it is easy to verify that, by construction, $\Omega$ is a connection with vanishing curvature. To sum up we have 
\begin{equation}
 D\eta^i=d\eta^i-{\O^i}_j \eta^j\, , \qquad D\eta_i=d\eta_i+ \eta_j{\O^j}_i\, , \qquad d\O-\O\wedge\O=0\, , 
\end{equation}
and similarly for $\theta$. Using equation \eqref{eq:S5vielbein} we can express also the $S^5$ vielbein in terms of $\O$ as 
\begin{equation}
e^A=-\frac{i}{2}\Tr(\g^A\O) \, .
\end{equation}
This is nothing else than a projection of the matrix $\O$, which in general admits the decomposition
\begin{equation}\label{eq:Omegadec}
 {\O^i}_j=\frac{i}{2} e_A (\g^A)^i_j+\frac14 \o_{AB} (\g^{AB})^i_j\, ,
\end{equation}
where $\o_{AB}$ is the spin connection in our parametrization of $S^5$. Therefore, $D$ assumes a precise geometrical interpretation as the covariant derivative of a spinor on $S^5$. Notice that the whole dependence on $U$ has now been reabsorbed and the Lagrangian depends only on $\O$, which carries all the information about the $S^5$ background. The explicit form of $\O$ is 
\begin{equation}
\O=\frac{i}{2} \g^A \left[y_A \frac{y^B d y_B}{|y|^2} \left(1-\frac{\sin |y|}{|y|}\right)+dy_A \frac{\sin |y|}{|y|}\right]-\frac14 \g^{AB} \left[(y_Ady_B-y_Bdy_A) \frac{1-\cos|y|}{|y|^2}\right],
\end{equation}
and the final form of the Lagrangian is 
\begin{equation}
 \mathcal{L}=\mathcal{L}_B+\mathcal{L}^{(2)}_F+\mathcal{L}^{(4)}_F\, ,
\end{equation}
where the bosonic component is simply the Polyakov lagrangian in our parametrization of $AdS_5\times S^5$
\begin{equation}
 \mathcal{L}_B=\g^{\a\b}\Big( 2\, e^{2\phi}(\pa_\a x^+ \pa_\b  x^- + \pa_\a x \pa_\b \bar x) +\pa_\a \phi \pa_\b \phi+ G_{AB}(y) \pa_\a y^A \pa_\b y^B\Big)\, .
\end{equation}
Notice that, as expected, the bosonic Lagrangian depends only on the vielbein of $S^5$. The dependence on the spin connection enters, through $\O$, in the fermionic interactions. The quadratic part of the fermion action reads
\begin{align}
\mathcal{L}_F^{(2)}=e^{2\phi} \pa_\a x^+ &\Big[\g^{\a\b}\left(i\, \eta^i D_\b \eta_i+i\, \theta^i D_\b \theta_i+e_\b^A\eta_i{(\g_A)^i}_j \eta^j\right) \nonumber \\
-&2\,\e^{\a\b}\eta^iC_{ij}\left(D_\b \theta^j-i\sqrt2 e^{\phi}\eta^j\pa_\b x\right)+h.c.\Big]\, .\label{eq:L2fWZ}
\end{align}
Finally the quartic fermionic term depends only on $\eta$
\begin{equation}\label{eq:L4fWZ}
 \mathcal{L}^{(4)}_F=-e^{4\phi} \g^{\a\b} \pa_\a x^+ \pa_\b x^+ \left[\left(\eta^2\right)^2-\left(\eta_i{(\g_A)^i}_j \eta^j\right)^2\right]\, .
\end{equation}
This Lagrangian, although still complicated by the presence of the connection $\O$, has the privilege of having a clear geometric interpretation. As we mentioned, a similar construction can be carried out also in the $AdS_4\times \mathbb{CP}^3$ case (see section \ref{sec:WZAdS4}).

\subsubsection{4+6 parametrization}\label{sec:4+6}
The choice of a light-cone gauge involving only coordinates in the $AdS$ part of the space suggests that the sphere is unaffected by this procedure and all the $SU(4)$ generators  simply commute with the generators of translations in the $x^+$ and $x^-$ directions. We therefore expect our Lagrangian to have an explicitly realized $SO(6)$ symmetry, which is clearly not there in \eqref{eq:Lkin} and \eqref{eq:LWZ}. It turns out that it is possible to find a change of variables which brings this Lagrangian to an explicitly $SO(6)$ invariant form. The idea is to use the coordinate $\phi$, together with the five $y^A$ coordinates, to build a $SO(6)$ vector
\begin{equation}\label{eq:46metric}
 z^A=e^{-\phi} \sin |y| n^A\, , \qquad z^6=e^{-\phi} \cos |y| \, , \qquad |z|^2=z^M z_M =e^{-2\phi}\, ,
\end{equation}
with $M=1,...,6$. 
The metric in the new coordinates looks extremely simple and has the standard 4+6 form from which we borrowed the name for the parametrization
\begin{equation}
 ds^2=\frac{dx^\m dx_\m +dz^M dz_M}{|z|^2}\, .
\end{equation}

To perform the change of variable in the fermionic Lagrangian, it is useful to introduce the $SO(6)$ matrices $\rho^M_{ij}$. As usual, they carry two indices changing in the fundamental and one in the vector representation of $SU(4)\sim SO(6)$. We also indicate by $(\rho^M)^{ij}$ the hermitian conjugate of $\rho^M_{ij}$. Therefore vectorial indices are raised and lowered by a six-dimensional identity matrix and there is no difference between $\rho^M$ and $\rho_M$. On the other hand, raising or lowering fundamental indices always implies some kind of complex conjugation as we have already observed for the fermions in \eqref{eq:ccferm}. 
The commutator of two $\rho$ matrices is abbreviated as
\begin{equation}\label{eq:rhomndef}
 {(\rho^{MN})^i}_j=\frac12\left[(\rho^M)^{il} \rho^N_{lj}-(\rho^N)^{il}\rho^M_{lj}\right]\, .
\end{equation}
Explicit expressions of all the matrices and additional relations among them are spelled out in appendix \ref{notation}.

The mapping between $SO(5)$ gamma matrices and $\rho$ matrices is provided by
\begin{equation}
 {(\g^A)^i}_j=i(\rho^A)^{il} (\rho^6)_{lj}\, , \qquad C_{ij}=\rho^6_{ij}\, ,
\end{equation}
as one can easily check looking at the explicit expressions or checking the defining properties.
It is interesting to note, using \eqref{eq:Uexpl}, that
\begin{equation} \label{eq:gammarhoid}
{(U^\dagger)^i}_j {(\g^A)^j}_k {U^k}_l= i\, n_M {(\rho^{MA})^i}_l-i\,\frac{1}{1+n_6}z^A n_M {(\rho^{M6})^i}_l \, , \quad   e^{-\phi} {U^i}_j C_{ik} {U^k}_l=\rho^M_{jl} z_M \, ,
\end{equation}
where we introduced the notation $n_M=\frac{z_M}{|z|}$.
Combining the first identity with the kinetic terms of the bosons, one notices a very powerful cancellation of all the terms that are not explicitly $SO(6)$ invariant. We can therefore rewrite the kinetic lagrangian \eqref{eq:Lkin} as
\begin{align}\label{eq:Lkinnew}
 \mathcal{L}_{kin}=\frac{\g^{\a\b}}{|z|^2}&\bigg[2\, \pa_\a x^+ \pa_\b x^-+2\, \pa_\a x \pa_\b \bar x+ D_\a z^M D_\b z_M \nonumber\\
 &+i\, \pa_\a x^+ \left(\eta^i\pa_\b \eta_i +\eta_i\pa_\b \eta^i+\theta^i \pa_\b \theta_i +\theta_i\pa_\b \theta^i\right)- \frac{\pa_\a x^+\pa_\b x^+}{|z|^2} (\eta^2)^2\bigg]\, ,
\end{align}
where the covariant derivative simply acts as follows
\begin{equation}\label{eq:covdevz}
 D_\a z^M= \pa_\a z^M+i\, \eta_i {(\rho^{MN})^i}_j \eta^j\frac{z_N}{|z|^2} \pa_\a x^+\, .
\end{equation}
The Lagrangian \eqref{eq:Lkinnew} is explicitly $SO(6)$ invariant, however it has the unpleasant feature of containing inverse powers of $|z|$, preventing us from expanding around the trivial vacuum where all the fields are set to zero. This is a common feature in integrable systems such as the $O(N)$ sigma model~\cite{Zamolodchikov:1977nu}. In that case the Lagrangian has explicit $O(N)$ invariance. Nevertheless, in order to perform a perturbative computation, one has to pick a vacuum which breaks the symmetry to $O(N-1)$. In that case the symmetry is restored non-perturbatively and we expect something similar to happen also for the string theory model.

The second identity in \eqref{eq:gammarhoid} allows to rewrite the WZ term \eqref{eq:LWZ} in the explicitly $SO(6)$ invariant form 
 \begin{align}\label{eq:LWZnew}
 \mathcal{L}_{WZ}=-\frac{2}{|z|^3} \e^{\a\b}\pa_\a x^+ z_M\eta^i \rho^M_{ij} \left(\pa_\b \theta^j-\frac{i}{|z|} \sqrt2 \eta^j\pa_\b x\right)+h.c.\, .
\end{align}
Notice that the dependence of the Lagrangian on the matrices $U$ has been completely reabsorbed by the identities \eqref{eq:gammarhoid}. This Lagrangian will be the starting point for some of the applications discussed in chapter \ref{GKP}.

\section{Superstring theory in \txpf{$AdS_4\times \mathbb{CP}^3$}{AdS4 x CP3}}\label{sec:stringAdS4}
The importance of constructing superstring theory for various examples of $AdS/CFT$ has been widely emphasized in the introduction of this review. Nevertheless, the construction of a Lagrangian for a superstring moving in $AdS_4\times \mathbb{CP}^3$ is not as straightforward as the higher dimensional counterpart, analyzed in section \ref{sec:stringAdS5}. At first sight this may seem counterintuitive, since in section \ref{sec:action} we described a general procedure to build an action for any supercoset target space and the coset $\frac{SO(2,3)\times SU(4)}{SO(1,3)\times U(3)}$, describing $AdS_4\times \mathbb{CP}^3$, allows for a supersymmetric extension to $\frac{OSp(2,2|6)}{SO(1,3)\times U(3)}$\footnote{Remember that $SO(2,3)\sim USp(2,2)$ and in our notation $OSp(2,2|6)$ has bosonic subgroup $USp(2,2)\times SO(6)$.}. This possibility was explored in~\cite{Stefanski:2008ik,Arutyunov:2008if} and the resulting action  can be interpreted as a partially gauge-fixed type 
IIA Green Schwarz action, where the $\kappa$-symmetry gauge-fixing sets to zero eight fermionic 
modes corresponding to the eight broken supersymmetries~\cite{Gomis:2008jt,Grassi:2009yj}. Indeed, unlike $AdS_5\times S^5$, the $AdS_4\times \mathbb{CP}^3$ background preserves only 24 of the 32 supersymmetries of type IIA supergravity.

Therefore, it looks like, up to this apparently irrelevant difference, we have a way to derive an action for the new background. However, as first argued in~\cite{Arutyunov:2008if} and 
 later clarified in~\cite{Gomis:2008jt}, this action is not suitable 
 to describe the dynamics of a string lying solely in the $AdS_4$ part~\footnote{The same is true 
 when the string forms a worldsheet instanton by wrapping a $\cp^1$ cycle in $\cp^3$~\cite{Cagnazzo:2009zh}.} 
 of the   \adscp superspace. In this case four of the eight modes set to zero 
 are in fact dynamical fermionic degrees of freedom of the superstring.  Any action willing to capture 
 the semiclassical dynamics on these 
classical string configurations should contain these physical fermions, and should therefore be found 
via another, sensible $\kappa$-symmetry 
 gauge-fixing of the full action. 
 This has been done 
 in~\cite{Uvarov:2009hf,Uvarov:2009nk}~\footnote{See also~\cite{Grassi:2009yj}.},  starting  from the $D=11$ membrane action~\cite{deWit:1998yu} based on the supercoset 
    $OSp(8|4)/\left(SO(7)\times SO(1, 3)\right)$, 
 performing double dimensional reduction and choosing
 a $\kappa$-symmetry light-cone gauge.
 The output is an action, at most  quartic in the fermions, which is the  \adscp counterpart of the 
 gauge-fixed action of section \ref{sec:stringAdS5}.
 
 Here we will be mostly interested in this second version of the action. Therefore, after quickly sketching the main features of the coset construction, we will review in some details the derivation of the action of~\cite{Uvarov:2009hf,Uvarov:2009nk}.
 
 \subsection{The coset approach}
 The Lie algebra $\mathfrak{osp}(2,2|6)$ can be realized by $10 \times 10$ matrices of the form
 \begin{equation}\label{eq:supermatrixAdS4}
 M=\left(\begin{array}{cc} m & \theta \\
         \eta & n
        \end{array}\right)\, ,
\end{equation}
where $m$ and $n$ are Grassmann even $4\times 4$ and $6 \times 6$ matrices respectively. The Grassmann odd matrix $\theta$ is $4 \times 6$ while $\eta$ is $6\times 4$. In order to belong to $\mathfrak{osp}(2,2|6)$, M has to satisfy two conditions. The first one singles out the complex algebra $\mathfrak{osp}(4|6)$ through the constraint
\begin{equation}\label{eq:cond1}
 M^{st}\left(\begin{array}{cc} C_4 & 0 \\
         0 & \mathbb{1}_6
        \end{array}\right)+
        \left(\begin{array}{cc} C_4 & 0 \\
         0 & \mathbb{1}_6
        \end{array}\right) M=0\, ,
\end{equation}
where $C_4$ is a charge conjugation matrix, which can be chosen to be real skew symmetric and satisfying $C_4^2=-\mathbb{1}_4$. $M^{st}$ indicates the supertrasposition introduced in \eqref{eq:omegaAdS5}. If we restrict equation \eqref{eq:cond1} to the bosonic matrices, we notice that the condition is translated into 
\begin{equation}
 m^t=C_4 m C_4\, , \qquad n^t=-n\, ,
\end{equation}
which tells us that the bosonic subalgebra is $\mathfrak{sp}(4,\mathbb{C})\oplus\mathfrak{so}(6,\mathbb{C})$. We can also pick a real section of that imposing
\begin{equation}\label{eq:cond2}
 M^\dagger \left(\begin{array}{cc} \S & 0 \\
         0 & -\mathbb{1}_6
        \end{array}\right)+
        \left(\begin{array}{cc} \S & 0 \\
         0 & -\mathbb{1}_6
        \end{array}\right)M=0\, ,
\end{equation}
where $\S$ was defined in \eqref{eq:HandSigma}. This last equation defines the algebra $\mathfrak{osp}(2,2|6)$ as a real section of $\mathfrak{osp}(4|6)$. As a consequence of \eqref{eq:cond1} and \eqref{eq:cond2}, we have the following relations among the fermionic components
\begin{equation}
 \eta=-\theta^t C_4\, , \qquad \theta^*=i\Gamma^3 \theta\, . \label{eq:rel3}
\end{equation}

Following the same lines of equation \eqref{eq:omegaAdS5} we can write down explicitly the automorphism $\O$ introduced in \eqref{eq:omega} as
\begin{equation}
 \O(M)=-\mathcal{K}M^{st} \mathcal{K}^{-1}, 
 \qquad \mathcal{K}=\left(\begin{array}{cc} K_4 & 0 \\
         0 & -K_6
        \end{array}\right)\, ,
\end{equation}
and, similarly to \eqref{eq:Kdef}, the matrices $K_4$ and $K_6$ can be chosen to be 
\begin{equation}
K_4= \left(\begin{array}{cc} \e & 0 \\
         0 & \e
        \end{array}\right)\, , \qquad K_6=\left(\begin{array}{ccc} \e & 0 & 0\\
         0 & \e & 0\\
         0 & 0 & \e
        \end{array}\right)\, .
\end{equation}
Since the structure of the automorphism $\O$ is exactly the same as in \eqref{eq:omegaAdS5}, we can just use formulae \eqref{eq:bospro} and \eqref{eq:ferpro} to find the $\mathbb{Z}_4$ projections of the supermatrix $M$ and build all the ingredients for the superstring action. 

This is the point where we would need to introduce a parametrization of the coset representative of $\frac{OSP(2,2|6)}{SO(1,3)\times U(3)}$ and write down the explicit form of the Lagrangian. However, as we mentioned in chapter \ref{GKP}, we will be interested in classical string configurations lying entirely in the $AdS$ part of the space and this description cannot be employed in that particular case. Therefore, instead of focusing on deriving a closed-form Lagrangian, we would rather study the properties of $\k$-symmetry in the coset description and analyze why this description is not suitable for the configurations we are interested in.

\subsubsection{\txpf{$\k$}{k}-symmetry}

As we have already done for $AdS_5\times S^5$ we start from equation \eqref{eq:lagvar}, which holds for a general supercoset sigma model, and we look for a change of parametrization which is particularly suitable for the supergroup we are dealing with. In this case one finds that the convenient change of variables is provided by
\begin{equation}
 \varepsilon^{(1)}=A_{\a,-}^{(2)}A_{\b,-}^{(2)}\kappa^{\a\b}
+\kappa^{\a\b}A_{\a,-}^{(2)}A_{\b,-}^{(2)}+A_{\a,-}^{(2)}\k^{\a\b}A_{\b,-}^{(2)}-\frac{1}{8}\,
{\rm str}(\Sigma A_{\a,-}^{(2)}A_{\b,-}^{(2)})\kappa^{\a\b}\,
, \label{eq:eps1}
\end{equation}
where $\kappa^{\a\b}$ is the
$\kappa$-symmetry parameter which is assumed to be independent on
the dynamical fields of the model. A similar change of variable can be performed also for the $\ve^{(3)}$ parameter introducing a new parameter $\varkappa^{\a\b}$. After some algebra, whose details can be found in~\cite{Arutyunov:2008if}, one can find the following variation for the worldsheet metric
\begin{equation}
 \delta \gamma^{\a\beta}=\frac{1}{2}\, {\rm str}\Big(\Sigma
A^{(2)}_{\delta,-}
[\kappa^{\a\b},A^{(1),\delta}_+]\Big)+\frac{1}{2}\, {\rm
str}\Big(\Sigma A^{(2)}_{\delta,+}
[\varkappa^{\a\b},A^{(3),\delta}_-]\Big)\, ,
\end{equation}
We stress once again that in our derivation of
$\kappa$-symmetry we used the fact that ${\rm P}_{\pm}^{\a\b}$ are
orthogonal projectors and, therefore, we required $\kappa\pm 1$.

To understand how many fermionic degrees of freedom we can fix using $\k$-symmetry, we follow a procedure similar to the $AdS_5\times S^5$ case and without loss of generality we consider a current of the form
\begin{equation}
A^{(2)}=i\, A_t \G^t+A_6 T^6\, ,
\end{equation}
where $t$ is the time direction in $AdS$ and $T^6$ is the generator of translation in one of the $\mathbb{CP}^3$ directions. As in $AdS_5\times S^5$ the Virasoro contraint demands $A_t=\pm A_6$ and, picking the first solution, we can use equations \eqref{eq:Aprop} and \eqref{eq:eps1} to write down the form of $\ve^{(1)}$ as 
\begin{equation}
 \varepsilon^{(1)}=x^2\left(\begin{array}{cc} 0 ~&~ \varepsilon
\\
-\varepsilon^tC_4 ~&~ 0
\end{array} \right)\, ,
\end{equation}
where $\varepsilon$ is the following matrix 
\begin{equation}
\nonumber\varepsilon={ \left(\begin{array}{rrrrrr}
 0 ~&~ 0 ~&~ i(i \k_{13}-\k_{16}) ~&~ i(i\k_{14}-\k_{15}) ~&~ i\k_{14}-\k_{15} ~&~
i \k_{13}-\k_{16}
\\
 0 ~&~ 0 ~&~ i(i\k_{23}-\k_{26}) ~&~ i(i\k_{24}-\k_{26}) ~&~ i\k_{24}-\k_{25} ~&~ i\k_{23}-\k_{26}\\
 0 ~&~ 0 ~&~-i(-i\k_{33}-\k_{36}) ~&~ -i(-i\k_{34}-\k_{35}) ~&~ -i\k_{34}-\k_{35} ~&~
-i\k_{33}-\k_{36}  \\
 0 ~&~ 0 ~&~ -i(-i\k_{43}-\k_{46}) ~&~-i( -i\k_{44}-\k_{45}) ~&~ -i\k_{44}-\k_{45} ~&~
-i\k_{43}-\k_{46}
\end{array} \right)\,
}
\end{equation}
and $\kappa_{ij}$ are the entries of
the matrix $\kappa$. As we see, the matrix $\varepsilon$
depends on 8 independent complex fermionic parameters (e.g. the
last two columns). The reality condition \eqref{eq:rel3} for
$\varepsilon$ reduces this number by half. Finally,
$\varepsilon^{(1)}$ must belong to the component $A^{(1)}$ which
further reduces the number of fermions by half. As a result,
$\varepsilon^{(1)}$ depends on four real fermionic parameters. A
similar analysis applies to $\varepsilon^{(3)}$. Thus, in total $\varepsilon^{(1)}$ and
$\varepsilon^{(3)}$ depend on 8 real fermions and these are those
degrees of freedom which can be gauged away by $\kappa$-symmetry.
The gauge-fixed coset model will therefore  involve 16 physical
fermions only.

It should be noted that the considerations above are applicable to
a generic case, where string motion occurs in both ${\rm AdS}_4$
and $\cp^3$ spaces. There is however a singular situation, when
string moves in the $AdS$ space only. One can show that for this case
the transformation (\ref{eq:eps1}) vanishes and only 12 fermionic equations (out of 24) are
independent. This suggests that $\kappa$-symmetry in this singular situation  becomes
capable of gauging away 12 over 24 fermions. The singular nature of
the corresponding bosonic background shows up in the fact that, as
soon as fluctuations along $\cp^3$ directions are switched on, the
rank of $\kappa$-symmetry gets reduced to 8~\cite{Arutyunov:2008if}. We therefore conclude that
singular backgrounds cannot be quantized semi-classically within
the coset sigma-model. Since in this review we are interested in a classical string configuration which lies entirely in $AdS$ (see chapter \ref{GKP}), we need to develop an alternative approach which includes this configuration. This is the aim of the following sections.

\subsection{String action from double dimensional reduction}
As we mentioned, the Lagrangian obtained via the coset construction can be interpreted as the full Green-Schwarz type IIA superstring Lagrangian in $AdS_4\times \mathbb{CP}^3$, after the $\k$-symmetry gauge has been partially fixed setting to zero the fermionic coordinates associated to the broken supersymmetries. Nevertheless, the action of GS superstring in curved background is known only up to quartic order in the fields and one may wonder whether there is a way of building the full Lagrangian for type IIA superstring on $AdS_4\times \mathbb{CP}^3$ with full $\k$-symmetry freedom (i.e. with the usual 32 fermionic degrees of freedom of the GS superstring). It turns out this can be done exploiting the fact that $S^7$ is a $U(1)$ Hopf fibration over $\mathbb{CP}^3$,
and therefore the  $AdS_4\times \mathbb{CP}^3$ solution of the type IIA supergravity bosonic equations of motion~\cite{Watamura:1983hj} is connected
to the Freund--Rubin $AdS_4\times S^7$ bosonic solution of
$D=11$ supergravity by reducing along the
$U(1)$--fiber direction of the sphere~\cite{Nilsson:1984bj,Sorokin:1985ap}. The superspace extension of this reduction is rather 
subtle and it was achieved in~\cite{Gomis:2008jt,Grassi:2009yj}, where the complete action for type IIA superstring in $AdS_4\times \mathbb{CP}^3$ was written down. It describes all possible superstring motions and allows a wider choice of $\kappa$-symmetry gauges
compared to the supercoset action. About the integrability 
of this string non-coset model, the standard analysis of section \ref{sec:ksymm}- 
which applies to the supercoset action - is not possible here. 
The  classical integrability of strings generically moving in 
the full \adscp superspace has been however shown by constructing
a Lax connection with zero curvature  up to  quadratic order in the 
fermions~\cite{Sorokin:2010wn}~\footnote{A study of classical 
integrability (prior to gauge-fixing) for general motion of the string in several backgrounds of 
interest for the $AdS/CFT$ correspondence is in~\cite{Wulff:2014kja}.}. In~\cite{Uvarov:2009hf} a $\k$-symmetry gauge particularly suitable for the $AdS$ light-cone gauge fixing was introduced. In the following we summarize the construction of~\cite{Uvarov:2009hf}.

\subsubsection{The membrane action in \txpf{$AdS_4\times S^7$}{AdS4 x S7}}
Actions for the M2-brane and the M5-brane in the $AdS_4\times S^7$ and $AdS_7\times S^4$
superbackgrounds respectively were derived in~\cite{deWit:1998tk,deWit:1998yu, Claus:1998fh, Pasti:1998tc}. Similarly to the case of $AdS_5\times S^5$ in ten dimensions, $AdS_4\times S^7$ and $AdS_7\times S^4$ are maximally supersymmetric backgrounds in eleven dimensions. Thus one can exploit a coset construction similar to the one described in section \ref{sec:action} to build the geometric ingredients entering the supermembrane action. $AdS_4\times S^7$ can be described as a coset $\frac{SO(2,3)}{SO(1,3)}\times \frac{SO(8)}{SO(7)}$ and the supersymmetric extension with 32 superspace directions is given by $\frac{OSp(4|8)}{SO(1,3)\times SO(7)}$. As a starting point, one can choose a set of generators for the algebra $\mathfrak{osp}(4|8)$ of the form $\{M_{\m\n},M_{\m},V_{IJ},V_{8I},Q_{A'}\}$, with $\m, \n=0,...,3$; $I,J=1,...,7$ and $A'=1,...,32$. The set of bosonic generators $\{M_{\m\n},M_{\m}\}$ span the subalgebra $\mathfrak{sp}(4)\sim \mathfrak{so}(2,3)$ and, since $M^{\m\n}$ alone generate $\mathfrak{
so}
(1,3)$, the generators $M^\m$ are associated to the coset degrees of freedom $\frac{SO(2,3)}{SO(1,3)}$. In a similar way $V_{IJ}$ are generators of $\mathfrak{so}(7)$ and, together with $V_{8I}$, they generate $\mathfrak{so}(8)$. All the fermionic generators are encoded in $Q_{A'}$. In this basis the Cartan form reads
\begin{equation}\label{eq:cartanformAdS4first}
 A= g^{-1}d g =\o^{\m\n} M_{\m\n}+E^\m M_{\m}+\O^{IJ}V_{IJ}+\O^{8I} V_{8I}+F^{A'} Q_{A'}\, .
\end{equation}
The geometric interpretation of the coefficients is the usual one, i.e. $E^\m$ and $\O^{8I}\equiv E^{I}$ are the supervielbeine of $AdS_4$ and $S^7$ respectively. In this setup, the M2-brane action in the $AdS_4\times S^7$ background reads
\begin{equation}\label{eq:membrane}
S=-\int\limits_{V} d^3\xi\sqrt{-g^{(3)}}+S_{WZ}\, .
\end{equation}
Here $g^{(3)}$ is the determinant of the induced world-volume metric 
\begin{equation}\label{eq:indmet}
g^{(3)}_{\hat \a \hat \b}=E^\m_{\hat \a} E_{\m\, \hat \b} +E^{I}_{\hat \a} E_{I\, \hat \b} \, ,\qquad \hat \a, \hat \b =0,1,2; 
\end{equation}
with the components of the Cartan form defined by $A=A_{\hat \a} d\xi^{\hat \a}$.
The Wess-Zumino (WZ) term
\begin{equation}
S_{WZ}=\frac14 \int\limits_{\mathcal M_4} H_{(4)}
\end{equation}
is the integral  of the closed 4-form
\begin{equation}\label{eq:4formWZ}
H_{(4)}=\frac{i}{2}F^{A'}\wedge ({\G}^{\hat \m\hat
\n})_{A'}{}^{B'}F_{B'} \wedge E_{\hat
\m}\wedge E_{\hat \n}+\varepsilon_{\m\n\r\l}E^{\m}\wedge
E^{\n}\wedge E^{\r}\wedge E^{\l}
\end{equation}
over the 4-dimensional auxiliary hypersurface $\mathcal M_4$, whose boundary coincides with the supermembrane world volume $V$.
In \eqref{eq:4formWZ} we introduced the eleven dimensional vector $E^{\hat \m}=(E^{\m},E^{I})$ and the matrix $\G^{\hat \m \hat \n}$, commutator of two $SO(1,10)$ gamma matrices.
The coefficient of the WZ term is fixed by requiring $\k$-symmetry invariance (see~\cite{Uvarov:2009hf} for details).

In the following we will be interesting in performing a double dimensional reduction and fixing an $AdS$ light-cone gauge on the superstring action. Both these tasks are best achieved in a different basis of generators with respect to the one we just introduced. As far as the light-cone gauge is concerned, we learned in section \ref{sec:adslccoset} that a convenient basis is provided by the interpretation of $SO(2,d)$ as the superconformal group in $d$ dimension. Therefore the set of bosonic generators for $\mathfrak{sp}(4)$ is exactly the same as in section \ref{sec:adslccoset}, but with one dimension less. Introducing light-cone coordinates
\begin{equation}\label{eq:lccoordAdS4}
 x^{\pm}=x^2\pm x^0\, ,
\end{equation}
one ends up with
 \begin{equation}\label{eq:lcbosgenAdS4}
 \{J^{+-},J^{+x}, P^\pm, P, K^\pm, K, D\}\, ,
\end{equation}
where $P$ generates translations along the coordinate $x^1$, which hereafter we simply label as $x$. The metric of the $AdS$ part of the space is naturally expressed in Poincar\'e coordinates
\begin{equation}
ds^2_{AdS_4}=R_{AdS}^2 \frac{dx^m dx_m +dz^2}{z^2}\, ,
\end{equation}
with $m=0,1,2$ (or equivalently $m=+,-,1$), and the $AdS$ components of the Cartan form \eqref{eq:cartanformAdS4first} are rearranged as (see appendix \ref{notation})
\begin{equation}
 \o^{\m\n} M_{\m\n}+E^\m M_{\m}=\o^{mn}J_{mn}+\Delta D+\omega^m P_m+c^m K_m\, .
\end{equation}
Working out the transformation of the first term in the induced metric \eqref{eq:indmet} one obtains
\begin{equation}\label{eq:indmetAdS4}
 g^{(3),AdS}_{\hat \a \hat \b} d\xi^{\hat \a} d\xi^{\hat \b}=\frac14 (\o_m+c_m)(\o^m+c^m)+\D^2\, .
\end{equation}
Notice the analogy between this equation and the first three terms in equation \eqref{eq:A2A2}.

The 32 supercharges can also be organized in a convenient light-cone representation  $\{Q^{\pm i},Q^{\pm}_i,S^{\pm i},S^{\pm}_i\}$, as in section \ref{sec:adslccoset}. These supercharges clearly describe a $\mathcal{N}=8$ superspace in three dimensions. However, the dimensional reduction preserves only $\mathcal{N}=6$ supersymmetry. For this reason, it is convenient to split the index $i=1,..,4$ of the supercharges as $Q^{\pm}_i=(Q^{\pm}_a,Q^{\pm}_4)$ and similarly for the antifundamental index. Now $a=1,...,3$ is an index in the (anti)fundamental of $SU(3)$, which is the symmetry we expect to be explicitly realized in the superstring action for $AdS_4\times\mathbb{CP}^3$.

Unlike the $AdS_5\times S^5$ case, the treatment of the $S^7$ part of the space is quite involved due to the dimensional reduction. Indeed, one needs to find a basis of generators which makes the Hopf fibration structure of $S^7$ manifest. First of all it is convenient to introduce a $\mathfrak{so}(6)\oplus \mathfrak{so}(2)\sim \mathfrak{su}(4)\oplus \mathfrak{u}(1)$ basis
\begin{equation}\label{eq:cfso8}
\Omega^{IJ}V^{IJ}+\Omega^{8I}V^{8I}=\Omega^{MN}V_{MN}+\Omega^{78}V_{78}+\Omega^{8M}V_{8M}+2\, \Omega^{7M}V_{7M}\, ,\qquad M,N=1,...,6\, . 
\end{equation}
Here $V_{MN}$ and $V_{78}$ are generators of $\mathfrak{so}(6)$ and $\mathfrak{so}(2)$ respectively and the remaining generators are associated to the coset directions $\frac{SO(8)}{SO(2)\times SO(6)}$. Using the $\rho$ matrices $\eqref{eq:rhomat}$, one can convert all the $SO(6)$ vector indices in $SU(4)$ (anti)fundamental ones. The detailed procedure is spelled out in appendix \ref{notation}. The final result is
\begin{align}
 \Omega^{8I}V^{8I}+2\Omega^{7I}V^{7I}&=\Omega_a T^a+\Omega^aT_a+\tilde\Omega_a\tilde T^a+\tilde\Omega^a\tilde T_a\, , \nonumber\\
\Omega^{IJ}V^{IJ}+\Omega^{78}V^{78}&=\Omega_a{}^b
V_b{}^a+\Omega_b{}^b
V_a{}^a+\Omega_a{}^4V_4{}^a+\Omega_4{}^aV_a{}^4+hH\, . 	\label{cfso6so2}
\end{align}
The relations between the coefficients of \eqref{eq:cfso8} and \eqref{cfso6so2} are given in appendix \ref{notation}. Here we just resume the role of the generators on the r.h.s. of \eqref{cfso6so2}. The 1-form $h$ in (\ref{cfso6so2}) corresponds to the fiber direction of $\mathbb{CP}^3\times U(1)$. The 8 generators $V_b{}^a$ span a $\mathfrak{su}(3)$ algebra, which is enhanced to $\mathfrak{u}(3)$ when $V^a_a$ is included. Including also the 6 generators $T^a$ and $T_a$, the full set of 15 generators $\{T^a,T_a,V_b{}^a,V^a_a\}$ span the $\mathfrak{su}(4)$ algebra. Finally, the remaining 12 generators $\{V_a{}^4,V_4{}^a,\tilde T_a,\tilde T^a\}$ are associated to the coset $\frac{SO(8)}{SU(4)\times U(1)}$. The relation between these coefficients and those given in \eqref{eq:cartanformAdS4first} is quite involved and, to understand which degrees of freedom are relevant for the construction of the 
Lagrangian, one has to transform the original vielbeine to the new ones in \eqref{cfso6so2} (see appendix \ref{notation} for details). The result of this change of coordinates is quite simple and yields the new induced metric
\begin{equation}\label{eq:indmetS7}
 g^{(3),S^7}_{\hat \a \hat \b} d\xi^{\hat \a} d\xi^{\hat \b}=\Omega^{8I'}\Omega^{8I'}=(h+\Omega_a{}^a)^2+(\Omega_a+\tilde\Omega_a)(\Omega^a+\tilde\Omega^a)\, .
\end{equation}
The aim of the following discussion is to find the explicit expressions for these supervielbeine using a specific coset representative for $\frac{OSp(4|8)}{SO(1,3)\times SO(7)}$ 

There is always some degree of arbitrariness in the choice of the coset representative. The idea here is to use a dressed version of the coset representative for $\frac{OSp(4|6)}{SO(1,3)\times U(3)}$, adding the fiber direction $y$ to parametrize $S^7$ and the superspace directions associated to the broken supersymmetries. As we have already experimented in section \ref{sec:adslccoset}, it is convenient to fix the $\k$-symmetry before deriving the current. Following the same strategy as for $AdS_5\times S^5$, we set to zero all the fermionic directions with negative\footnote{For $AdS_5\times S^5$ we chose the positively charged fermions, however here we switch convention to be consistent with the literature.} charge under the $J^{+-}$ generator. This implies
\begin{equation}
 \theta^-_i=\theta^{-\, i}=0\, ,\qquad \eta^-_i=\eta^{-\, i}=0\, .
\end{equation}
We also set  $\theta^+_i\equiv \theta_i$ and $\theta^{+\, i}\equiv \theta^i$, and similarly for $\eta$. The $\k$-symmetry gauge fixed coset representative will be given by
\begin{equation}
 g=g(x,\theta^a)\, g(\eta^a)\, g(z)\, g(\phi)\, g(y)\, g(\theta^4)\, g(\eta^4)\, ,
\end{equation}
where the first four factors are the precise analogue of the higher dimensional counterpart \eqref{eq:cosetrep}
\begin{align}
 g(x,\theta^a)&=e^{x^m P_m+\theta^a Q^-_a+\theta_a Q^{-\, a}}, & g(\eta)&=e^{\eta_a S^{-\, a}+\eta^a S^-_a}, & g(z)&=e^{z^a T_a+z_a T^a}, & g(\phi)&=e^{\phi D},
\end{align}
while the last three factors read
\begin{equation}
 g(y)=e^{yH}\, , \qquad g(\theta^4)=e^{\theta^4 Q^-_4+\theta_4 Q^{-\, 4}}\, , \qquad g(\eta^4)=e^{\eta^4 S^-_4+\eta_4 S^{-\, 4}}\, .
\end{equation}
The position of the coordinate $y$, on the left of the supercharges associated to broken supersymmetries, allows to have no dependence on $y$ in the vielbein. This is necessary to perform the dimensional reduction, as we will point out in the following. Before that, let us compute the relevant components of the Cartan form.

In general the current $A$ can be decomposed as
\begin{align}
 A&=g^{-1}dg=A_{AdS_4}+A_{S^7}+A_{\text{ferm}}\, ,\\
 A_{AdS_4}&=\o^{mn}J_{mn}+\Delta D+\omega^m P_m+c^m K_m\, , \label{eq:AAds4} \\
 A_{S^7}&=\Omega_a T^a+\Omega^aT_a+\tilde\Omega_a\tilde T^a+\tilde\Omega^a\tilde T_a\nonumber\\
&+\Omega_a{}^bV_b{}^a+\Omega_b{}^b
V_a{}^a+\Omega_a{}^4V_4{}^a+\Omega_4{}^aV_a{}^4+hH\, , \label{eq:AS7}\\
A_{\text{ferm}}&=\omega^-_i Q^{+\, i}+\omega^{- i} Q^+_{ i}+\omega^+_i Q^{-\, i}+\omega^{+ i} Q^-_{ i} \nonumber\\
&+\chi^-_i S^{+\, i}+\chi^{- i} S^+_{ i}+\chi^+_i S^{-\, i}+\chi^{+ i} S^-_{ i}\, .\label{eq:Aferm}
\end{align}
We stress that in the last line we grouped the different contributions in a $SU(4)$ notation, but in the following we will always deal with the $\o_a$ and $\o_4$ separately. We report here only the components that are relevant for the construction of the induced metrics \eqref{eq:indmetAdS4} and \eqref{eq:indmetS7}. We start from the $S^7$ components
\begin{align}
 \Omega_{a}&=d\bar z_a\frac{\sin{|z|}}{|z|}+\bar z_a\frac{\sin{|z|}(1-\cos{|z|})}{2|z|^3}(dz^c\bar z_c-z^cd\bar z_c)+\bar z_a\left(\frac{1}{|z|}-\frac{\sin{|z|}}{|z|^2}\right)d|z|,\label{eq:cp3vielbein1}\\
\Omega^{a}&=dz^a\frac{\sin{|z|}}{|z|}+z^a\frac{\sin{|z|}(1-\cos{|z|})}{2|z|^3}(z^cd\bar
z_c-dz^c\bar
z_c)+z^a\left(\frac{1}{|z|}-\frac{\sin{|z|}}{|z|^2}\right)d|z|\, ,\label{eq:cp3vielbein2}
\end{align}
where $|z|^2 \equiv z^a\, \bar z_a$.
From the decomposition \eqref{eq:AS7} and the interpretation of $T^a$ as coset generators of $\frac{SU(4)}{U(3)}$, it is clear that $\Omega_a$ inherits the geometrical interpretation of vielbein of $\mathbb{CP}^3$. Indeed
\begin{align}
 ds^2_{\mathbb{CP}^3} = \Omega_a\Omega^a = g_{ab}\, dz^a\, dz^b + g^{ab}\, d\bar z_a\, d\bar z_b + 2\, 
g_a^{\phantom{a}b}\, dz^a\,d\bar z_b\,,
\end{align}
with
\begin{align}
 g_{ab} &= \frac{1}{4|z|^4} \left( |z|^2 - \sin^2{|z|} + \sin^4{|z|} \right)\bar z_a\,\bar z_b\, ,  \nonumber\\
 g_a^{\phantom{a}b} &= \frac{\sin^2{|z|}}{2|z|^2}\, \delta_a^b + \frac{1}{4|z|^4} \left( |z|^2 - \sin^2{|z|} - \sin^4{|z|} \right)\bar z_a\, z^b \, ,
\end{align}
and $g^{ab}$ is simply obtained by $g_{ab}$ replacing $z$ with $\bar z$.
The other relevant components for \eqref{eq:indmetS7} read
\begin{align}
\tilde \Omega_a&=\varepsilon_{abc}\hat{\eta}{}^b\hat{\eta}{}^cdx^+-2e^{-\varphi}\hat\eta_a\eta_4dx^+\, , &
\tilde \Omega^a&=-\varepsilon^{abc}\hat\eta_b\hat\eta_cdx^++2e^{-\varphi}\hat{\eta}{}^a\eta^4dx^+\, ,\\
h&=dy-e^{-2\varphi}\eta_4\eta^4dx^+,&
\Omega_a^{\phantom{a}a} &= i\, \frac{\sin^2|z|}{|z|^2} \left( dz^a\, \bar z_a - z^a\, d z_a 
\right)\,,
\end{align}
where we introduced the notation 
 \begin{equation}\label{eq:rotfermAdS4}
\hat\eta_a=T_a{}^b\eta_b+T_{ab}\eta^b,\qquad\hat{\eta}{}^a=T^a{}_b\eta^b+T^{ab}\eta_b\, ,
\end{equation}
in the same spirit of \eqref{eq:tildedef}. The origin of the matrix $T$ is the same as the matrix $U$ in \eqref{eq:tildedef}, with the important difference of containing non-diagonal terms\footnote{Here non-diagonal means that the rotation of $\eta_a$ involves also the complex conjugate $\eta^a$.}. Indeed, the definition of $T$ is 
\begin{equation}
 {T_{\hat a}}^{\hat b}=\left(\begin{array}{cc}
                        {T_a}^b & T_{ab}\\
                        T^{ab} & {T^b}_a
                       \end{array}\right)=\exp\left(\begin{array}{cc}
                        0 & i\, \e_{acb}z^c\\
                        -i\, \e^{acb}\bar z_c & 0
                       \end{array}\right)\, ,
\end{equation}
with the hatted index defined such that $z^{\hat a}=(z^a,z_a)$. Using the properties of the matrix in the exponent, one can find an explicit expression of $T$ as
\begin{equation}\label{matrixT}
{T_{\hat{a}}}^{\hat b}
= \left(
 \begin{array}{cc}
  \delta_a^b\, \cos{|z|} + \bar z_a\, z^b\, \frac{1-\cos{|z|}}{|z|^2} & i\,\varepsilon_{acb}\, z^c\, \frac{\sin{|z|}}{|z|} \\
  -i\,\varepsilon^{acb}\, \bar z_c\, \frac{\sin{|z|}}{|z|} & \delta^a_b\, \cos{|z|} + z^a\, \bar z_b\, \frac{1-\cos{|z|}}{|z|^2}
 \end{array}\right)\,.
\end{equation} 
It is worthwhile noticing that the vielbeine are independent of $y$. As we will see, this is an essential property for performing the dimensional reduction and it is a consequence of our choice of the coset representative. 

This feature is present also in the $AdS$ components, although in this case we notice the appearance of a non-trivial dependence on $dy$
\begin{align}
\omega^-&=e^{-2\varphi}(dx^- +i\, d\theta_a \theta^a -i\, \theta_a d \theta^a)+i\, d\theta_4\theta^4-i\, \theta_4d\theta^4-4\theta_4\theta^4 dy\, , \\
\omega^+&=e^{-2\varphi}dx^+\phantom{0}\, , \qquad  \omega^1= e^{-2\varphi} dx \phantom{0},\nonumber\\
 c^-&=e^{2\varphi}(i\, d\eta_a\eta^a-i\, \eta_a d\eta^a)+ i\, d\eta_4\eta^4- i\, \eta_4 d \eta^4-4\eta_4\eta^4 dy\, ,\\
c^+&=0\, ,\phantom{e^{-2\varphi}dx^+}  \qquad ~ c^1=0,\\
\Delta&=d\varphi\, .
\label{omega}
\end{align}

In the fermionic components of the Cartan form associated to the unbroken supersymmetries there is no dependence on $y$. The non-vanishing components are
\begin{align}
\omega^+_a&=e^{-\varphi}(\hat d\theta_a+dx\, \hat\eta_a), & \omega^{+a}&=e^{-\varphi}(\hat d\theta^a+dx\, \hat\eta^a),  \\
\omega^-_a&=e^{-\varphi}dx^+\hat\eta_a, & \omega^{-a}&=e^{-\varphi}dx^+\hat\eta^a, \\
\chi^+_a&=e^{\vf} \hat d \eta_a\, , & \chi^{+ a}&=e^{\vf} \hat d \eta^a\, . \label{eq:chiplus}
\end{align} 
On the other hand, the coefficients of the generators associated to the broken supersymmetries exhibit an explicit dependence on $dy$
\begin{align}
\omega^+_4&=d\theta_4 +d\vf \theta_4+e^{-2\vf} dx\, \eta_4 +2i\theta_4\, dy, & \omega^{+4}&=d\theta^4+d\vf \theta^4+e^{-2\vf} dx\, \eta^4- 2i\theta^4\, dy, \\
 \omega^-_4&=e^{-2\varphi}dx^+\eta_4, & \omega^{-4}&=e^{-2\varphi}dx^+\eta^4, \\
\chi^+_4&= d \eta_4-d\vf \eta_4+2i\eta_4\, dy\, , & \chi^{+ 4}&= d \eta^4-d\vf \eta^4-2i\eta^4\, dy\, . \label{eq:chi4plus}
\end{align} 
We have now collected all the necessary ingredients to build the supermembrane action in eleven dimensions and we can move to the description of the dimensional reduction procedure.

\subsubsection{Dimensional reduction}

Dimensional reduction of the $D=11$ supermembrane action to the
$D=10$ Type IIA superstring was described for general
superbackground in~\cite{Duff:1987bx}. One crucial requirement for being able to perform such a reduction is that the first 10 components of the bosonic supervielbeine are independent of both $y$ and $dy$, while the eleventh component should appear in the Kaluza-Klein Ansatz form
\begin{equation}\label{eq:KKAnsatz}
\mathrm E^{11}=\Phi(dy+A)\, ,
\end{equation}
where $\Phi$ is related to the dilaton and $A$ is the RR 1-form potential. However we noticed that, for the case at hand, this requirement is not satisfied since the bosonic vielbeine $\omega^m$ and $c^m$ in \eqref{omega} depend explicitly on $dy$. To remove this dependence one has to perform a local Lorentz rotation in the tangent
space
\begin{equation}
E^{\hat \m}\rightarrow L^{\hat \m}{}_{\hat \n}E^{\hat \n},\quad
F^{A'}\rightarrow
L^{A'}{}_{B'}F^{B'},\quad L^{\hat \m}{}_{\hat
\n}\in SO(1,10),\quad L^{A'}{}_{B'}\in \text{Spin}(1,10)\, ,
\end{equation} 
where $E^{\hat \m}$ and $F^{A'}$ are the bosonic and fermionic components of the supervielbeine entering \eqref{eq:4formWZ}. We should stress that
such a transformation is not part of the isometry of the $AdS_4\times S^7$
solution and should be
regarded as an appropriate choice of a different supervielbein basis of $\frac{OSp(8|4)}{SO(7)\times
SO(1,3)}$, which has the Kaluza–Klein form compatible with the Hopf fibration. In our case,
since the
$\mathbb{CP}^3$ vielbein components do not
contain any contribution proportional to $dy$, the necessary
frame rotation $L$ involves only the directions tangent to $AdS_4$ and
the one tangent to the U(1)-fiber direction on $S^7$
\begin{equation}
\left(\begin{array}{c}
 {\hat E}^{\m}\\
 \hat E^{11}
\end{array}\right)=
L\left(\begin{array}{c}
E^{\mu}\\
\Omega^{78}
\end{array}\right),
\end{equation}
where $E^\mu$ is defined in \eqref{eq:cartanformAdS4first} and, when translated to light-cone coordinates, is given by
\begin{equation}
 E^{\mu}=\left(\tfrac12(\o^m+c^m),\D\right)\, .
\end{equation}
The entries of the matrix $L$
\begin{equation}
L=\left(\begin{array}{cc}
L^{\m}{}_{\n}& L^{\m}{}_{7}\\
L^7{}_{\m}&L^7{}_7
\end{array}\right)\in SO(1,4)
\end{equation}
are fixed by the requirement that the transformed vielbein ${\hat E}^\m$ does not depend on $dy$
\begin{equation}
L^{\m}{}_{\n}=\delta^{\m}_{\n}-\frac12 E^{\m}_y E_{y\n},\quad
L^{\m}{}_7=-E^{\m}_y,\quad L^{7}{}_{\m}=E_{y\m},\quad
L^7{}_7=1,
\end{equation}
where
\begin{equation}
E^{\m}_y=2\Theta(1,0,-1,0)
\end{equation}
is a light-like vector, expressed in terms of $\Theta=\theta_4\theta^4+\eta_4\eta^4$.  The corresponding Lorentz rotation acting on the
supervielbein fermionic components is generated by the matrix
\begin{equation}
L^{A'}{}_{B'}=\delta^{A'}_{B'}-\frac12 E_{y\m} (\G^\m)^{A'}{}_{C'} (\G^{11})^{C'}{}_{B'},
\end{equation}
where $\G^\m$ and $\G^{11}$ are $SO(1,10)$ gamma-matrices. One can split the eleven dimensional spinor indices in $4d$ and $7d$ spinor indices and then, using the identity
$F^{A'}Q_{A'}=A_{\text{ferm}}$ with $A_{\text{ferm}}$ defined in \eqref{eq:Aferm}, one can find the action of $L$ on the fermionic components $\omega$ and $\chi$ in \eqref{eq:Aferm}.

After the Lorentz transformation, the bosonic components of the $D=11$ supervielbein in the light-cone basis equal
\begin{align}\label{tbv}
\hat E^{-}&=\frac12e^{-2\varphi}dx^-+\hat \omega -2e^{-2\varphi}\Theta^2dx^+ +4\Theta(\Omega_a{}^a-e^{-2\varphi}\eta_4\eta^4dx^+),\\
\hat E^{+}&=\frac12e^{-2\varphi}dx^+ ,  \qquad \hat E^1=\frac12e^{-2\varphi}dx, \qquad \hat E^3=-d\varphi,\qquad 
\hat E^{11}=dy+A\, ,
\end{align}
where
\begin{align}
 \hat \o&=ie^{-2\varphi}(d\theta_a\theta^a-\theta_ad\theta^a)+i(d\theta_4\theta^4-\theta_4d\theta^4)+ie^{2\varphi}(d\eta_a\eta^a-\eta_ad\eta^a)+i(d\eta_4\eta^4-\eta_4d\eta^4), \nonumber \\
 A&=\Omega_a{}^a-e^{-2\varphi}\eta_4\eta^4dx^+ -e^{-2\varphi}\Theta dx^+\, .
\end{align}
We notice that the new vielbein $\hat E^{11}$ contains all the dependence on $dy$ and has the required form \eqref{eq:KKAnsatz} with $\Phi=1$. Therefore, identifying the direction $y$ with the world-volume compact direction we obtain
\begin{equation}
\int\limits_V d^3\xi\sqrt{-g^{(3)}}\rightarrow\int\limits_\Sigma
d\tau d\sigma\sqrt{-g^{(2)}},
\end{equation}
or alternatively in the Polyakov form
\begin{equation}
S_{kin}=-\frac12 \int \g^{\a\b} \left(g^{(2),AdS_4}_{\a\b}+g^{(2),\cp^3}_{\a\b}\right)\, ,
\end{equation}
where $g^{(2)}$ stands for the induced worldsheet metric. Explicitly
\begin{align}
g^{(2),AdS}_{\a\b}&=\hat E^+_\a\hat E^-_{\b}+\hat E^1_\a\hat E^1_\b+\hat
E^3_\a\hat
E^3_\b=\frac14 e^{-4\varphi}(\partial_\a x^+\partial_\b x^- +\partial_\a x\partial_\b x)
+\partial_\a\varphi\partial_\b\varphi\\
&+\frac12e^{-2\varphi}\partial_\a x^+(\hat \o_\b+4\, \Theta\,  \Omega_{\b a}{}^a)-2e^{-4\varphi}\Theta^2\partial_\a x^+\partial_\b x^+\, , \nonumber\\
g^{(2),\cp^3}_{\a\b}&=(\O_{a}+\tilde \O_{a})_\a (\O^a+\tilde \O^a)_\b\\
&=\left[\O_{a\, \a}+\pa_\a x^+(\varepsilon_{abc}\hat{\eta}{}^b\hat{\eta}{}^c-2e^{-\varphi}\hat\eta_a\eta_4)\right]\left[\O^a_\b-\pa_\b x^+(\varepsilon^{abc}\hat\eta_b\hat\eta_c+2e^{-\varphi}\hat{\eta}{}^a\eta^4) \right]\, ,\nonumber
\end{align}
where the vielbein $\Omega^a_\a$ are defined in the natural way 
${\Omega^a}=\Omega^a_\a\, d\s^\a$ with $\s^\a=(\t,\s)$. 

As far as the fermionic components of the supervielbeine are concerned, let us separate them as
\begin{equation}\label{fermviel}
F^{A'}=f^{A'}+dyF^{A'}_y\, ,
\end{equation}
where the second term contains the whole dependence on $dy$. After Lorentz rotation the transformed fermionic vielbeine assume the form
 \begin{equation}\label{kkfermviel}
(LF)^{A'}=E^{A'}+\chi^{A'}(dy+A)\,\qquad
E^{A'}=(Lf)^{A'}-(LF_y)^{A'} A,\qquad
\chi^{A'}=(LF_y)^{A'},
\end{equation}
where $E^{A'}$ are the $D=10$ supervielbein fermionic
components and $\chi^{A'}$ is the dilatino superfield. With the fermionic components organized in this way, the dimensional reduction of the WZ term implies
\begin{equation}\label{wzred}
\int\limits_{\mathcal M_4} H_{(4)}\rightarrow \int\limits_{\mathcal M_3}H_{(3)},
\end{equation}
where $H_{(3)}$ is the NS-NS 3-form
\begin{equation}\label{eq:3form}
H_{(3)}=\frac{i}{4}(E^{A'}\G^{\hat \m\hat
\n}{}_{A'}{}^{B'}\chi_{B'}\wedge\hat
E_{\hat \m}\wedge\hat E_{\hat
\n}+E^{A'}\G^{\hat
\m\, 11}{}_{A'}{}^{B'}\wedge
E_{B'}\wedge\hat E_{\hat \m})-\epsilon_{\m\n\k\l}\hat E^\m \wedge\hat
E^{\n}\wedge\hat E^{\k}L^{\l}{}_7.
\end{equation}
This is a closed 3-form which can be expressed locally as the differential of a 2-form. Nevertheless it is not always easy to find a general expression for this 2-form. In our case the effect of the Lorentz rotation affects only the $\chi$ components \eqref{eq:chiplus} and \eqref{eq:chi4plus}
\begin{align}
 (L\chi)^+_a&=e^{\varphi}\hat d\eta_a+2\, i\, e^{-\varphi}\Theta\,\hat\eta_a dx^+ & (L\chi)^{+,a}&=e^{\varphi}\hat d\eta^a-2\, i\, e^{-\varphi}\Theta\, \hat\eta^a dx^+,\\
 (L\chi)^+_4&=d\eta_4-d\varphi\eta_4+2\, i\, e^{-2\varphi}\Theta\, \eta_4dx^+ ,& (L\chi)^{+,4}&=d\eta^4-d\varphi\eta^4+2\, i\, e^{-2\varphi}\Theta\, \eta^4dx^+ ,
\end{align}
and the 3-form can be expressed as the total differential of the rather lengthy 2-form
\begin{align}
B_{(2)}&=\frac{1}{2}e^{-4\varphi}\Theta\,  dx\wedge
dx^+ + \frac14e^{-2\varphi}(d\theta_4\eta^4-d\eta_4\theta^4+\eta_4d\theta^4-\theta_4d\eta^4)\wedge
dx^+ \nonumber\\
&+ie^{-2\varphi}\tilde \Theta\,  dx^+\wedge\Omega_{a}{}^a
+ie^{-\varphi}\hat\eta_a\theta_4dx^+\wedge\Omega^a+ie^{-\varphi}\hat{\eta}{}^a\theta^4dx^+\wedge\Omega_a \nonumber\\
&+e^{-2\varphi}\hat\eta_a\hat{\eta}{}^adx\wedge
dx^+ +\frac12e^{-2\varphi}(\hat\eta_a\hat d\theta^a+\hat
d\theta_a\hat{\eta}{}^a)\wedge dx^+\, ,
\end{align}
where we defined $\tilde \Theta =\theta_4{\eta}^4-\eta_4{\theta}^4$. We can now put all the terms together and, after the following rescaling 
\begin{equation}
\theta_{a}  \rightarrow \sqrt{2}\, \theta_{a}\, , \qquad
\theta_{4} \rightarrow \sqrt{2}\, e^{-\varphi}\theta_{4}\, , \qquad
\eta_{a} \rightarrow \sqrt{2}\, e^{-2\varphi}\eta_{a}\, , \qquad
\eta_{4} \rightarrow \sqrt{2}\, e^{-\varphi}\eta_{4}\, ,
\end{equation}
and similar ones for the complex conjugates, we get the $\k$-symmetry light-cone gauge fixed action for the superstring in $AdS_4\times \cp^3$ background
\begin{align}\label{eq:lagrangian}
S &= -\frac{T}{2} \int\, d\tau\, d\sigma\, \mathcal{L} \qquad\qquad \\ %T = \frac{R^2}{2\pi\alpha'}\\ 
\mathcal{L} = \gamma^{\a\b }\Big[&e^{-4\varphi} \frac{\partial_\a x^+\partial_\b x^-+
\partial_\a x\partial_\b x}{4} + \partial_\a \varphi
\partial_\b \varphi + \O^a_\a \O_{a \b}\nonumber\\
\phantom{=\gamma^{\a\b }} +&e^{-4\varphi}\partial_\a x^+\left(\varpi_\b + h_\b+e^{-4\vf} B\, \partial_\b x^+\right)\Big]\nonumber\\
+\epsilon^{\a\b }2\, &e^{-4\varphi}\partial_\a x^+ \left(~{\omega}_\b -\ell_\b+e^{-2\varphi}C\,\partial_\b x
~\right),\nonumber
\end{align}
where $T$ is the string tension and   the following quantities
\begin{align}
\varpi_\a &=i\left(\partial_\a \theta_a{\theta}^a-\theta_a\partial_\a {\theta}^a
+\partial_\a \theta_4{\theta}^4-\theta_4\partial_\a {\theta}^4
+\partial_\a \eta_a{\eta}^a-\eta_a\partial_\a {\eta}^a
+\partial_\a \eta_4{\eta}^4-\eta_4\partial_\a {\eta}^4\right),\nonumber\\ 
\omega_\a &=\hat{\eta}_a\hat\partial_\a {\theta}^a
+\hat\partial_\a \theta_a\hat{{\eta}}^a+\frac12\left(\partial_\a \theta_4{\eta}^4-\partial_\a \eta_4
{\theta}^4+\eta_4\partial_\a {\theta}^4-\theta_4
\partial_\a {\eta}^4\right) \, , \nonumber\\ 
B&=8\,  \left[(\hat{\eta}_a\hat{{\eta}}^a)^2+\varepsilon_{abc}\hat{{\eta}}^a\hat{{\eta}}^b
\hat{{\eta}}^c{\eta}^4+\varepsilon^{abc}
\hat{\eta}_a\hat{\eta}_b\hat{\eta}_c\eta_4+2
\eta_4{\eta}^4 \left(\hat{\eta}_a\hat{{\eta}}^a-\theta_4{\theta}^4\right)\right],\nonumber\\ 
C&=2\, \hat{\eta}_a\hat{{\eta}}^a+
\theta_4{\theta}^4+\eta_4{\eta}^4 \, ,\nonumber\\ 
h_\b&=2\, \left[\Omega^a_\b\varepsilon_{abc}\hat{{\eta}}^b\hat{
{\eta}}^c-\Omega_{a\b}\varepsilon^{abc}\hat{\eta}_b
\hat{\eta}_c + 2\left(\Omega_{a\b}\hat{{\eta}}^a
{\eta}^4-\Omega^a_\b\hat{\eta}_a\eta_4\right) + 2
\left(\theta_4{\theta}^4+\eta_4{\eta}^4\right)
{\Omega}_{a\ \b}^{\ a}\right],\nonumber\\ 
\ell_\b&=2\, i\, \left[\Omega_{a\b}\hat
{{\eta}}^a{\theta}^4+\Omega^a_\b\hat{\eta}_a\theta_4
+ \left(\theta_4{\eta}^4-\eta_4{\theta}^4\right)
{\Omega}_{a\ \b}^{\ a}\right] \label{eq:pieces}
\end{align}
include fermions up to the fourth power. It is clear from \eqref{eq:pieces} that, despite the result has the same structure as \eqref{eq:Lkin} and \eqref{eq:LWZ}, the expressions are definitely more involved due to the non maximally supersymmetric background. To facilitate the comparison with $AdS_5$ and to eliminate the tedious rotations \eqref{eq:rotfermAdS4}, we introduce the Wess Zumino parametrization as we did in section \ref{sec:WZpar} for $AdS_5\times S^5$.

\subsubsection{WZ parametrization}\label{sec:WZAdS4}
We first introduce a collective index for upper and lower indices so that  
\begin{equation} 
 \eta_{\hat{a}}=\left(\begin{array}{c}
                      \eta_a \\
                      \eta^a 
                      \end{array}\right)\,.\, 
\end{equation}
In this notation the action of the matrix $T$ on the fermions \eqref{eq:rotfermAdS4} can be rewritten as 
\begin{equation}
\hat{\eta}_{\hat a}= {T_{\hat{a}}}^{\hat b}\eta_{\hat b}\, ,
\end{equation}
where the matrix ${T_{\hat{a}}}^{\hat b}$ is given in \eqref{matrixT}. 
We also  introduce the shorthand notation
\begin{equation}
 \partial_i\eta_a{\eta}^a-\eta_a\partial_i{\eta}^a=-\eta^{\hat{a}}\partial_i\eta_{\hat{a}}~,
\end{equation}
where $\eta^{\hat{a}}=({\eta}^a,\eta_a)$. In \eqref{eq:rot} a recipe for eliminating the rotation of the fermions was given. 
This generates additional terms coming from derivatives that can be reabsorbed into a covariant derivative. 
In particular, we apply the transformation
\begin{equation}
 \eta_{\hat a}\to \big(T^{-1}\big)_{\hat a}^{\hat b} \,\eta_{\hat b}\,.
\end{equation}

In contrast with the $AdS_5\times S^5$ case the matrix $T$ is not block diagonal, therefore one has $\eta^{\hat{a}}\partial_i\eta_{\hat{a}}=\hat{\eta}^{\hat{a}}\hat\partial_i {\eta}_{\hat{a}}$, where 
it is crucial to use  hatted indices. This transformation removes all the hats from fermions, at the price of introducing the covariant derivative 
\begin{equation}
 D=d-\Omega\,,
\end{equation}
where $\Omega\equiv{\Omega_{\hat a}}^{\hat b}=d {{T}_{\hat a}}^{\hat c}\, {(T^{-1})_{\hat c}}^{\hat b}$ and $d\Omega-\Omega\wedge\Omega=0$. More explicitly\footnote{
The matrix $\Omega$ was already introduced in~\cite{Uvarov:2008yi}, 
however there it was defined as 
${\Omega_{\hat a}}^{\hat{b}}=i {T_{\hat a}}^{\hat c}d  {{T^{-1}}_{\hat c}}^{\hat b}
=-i d{T_{\hat a}}^{\hat c} {{T^{-1}}_{\hat c}}^{\hat b}$, differing from ours by a factor of $i$. 
To make contact with the expressions of~\cite{Uvarov:2008yi} 
we add such a factor in formula \eqref{Omega}.},
\begin{equation}\label{Omega}
 {\Omega_{\hat a}}^{\hat{b}}=i\, \left(\begin{array}{cc}
                              \frac12(\Omega_{a}^{\phantom{a}b} - \d_a^b \Omega_c^{\phantom{c}c})& \e_{acb}\Omega^c\\
                              -\e^{acb}\Omega_{c} & -\frac12(\Omega^a_{\phantom{a}b}-\d_b^a \Omega_c^{\phantom{c}c})
                             \end{array}\right)\,,
\end{equation}
where the components ${\Omega_a}^b$ already appeared in \eqref{cfso6so2} and they read
\begin{align}\label{eq:forms}
\Omega_{a}^{\phantom{a}b}&=2\,i\frac{(1-\cos{|z|})}{|z|^2}(\bar z_adz^b-d\bar z_az^b)-i\bar z_az^b\frac{(1-\cos{|z|})^2}{|z|^4}(dz^c\bar z_c-z^cd\bar z_c),
\end{align}
On the other hand the components $\O^a$ and $\O_a$ are simply the $\cp^3$ vielbeine \eqref{eq:cp3vielbein1} and \eqref{eq:cp3vielbein2}.

We can also decompose the matrix $\Omega$ in order to separate the contributions from the vielbein and from the spin connection\footnote{A similar procedure was applied in \eqref{eq:Omegadec}, where in that case the decomposition is expressed in terms of the $SO(5)$ $\gamma$-matrices.} 
\begin{equation}\label{decomposition}
 {\Omega_{\hat a}}^{\hat{b}}=\Omega^{\hat c} {(E_{\hat c})_{\hat a}}^{\hat b}+ \Omega^c_{\phantom{c}d} {(J_c^d)_{\hat a}}^{\hat b}\, ,
\end{equation}
with\footnote{Let us stress that the meaning of the first term of equation \eqref{decomposition} in matrix form is the following 
\begin{equation}
\Omega^{\hat c}{(E_{\hat c})_{\hat a}}^{\hat{b}}=\left(\begin{array}{cc}
                             \Omega^c{(E_{c})_a}^b+\Omega_c{(E^{c})_a}^b&  \Omega^c (E_{c})_{ab}+\Omega_c (E^{c})_{ab}\\
                               \Omega^c (E_{c})^{ab}+\Omega_c(E^{c})^{ab} &  \Omega^c{(E_{c})^a}_b+\Omega_c{(E^{c})^a}_b
                             \end{array}\right)
\end{equation}
and the explicit expression of ${(E_{\hat c})_{\hat a}}^{\hat{b}}$ shows that the only non-vanishing elements are $(E_{c})_{ab}$ and $(E^{c})^{ab}$.}
\begin{align}
 {(E_{\hat c})_{\hat a}}^{\hat{b}}&=i\, \left(\begin{array}{cc}
                             0& \e_{acb}\\
                              -\e^{acb} & 0
                             \end{array}\right),
&{(J_c^d)_{\hat a}}^{\hat b}&=\frac{i}{2}\, \left(\begin{array}{cc}
                             \d_a^d \d_c^b-\d_a^b \d_c^d& 0\\
                              0 & -\d_b^d \d_c^a+\d_b^a \d_c^d
                             \end{array}\right) \,.
\end{align}
This decomposition provides a way to project out the spin connection and find the exact relation between the vielbein $\Omega_{\hat a}$ and the matrix $\Omega$
\begin{equation}
 \Omega_{\hat c}=\frac12 \Tr(E_{\hat c}\,\Omega)~.
\end{equation}
After having introduced all the necessary ingredients, we are ready to rewrite the Lagrangian in a form which resembles the $AdS_5\times S^5$ case. We separate it into
\begin{equation}
 \mathcal{L} = \mathcal{L}_B + \mathcal{L}_F^{(2)} + \mathcal{L}_F^{(4)}\, ,
\end{equation}
where the bosonic contribution is simply given by the standard bosonic sigma model with $AdS_4\times \mathbb{CP}^3$ as target space
\begin{equation}
\mathcal{L}_B = \gamma^{\a\b}\left[\frac{e^{-4\varphi}}{4}\left(\partial_\a x^+\partial_\b x^-+
\partial_\a x^1\partial_\b x^1\right)+\partial_\a \varphi
\partial_\b \varphi+{\Omega^a}_\a{\Omega_a}_\b\right]\, .
\end{equation}
Notice that ${\Omega^{\hat a}}_\a\,{\Omega_{\hat a}}_\b=2\, {\Omega^a}_\a\,{\Omega_a}_\b$
 for the symmetry of the worldsheet metric. The quadratic part in the fermion fields can be expressed 
 as
\begin{align}\label{lagnew}
\mathcal{L}_F^{(2)} = -e^{-4\vf} \pa_\a x^+ \Big[&i \g^{\a\b} \big(\eta^{\hat{a}}D_\b \eta_{\hat{a}}+
\theta^{\hat{a}}D_\b \theta_{\hat{a}}-2\, \Omega_\b ^{\hat c}\, \eta E_{\hat c}\eta \big)\\
+&2\epsilon^{\a\b} 
\eta^{\hat{a}} {C_{\hat a\hat{b}}} \big(D_\b  \theta^{\hat b}+e^{-2\vf} \eta^{\hat b} \pa_\b  x) \nonumber\\
 +&i \g^{\a\b} \big(\eta^{4}\pa_\b \eta_{4}+\theta^{4}\pa_\b \theta_{4}
 -4\, i\, \eta_a\Omega^{a}_\b \eta_4+ 2\, i\, \Omega_{a\phantom{a}j}^{\phantom{a}a} \Theta-h.c.\big)\nonumber \\
 +&\phantom{2}\epsilon^{\a\b}\big( \eta^4 \pa_\b  \theta_4- \theta^4 \pa_\b  \eta_4 +4\, i\, \eta_a\Omega^{a}_\b \theta_4 +2\,i\, \Omega_{a\phantom{a}j}^{\phantom{a}a} \tilde\Theta-e^{-2\vf} \Theta \pa_\b  x+h.c.\big)
 \Big]\nonumber
\end{align}
where we have introduced the charge conjugation matrix $C$, 
\begin{equation}
 C_{\hat a\hat b}=\left(\begin{array}{cc}
                              0 & \d_a^b \\
                              -\d^b_a & 0
                             \end{array}\right).
\end{equation}
The first two lines of the Lagrangian \eqref{lagnew} closely resembles expression \eqref{eq:L2fWZ}, that is the $AdS_5\times S^5$ Lagrangian in Wess-Zumino type parametrization. 
This is the part of the Lagrangian that does not contain the fermions $\eta_4$ and $\theta_4$, which emerge when 
obtaining the \adscp action from dimensional reduction.
%This is in contrast with the $OSp(4|6)/\left( SO(1,3)\times U(3) \right)$ supercoset formulation, 
%where only 24 fermionic coordinates arise. However, in the latter setting the $\kappa$-symmetry 
%light-cone gauge fixing for strings moving in the $AdS$ space only (as for the light-like cusp) is not viable. 
%On the contrary the former approach~\cite{Uvarov:2009hf} allows to impose a $\kappa$-symmetry 
%gauge condition that consists of setting to zero all the coordinates associated to fermionic generators 
%with negative charge with respect to the $SO(1,1)$ generator $M^{+-}$ (see~\cite{Uvarov:2009hf} and~\cite{Metsaev:2000yf} for the $AdS_5\times S^5$ case). 
%The main difference with respect to $AdS_5\times S^5$ is that the $SU(4)$ R-symmetry 
%is not explicitly realized on the fermionic Lagrangian \eqref{lagnew}. 
%This feature is inherited by the quantum fluctuations around the light-like cusp. 
%As a result of the broken symmetry, the spectrum contains fermionic degrees of freedom with 
%different masses (one gets $6$ massive and $2$ massless excitations). 
%Our one- and two-loop calculations have explicitly shown that the role of the 
%massless fermions ($\tilde\eta_4$ and $\tilde\theta_4$) is crucial for compensating 
%the bosonic degrees of freedom, making the one-loop partition function UV-finite. 
%At two loops their interactions with the other excitations would in principle 
%start playing a part. Nevertheless it turns out that the massless fermions 
%decouple from the computation and do not contribute to the two-loop result.

The last term of the superstring Lagrangian is quartic in fermions
\begin{equation}
\mathcal{L}_F^{(4)} = 4\, e^{-8\vf} \g^{\a\b} \pa_\a x^+\pa_\b x^+   [(\eta_a{\eta}^a)^2+2\,\varepsilon^{abc}
\eta_a\eta_b\eta_c\eta_4+2
\eta_4{\eta}^4\eta_a{\eta}^a-\Theta^2+h.c.] \,.
\end{equation}
As discussed for the quadratic part, the first terms clearly reminds the expression for $AdS_5\times S^5$ (equation \eqref{eq:L4fWZ}), whereas the others contain the non-trivial interactions of $\eta_4$ and $\theta_4$.

\section{Superstring theory in \txpf{$AdS_3\times S^3 \times M^4$}{AdS3 x S3 x M4}}\label{sec:stringAdS3}

The last example of $AdS/CFT$ we are going to analyze in this thesis is the $AdS_3/CFT_2$ one. We focus on supergravity backgrounds preserving 16 real supercharges, i.e. $AdS_3\times S^3\times T^4$ and $AdS_3\times S^3\times S^3\times S^1$. The two models are not entirely independent. Indeed the second can be seen as a deformation of the first, where the deformation parameter $\a$ is introduced by the following triangle equality imposed by the supergravity equations of motion
\begin{equation}\label{supgraradii}
 \frac{1}{R_+^2}+\frac{1}{R_-^2}=\frac{1}{R^2}\,.
\end{equation}
This is a relation among the radii of the two spheres ($R_+$ and $R_-$) and the radius of $AdS$ ($R$). We can re-express this relation as
\begin{equation}
 \frac{R^2}{R_+^2}=\a\, , \qquad \frac{R^2}{R_-^2}=1-\a\, .
\end{equation}
Hence the superstring action on this background will be a function of $\a$. The same triangle equality appears in the invariant bilinear form of
the exceptional Lie superalgebra $\mathfrak{d}(2,1;\alpha )$~\cite{Kac:1977qb,Frappat:1996pb}; this is because the
super-isometries of the $AdS_3\times S^3\times S^3$ background form
two copies of $\mathfrak{d}(2,1;\alpha )$~\cite{Gauntlett:1998kc}. In the limit $\a\to 0$ (or equivalently $\a\to 1$) one of the two spheres assumes the same radius of $AdS_3$ and the other one blows up into a plane. Up to compactifying this plane to a $T^3$, this limit is equivalent to considering $AdS_3\times S^3\times T^4$. In this case the algebra of superisometries of $AdS_3\times S^3$ consists of two copies of $\mathfrak{psu}(1,1|2)$ and this hints at a similarity with the $AdS_5\times S^5$ example. On the other hand, in the limit $\a\to \frac12$ the 
exceptional superalgebra $\mathfrak{d}(2,1;\alpha )$ coincides with the classical $\mathfrak{osp}(4|2)$,
superalgebra hinting to similarities with the $AdS_4\times \mathbb{CP}^3$ case. We will comment further on those similarities in chapter \ref{BMN}.

As we mentioned in chapter \ref{introduction}, the $AdS_4\times S^3\times T^4$ background, as well as other $AdS_3$ backgrounds, support both NSNS and RR fluxes. The NSNS flux theory admits a NSR description and it can be formulated as a supersymmetric extension of the $SL(2)\times SU(2)$ WZW model. It is then solvable by representation theory of chiral algebras~\cite{Giveon:1998ns,deBoer:1998pp,Kutasov:1999xu, Maldacena:2000hw, Maldacena:2000kv, Maldacena:2001km}. On the other hand, this path is not viable in presence of RR flux. Nevertheless, the GS formulation for pure RR flux~\cite{Pesando:1998wm,Rahmfeld:1998zn,Park:1998un,Metsaev:2000mv} can be deformed by the introduction of a parameter $q$~\cite{Pesando:1998wm}, interpolating between pure RR and pure NSNS. The corresponding supergravity background is the near-horizon limit of the mixed NS5-NS1+D5-D1 solution and it is invariant under S-duality transformation. The latter transforms NSNS into RR flux, so that if the coefficients of the NSNS and RR fluxes are chosen as $q$ and $q'$, respectively, 
then they enter symmetrically into the supergravity equations, e.g., as $q^2+{q'}^2= 1$ (we set the
curvature radius to $R_{AdS_3}^2=R_{S^3}^2=1$). Nevertheless the free (i.e. no target space interaction) superstring theory is not invariant under
S-duality and should thus depend non-trivially on the parameter $q$. In particular we assume $0<q<1$, with $q=0$ corresponding to pure RR flux and $q=1$ to pure NSNS flux.

The GS formulation for pure RR and mixed flux has received a lot of attention in the last years because of its integrable properties~\cite{Babichenko:2009dk,Sundin:2012gc,Sundin:2013uca,Cagnazzo:2012se}.
As it happens for the $AdS_4\times \cp^3$ superstring, the coset formulation~\cite{Babichenko:2009dk,Cagnazzo:2012se} suffers from a serious drawback. Indeed, the action obtained by the supercoset construction contains only 16 fermionic degrees of freedom and can be interpreted as a $\k$-symmetry gauge-fixed version of the full GS superstring Lagrangian~\cite{Babichenko:2009dk,Rughoonauth:2012qd}. As discussed for the $AdS_4\times \cp^3$ case in section \ref{sec:stringAdS4}, this $\k$-symmetry gauge fixing may be uncompatible with some particular string configurations. Compared to $AdS_4\times \cp^3$, however, there is no known way to obtain the full superstring theory action, and the path that has been followed so far consists in expanding the GS action for curved backgrounds in higher powers of fermions~\cite{Sundin:2012gc,Rughoonauth:2012qd,Sundin:2013ypa,Wulff:2013kga,Wulff:2014kja}. 

In the following, we will sketch the coset formulation of $AdS_3$ with the aim of proving classical integrability in the particular $\k$-symmetry gauge fixing implied by the construction. This $\k$-symmetry gauge is not compatible with any light-cone gauge fixing. Since this is the main requirement to be able to expand the action around the BMN vacuum, the starting point for any perturbative computation is the expansion of the general GS superstring action for curved backgrounds supported by RR flux.

\subsection{The coset approach}
 One important feature of the $AdS_3$ backgrounds is that the group of isometries of $AdS_3$ is the conformal group in two dimensions. The latter is a two-fold tensor product, with two factors acting independently on the left and right
movers. Therefore the cosets entering the $AdS_3/$CFT${}_2$
correspondence are of the form $\frac{H\times H}{H_0}$. The immediate implication of this fact is that, if $H$ is a
supergroup, such a coset will inherit a $\mathbb{Z}_4$
structure. Indeed, one can define a $\mathbb{Z}_4$ automorphism by combining the
fermion parity with the permutation of the two factors:
\begin{equation}\label{omegad}
 \Omega =
 \begin{pmatrix}
   0  & \mathbb{1}  \\
    (-1)^F & 0  \\
 \end{pmatrix},
\end{equation}
where the supermatrix is acting on a superalgebra element $(X_1,X_2)$, with $X_1$ in the first copy of $\mathfrak{h}$ and $X_2$ in the second one. This map 
squares to $(-1)^F$ and its forth power is the identity:
$\Omega^4=\mathbb{1}$.  Given this particular structure, one can define the two Cartan forms
\begin{equation}
A_{L,R}=g_{L,R}^{-1} dg_{L,R}\, ,
\end{equation}
where $g_L$ is an element of the first $H$ factor and $g_R$ of the second one.
The $\mathbb{Z}_4$ grading of the Cartan current is simply
\begin{align}
 A^{(0)}&=\frac12(A_L^{\text{even}}+A_R^{\text{even}})\,, & A^{(2)}&=\frac12(A_L^{\text{even}}-A_R^{\text{even}})\, ,\\
 A^{(1)}&=\frac12(A_L^{\text{odd}}+A_L^{\text{odd}})\, , & A^{(3)}&=\frac12(A_L^{\text{odd}}-A_R^{\text{odd}})\, .
\end{align}
Notice also that the invariant subspace of $\O$ is the diagonal bosonic
subalgebra. Therefore we can conclude that, for any superalgebra $H$, one can construct a $\mathbb{Z}_4$
invariant coset sigma-model with the global  $H\times H$ symmetry. The
bosonic part of the action is the sigma-model on
$H_{\rm bos}\times H_{\rm bos}/H_{\rm diag}$ isomorphic
to $H_{\rm bos}$. This sigma-model will be automatically integrable, as we showed in section \ref{sec:ksymm}. 

This general construction naturally applies also to $AdS_3\times S^3\times T^4$ and $AdS_3\times S^3\times S^3\times S^1$. In the former case $H=PSU(1,1|2)$ while in the latter $H=D(2,1;\a)$. To be precise, the supergroup $PSU(1,1|2)$ describes only the $AdS_3\times S^3$ part of the background and the additional abelian factors associated to $T^4$ needs to be added in by hand. A similar argument applies to the last $S^1$ in $AdS_3\times S^3\times S^3\times S^1$. Therefore the question arises as to whether and how the sigma model obtained this way is related to the GS action for superstrings in this backgrounds. In~\cite{Babichenko:2009dk} it was proven that the action obtained via the coset sigma model for $AdS_3\times S^3 \times S^3\times S^1$ is equivalent to GS superstring in some particular $\k$-symmetry gauge. This is equivalent to prove that there exists a $\k$-symmetry gauge that decouples the $S^1$ direction from the other degrees of freedom. A similar mechanism works for $AdS_3\times S_3\times T^4$, 
i.e. the coset action on $AdS_3\times S^3$,
supplemented with four  free bosons, describes
ten-dimensional Type IIB GS strings on $AdS_3\times S^3\times T^4$
in a fully fixed kappa-symmetry gauge. One may wonder how could this be possible since the
six-dimensional coset
action for $\frac{PSU(1,1|2)\times PSU(1,1|2)}{SL(2,\mathbb{R})\times SU(2)}$~\cite{Rahmfeld:1998zn,Park:1998un,Metsaev:2000mv} has only 8
physical fermions, a factor of two short
of the 16 fermions required in ten dimensions. However, rather surprisingly
the extra $T^4$ factor in the action changes the number of
physical degrees of freedom in the coset sector. This is a consequence of the interaction of the four bosons of
$T^4$ with the coset fermions through the 2d metric
coupling or, in the conformal gauge, through the Virasoro
constraints.  In other words the Virasoro constraints are modified by the addition of the four free bosons and consequently the kappa-symmetry of the six-dimensional action is not a symmetry of the ten-dimensional
action. We can then conclude that the coset + $T^4$ model has 16 physical fermions.

The extension of this construction to the mixed flux case involves the addition of a WZ term to the coset action, due to the presence of a B-field
\begin{align}
 S&=\frac12 \int_M \Str(A^{(2)}\wedge * A^{(2)}+ \sqrt{1-q^2}A^{(1)}\wedge A^{(3)})\\
 &+q\int_B\Str(\frac23 A^{(2)}\wedge A^{(2)}\wedge A^{(2)}+A^{(3)}\wedge A^{(1)}\wedge A^{(2)}+A^{(1)}\wedge A^{(3)}\wedge A^{(2)})\, ,
\end{align}
where the first line is just another way to rewrite \eqref{eq:action} with $\k=\sqrt{1-q^2}$. The coefficient of the new WZ term is fixed by requiring $\k$-symmetry, classical integrability and conformal invariance~\cite{Cagnazzo:2009zh}.

% Chapter Template

\chapter{Near-BMN string and worldsheet scattering} % Main chapter title

\label{BMN} % Change X to a consecutive number; for referencing this chapter elsewhere, use \ref{ChapterX}

\lhead{Chapter 3. \emph{BMN string}} % Change X to a consecutive number; this is for the header on each page - perhaps a shortened title

%--------------------------------------------------------------------------------------------------------------

The light-cone gauge fixed worldsheet sigma model is a two-dimensional quantum field theory with interactions vertices involving an arbitrary number of fields. The quadratic Lagrangian is that of the light-cone gauge fixed string theory in a plane-wave Ramond-Ramond background. The latter, together with flat space and $AdS_5\times S^5$, constitutes the set of all the maximally supersymmetric backgrounds for type IIB superstring. Moreover, it was shown in~\cite{Blau:2001ne,Blau:2002dy} that the parallel-plane ($pp$) wave background can be obtained as a limit of $AdS_5\times S^5$, when considering the geometry seen by a point-like string (i.e. a particle) moving very fast along $S^5$. It is clear that such a motion can be very conveniently described in light-cone gauge. Indeed, considering an angular coordinate $\phi$ on the sphere and the $AdS$ time $t$, we can describe the trajectory of a light-like particle as $t=\phi=\tau/2$, which is clearly very well suited for the light-cone gauge condition
\begin{equation}
x^+=t+\phi=\tau\, .
\end{equation} 
As a consequence the light-cone gauge fixed superstring action describes the quantum fluctuations around this classical solution, and in $pp$-wave background is simply given by a free worldsheet theory with eight massive bosons and eight massive fermions~\cite{Metsaev:2001bj}. This free Lagrangian can be easily quantized and the spectrum is known exactly~\cite{Metsaev:2002re}. The precise relation with the corresponding operators in $\mathcal{N}=4$ SYM was found by Berenstein, Maldacena and Nastase (BMN) in~\cite{Berenstein:2002jq}. The expansion of the ligth-cone gauge fixed superstring action in higher powers of the fields can be seen as a perturbation of the $pp$-wave background and it is often referred to as near-BMN expansion.

In the following, we will briefly sketch the procedure of uniform light-cone gauge fixing (see also section \ref{sec:unilccoset}) and expand the corresponding action up to quartic order. This action describes a closed superstring and therefore it is defined on a worldsheet with one compact direction. In principle this would prevent us from defining any kind of scattering among the worldsheet quantum fluctuations. Nevertheless, one can formally define a decompactification limit such that the radius of the compactified direction becomes very large and massive, asymptotic states with arbitrary momentum arise. This limit, though yielding a non-physical string action, is very important for the comparison with integrability since it allows for a proper definition of the scattering matrix. In section \ref{sec:decvir} we will describe this limit and relax the Virasoro condition in order to deal with non-vanishing worldsheet momenta.

In this setup the asymptotic states are well defined and a natural observable is the S-matrix for the worldsheet excitations~\cite{Klose:2007rz}. This S-matrix is clearly not a physical object, since we gave up the level-matching condition and we took the decompactification limit. Nonetheless, it still contains all the information about the spectrum of the theory, due to the expected integrability properties. Indeed the S-matrix is the main building block of the Bethe Ansatz (either asymptotic or thermodynamic) whose solution yield the Hamiltonian eigenstates, i.e. the spectrum of the full superstring theory. It turns out that the symmetry of the $AdS_5\times S^5$ background is large enough to fix completely the structure of the S-matrix up to an overall factor which contains most of the dynamical information about the scattering process. In this chapter we follow a different strategy and we study the S-matrix perturbatively, first reviewing the tree-level calculation and then introducing the so-called unitarity techniques, which dramatically simplify the one-loop computation.

In section \ref{sec:AdS3} we apply the same technique to the worldsheet scattering in $AdS_3\times S^3\times M^4$, where an additional obstacle for the standard Feynman diagram technique comes from the computational difficulty in expanding the GS superstring action beyond quartic order. We will see that the only ingredient for a one-loop computation by unitarity is the tree-level S-matrix and this provides a drastic simplification for theories with many interactions, such as the string sigma models on curved backgrounds. 

\section{Uniform light-cone gauge fixing}
The construction of the action in uniform light-cone gauge is slightly involved and here we only describe the main steps of the procedure. We first focus on the bosonic part and subsequently we report the results for the fermionic Lagrangian, whose explicit derivation can be found in~\cite{Frolov:2006cc,Arutyunov:2009ga}. 

\subsection{Bosonic strings in light-cone gauge}
Let us start from the bosonic part of the superstring action in $AdS_5\times S^5$ \eqref{eq:ulcLbos}. Consider the momenta canonically conjugated to the coordinates $X^{\hat m}=\{t,\phi,y^i,z^i\},$\footnote{Notice that the same procedure cannot be straightforwardly extended to the full superstring case due to the contributions to the momenta $p_{\hat m}$ coming from the WZ term.} 
\begin{equation}
p_{\hat m} =\frac{\d S}{\d \dot{X}^{\hat m}} = -T\, \g^{0\b}\partial_\b X^{\hat n}\,
G_{{\hat m}{\hat n}}\,,\qquad \dot{X}^{\hat m}\equiv \pa_0 X^{\hat m} \,, 
\end{equation} 
and rewrite the
string action (\ref{eq:ulcLbos}) in the first-order form 
\begin{equation} \label{S2} 
S=
\int\, {\rm d}\s{\rm d}\tau\, \left( p_{\hat m} \dot{X}^{\hat m} +
\frac{\g^{01}}{\g^{00}} C_1+ \frac{1}{2T\, \g^{00}}C_2\right)\,,
\end{equation}
where $C_1$ and $C_2$ represent the two Virasoro constraints
\begin{equation} C_1=p_{\hat m}\acute{X}^{\hat m}\,,\quad
C_2=G^{{\hat m}{\hat n}} p_{\hat m} p_{\hat n} + T^2\, \acute{X}^{\hat m} \acute{X}^{\hat n} G_{{\hat m}{\hat n}}\,,\qquad \acute{X}^{\hat m}\equiv
\pa_1 X^{\hat m}\,, \end{equation}
which need to be solved after the gauge fixing. 

To impose a uniform gauge we introduce the ``light-cone''
coordinates and momenta
\begin{align}
 \label{lcc} 
 x_- &=\phi \,-\,t\ , & x_+ &=(1-a)\,t\, +\,a\,\phi\ ,\\
 p_- &= p_\phi\,+\,p_t\ ,& p_+ &=
(1-a) p_\phi \,-\,a\,p_t\ ,
\end{align}
where the parameter $a$ is a residual gauge freedom which parametrizes the most general uniform gauge such
that the light-cone momentum $p_-$ is equal to $p_\phi\,+\,p_t$. To better understand the role of $a$ let us introduce the conserved charges
\begin{equation}
\label{charges} E = - \int d\s\, p_t\ \, , \qquad J= \int d\s\, p_\phi\ ,
\end{equation}
which are related to the light-cone momenta by\footnote{Here $P_+$ and $P_-$ are not to be confused with the translation generators introduced in \eqref{eq:lcbosgen}.}
\begin{equation}
 \label{charges2}
P_- \,=\,\int\, {\rm d}\s\, p_-\,= \, J\,-\,E\ \, ,
\qquad P_+ \,=\, \int\, {\rm d}\s\, p_+\,=\, (1-a)\, J
\,+\, a\,E\, . 
\end{equation}
The second relation relates the momentum $P_+$ to some combination of $E$ and $J$. We can observe that there are three natural choices for the value of the parameter $a$. If $a=0$ we
have the temporal gauge $t = \tau\,,\ P_+=J $, if
$a=\frac12$, we obtain the usual light-cone gauge $x_+ =\frac12(t +\phi) =\tau\,,\ P_+ = \frac12(E+J)$, while for $a=1$ the uniform gauge
reduces to $x_+=\phi =\tau\,,\ P_+=E$, where the angle variable $\phi$
is identified with the world-sheet time $\tau$, and the energy $E$
is distributed uniformly along the string. 

In general we can consider the variable $x^+$ in \eqref{lcc} and impose $x^+=\t$. Nevertheless, one has to take into account that the light-cone direction $\phi$ is compact and the closed string may have a non-trivial winding in that direction. In particular, for $\s_i<\s<\s_f$, the condition 
\begin{equation}
 \phi(\s_f)-\phi(\s_i)=2\pi m
\end{equation}
has to hold. Consequently a consistent gauge choice is 
\begin{equation}
 \label{ulc} x_+ \,=\, \tau \,+\, a\, m\,\s\
,\quad\quad p_+ \,=\, 1\ \, ,
\end{equation}
where the winding (with $m$ labeling the winding number) correctly vanishes in the temporal gauge. The second condition in \eqref{ulc} states
that the light-cone momentum is distributed uniformly along the
string, and this explains the word ``uniform'' in the name of the
gauge. Our particular gauge choice fixes the value of the total momentum $P_+$ in \eqref{charges2} as
\begin{equation}
 P_+=|\s_f-\s_i|\, .
\end{equation}
To find the gauge fixed action one can solve the Virasoro constraint for $x_-$ and $p_-$, such that the action assumes the form
\begin{equation} \label{S4} 
S
=\int {\rm d}\s{\rm d}\tau\, \left( p_{\hat i} \dot{x}^{\hat i}
\,-\, \mathcal{H} \right)\, , \qquad \mathcal{H} \,=\, -p_-(p_{\hat i}, x^{\hat i} ,{\acute{x}}^{\, \hat i} ) \, ,
 \end{equation}
 where the vector $x^{\hat i}=(y^i,z^i)$.
It is worth noticing that the whole dependence on $P_+$ is contained in the integration bounds on $\s$. In other words the theory is defined on a cylinder of circumference $P_+$. 

For simplicity let us now restrict to the $m=0$ case, so that invariance under translations in the $\s$ direction implies that the total worldsheet momentum of the string
\begin{equation}\label{pws} p_{{\rm ws}} = -\int_{-\frac{P_+}{2}}^{\frac{P_+}{2}}\, {\rm d}\s\, p_{\hat i} {\acute{x}}^{\, \hat i}\,
\end{equation}
is conserved. Furthermore, a closed string should satisfy the level-matching condition, which for $m=0$ imposes
\begin{equation}
 \label{LM} \D x_-=
\int_{-\frac{P_+}{2}}^{\frac{P_+}{2}}\, {\rm d}\s\, \acute{x}_- = 0 .
\end{equation}
When this condition is combined with the solution of the Virasoro constraint $C_1=0$, one finds that
\begin{equation}
\D x_-=
\int_{-\frac{P_+}{2}}^{\frac{P_+}{2}}\, {\rm d}\s\, \acute{x}_- = -\int_{-\frac{P_+}{2}}^{\frac{P_+}{2}}\, {\rm
d}\s\, p_{\hat i} {\acute{x}}^{\, \hat i}=p_{{\rm ws}} ,
\end{equation}
which implies that the physical string states have vanishing worldsheet momentum
\begin{equation}
 \label{LM2} \D x_-= p_{{\rm ws}} =0\,,\qquad
m=0\,.  
\end{equation}
Nevertheless, it is worth remembering that a proper quantization of superstring in light-cone gauge requires considering all states with periodic target space coordinates and imposing the level-matching condition only
at the very end. Hence, before imposing the level matching condition, the string states are not physical and in a uniform light-cone gauge the target spacetime image is an open string with end points
moving in unison so that $\D x_-$ remains constant (this is because $p_{{\rm ws}}=\D x_-$ is conserved). Moreover, in general, string configurations which
violate the level-matching condition may depend on $a$. We will see this gauge dependence appearing explicitly in the main object we study in this chapter, i.e. the S-matrix for the scattering of worldsheet excitations.

Solving the second Virasoro condition $C_2=0$ for the action \eqref{eq:ulcLbos}, we can find the explicit expression for the Hamiltonian density $\mathcal{H}$ 
\begin{align}
\mathcal{H} &= \frac{\sqrt{G_{\phi\phi}G_{tt}\left( 1+ \left((a-1)^2G_{\phi\phi} - a^2G_{tt}\right) \mathcal{H}_x  + T^2\left((a-1)^2G_{\phi\phi} - a^2G_{tt}\right)^2 \acute{x}_-^2  \right)}}{(a-1)^2G_{\phi\phi} - a^2G_{tt}}~~~\nonumber\\
\label{Hb} &+  \frac{(a-1)G_{\phi\phi} - a G_{tt}}{(a-1)^2G_{\phi\phi} -
a^2G_{tt}}\,, 
\end{align}
where $\mathcal{H}_x$ depends only on the transverse coordinates
\begin{equation} 
\mathcal{H}_x = G^{\hat i \hat j}p_{\hat i}  p_{\hat j} + g^2\, \acute{x}^{\hat i} \acute{x}^{\hat j}\,
G_{\hat i\hat j}\,. 
\end{equation}
Let us stress that, using the relation 
\begin{equation}
\label{He} H \,=\, \int_{-\frac{P_+}{2}}^{ \frac{P_+}{2}}\, {\rm d}\s\, \mathcal{H} =-P_-= E-J\,, 
\end{equation}
one can relate the worldsheet Hamiltonian to the target space energy (notice that $E$ appears also on the l.h.s. of \eqref{He} through the dependence on $P_+$), and therefore the knowledge of the spectrum of $H$ would give an algebraic equation for $E$. This is particularly relevant in the context of the AdS/CFT correspondence since it would yield the anomalous dimension of all the single-trace local gauge invariant operators of the CFT.
Unfortunately this cannot be achieved because the Hamiltonian, even without fermions, has a complicated non-polynomial dependence and it is not suitable for a direct canonical quantization. The best we can do is to quantize the theory perturbatively around some particular vacuum. Before doing that, let us include the fermionic part of the action.

\subsection{Full superstring action}
As we mentioned, the inclusion of the fermions in the previous construction is not completely straightforward due to the non-trivial interaction between the bosonic and fermionic fields. Here we simply state some intermediate results of the involved derivation (for details see~\cite{Frolov:2006cc,Arutyunov:2009ga}). We start from the currents derived in \eqref{eq:ulcAeven} and \eqref{eq:ulcAodd} and we conveniently fix $a=\frac12$. In order to extract the conjugated momenta, it is useful to introduce a Lie-algebra valued
auxiliary field $\bp$, and rewrite the superstring action
(\ref{eq:action}) in the form
\begin{eqnarray}
S =\int d\t d\s \left[ -\Str \left( \bp \, A_0^{(2)}\, + \, \kappa \frac{T}{
2}\epsilon^{\a\b}A_\a^{(1)} A_\b^{(3)}\right)\, -\,  \frac{\gamma^{01}}{\gamma^{00}} \, C_1
 + \, \frac{1}{ 2 T \gamma^{00}} C_2\right]
\,, \label{Lang}
\end{eqnarray}
where the Virasoro contraints in this case are
\begin{align}
 \label{C1}
C_1 &= \Str\,\bp \, A_1^{(2)}=0\,,\\
\label{C2} C_2 &= \Str\left( \bp^2 + g^2 (A_1^{(2)})^2 \right) =0\,.
\end{align}
In this way, one can easily express the Lagrangian in first-order formalism and impose the condition \eqref{ulc}. Afterwards, one has to solve the Virasoro constraint and replace the solutions in the Lagrangian. Here we omit the full derivation and we only quote the final result in first-order formalism as
\begin{equation} 
S=\int d\t d\s \mathcal{L}_{gf}\, , \qquad  \qquad \label{Lgf} \mathcal{L}_{gf} = \mathcal{L}_{kin} - \mathcal{H} \,.
\end{equation}
The kinetic term $\mathcal{L}_{kin}$ depends on the time derivatives of the
physical fields, and determines the Poisson structure of the
theory. It can be cast in the form
\begin{eqnarray} \label{Lkin} \mathcal{L}_{kin} &=&
p_{\hat i}\dot{x}^{\hat i} -
\frac{i}{2}\Str\left(\Sigma_+\chi\pa_\tau\chi\right)
+\frac{1}{2} g^{\hat j}\bp^{\hat i}\, \Str\left(\left[\S_{\hat j},\Sigma_{\hat i}\right] B_\tau\right)\\
\nonumber &-& i\kappa\frac{g}{2}(G_+^2-G_-^2)\, \Str\left( F_\tau
\mathcal{K} F_\s^{st} \mathcal{K}\right) +i\kappa\frac{g}{2} G^{\hat i} G^{\hat j}\,
\Str\left(\S_{\hat j} F_\tau\S_{\hat i} \mathcal{K} F_\s^{st} \mathcal{K}\right) \,,~~~~~~~
\end{eqnarray} where  we  use the following decompositions
 \begin{align}\label{g}\nonumber g(x)
= g_+ \mathbb{1}_8 + g_-\Upsilon + g^{\hat i} \Sigma_{\hat i}\, &,\qquad
 g(x)^2 = G_+\mathbb{1}_8 +G_-\Upsilon +G^{\hat i}\S_{\hat i} \, , \\
 \bp = \frac{i}{ 2}\bp_+ \Sigma_+ +&
 \frac{i}{4}\bp_- \Sigma_- + \frac{1}{2}\bp^{\hat{i}} \Sigma_{\hat i}
 + \bp_{\mathbb{1}} i\mathbb{1}_8 \, ,
\end{align} with the $8\times 8$ matrices $\Upsilon$ and $\S_{\hat i}=(\S_i,\tilde \S_i)$ defined by
\begin{equation}
 \Upsilon=\left(\begin{array}{cc} \mathbb{1}_4 & 0\\
                 0 & -\mathbb{1}_4
                \end{array}\right)\, ,\qquad \S_i=\left(\begin{array}{cc} \g_i & 0\\
                 0 & 0
                \end{array}\right)\, , \qquad \tilde\S_i=\left(\begin{array}{cc} 0 & 0\\
                 0 & i\g_i
                \end{array}\right)\, ,
\end{equation}
and $\S_\pm$ given in \eqref{eq:sigmapm}.
The functions $B_\alpha$ and $F_\alpha$ refer to the even
and odd components of $g^{-1}(\chi)\,\partial_\alpha  g(\chi)$
 \begin{align}\label{BF}
g^{-1}(\chi)\pa_\a g(\chi) &= B_\a +
F_\a\,,\\\nonumber B_\a &= -\frac12\chi\pa_\a\chi +
\frac12\pa_\a\chi\chi + \frac12\sqrt{1+\chi^2}\pa_\a\sqrt{1+\chi^2} - \frac12 \pa_\a\sqrt{1+\chi^2}\sqrt{1+\chi^2}\, ,\\\nonumber F_\a
&=
  \sqrt{1+\chi^2}\pa_\a\chi - \chi\pa_\a\sqrt{1+\chi^2}\, .
\end{align}
The Hamiltonian density $\mathcal{H}$ reads
\begin{align} 
 \mathcal{H}&=-\frac{i}{2}\, \Str
\left( \bp \Sigma_+ g(x)(1+2\chi^2)g(x) \right)-\kappa\frac{T}{2}(G_+^2-G_-^2)\,
\Str\left(\S_+ \chi\sqrt{1+\chi^2}\mathcal{K} F_\s^{st} \mathcal{K}\right)
 \nonumber\\
&- \kappa\frac{T}{2}G_{\hat i} G_{\hat j}\,
\Str\left(\S_+\S_{\hat j}\chi\sqrt{1+\chi^2} \S_{\hat i} \mathcal{K} F_\s^{st}
\mathcal{K}\right) \, .
\end{align}
This form of the action is still very implicit and not particularly suitable for explicit computations.
We now move to the perturbative quantization of this action around the BMN vacuum.

\subsection{Near-BMN action}
The BMN limit is defined by 
\begin{equation}
T\to\infty\, ,\qquad P_+ \to \infty\, , \qquad T/P_+\  \text{fixed}.
\end{equation}
The near-BMN expansion is then obtained considering subleading corrections in the large $T$ limit. Detailed expressions of the expanded action in first-order formalism are given in~\cite{Frolov:2006cc,Arutyunov:2009ga}. Here we only mention that, in order to obtain a canonical kinetic term, one has to perform a non-linear field redefinition of the fermionic fields $\chi \mapsto \chi + \Phi(p,x,\chi)$. In~\cite{Klose:2006zd} the action was converted to second-order formalism and, after rescaling $X\to\sqrt{P_+}\,X$, $\chi\to\sqrt{\frac{P_+}{2}}\,\chi$,
$\sigma\to \frac1{2T} \, \sigma$ and fixing $\kappa=1$, the action up to quartic order in the fields can be expressed as
\[ \label{eq:action-quadratic-matrix}
 S = T
           \int\limits_{-\infty}^{\infty} \!d\tau
           \int\limits_{-\frac{P_+}{4T}}^{\frac{P_+}{4T}} d\sigma\: \left(\mathcal{L}_2+\mathcal{L}_4+...\right)\, ,
\]
with
\[ \label{eqn:lagr-quadratic-matrix}
\begin{split}
\mathcal{L}_2 &=   \Str \left(
   \frac{1}{4} \dot{X} \dot{X}
 - \frac{1}{4} \acute{X} \acute{X}
 - \frac{1}{4}   {X}     {X}
 - \frac{i}{2} \Sigma_+ \chi \dot{\chi}
 - \frac{1}{2} \Sigma_+ \chi \acute{\chi}^\natural
 - \frac{1}{2} \chi \chi
 \right) \; ,
\\[3mm]
\mathcal{L}_4 &=  
  \frac{1}{8} \Str \Sigma_+\Sigma_- X X \Str \acute{X} \acute{X}
\\ &
 + \frac{1}{8}  \Str \chi \acute{\chi} \chi \acute{\chi}
 + \frac{1}{8}  \Str \chi \chi \acute{\chi} \acute{\chi}
 + \frac{1}{16} \Str \comm{\chi}{\acute{\chi}} \comm{\chi^\natural}{\acute{\chi}^\natural}
 + \frac{1}{4}  \Str \chi \acute{\chi}^\natural \chi \acute{\chi}^\natural
\\ &
 - \frac{1}{8} \Str \Sigma_+\Sigma_- X X \Str \acute{\chi}\acute{\chi}
 + \frac{1}{4} \Str \comm{X}{\acute{X}} \comm{\chi}{\acute{\chi}}
 + \Str X \acute{\chi} X \acute{\chi}
\\ &
 + \frac{i}{8} \Str \comm{X}{\dot{X}} \comm{\chi^\natural}{\acute{\chi}}
 - \frac{i}{8} \Str \comm{X}{\dot{X}} \comm{\chi}{\acute{\chi}^\natural}
 \; ,
\end{split}
\]
where the matrices $X$ and $\chi$ are given in \eqref{eq:X} and \eqref{eq:chi}, $\S_\pm$ appeared already in \eqref{eq:sigmapm} and the charge conjugation $()^\natural$ is defined in terms of the matrices $\mathcal{K}$ and $\tilde{\mathcal{K}}$ introduced in section \ref{sec:stringAdS5} as
\begin{equation}
 X^\natural\equiv\mathcal{K} X^t \mathcal{K}\, , \qquad  \chi^\natural\equiv\tilde{ \mathcal{K}} \chi^t \mathcal{K}\, .
\end{equation}
Notice that the action depends on the string tension only through an overall factor. The fixed ratio $\frac{T}{P_+}$ appears in the integration bounds of $\s$, which is a compactified worldsheet direction of circumference $\frac{P_+}{2T}$. This clearly constitutes an obstacle for the definition of worldsheet asymptotic states. In the next section we describe in detail how to relax this restriction.

\subsection{Decompactification limit and level-matching condition}\label{sec:decvir}
The boundaries of the integral over $\s$ in equation \eqref{eq:action-quadratic-matrix} are $-\frac{P_+}{4T}$ and $\frac{P_+}{4T}$. Since we are studying a closed string, the fields $X$ and $\chi$ are periodic in the coordinate $\s$ and this does not allow for a straightforward definition of the asymptotic states. Nevertheless, the whole dependence on $P_+$ is contained in the integration bounds and we can circumvent this difficulty taking the limit $P_+\to \infty$. In this limit we are left with an interacting theory on the plane, whose asymptotic states are eight massive bosons and eight massive fermions. Their interaction appears not to be Lorentz invariant, but their S-matrix is well defined and one can compute it perturbatively. 

Actually, it turns out one can do better than that. Using the symmetries of the theory in the BMN vacuum ($SU(2|2)^2$) one can fix the dispersion relation of the excitations and the two-body S-matrix for any value of the string tension up to an overall phase~\cite{Beisert:2005tm}. Moreover, the theory is believed to be integrable at the quantum level, which would imply that higher point S-matrices are fully determined by the $2\to2$ one. We will discuss thoroughly these issues in section \ref{sec:Smat}.

On a decompactified worldsheet we can also give up the level-matching condition and, according to \eqref{LM2}, this allows us to consider particles with arbitrary worldsheet momenta, i.e.
unphysical configurations that do not correspond to closed
strings. As a result the world-sheet
S-matrix, as well as other quantities, acquires a mild gauge dependence.

\subsection{Quadratic action and quantization}\label{sec:quant}
In the decompactification limit it is straightforward to quantize the theory perturbatively for large values of the string tension. The quadratic Lagrangian in terms of the elementary excitations reads
\[\label{free-Lag-uniform}
\begin{split}
\mathcal{L}_2 =&   \tfrac{1}{2} \dot{Y}_{a\dot{a}} \dot{Y}^{a\dot{a}}
- \tfrac{1}{2} \acute{Y}_{a\dot{a}} \acute{Y}^{a\dot{a}} -
\tfrac{1}{2}     {Y}_{a\dot{a}}     {Y}^{a\dot{a}}\\
+& \tfrac{1}{2} \dot{Z}_{\a\dot {\a}} \dot{Z}^{\a\dot {\a}} -
\tfrac{1}{2} \acute{Z}_{\a\dot {\a}} \acute{Z}^{\a\dot {\a}} - \tfrac{1}{2}
{Z}_{\a\dot {\a}}     {Z}^{\a\dot {\a}}
\\ 
+&
i\,\eta_{\a\dot {a}}^\dagger\dot{\eta}^{\a\dot {a}} + \tfrac12 \left(\eta^{\dagger\a\dot {a}}
\acute{\eta}_{\a\dot {a}}^\dagger-\eta^{\a\dot {a}}\acute{\eta}_{\a\dot {a}} \right) -
\eta_{\a\dot {a}}^\dagger\eta^{\a\dot {a}}  \\
+&
i\,\theta_{a\dot{\a}}^\dagger\dot{\theta}^{a\dot{\a}}  +
\tfrac12\left(\theta^{\dagger a\dot{\a}} \acute{\theta}_{a\dot{\a}}^\dagger-\theta^{a\dot{\a}}\acute{\theta}_{a\dot{\a}} \right)- \theta_{a\dot{\a}}^\dagger\theta^{a\dot{\a}}\, ,
\end{split}
\]
where we lower and raise the indices by using the
$\e$-tensor \begin{equation}
 Y_{a\dot{a}}
=\e_{ab}\e_{\dot{a}\dot{b}}Y^{b\dot{b}}\,,\quad \eta_{\a\dot{a}}
=\e_{\a\b}\e_{\dot{a}\dot{b}}\eta^{\b\dot{b}}\,,\quad \eta^{\dagger\a\dot{a}}
=\e^{\a\b}\e^{\dot{a}\dot{b}}\eta^\dagger_{\b\dot{b}}\,, 
            \end{equation}
 and similar
formulae for $Z_{\a\dot{\a}}\,,\  \theta_{a\dot{\a}}\,,\
\theta^{\dagger a\dot{\a}} $.            
The expression \eqref{free-Lag-uniform} is clearly a free relativistic action describing eight bosons and eight fermions of mass $1$. The corresponding free equations of motion can be solved by the mode decomposition 
 \begin{align}
Y_{a\dot{a}}(\t,\s)      & = \int\frac{dp}{2\pi}
\frac{1}{\sqrt{2e}} \:
                           \left( a_{a\dot{a}}(p)      \, e^{-i(e \t +p \s)}
                                  + a^\dagger_{a\dot{a}}(p) \, e^{+i(e \t +p \s)} \right) \; , \\
Z_{\a\dot{\a}}(\t,\s)      & = \int\frac{dp}{2\pi}
\frac{1}{\sqrt{2e}} \:
                           \left( a_{\a\dot{\a}}(p)      \, e^{-i(e \t + p \s)}
                                  + a^\dagger_{\a\dot{\a}}(p) \, e^{+i(e \t +p \s)} \right) \; , \\
\theta_{a\dot{\a}}(\t,\s)   & = e^{-i \frac{\p}{4}}\int\frac{dp}{2\pi}
\frac{1}{\sqrt{e}} \:
                           \left( b_{a\dot{\a}}(p)      \, u(p) \, e^{-i(e \t +p \s)}
                                  + b^\dagger_{a\dot{\a}}(p) \, v(p) \, e^{+i(e \t +p \s)} \right) \; , \\
\eta_{\a\dot{a}}(\t,\s)   & = e^{-i \frac{\p}{4}}\int\frac{dp}{2\pi}
\frac{1}{\sqrt{e}} \:
                           \left( b_{\a\dot{a}}(p)      \, u(p) \, e^{-i(e \t +p \s)}
                                  + b^\dagger_{\a\dot{a}}(p) \, v(p) \, e^{+i(e \t +p \s)} \right) \; ,
\end{align}
where the energy is $e = \sqrt{1 + p^2}$, the wave functions
are
\[
  u(p) = \cosh{\tfrac{\theta}{2}}\, ,
  \qquad
  v(p) = \sinh{\tfrac{\theta}{2}}\, ,
\]
and the rapidity $\theta$ is defined through $p = \sinh\theta$. The creation and annihilation operators satisfy the canonical commutation relations
\begin{align}
 & \comm{ a^{a\dot{a}}(p)}{a^\dagger_{b\dot{b}}(p')} = 2\pi \, \delta^{a}_{b} \delta^{\dot{a}}_{\dot{b}} \, \delta(p-p') \; , &
 & \acomm{b^{a\dot{\a}}(p)}{b^\dagger_{b\dot{\b}}(p')} = 2\pi \, \delta^{a}_{b} \delta^{\dot{\a}}_{\dot{\b}} \, \delta(p-p') \; , \nonumber \\
 & \comm{ a^{\a\dot{\a}}(p)}{a^\dagger_{\b\dot{\b}}(p')} = 2\pi \, \delta^{\a}_{\b} \delta^{\dot{\a}}_{\dot{\b}} \, \delta(p-p') \; , &
 & \acomm{b^{\a\dot{a}}(p)}{b^\dagger_{\b\dot{b}}(p')} = 2\pi \, \delta^{\a}_{\b} \delta^{\dot{a}}_{\dot{b}} \, \delta(p-p') \; .
\end{align}
Let us stress that this decomposition allows to consider particles and anti-particles at once, without any notational difference. If one considers, for instance, the field $Y_{a\dot{a}}$ clearly the oscillator $a^\dag_{a\dot a}$ creates the
``anti-particle'' of the ``particle'' that is destroyed by the
oscillator $a_{a\dot a}$. These two oscillators appear in the decomposition of the same field $Y_{a\dot{a}}$, but clearly they do not form a canonical pair. Rather $a^\dagger_{a\dot{a}}$ and $a^{a\dot{a}}=\e^{ab}\e^{\dot{a}\dot{b}}a_{b\dot{b}}$ are conjugated to each other as one can see from the commutation relations. 

Interpreting the higher order corrections in the Lagrangian as perturbations of this free action for large string tension, it is a straightforward exercise to compute the scattering process at tree-level. Before showing the explicit expressions, let us recall some generalities about S-matrices of two-dimensional systems.

\section{Worldsheet scattering in \txpf{$AdS_5\times S^5$}{AdS5 x S5}}\label{sec:Smat}
As usual in scattering theory, the two-particle asymptotic states are simply the tensor product of two one-particle states with different momenta $p$ and $p'$. In general, the S-matrix of a $2\to n$ process can be seen as an operator from an asymptotic two-particle state with arbitrary momenta (rapidities) to an asymptotic $n$-particle state with arbitrary momenta (rapidities). Nevertheless, in an integrable theory, the S-matrix satisfies a number of
additional kinematic constraints~\cite{Zamolodchikov:1978xm} (see also~\cite{Dorey:1996gd} for a review), as a consequence of the infinite number of conserved charges:
\begin{itemize}
 \item there is no-particle production, i.e. the number of ingoing particles is equal to the number of outgoing particles;
 \item the set of outgoing momenta is equal to the set of ingoing momenta;
 \item the many-body S-matrix factorizes into the
products of two-body S-matrices.
\end{itemize}
This final requirement implies the Yang-Baxter equation, a
consistency condition for equivalent orderings of scattering of three-particle states, which can be represented diagrammatically as follows:
\begin{equation}
\begin{tikzpicture}[line width = 2pt,scale=0.6,rotate=90,baseline=5.5em,yscale=-1]
\draw[->] (0,{16*(-5+4*sqrt(3))/23}) -- (8,{24*(12-5*sqrt(3))/23});
\node at (-0.4,{16*(-5+4*sqrt(3))/23-0.2}) {{\large $\vartheta_2$}};
\draw[->] ({16*(-5+4*sqrt(3))/23},0) -- ({24*(12-5*sqrt(3))/23},8);
\node at ({16*(-5+4*sqrt(3))/23},-0.45) {{\large $\vartheta_1$}};
\draw[->] ({8*(29-14*sqrt(3))/23},{24*(2+3*sqrt(3))/23}) -- ({24*(2+3*sqrt(3))/23},{8*(29-14*sqrt(3))/23});
\node at ({8*(29-14*sqrt(3))/23-0.3},{24*(2+3*sqrt(3))/23+0.3}) {{\large $\vartheta_3$}};
\node[circle,draw=black,fill=black!02,line width = 2pt,opacity=.90] at ({8*(7-sqrt(3))/23},{8*(7-sqrt(3))/23}) {{\Large $\theta_{12}$}};
\node[circle,draw=black,fill=black!10,line width = 2pt,opacity=.90] at ({8*(19-6*sqrt(3))/23},{8*(16+sqrt(3))/23}) {{\Large $\theta_{13}$}};
\node[circle,draw=black,fill=black!18,line width = 2pt,opacity=.90] at ({8*(16+sqrt(3))/23},{8*(19-6*sqrt(3))/23}) {{\Large $\theta_{23}$}};
\end{tikzpicture}
\qquad
\mathbf{=}
\qquad
\begin{tikzpicture}[line width = 2pt,scale=0.6,rotate=90,baseline=21.5em,yscale=-1]
\draw[->] (11,{8-24*(12-5*sqrt(3))/23}) -- (11+8,{8-16*(-5+4*sqrt(3))/23});
\node at (11-0.4,{8-24*(12-5*sqrt(3))/23-0.2}) {{\large $\vartheta_2$}}; 
\draw[->] ({11+8-24*(12-5*sqrt(3))/23},0) -- ({11+8-16*(-5+4*sqrt(3))/23},8);
\node at ({11+8-24*(12-5*sqrt(3))/23},-0.45) {{\large $\vartheta_1$}};
\draw[->] ({11+8-24*(2+3*sqrt(3))/23},{8-8*(29-14*sqrt(3))/23}) -- ({11+8-8*(29-14*sqrt(3))/23},{8-24*(2+3*sqrt(3))/23});
\node at ({11+8-24*(2+3*sqrt(3))/23-0.3},{8-8*(29-14*sqrt(3))/23+0.3}) {{\large $\vartheta_3$}}; 
\node[circle,draw=black,fill=black!02,line width = 2pt,opacity=.90] at ({11+8-8*(7-sqrt(3))/23},{8-8*(7-sqrt(3))/23}) {{\Large $\theta_{12}$}};
\node[circle,draw=black,fill=black!10,line width = 2pt,opacity=.90] at ({11+8-8*(19-6*sqrt(3))/23},{8-8*(16+sqrt(3))/23}) {{\Large $\theta_{13}$}};
\node[circle,draw=black,fill=black!18,line width = 2pt,opacity=.90] at ({11+8-8*(16+sqrt(3))/23},{8-8*(19-6*sqrt(3))/23}) {{\Large $\theta_{23}$}};
\end{tikzpicture}\label{ybe}
\end{equation}
where the grey blobs stand for $2\to2$ S-matrices. Let us stress that these constraints are applicable to a theory whose integrability is preserved at the quantum level (the bosonic $\mathbb{CP}^n$ models~\cite{D'Adda:1978kp,Witten:1978bc} constitute a well known example of classical integrability broken by quantum anomalies~\cite{Abdalla:1980jt}). Although nowadays we have many different and strong indications for the quantum integrability of the string sigma model in $AdS_5\times S^5$, we do not have an explicit proof of that. Furthermore, in general, the preservation of quantum integrability requires a careful choice of regularization or addition of local counterterms. Indeed the quantization of a classical theory is not unique and one is to impose Ward identities or use a particular regularization to preserve a classical symmetry at the quantum level (see for instance~\cite{deVega:1981ka,deVega:1982sh,Hoare:2010fb}). One of our aims in the following discussion will be to provide non-trivial quantum checks of the expected quantum integrability of the model and suggest a possible regularization scheme preserving the classical symmetries.

The properties listed above clearly indicate that, for an integrable theory, the two-particle S-matrix is the fundamental building block for the construction of many body S-matrices. From now on we focus on the two-particle S-matrix in two dimensions. It is interesting to note that for a theory with a single mass scale (as is the case for the light-cone gauge fixed string theory in $AdS_5\times S^5$) the scattering of two relativistic excitations of momenta $p$ and $p'$ automatically yields two excitations with the same momenta. One can easily see that introducing light-cone 2d coordinates $\xi^{\pm}=\t\pm\s$ and implementing the mass-shell condition for a relativistic particle
\begin{equation}
 p_+ p_-=m^2\,,
\end{equation}
the latter can be parametrized in terms of rapidities as $p_\pm=m\, e^{\pm \theta}$, and momentum conservation would read
\begin{equation}
\left\{\begin{array}{ll}
        e^{\theta_1}+e^{\theta_2}=e^{\theta_3}+e^{\theta_4}\\
        e^{-\theta_1}+e^{-\theta_2}=e^{-\theta_3}+e^{-\theta_4}\, .\\
       \end{array}\right.
\end{equation}
It is easy to see that this system admits a discrete set of solutions
\begin{equation}
\left\{\begin{array}{ll}
         \theta_1=\theta_3\\
        \theta_2=\theta_4
       \end{array}\right. \qquad \text{or} \qquad
       \left\{\begin{array}{ll}
         \theta_1=\theta_4\\
        \theta_2=\theta_3\, .
       \end{array}\right.
\end{equation}
As we mentioned, in general, the light-cone gauge fixed string theory in $AdS_5 \times S^5$ is not invariant under worldsheet Lorentz transformations. As a consequence, the dispersion relation of the fundamental excitations is non-relativistic. However, at quadratic order in the near-BMN expansion (i.e.
for the free states in perturbation theory) it is relativistic. As the symmetry breaking terms appear at quartic order, the first non-relativistic correction to the dispersion relation would appear in the two-loop two-point functions and are irrelevant for the tree-level calculation. We can then conclude that the simple kinematical constraint we just derived is enough to fix the kinematics of the tree-level S-matrix. Let us stress that for higher loop corrections and for theories with different masses this argument does not apply, and we have to rely either on integrability or on an explicit computation to fix the set of outgoing momenta equal to the set of ingoing momenta. In general we will use $e$ to denote the relativistic energies of the free states, and
$\o$ to denote their all-order form.

In this setup, we can interpret the S-matrix as an operator mapping a two-particle state with momenta $p$ and $p'$ to a different two-particle state with the same momenta $p$ and $p'$
 \[\label{smatrix_components}
 \mathbb{S} \, \ket{\Phi_{A\dot{A}}(p) \Phi_{B\dot{B}}(p')}
 = \ket{\Phi_{C\dot{C}}(p) \Phi_{D\dot{D}}(p')} \,
S_{A\dot{A}B\dot{B}}^{C\dot{C}D\dot{D}}(p,p')\, ,
\]
with the index $A$ taking values $(a|\a)$ and similarly for $\dot{A}$. The field $\Phi$ corresponds to $Y,Z,\theta,\eta$ according to the values of its indices. Modulo anomalies, the S-matrix should enjoy the symmetries that are explicit in the Lagrangian. In this case the symmetry is a centrally extended PSU$(2|2)\times$ PSU$(2|2)$~\cite{Arutyunov:2006ak}, the same appearing in the dual gauge theory~\cite{Beisert:2005tm}. The invariance of the S-matrix under a non-simple group, such as PSU$(2|2)\times$ PSU$(2|2)$, with the constraints coming from the YBE \eqref{ybe}, lead to the group-factorization~\footnote{This can also be interpreted as  the requirement that the Faddeev-Zamolodchikov algebra, used in describing the Hilbert space of the asymptotic states, is a direct product~\cite{Arutyunov:2006yd,Klose:2006zd}.}
\begin{equation}\label{s2smatfact_string}
S_{A\dot A,B\dot B}^{C \dot C, D \dot D}(p,p') = (-1)^{[\dot A][B]+[\dot C][D]} S_{AB\vphantom{\dot B}}^{CD \vphantom{\dot D}}(p,p') S_{\dot A \dot B}^{\dot C \dot D}(p,p')~,
\end{equation}   
which has indeed been verified at tree level~\cite{Klose:2006zd}.
Since only $SU(2)\times SU(2)\subset PSU(2|2)$
is a manifest symmetry of the gauge-fixed worldsheet theory,
$S$ may be parametrized in terms of ten unknown functions
of the momenta $p$ and $p'$ of the two incoming particles:%
\begin{align}
S_{ab}^{cd} & = A \,\delta_a^c \delta_b^d + B \,\delta_a^d \delta_b^c &&
	    \begin{tikzpicture}
                 \draw [-] (0.2,0.2)--(0.2,-0.2);
                 \draw [-] (-0.2,-0.2)--(-0.2,0.2);
                 \draw [-] (1.2,-0.2)--(0.8,0.2);
                 \draw [-] (1.2,0.2)--(0.8,-0.2);
            \end{tikzpicture} 
 &
S_{ab}^{\g\d} & = C \,\e_{ab} \e^{\g\d} &&
\begin{tikzpicture}
                 \draw [-,dashed] (-0.2,0.2)--(0,0);
                 \draw [-] (0,0)--(0.2,-0.2);
                 \draw [-] (-0.2,-0.2)--(0,0);
		 \draw [-,dashed] (0,0)--(0.2,0.2);
                 \end{tikzpicture} \label{eq:comps1}  \\
S_{\a\b}^{\g\d} & = D \,\delta_\a^\g \delta_\b^\d + E \,\delta_\a^\d \delta_\b^\g &&
\begin{tikzpicture}
                 \draw [-,dashed] (0.2,0.2)--(0.2,-0.2);
                 \draw [-,dashed] (-0.2,0.2)--(-0.2,-0.2);
                 \draw [-,dashed] (1.2,-0.2)--(0.8,0.2);
                 \draw [-,dashed] (0.8,-0.2)--(1.2,0.2);
            \end{tikzpicture} 
 &
S_{\a\b}^{cd} & = F \,\e_{\a\b} \e^{cd} &&
\begin{tikzpicture}
                 \draw [-] (-0.2,0.2)--(0,0);
                 \draw [-,dashed] (0,0)--(0.2,-0.2);
                 \draw [-,dashed] (-0.2,-0.2)--(0,0);
		 \draw [-] (0,0)--(0.2,0.2);
                 \end{tikzpicture} \label{eq:comps2}  \\
S_{a\b}^{c\d} & = G \,\delta_a^c \delta_\b^\d &&
\begin{tikzpicture}
                 \draw [-,dashed] (0.2,0.2)--(0.2,-0.2);
                 \draw [-] (-0.2,-0.2)--(-0.2,0.2);
            \end{tikzpicture}  &
S_{\a b}^{\g d} & = L \,\delta_\a^\g \delta_b^d &&
\begin{tikzpicture}
                 \draw [-] (0.2,0.2)--(0.2,-0.2);
                 \draw [-,dashed] (-0.2,-0.2)--(-0.2,0.2);
            \end{tikzpicture} \label{eq:comps3}
 \\
S_{a\b}^{\g d} & = H \,\delta_a^d \delta_\b^\g &&
\begin{tikzpicture}
                 \draw [-,dashed] (-0.2,0.2)--(0.2,-0.2);
                 \draw [-] (-0.2,-0.2)--(0.2,0.2);
            \end{tikzpicture} 
 &
S_{\a b}^{c\d} & = K\,\delta_\a^\d \delta_b^c &&
\begin{tikzpicture}
                 \draw [-] (-0.2,0.2)--(0.2,-0.2);
                 \draw [-,dashed] (-0.2,-0.2)--(0.2,0.2);
            \end{tikzpicture} \label{eq:comps4} 
 \ \ . 
\end{align}

\subsection{Tree-level S-matrix}
The S-matrix can be expanded perturbatively in powers of the inverse string tension
\begin{equation}
\mathbb{S}=\mathbb{1}+i \co \mathbb{T}^{(0)}+ i \co^2 \mathbb{T}^{(1)}+\mathcal{O}(\co^3) \, ,
\end{equation}
with 
\begin{equation}
\co^{-1} \equiv T = \frac{\sqrt{\lambda}}{2\pi} \ .
\end{equation} 
This kind of expansion can be performed either for the total S-matrix in the l.h.s. of \eqref{s2smatfact_string} or for the two factors in the r.h.s. of \eqref{s2smatfact_string}. 
The relation between the two expansions at tree-level reads
\begin{equation}
 T^{(0)}{}_{A\dot{A}B\dot{B}}^{C\dot{C}D\dot{D}}(p,p')=T^{(0)}{}_{AB}^{CD}(p,p')\d_{\dot{A}}^{\dot{C}}\d_{\dot{B}}^{\dot{D}}+\d_{A}^{C}\d_{B}^{D} T^{(0)}{}_{\dot{A}\dot{B}}^{\dot{C}\dot{D}}(p,p')\, .
\end{equation}
This property has been checked to hold at tree-level in~\cite{Klose:2006zd}, where explicit expressions for the leading order expansion of the S-matrix were given. They can be computed in a straightforward way starting from the action \eqref{eqn:lagr-quadratic-matrix} expanded in terms of the physical degrees of freedom, as we did for the quadratic action in \eqref{free-Lag-uniform}. As we anticipated, the tree-level S-matrix shows a mild dependence on the gauge parameter $a$. It turns out that this dependence has the following exact form
\begin{equation}\label{gaugeAdS5}
\exp\big[\frac{i}{2}(a-\tfrac12)(\o'p-\o p')\big] \ ,
\end{equation}
and in order not to clutter the equations we display results only for $a=\frac12$.
The tree-level S-matrix reads
\begin{align}
\!\!A^{(0)} & =   \frac{1}4 \frac{(p-p')^2}{e' p - e p'} \ , \nonumber
& \nonumber
\!\!C^{(0)} & =   \frac{1}2 \sqrt{(e+1)(e'+1)} \frac{e' p - p' e - p + p'}{e' p - e p'} \ , \nonumber
\\ \nonumber
\!\!D^{(0)} & =  \frac{1}4 \frac{(p-p')^2}{e' p - e p'} \ , 
&
\!\!F^{(0)} & =   \frac{1}2 \sqrt{(e+1)(e'+1)} \frac{e' p - p' e - p + p'}{e' p - e p'} \ , \nonumber
 \nonumber
\\
\!\!B^{(0)} & =   \frac{p p'}{e' p - e p'} \ ,\nonumber
&
\!\!H^{(0)} & =   \frac{1}2 \frac{p p'}{e' p - e p'} \frac{(e+1)(e'+1)- pp'}{\sqrt{(e+1)(e'+1)}} \ ,
\\
\!\!E^{(0)} & = - \frac{p p'}{e' p - e p'} \ , \nonumber
&
\!\!K^{(0)} & =   \frac{1}2 \frac{p p'}{e' p - e p'} \frac{(e+1)(e'+1)- pp'}{\sqrt{(e+1)(e'+1)}} \ , \nonumber
\\
\!\!G^{(0)} & =  - \frac{1}4 \frac{p^2 - p'^2}{e' p - e p'} \ , 
&
\!\!L^{(0)} & =   \frac{1}4 \frac{p^2 - p'^2}{e' p - e p'} \ .
\end{align}
where $e$ and $e'$ are the relativistic energies $e=\sqrt{1+p^2}$. As one can see from equations \eqref{eq:comps1}--\eqref{eq:comps4}, the components $A$, $D$, $G$ and $L$ correspond to the contributions proportional to the identity, and for $a\neq \frac12$ they would contain the $a$ dependence from the phase \eqref{gaugeAdS5}. Notice that the tree-level S-matrix is not Lorentz invariant, as we could expect due to the lack of Lorentz symmetry in the quartic action \eqref{eqn:lagr-quadratic-matrix}. 

\subsection{One-loop S-matrix}

The computation of the one-loop correction is definitely more involved because of the complicated structure of the interactions. Indeed, before~\cite{Bianchi:2013nra,Engelund:2013fja}, the perturbative S-matrix was known beyond the leading order~\cite{Klose:2007wq,Klose:2007rz} only in the   kinematic truncation 
known as near-flat-space limit~\cite{Maldacena:2006rv}. In~\cite{Bianchi:2013nra,Engelund:2013fja} the logarithmic part of the one-loop result was computed using the so-called unitarity techniques and in~\cite{Bianchi:2013nra} a prescription was given to fix the remaining rational terms. The latter turned out to be successful for a number of integrable models and, for the light-cone gauge fixed string in $AdS_5\times S^5$, it yields a result which agrees with the prediction from integrability~\cite{Beisert:2005tm}. The same result was then re-derived in~\cite{Roiban:2014cia} using standard Feynman diagrams techniques, although with a fairly unusual regularization, which allows to perform the computation in strictly two dimensions. 

Section \eqref{sec:unitarity} contains a very detailed description for the construction of the one-loop S-matrix using unitarity. For clarity, here we simply report the final result and we anticipate some important observations. The result can be written as follows 
\begin{equation}\begin{split}\label{Sstrings_cut}
S_{AB}^{CD}(p,p') & = e^{ i\co^2 \varphi (p,p')} \ \tilde S_{AB}^{CD}(p,p') + \mathcal{O}(\co^3) \ , 
\end{split}\end{equation}
where we have pulled out a factor that to the one-loop order can be resummed as an overall phase. Expanding for large string tension we get
\begin{equation}\begin{split}
S_{AB}^{CD}(p,p') & =\d_A^C \d_B^D+ i \co T^{(0)}{}_{AB}^{CD}(p,p')+  i\co^2 \left(\varphi (p,p')\d_A^C \d_B^D+  \tilde T^{(1)}{}_{AB}^{CD}(p,p')\right) + \mathcal{O}(\co^3) \ .
\end{split}\end{equation}
The one-loop contribution $\tilde T^{(1)}{}_{AB}^{CD}(p,p')$ has the same structure as in \eqref{eq:comps1}--\eqref{eq:comps4}
with parametrizing functions given by
\allowdisplaybreaks{
\begin{align}\nonumber
\tilde A^{(1)} & =- \frac{i}{4} \left(pp'-\frac{(p+p')^4}{8(e' p - e p')^2}\right) \ ,
&\nonumber
\tilde B^{(1)} & =   \frac{i}4 p p' \ ,
\\\nonumber
\tilde D^{(1)} & =  - \frac{i}{4} \left(pp'-\frac{(p+p')^4}{8(e' p - e p')^2}\right) \ ,
&\nonumber
\tilde E^{(1)} & =    \frac{i}4 p p' \ ,
\\\nonumber
\tilde C^{(1)} & = 0 \ ,
&\nonumber
\tilde F^{(1)} & = 0 \ ,
\\\nonumber
\tilde H^{(1)} & = 0 \ ,
&\nonumber
\tilde K^{(1)} & = 0 \ ,
\\\nonumber
\tilde G^{(1)} & = - \frac{i}{8} \left(pp'-\frac{(p+p')^4}{4(e' p - e p')^2}\right) \ ,
&\label{functions_strings_cuts}
\tilde  L^{(1)} & =- \frac{i}{8} \left(pp'-\frac{(p+p')^4}{4(e' p - e p')^2}\right) \ ,
\end{align}}
and 
\begin{equation}\label{phase_strings_cuts}
\varphi(p,p') = \frac{1}{2\pi}\frac{p^2 p'^2 \left((e' p - e p') - (e e' - p p')\arsinh [e' p - e p'] \right)}{(e' p - e p')^2}\ . 
\end{equation}
A few comments about this result are in order. First of all, one should notice that the real part of the one-loop S-matrix is fully contained in the phase factor $\varphi(p,p')$. The matrix part is purely imaginary and can be fully reproduced by the optical theorem. We will see in section \ref{sec:unitarity} that the unitarity computation separates nicely the real and the imaginary contributions. It is also interesting to note that all the logarithmic dependence on the kinematical variables is contained in the phase factor $\varphi(p,p')$. This is an essential requirement of integrability. Indeed the matrix structure of the S-matrix is completely fixed by symmetries and it is a rational function of the Zhukovsky variables (see appendix \ref{app:exactSmat} for details). This implies that the whole logarithmic dependence must appear in the overall phase factor that cannot be fixed by symmetries. The latter usually goes under the name of dressing phase or dressing factor and its exact determination exploited 
the non-relativistic generalization of the crossing symmetry~\cite{Janik:2006dc,Volin:2009uv}
as well as perturbative data both from the string and gauge theory sides~\cite{Beisert:2006ib,Beisert:2006ez}. 
In appendix \ref{app:exactSmat} we provide the detailed expressions of the exact S-matrix and explain how to expand it to reproduce the one-loop result quoted here. We now move to the derivation of a general formula for the one-loop S-matrix in terms of the tree-level one.  

\section{Unitarity techniques}\label{sec:unitarity}

The remarkable efficiency of unitarity-based methods~\cite{Bern:1994zx,Bern:1994cg,Roib_Review,Elvang:2013cua,Henn:2014yza} for the calculation of space-time scattering amplitudes in non-abelian gauge theories motivates the application of 
similar techniques 
to perturbative regimes of other interesting models. Here we focus on two-dimensional models whose integrability has been proven at the classical level and whose tree-level S-matrix satisfies all the requirements coming from integrability (see section \ref{sec:Smat}). We first outline the general construction of~\cite{Bianchi:2013nra,Engelund:2013fja,Bianchi:2014rfa} and then apply it to the light-cone gauge fixed superstring theory in $AdS_5 \times S^5$ and in $AdS_3 \times S^3 \times M^4$.

\subsection{Theories with a single mass}\label{singlemass}
%%%%%%%%%%%%%%%%%%%%%%%%%%%%%%%%%

The object of interest is the two-particle S-matrix \eqref{smatrix_components}. The latter is related to the
four-point scattering amplitude by
\begin{equation}\label{eqn:4ptamp}
\langle\Phi^\IndC(\rmq)\Phi^\IndD(\rmq')\,|\mathbb{S}|\,\Phi_\IndA(\rmp)\Phi_\IndB(\rmp')\rangle=\mathcal{A}_{\IndA\IndB}^{\IndC\IndD}(\rmp,\rmp',\rmq,\rmq')~.
\end{equation}
Here $\IndA,\IndB,\ldots$ are indices
running over the particle content of the theory and $\rmp,\rmp',\rmq,\rmq'$ are
the on-shell two-momenta of the fields. For now we will restrict to the case where
all the particles have equal non-vanishing mass, which we set to one. As a
consequence of momentum conservation, the four-point amplitude takes the form
\begin{equation}\label{eqn:ampcons}
\mathcal{A}_{\IndA\IndB}^{\IndC\IndD}(\rmp,\rmp',\rmq,\rmq')=(2\pi)^2 \delta^{(2)}(\rmp+\rmp'-\rmq-\rmq')\, \widetilde{\mathcal{A}}_{\IndA\IndB}^{\IndC\IndD}(\rmp,\rmp',\rmq,\rmq')~.
\end{equation}
Furthermore, as we derived in section \ref{sec:Smat}, two-dimensional kinematics implies that the set of
initial momenta is preserved in the scattering process. This property is translated in the following distribution identity
\begin{equation}\label{delta2d}
\delta^{(2)}(\rmp+\rmp'-\rmq-\rmq')=\frac{\Jc(p,p')}{4\o\o'}\,\big(2\o\,\delta(p - q)\,2\o'\delta(p' - q') + 2\o\, \delta(p - q')\,2\o'\delta(p' - q) \big)~,
\end{equation}
where $p,p',q,q'$ are the spatial momenta and the Jacobian
$\Jc(p,p')=1/(\partial \o/\partial p-\partial\o'/\partial p')$ depends on the
on-shell energies $\o(p),\o'(p')$. Note that we have assumed the particle
velocities ordered as $v = \partial \o/\partial p > \partial \o'/\partial
p'= v'$, and for the spatial momentum $\delta$-functions we have used a
normalization that becomes the standard Lorentz-invariant one in the
relativistic case.

Substituting \eqref{delta2d} into \eqref{eqn:ampcons} we find two terms.
Without loss of generality we can consider just the amplitude associated to the
first product of $\delta$-functions, $2\o\, \delta(p - q)\,2\o'\delta(p' - q')$.
The two-particle S-matrix is then defined as
\begin{equation}\label{AandS}
S_{\IndA\IndB}^{\IndC\IndD}(p,p')\equiv \frac{\Jc(p,p')}{4\e\e'}\widetilde{\mathcal{A}}_{\IndA\IndB}^{\IndC\IndD}(\rmp,\rmp',\rmp,\rmp')~.
\end{equation}
We will be interested in computing the cut-constructible part of
$T^{(1)}$ from the tree-level S-matrix $T^{(0)}$.
Once more, the correction to the dispersion relation would affect the pre-factor in \eqref{AandS} starting from $\mathcal{O}(\co^2)$ corrections, and for our purposes the Jacobian in \eqref{AandS} is just given by
\begin{equation}\label{jdef}
J(p,p') = \frac{1}{4(e'p-e p')} \ , \qquad e = \sqrt{p^2 + 1} \ , \qquad e' = \sqrt{p'^2 + 1} \ .
\end{equation}

\begin{figure}[t]
\begin{center}
\begin{tikzpicture}[line width=1pt,scale=1,baseline=-184]
\draw[-] (-5,-3) -- (-4,-4);
\draw[-] (-5,-5) -- (-4,-4);
\draw[-] (-2,-4) -- (-1,-3);
\draw[-] (-2,-4) -- (-1,-5);
\draw (-3,-4) circle (1cm);
\draw[|-|,dashed,blue!70,line width=1pt] (-3,-2.5) -- (-3,-5.5);
\draw[->] (-4.7,-3.0) -- (-4.3,-3.4);
\node at (-4.7,-2.8) {$\rmp$};
\draw[->] (-4.7,-5.0) -- (-4.3,-4.6);
\node at (-4.7,-5.2) {$\rmp'$};
\draw[<-] (-1.3,-3.0) -- (-1.7,-3.4);
\node at (-1.3,-2.8) {$\rmq'$};
\draw[<-] (-1.3,-5.0) -- (-1.7,-4.6);
\node at (-1.3,-5.2) {$\rmq$};
\draw[<-] (-3.1,-3.2) -- (-3.5,-3.35);
\node at (-3.2,-3.5) {$\rml_1$};
\draw[->] (-2.9,-4.8) -- (-2.5,-4.65);
\node at (-2.8,-4.5) {$\rml_2$};
\node at (-3.4,-2.85) {$\IndE$};
\node at (-2.6,-5.15) {$\IndF$};
\node at (-5.2,-3) {$\IndA$};
\node at (-5.2,-5) {$\IndB$};
\node at (-0.8,-5) {$\IndC$};
\node at (-0.8,-3) {$\IndD$};
\node [circle,draw=black!100,fill=black!5,thick,opacity=.95] at (-4,-4) {\footnotesize$\mathcal{A}^{(0)}$\normalsize};
\node [circle,draw=black!100,fill=black!5,thick,opacity=.95] at (-2,-4) {\footnotesize$\mathcal{A}^{(0)}$\normalsize};
\end{tikzpicture}
\begin{tikzpicture}[line width=1pt,scale=1,rotate=90]
\draw[-] (-5,-3) -- (-4,-4);
\draw[-] (-5,-5) -- (-4,-4);
\draw[-] (-2,-4) -- (-1,-3);
\draw[-] (-2,-4) -- (-1,-5);
\draw (-3,-4) circle (1cm);
\draw[|-|,dashed,blue!70,line width=1pt] (-3,-2.5) -- (-3,-5.5);
\draw[->] (-4.7,-3.0) -- (-4.3,-3.4);
\node at (-4.7,-2.8) {$\rmp'$};
\draw[<-] (-4.7,-5.0) -- (-4.3,-4.6);
\node at (-4.7,-5.2) {$\rmq'$};
\draw[->] (-1.3,-3.0) -- (-1.7,-3.4);
\node at (-1.3,-2.8) {$\rmp$};
\draw[<-] (-1.3,-5.0) -- (-1.7,-4.6);
\node at (-1.3,-5.2) {$\rmq$};
\draw[<-] (-3.1,-3.2) -- (-3.5,-3.35);
\node at (-3.2,-3.5) {$\rml_1$};
\draw[<-] (-2.9,-4.8) -- (-2.5,-4.65);
\node at (-2.8,-4.5) {$\rml_2$};
\node at (-3.4,-2.85) {$\IndE$};
\node at (-2.6,-5.15) {$\IndF$};
\node at (-5.2,-3) {$\IndB$};
\node at (-5.2,-5) {$\IndD$};
\node at (-0.8,-5) {$\IndC$};
\node at (-0.8,-3) {$\IndA$};
\node [circle,draw=black!100,fill=black!5,thick,opacity=.95] at (-4,-4) {\footnotesize$\mathcal{A}^{(0)}$\normalsize};
\node [circle,draw=black!100,fill=black!5,thick,opacity=.95] at (-2,-4) {\footnotesize$\mathcal{A}^{(0)}$\normalsize};
\end{tikzpicture}
\hspace{10pt}
\begin{tikzpicture}[line width=1pt,scale=1,rotate=90]
\draw[-] (-5,-3) -- (-4,-4);
\draw[-] (-5,-5) -- (-4,-4);
\draw[-] (-2,-4) -- (-1,-3);
\draw[-] (-2,-4) -- (-1,-5);
\draw (-3,-4) circle (1cm);
\draw[|-|,dashed,blue!70,line width=1pt] (-3,-2.5) -- (-3,-5.5);
\draw[->] (-4.7,-3.0) -- (-4.3,-3.4);
\node at (-4.7,-2.8) {$\rmp'$};
\draw[<-] (-4.7,-5.0) -- (-4.3,-4.6);
\node at (-4.7,-5.2) {$\rmq$};
\draw[->] (-1.3,-3.0) -- (-1.7,-3.4);
\node at (-1.3,-2.8) {$\rmp$};
\draw[<-] (-1.3,-5.0) -- (-1.7,-4.6);
\node at (-1.3,-5.2) {$\rmq'$};
\draw[<-] (-3.1,-3.2) -- (-3.5,-3.35);
\node at (-3.2,-3.5) {$\rml_1$};
\draw[<-] (-2.9,-4.8) -- (-2.5,-4.65);
\node at (-2.8,-4.5) {$\rml_2$};
\node at (-3.4,-2.85) {$\IndE$};
\node at (-2.6,-5.15) {$\IndF$};
\node at (-5.2,-3) {$\IndB$};
\node at (-5.2,-5) {$\IndC$};
\node at (-0.8,-5) {$\IndD$};
\node at (-0.8,-3) {$\IndA$};
\node [circle,draw=black!100,fill=black!5,thick,opacity=.95] at (-4,-4) {\footnotesize$\mathcal{A}^{(0)}$\normalsize};
\node [circle,draw=black!100,fill=black!5,thick,opacity=.95] at (-2,-4) {\footnotesize$\mathcal{A}^{(0)}$\normalsize};
\end{tikzpicture}
\caption{Diagrams representing s-, t- and u-channel cuts contributing to the four-point one-loop amplitude.}
\label{stu}\nonumber
\end{center}
\end{figure}
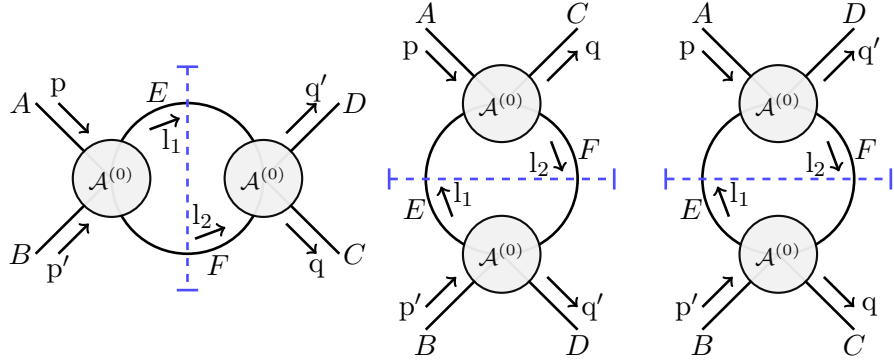

In general, there are three possible contributions (shown in figure~\ref{stu})  that can arise in a
unitarity computation. We ignore tadpoles
and graphs built from a three- and five-point amplitude. In the standard
unitarity procedure such graphs have no physical two-particle cuts and
therefore they can safely be ignored. However, in higher dimensions a recipe to
deal with tadpole diagrams in the context of generalized unitarity for massive
theories was given in~\cite{Britto:2010um}. In two dimensions the situation is
slightly different.
%v2-start
In particular, tadpole diagrams require the introduction of a regularization
since they develop a logarithmic divergence.
%v2-end
Since our procedure is inherently finite it is not clear
how tadpoles should be included, but it appears that they do not need to be to
construct the one-loop S-matrix (up to possible shifts in the coupling), as we
have explicitly checked in all the cases under consideration. 

The explicit expression for the three contributions in figure \ref{stu} is
 \begin{align}\nonumber
\mathcal{A}^{(1)}{}^{\IndC\IndD}_{\IndA\IndB}(\rmp,\rmp',\rmq,\rmq')|_{s-cut}&=\frac12\int\frac{d^2 \rml_1}{(2\pi)^2}\int\frac{d^2 \rml_2}{(2\pi)^2}\ i\pi\delta^+({\rml_1}^2-\ms)\ i\pi\delta^+(\rml_2^2-\ms)\\\label{eqn:sch1}%BHsymfac
&\times\,\mathcal{A}^{(0)}{}_{\IndA\IndB}^{\IndE\IndF}({\rmp,\rmp',\rml_1,\rml_2})\mathcal{A}^{(0)}{}_{\IndF\IndE}^{\IndC\IndD} ({\rml_2,\rml_1,\rmq,\rmq'})\, ,
\\\nonumber
\mathcal{A}^{(1)}{}^{\IndC\IndD}_{\IndA\IndB}(\rmp,\rmp',\rmq,\rmq')|_{t-cut}&=\frac12\int\frac{d^2 \rml_1}{(2\pi)^2}\int\frac{d^2 \rml_2}{(2\pi)^2}\ i\pi\delta^+({\rml_1}^2-\ms)\ i\pi\delta^+({\rml_2}^2-\ms)\\\label{eqn:tch1}%BHsymfac
&\times\,\mathcal{A}^{(0)}{}_{\IndA\IndE}^{\IndF\IndC}({\rmp,\rml_1,\rml_2,\rmq})\mathcal{A}^{(0)}{}_{\IndF\IndB}^{\IndE\IndD}({\rml_2,\rmp',\rml_1,\rmq'})\, ,\\\nonumber
\mathcal{A}^{(1)}{}^{\IndC\IndD}_{\IndA\IndB}(\rmp,\rmp',\rmq,\rmq')|_{u-cut}&=\frac12\int\frac{d^2 \rml_1}{(2\pi)^2}\int\frac{d^2 \rml_2}{(2\pi)^2}\ i\pi\delta^+({\rml_1}^2-\ms)\ i\pi\delta^+({\rml_2}^2-\ms)\\\label{eqn:uch1}%BH-symfac
&\times\,\mathcal{A}^{(0)}{}_{\IndA\IndE}^{\IndF\IndD}({\rmp,\rml_1,\rml_2,\rmq'})\mathcal{A}^{(0)}{}_{\IndF\IndB}^{\IndE\IndC}({\rml_2,\rmp',\rml_1,\rmq})\, ,
\end{align}
where $\mathcal{A}^{(0)}$ are tree-level amplitudes and a sum over the complete set of intermediate states $\IndE,\IndF$ (all allowed particles for the cut lines) is understood. The on-shell propagator is given in terms of $\delta^+(k^2 - 1) = \theta(k^0) \delta(k^2-1)$ and we have included a symmetry factor of $\tfrac12$.%BHsymfac

To proceed, in each case we use (\ref{eqn:ampcons}) and the two-momentum conservation at the vertex involving the momentum $\rmp$ to integrate over $\rml_2$
\begin{align} 
 \mathcal{\widetilde{A}}^{(1)}{}^{\IndC\IndD}_{\IndA\IndB}(\rmp,\rmp',\rmq,\rmq')|_{s-cut}&=\frac12\int\frac{d^2 \rml_1}{(2\pi)^2}\,i\pi\delta^+({\rml_1}^2-\ms)\, i\pi\delta^+(({\rml_1}-{\rmp}-{\rmp'})^2-\ms)\label{2_8}\\%BHsymfac
&\times \,\,\widetilde{\mathcal{A}}^{(0)}{}_{\IndA\IndB}^{\IndE\IndF}({\rmp,\rmp',\rml_1,-\rml_1+\rmp+\rmp'}) \,\widetilde{\mathcal{A}}^{(0)}{}_{\IndF\IndE}^{\IndC\IndD}({-\rml_1+\rmp+\rmp',\rml_1,\rmq,\rmq'})\,,\nonumber \\
 \mathcal{\widetilde{A}}^{(1)}{}^{\IndC\IndD}_{\IndA\IndB}(\rmp,\rmp',\rmq,\rmq')|_{t-cut}&=\frac12\int\frac{d^2 \rml_1}{(2\pi)^2}\,i\pi\delta^+({\rml_1}^2-\ms)\,i\pi\delta^+(({\rml_1}+{\rmp}-{\rmq})^2-\ms) \label{2_9}  \\%BHsymfac
&\times\,\, \widetilde{\mathcal{A}}^{(0)}{}_{\IndA\IndE}^{\IndF\IndC}({\rmp,\rml_1,\rml_1+\rmp-\rmq,\rmq})\,  \widetilde{\mathcal{A}}^{(0)}{}_{\IndF\IndB}^{\IndE\IndD}({\rml_1+\rmp-\rmq,\rmp',\rml_1,\rmq'})\, , \nonumber \\
 \mathcal{\widetilde{A}}^{(1)}{}^{\IndC\IndD}_{\IndA\IndB}(\rmp,\rmp',\rmq,\rmq')|_{u-cut}&=
\frac12\int\frac{d^2 \rml_1}{(2\pi)^2}\,i\pi\delta^+({\rml_1}^2-\ms)\, i\pi\delta^+(({\rml_1}+{\rmp}-{\rmq'})^2-\ms)\label{2_10}\\%BHsymfac
&\times\,\, \widetilde{\mathcal{A}}^{(0)}{}_{\IndA\IndE}^{\IndF\IndD}({\rmp,\rml_1,\rml_1+\rmp-\rmq',\rmq'})\,\widetilde{\mathcal{A}}^{(0)}{}_{\IndF\IndB}^{\IndE\IndC}({\rml_1+\rmp-\rmq',\rmp',\rml_1,\rmq})\,.\nonumber
\end{align}
In each of these integrals the set of zeroes of the $\delta$-functions are discrete. This allows us to pull out the tree-level amplitudes with the loop-momenta evaluated at those zeroes, leaving scalar bubbles~\footnote{Note that if one first uses the $\delta$-function identity \eqref{delta2d} to fix, for example, $\rmp = \rmq$ and $\rmp' = \rmq'$ the $t$-cut integral is ill-defined. Furthermore, the procedure of fixing $\rml_1 = \rmq$ no longer follows. Therefore, to avoid this ambiguity we follow the prescription that we should only impose the $\delta$-function identity \eqref{delta2d} at the end. In some sense this is natural as, in general dimensions, QFT amplitudes have the form \eqref{eqn:ampcons}, while the $\delta$-function identity \eqref{delta2d} is specific to two dimensions.}. Following standard unitarity computations~\cite{Bern:1994zx}, we apply the following replacement in the imaginary part of the amplitude \eqref{2_8}--\eqref{2_10} to the internal on-shell propagators:
$i\pi \delta^+(\rml^2-1) \longrightarrow \tfrac{1}{\rml^2-1}$.
This allows us to rebuild, from its imaginary part, the cut-constructible piece of the amplitude
\begin{align}\nonumber
 \mathcal{\widetilde{A}}^{(1)}{}^{\IndC\IndD}_{\IndA\IndB}(\rmp,\rmp',\rmq,\rmq')&=\frac{I((\rmp+\rmp')^2,1,1)}{4}%\frac{ I(\rmp+\rmp')}{2}%BHsymfac
\Big[ \widetilde{\mathcal{A}}^{(0)}{}_{\IndA\IndB}^{\IndE\IndF}({\rmp,\rmp',\rmp,\rmp'}) \widetilde{\mathcal{A}}^{(0)}{}_{\IndF\IndE}^{\IndC\IndD}({\rmp',\rmp,\rmq,\rmq'})\\ \nonumber
&\qquad\qquad\qquad~~~~~+\,  \widetilde{\mathcal{A}}{}^{(0)}{}_{\IndA\IndB}^{\IndE\IndF}({\rmp,\rmp',\rmp',\rmp}) \widetilde{\mathcal{A}}^{(0)}{}_{\IndF\IndE}^{\IndC\IndD}({\rmp,\rmp',\rmq,\rmq'}) \Big] \\\nonumber
%\vphantom{\frac{ I(\rmp+\rmp')}{2}} %BHsymfac
& +\frac{I((\rmp-\rmq)^2,1,1)}{2}   \widetilde{\mathcal{A}}{}^{(0)}{}_{\IndA\IndE}^{\IndF\IndC}({\rmp,\rmq,\rmp,\rmq}) \widetilde{\mathcal{A}}^{(0)}{}_{\IndF\IndB}^{\IndE\IndD}({\rmp,\rmp',\rmq,\rmq'})\\ %\vphantom{\frac{ I(\rmp+\rmp')}{2}} %BHsymfac
& + \frac{ I((\rmp-\rmq')^2,1,1)}{2}   \widetilde{\mathcal{A}}{}^{(0)}{}_{\IndA\IndE}^{\IndF\IndD}({\rmp,\rmq',\rmp,\rmq'}) \widetilde{\mathcal{A}}^{(0)}{}_{\IndF\IndB}^{\IndE\IndC}({\rmp,\rmp',\rmq',\rmq})\, ,\label{2_11}%BHsymfac
\end{align}
where we have introduced the bubble integral
\begin{equation}\label{sbi}
I(\rmP^2,m,m')=\int \frac{d^2 \rmk}{(2\pi)^2} \frac{1}{(\rmk^2-m^2+i\e) ((\rmk-\rmP)^2-m'^2+i\e)}~.
\end{equation}
The structure of \eqref{2_11} shows the difference between the $s$-channel, for which there are two solutions of the $\delta$-function constraints in \eqref{2_8} (for positive energies), and the $t$- and $u$-channels, for which there is only one. 

Choosing  $\rmq=\rmp$, $\rmq'=\rmp'$, 
which corresponds to  considering the amplitudes associated to the first product of $\delta$-functions $\delta(p - q)\delta(p' - q')$, 
 it then follows that a candidate expression for the
one-loop S-matrix elements is given by the following simple sum of products of two tree-level amplitudes weighted by scalar bubble integrals.
\begin{align}
 {T^{(1)}}{}_{\IndA\IndB}^{\IndC\IndD}(p,p')=\frac{1}{8 (e'\,p-e\,p')}\,\Big[&{\tilde T^{(0)}}{}_{\IndA\IndB}^{\IndE\IndF}(p,p'){\tilde T^{(0)}}{}_{\IndE\IndF}^{\IndC\IndD}(p,p')I((\rmp+\rmp')^2,1,1)\nonumber\\ \vphantom{\frac{1}{4 (\e_2\,p-\e_1\,p')}} %BHsymfac
 +& {\tilde T^{(0)}}{}_{\IndA\IndE}^{\IndF\IndC}(p,p){\tilde T^{(0)}}{}_{\IndF\IndB}^{\IndE\IndD}(p,p')I(0,1,1) \nonumber\\
 \vphantom{\frac{1}{4 (\e_2\,p-\e_1\,p')}}
 +&{\tilde T^{(0)}}{}_{\IndA\IndE}^{\IndF\IndD}(p,p'){\tilde T^{(0)}}{}_{\IndF\IndB}^{\IndC\IndE}(p,p')I((\rmp-\rmp')^2,1,1)\,\Big]
 \label{eqn:final}~,
\end{align}
where $\tilde T^{(0)}(p,p')=4 (e'p-ep') T^{(0)}(p,p')$ and the scalar bubble integrals are
\begin{align}\label{ints}
I_s & \equiv I((\rmp + \rmp')^2,1,1) = \frac{1}{4(e'p-e p')}(1-\frac{\as(e'p-e p')}{i\pi}) = \frac{J}{i\pi}(i\pi-\theta) \ ,
\\\label{intt}
I_t & \equiv I(0,1,1) = \frac{1}{4\pi i} \ ,
\\\label{intu}
I_u & \equiv I((\rmp - \rmp')^2,1,1) = \frac{1}{4(e'p-e p')}\frac{\as(e'p-e p')}{i\pi} = \frac{J\theta}{i\pi} \ ,
\end{align}
where we have used \eqref{jdef} and defined
\begin{equation}\label{thetadef}
\theta \equiv \as(e'p-e p')\ .
\end{equation}

Let us stop for a second, and notice that there is a potential ambiguity in the way we proceeded. In particular, the t-channel contraction is rather subtle as there are two possible choices for
freezing the loop momenta (i.e. in terms of $p$ and $q$ or $p'$ and $q'$) giving
potentially different results. If we choose the alternative solution of the conservation $\delta$-function in (\ref{eqn:tch1}), the coefficient of $I(0)$ in (\ref{eqn:final}) would read
\begin{equation}
{\tilde T^{(0)}}{}_{\IndA\IndE}^{\IndC\IndF}(p,p'){\tilde T^{(0)}}{}_{\IndF\IndB}^{\IndD\IndE}(p',p')~.
\end{equation}
Therefore, consistency between the two expressions requires the following condition on the tree-level S-matrix 
\begin{equation}\label{consistency}
{\tilde T^{(0)}}{}_{\IndA\IndE}^{\IndF\IndC}(p,p)\,{\tilde T^{(0)}}{}^{\IndE\IndD}_{\IndF\IndB}(p,p')\,=\,{\tilde T^{(0)}}{}^{\IndC\IndF}_{\IndA\IndE}(p,p')\,{\tilde T^{(0)}}{}^{\IndD\IndE}_{\IndF\IndB}(p',p')~.
\end{equation}
Clearly this is a non-trivial constraint on the form of the tree-level S-matrix, and it turns out there are some non-relativistic models where this condition is not fulfilled. More specifically, for the light-cone gauge fixed string in $AdS_5\times S^5$ eq.~\eqref{consistency} still holds, despite the model being non-relativistic, however for $AdS_3 \times S^3 \times M^4$ this is no longer the case. This can be traced back to the fact that the function $\widetilde T^{(0)}(p,p)$ cannot have any momentum dependence
in a relativistic theory,\footnote{Let us recall that in a relativistic theory
the S-matrix depends only on the difference of rapidities, which vanishes for
$p'=p$.} whereas in a non-relativistic theory it can depend on $p$,
generating an asymmetry between $p$ and $p'$.  Hence it is natural to
conjecture that we should take the average of the two contractions.

For theories including fermionic fields, the above derivation holds up to signs. To display the general result in a compact fashion it is useful to define the following tensor contractions
\begin{align}
(A \cc{s} B)_{\IndA\IndB}^{\IndC\IndD}(p,p') &= A_{\IndA\IndB}^{\IndE\IndF}(p,p')B_{\IndE\IndF}^{\IndC\IndD}(p,p')\ ,\label{scont}
\\ (A \cc{u} B)_{\IndA\IndB}^{\IndC\IndD}(p,p') &= (-1)^{([\IndC]+[\IndF])([\IndD]+[\IndE])}A_{\IndA\IndE}^{\IndF\IndD}(p,p')B_{\IndF\IndB}^{\IndC\IndE}(p,p')\ ,\label{ucont}
\\ (A \cct{t}{\la} B)_{\IndA\IndB}^{\IndC\IndD}(p,p') &= (-1)^{[\IndC][\IndF] + [\IndE][\IndF]} A_{\IndA\IndE}^{\IndF\IndC}(p,p) B_{\IndF\IndB}^{\IndE\IndD}(p,p')\ ,\label{tcont1}
\\ (A \cct{t}{\ra} B)_{\IndA\IndB}^{\IndC\IndD}(p,p') &= (-1)^{[\IndD][\IndE] + [\IndE][\IndF]} A_{\IndA\IndE}^{\IndC\IndF}(p,p') B_{\IndF\IndB}^{\IndD\IndE}(p',p')\ ,\label{tcont2}
\end{align}
where $[\IndA] = 0$ for a boson and $1$ for a fermion. The two contractions \eqref{tcont1} and \eqref{tcont2} correspond to the two possible choices we discussed above \eqref{consistency}. In this notation the one-loop S-matrix reads\footnote{For clarity we have suppressed the flavour indices.}
\begin{equation}
T^{(1)}= \frac {iJ}2 (C_s I_s + C_t I_t + C_u I_u) \ ,
\end{equation}
with the matrices
$C_{s,u}$ given by
\begin{align}
C_{s} &= \widetilde T^{(0)} \cc{s} \widetilde T^{(0)} \ ,
\qquad
C_{u} = \widetilde T^{(0)} \cc{u} \widetilde T^{(0)} \ .
\end{align}
Equation \eqref{consistency} now reads
 \begin{equation}
\widetilde T^{(0)} \cct{t}{\la} \widetilde T^{(0)}=\widetilde T^{(0)} \cct{t}{\ra} \widetilde T^{(0)}, \label{cons}
\end{equation}
and since, as we discussed above \eqref{consistency}, this relation does not hold in general, for the coefficient of $I(0)$ we consider the average of the two contractions.
Therefore
\begin{equation}
C_t=\frac12(\widetilde T^{(0)} \cct{t}{\la} \widetilde T^{(0)}+\widetilde T^{(0)} \cct{t}{\ra} \widetilde T^{(0)})\ .
\end{equation}
To conclude the construction, we can use the explicit expressions of the
integrals $I_{s,t,u}$ in eqs.~\eqref{ints} to \eqref{intu} and the relation
between $T^{(0)}$ and $\widetilde T^{(0)}$ to rewrite the
one-loop result as
\begin{equation}\label{result}
T^{(1)}= \frac{\q}{2\p} (T^{(0)}\cc{u} T^{(0)}-T^{(0)} \cc{s} T^{(0)})+\frac i2 T^{(0)} \cc{s} T^{(0)} +\frac1{16\p} (\widetilde T^{(0)} \cct{t}{\la} T^{(0)}+T^{(0)} \cct{t}{\ra} \widetilde T^{(0)})\ ,
\end{equation}
where, under the assumption that $T^{(0)}$ is real, there is a natural split of
the result into three pieces; a logarithmic part, an imaginary rational part,
and a real rational part.

%%%%%%%%%%%%%%%%%%%%%%%%%%%%%%%%%
\subsection{Theories with multiple masses}\label{2mass}
%%%%%%%%%%%%%%%%%%%%%%%%%%%%%%%%%

We will now generalize the above construction to the case where the asymptotic
spectrum contains particles of different mass. In this derivation we will
restrict to theories whose tree-level S-matrix is integrable, in particular,
using the consequence that the set of outgoing momenta is a permutation of the
set of incoming momenta. This means that, for the reasons explained in section
\ref{singlemass}, tadpoles and one-loop graphs built from a three- and
five-point amplitude will be ignored in the unitarity computation. Therefore we
are again left with the three contributions given in figure~\ref{stu}.

We consider the configuration in which the external legs with indices $A$ and
$C$ have mass $m$ and the associated momenta are equal ($p=q$) and $B$ and $D$
have mass $m'$ with $p'=q'$.\footnote{Our procedure implies that if we assume
the set of outgoing momenta is equal to a permutation of the set of incoming
momenta at tree level, this property automatically extends to one loop.\vspace{2pt}}
For the s- and u-channels the story is then largely the same as the single-mass
case. It follows from the assumptions outlined in the previous paragraph that
when the two propagators are cut the internal loop momenta are frozen to the
values of the external momenta. The tree-level amplitudes on either side of the
cut can then be pulled out of the integral and we are left with scalar bubble
integrals with coefficients given by contractions of tree-level amplitudes.
Working through the remaining steps, which are essentially identical to the
single-mass case, it is clear that the contribution from these graphs is given
by
\begin{equation}\label{resultsumm}
T^{(1)}_{s,u} = \frac{\q}{2\p} (T^{(0)}\cc{u} T^{(0)}-T^{(0)} \cc{s} T^{(0)})+\frac i2 T^{(0)} \cc{s} T^{(0)}\ ,
\end{equation}
where
\begin{equation}\begin{split}
& \theta \equiv \as \big(\frac{e'p-e p'}{mm'}\big)\ , \qquad e = \sqrt{p^2 + m^2} \ , \qquad e' = \sqrt{p'^2 + m'^2} \ .
\\ & I_s \equiv I((\rmp + \rmp')^2,m,m') = \frac{1}{4(e'p-e p')}(1-\frac{\as(\frac{e'p-e p'}{mm'})}{i\pi}) = \frac{J}{i\pi}(i\pi-\theta) \ ,
\\ & I_u \equiv I((\rmp - \rmp')^2,m,m') = \frac{1}{4(e'p-e p')}\frac{\as(\frac{e'p-e p'}{mm'})}{i\pi} = \frac{J\theta}{i\pi} \ ,
\end{split}\end{equation}
Here $m$ and $m'$ are the masses of the two particles being scattered and the
scalar bubble integral $I(\rmP^2,m,m')$ is defined in eq.~\eqref{sbi}.
Eq.~\eqref{resultsumm} therefore fixes the logarithmic and imaginary rational
parts of the one-loop result.

The real rational part, which comes from the t-channel contribution, is, as
before, more subtle. In the single-mass case, the guiding principle for
computing the t-channel cuts was to only fix $q=p$ and $q'=p'$ at the end in
order to avoid ill-defined expressions in the intermediate steps. Therefore,
let us consider the t-channel graph in figure~\ref{stu} with the external legs
with indices $A$ and $C$ having mass $m$, $B$ and $D$ mass $m'$ and the loop
legs mass $m_l$, but $p$, $q$, $p'$ and $q'$ kept arbitrary, i.e. we do
{\em not} fix $q = p$ and $q'=p'$.

After putting the loop legs on-shell the loop momenta are fixed by the momentum
conservation delta functions in terms of the external momenta. Solving
in terms of $\rmp$ and $\rmq$ we find
\begin{equation}\begin{split}\label{sol1}
l_{1\pm}^{\uparrow} = & \frac12\big[ q_\pm - p_\pm + \sqrt{(q_\pm-p_\pm)^2 + 4 \frac{m_l^2}{m^2} q_\pm p_\pm}\big] \ ,
\\
l_{2\pm}^{\uparrow} = & \frac12\big[ p_\pm - q_\pm + \sqrt{(p_\pm-q_\pm)^2 + 4 \frac{m_l^2}{m^2} p_\pm q_\pm}\big]\ ,
\end{split}\end{equation}
while solving in terms of $\rmp'$ and $\rmq'$ gives
\begin{equation}\begin{split}\label{sol2}
l_{1\pm}^{\downarrow} = & \frac12\big[ p'_\pm - q'_\pm + \sqrt{(p'_\pm-q'_\pm)^2 + 4 \frac{m_l^2}{m^2} p'_\pm q'_\pm}\big]\ ,
\\
l_{2\pm}^{\downarrow} = & \frac12\big[ q'_\pm - p'_\pm + \sqrt{(q'_\pm-p'_\pm)^2 + 4 \frac{m_l^2}{m'^2} q'_\pm p'_\pm}\big] \ ,
\end{split}\end{equation}
where the light-cone momenta are defined in appendix \ref{notation}. The
first solution \eqref{sol1} then gives a contribution proportional to
\begin{equation}\label{expr1}
(-1)^{[C][F]+[E][F]}\widetilde{\mathcal{A}}^{(0)}{}_{AE}^{FC}(\rmp,\rml_1^\ua,\rml_2^\ua,\rmq)
\widetilde{\mathcal{A}}^{(0)}{}_{FB}^{ED}(\rml_2^\ua,\rmp',\rml_1^\ua,\rmq')\ .
\end{equation}
The arguments of the second factor of $\widetilde{\mathcal{A}}^{(0)}$ contain
all four of the external momenta and therefore this part is well-defined when
we fix $q=p$ and $q'=p'$. Therefore, let us focus on the first factor of
$\widetilde{\mathcal{A}}^{(0)}$, whose arguments only depend on two of the
momenta. Recalling that in an integrable theory the amplitude should vanish
unless the set of outgoing momenta is a permutation of the set of incoming
momenta, it follows that this first factor vanishes unless $m_l = m$. In this
case \eqref{expr1} reduces to
\begin{equation}
(-1)^{[C][F]+[E][F]}\widetilde{\mathcal{A}}^{(0)}{}_{AE}^{FC}(\rmp,\rmq,\rmp,\rmq)
\widetilde{\mathcal{A}}^{(0)}{}_{FB}^{ED}(\rmp,\rmp',\rmq,\rmq')\ .
\end{equation}
Finally setting $q=p$ and $q'=p'$ this expression can then be written in terms
of tree-level S-matrices. A similar logic follows for the second solution
\eqref{sol2}, except that here the contribution vanishes unless $m_l = m'$.

It therefore follows that the contribution from the t-channel is given by
\begin{equation}\label{resultsummt}
T^{(1)}_t = \frac1{16\p}( \frac1{m^2} \widetilde T^{(0)} \cct{t}{\la} T^{(0)}+ \frac{1}{m'^2}T^{(0)} \cct{t}{\ra} \widetilde T^{(0)})\ ,
\end{equation}
where $\widetilde T^{(0)}$ in the first term is built from the tree-level
S-matrix for the scattering of two excitations of mass $m$, while in the second
term it is built from the tree-level S-matrix for two excitations of mass $m'$.
We have included an additional factor of $\tfrac12$, as we should still use both
vertices to solve for the loop momenta and take the average.

Combining eqs.~\eqref{resultsumm} and \eqref{resultsummt} we find that the
one-loop result, in the case where an excitation of mass $m$ is scattered with
an excitation of mass $m'$, is given by
\begin{equation}\label{resulttwo}
T^{(1)}= \frac{\q}{2\p} (T^{(0)}\cc{u} T^{(0)}-T^{(0)} \cc{s} T^{(0)})
+ \tfrac i2 T^{(0)} \cc{s} T^{(0)}
+ \tfrac1{16\p} (\tfrac{1}{m^2}\widetilde T^{(0)} \cct{t}{\la} T^{(0)}+\tfrac{1}{m'^2}T^{(0)} \cct{t}{\ra} \widetilde T^{(0)})\ ,
\end{equation}
where, again under the assumption that $T^{(0)}$ is real, there is a natural
split of the result into three pieces: a logarithmic part, an imaginary
rational part, and a real rational part. Setting $m=m'=1$ we see that this
formula reduces to, and hence incorporates, the single-mass case given in
eq.~\eqref{result}.

A key consequence of the results in this section is that the cut-constructible
one-loop S-matrix for the scattering of a particle of mass $m$ with one of mass
$m'$ is built from the corresponding tree-level S-matrix along with the
tree-level S-matrices for the scattering of two particles of mass $m$ and for
two particles of mass $m'$, both evaluated at equal momenta. In particular
there are no contributions containing tree-level S-matrices for particles of
masses other than $m$ and $m'$. This will be important in later sections, as it
allows us to construct the one-loop cut-constructible S-matrix for various
sectors without knowing the full tree-level S-matrix.

\subsection{Relation to Yang Baxter equation}

The result \eqref{resulttwo} deserves a comment regarding its relation to
integrability and the Yang-Baxter equation (YBE) \eqref{ybe}. Up to signs related to fermions, which we are not
concerned with for this schematic discussion, the YBE can be written as
\begin{equation}
\mathbb{S}_{12} \mathbb{S}_{13} \mathbb{S}_{23} =
\mathbb{S}_{23} \mathbb{S}_{13} \mathbb{S}_{12} \ ,
\end{equation}
where these operators are acting on a three-particle state and the indices
denote the particles that are being scattered. The first non-trivial order in
its perturbative expansion is called the classical Yang-Baxter equation and is
a relation that is quadratic in the tree-level S-matrix,
\begin{equation}\label{cybe}
[\mathbb{T}_{12}^{(0)},\mathbb{T}^{(0)}_{13}] +
[\mathbb{T}_{12}^{(0)},\mathbb{T}^{(0)}_{23}] +
[\mathbb{T}_{13}^{(0)},\mathbb{T}^{(0)}_{23}] = 0 \ .
\end{equation}
At the next order we find the following relation
\begin{equation}\begin{split}\label{olybe}
&[\mathbb{T}_{12}^{(0)},\mathbb{T}^{(1)}_{13}] +
[\mathbb{T}_{12}^{(0)},\mathbb{T}^{(1)}_{23}] +
[\mathbb{T}_{13}^{(0)},\mathbb{T}^{(1)}_{23}] -
[\mathbb{T}_{13}^{(0)},\mathbb{T}^{(1)}_{12}] -
[\mathbb{T}_{23}^{(0)},\mathbb{T}^{(1)}_{12}] -
[\mathbb{T}_{23}^{(0)},\mathbb{T}^{(1)}_{13}] =
\\
&\hspace{275pt}\mathbb{T}^{(0)}_{23} \mathbb{T}^{(0)}_{13} \mathbb{T}^{(0)}_{12} -
\mathbb{T}^{(0)}_{12} \mathbb{T}^{(0)}_{13} \mathbb{T}^{(0)}_{23} \ .
\end{split}\end{equation}
One can check that, assuming that the tree-level S-matrix satisfies the
classical Yang-Baxter equation \eqref{cybe}, the rational s-channel
contribution to the cut-constructible one-loop S-matrix precisely cancels the
terms cubic in the tree-level S-matrix on the right-hand side of
eq.~\eqref{olybe}. Therefore, for the one-loop cut-constructible S-matrix to
respect integrability, the remaining terms should satisfy \eqref{olybe} with the
right-hand side set to zero. In general, this condition is not easy to solve,
but two solutions are clear. The first is the tree-level S-matrix itself (which
amounts to a shift in the coupling), and the second is any contribution
that can be absorbed into the overall phase factors.

%%%%%%%%%%%%%%%%%%%%%%%%%%%%%%%%%
\subsection{External leg corrections}\label{sec:elc}
%%%%%%%%%%%%%%%%%%%%%%%%%%%%%%%%%

In the construction outlined thus far we have not included any discussion of
corrections to the external legs. As shall become apparent, for the $AdS_3
\times S^3 \times S^3 \times S^1$ background these will be important even at
one loop. These corrections will give a rational contribution to the S-matrix
and can follow from the three types of Feynman diagrams in figure \ref{graphs}.
\begin{figure}[t]
\begin{center}
\begin{tikzpicture}[line width=1pt,scale=0.9]
\draw[-] (-5.4,-4) -- (-4,-4);
\draw[-] (-2,-4) -- (-0.6,-4);
\draw (-3,-4) circle (1cm);
\draw[->] (-5.2,-4.2) -- (-4.6,-4.2);
\node at (-5.2,-4.4) {$\rmp$};
\draw[->] (-1.4,-4.2) -- (-0.8,-4.2);
\node at (-1.4,-4.4) {$\rmp$};
\draw[<-] (-3.1,-3.2) -- (-3.5,-3.35);
\node at (-3.2,-3.6) {$\rml_1$};
\draw[<-] (-2.9,-4.8) -- (-2.5,-4.65);
\node at (-2.7,-4.4) {$\rml_2$};
\end{tikzpicture}
\qquad\qquad
\begin{tikzpicture}[line width=1pt,scale=0.9]
\draw[-] (-5.4,-5) -- (-0.6,-5);
\draw (-3,-4) circle (1cm);
\draw[->] (-5.2,-5.2) -- (-4.6,-5.2);
\node at (-5.2,-5.4) {$\rmp$};
\draw[->] (-1.4,-5.2) -- (-0.8,-5.2);
\node at (-1.4,-5.4) {$\rmp$};
\draw[<-] (-3.1,-3.2) -- (-3.5,-3.35);
\node at (-3.2,-3.6) {$\rml_{\hphantom{1}}$};
\end{tikzpicture}
\qquad\qquad
\begin{tikzpicture}[line width=1pt,scale=0.9]
\draw[-] (-5.4,-5) -- (-0.6,-5);
\draw[-] (-3,-5) -- (-3,-4.5);
\draw (-3,-3.75) circle (0.75cm);
\draw[->] (-5.2,-5.2) -- (-4.6,-5.2);
\node at (-5.2,-5.4) {$\rmp$};
\draw[->] (-1.4,-5.2) -- (-0.8,-5.2);
\node at (-1.4,-5.4) {$\rmp$};
\draw[<-] (-3.1,-3.2) -- (-3.45,-3.35);
\node at (-3.2,-3.6) {$\rml_{\hphantom{1}}$};
\end{tikzpicture}
\caption{Diagrams contributing to external leg corrections at one-loop.}
\label{graphs}
\end{center}
\end{figure}

We will be interested in external leg corrections at one loop that are caught
by unitarity. In order to approach this problem let us first review how
external leg corrections are usually dealt with in a standard Feynman diagram
calculation. We denote the sum of all one particle irreducible insertions into
a scalar propagator as $-i\Sigma(\rmp) = -i \co\Sigma^{(1)}(\rmp) +
\mathcal{O}(\co^2)$, where $-i \co\Sigma^{(1)}(\rmp)$ is the one-loop
contribution. After re-summing one finds
\begin{equation}
\begin{tikzpicture}[line width=1pt,baseline=-3]
\draw[-] (-1,0)--(0,0);
\draw [fill=black!20] (0.5,0) circle (0.5);
\draw[-] (1,0)--(2,0);
\end{tikzpicture}
=\frac{i}{\rmp^2-m^2-\Sigma(\rmp)}
%v2-start
\end{equation}
Expanding $\Sigma(\rmp)$ around the on-shell condition,
$\Sigma(\rmp)=\Sigma_0(p)+\Sigma_1(p)(\rmp^2-m^2)+\mathcal{O}((\rmp^2-m^2)^2)$,
one obtains a spatial momentum dependent shift in the pole and a non-vanishing
residue $Z(p)$ such that
\begin{equation}
\begin{tikzpicture}[line width=1pt,baseline=-3]
\draw[-] (-1,0)--(0,0);
\draw [fill=black!20] (0.5,0) circle (0.5);
\draw[-] (1,0)--(2,0);
\end{tikzpicture}
=\frac{iZ(p)}{\rmp^2-m^2-\Sigma_0(p)} + \ldots\ .
\end{equation}
where $Z=1+\co\Sigma^{(1)}_1(p)+\mathcal{O}(\co^2)$
and $\Sigma_0(p) = \co\Sigma_0^{(1)}(p) + \mathcal{O}(\co^2)$.
%v2-end
It is well-known that
the same quantity also appears in the LSZ reduction and the prescription to
take these contributions into account is to include a factor of $\sqrt{Z}$ for
the external legs of the scattering process. When inserting this into the
S-matrix of a $2\to2$ process one therefore gets an additional contribution to
$T^{(1)}$ of the form
\begin{equation}\begin{split}\label{important}
T^{(1)}_{ext} = (\Sigma^{(1)}_1(p) + \Sigma^{(1)}_1(p'))T^{(0)}\ ,
\end{split}
\end{equation}
where we recall that we are working in the configuration in which $q =p$ and
$q'=p'$. Here we can already make the observation that given that
$\Sigma^{(1)}_1(p)$ is real (and assuming that $T^{(0)}$ is real) the
contribution from external legs should contribute to the real rational part of
$T^{(1)}$.

The contribution $\Sigma^{(1)}_1(p)$ is a subleading contribution in the
expansion of the self-energy around the on-shell condition and in a standard
Feynman diagram computation would be regularization dependent. Since in the
unitarity computation we did not assume any explicit regularization we may
encounter problems combining the two results. For this reason, we will choose to
follow a rather different approach and compute this subleading contribution via
unitarity.

As we are considering a unitarity computation, we will only consider
contributions from the graphs in figure \ref{graphs} when they have a physical
two-particle cut. In particular, to be consistent with our approach for the
S-matrix we ignore the latter two tadpole diagrams and restrict our attention
to the first diagram. It therefore follows that, in the unitarity computation,
external leg corrections will only play a role at one loop in theories with
cubic vertices. In the context of generalized unitarity, as we discussed in
section \ref{singlemass}, tadpole diagrams may not be negligible and therefore
there is no guarantee that our procedure will provide the whole result. However,
the precise cancellation we observe in the specific example we discuss later is
a clear indication of the validity of our result up to a shift in the
coupling (for a more detailed discussion see section \ref{extlegs}).

\begin{figure}[t]
\begin{center}
\begin{tikzpicture}[line width=1pt,scale=1.3,baseline = -100]
\draw[-] (-5,-4.05) -- (-4,-4.05);
\draw[-] (-5,-3.95) -- (-4,-3.95);
\draw[-] (-2,-4.05) -- (-1,-4.05);
\draw[-] (-2,-3.95) -- (-1,-3.95);
\draw (-3,-4) circle (1cm);
\draw[|-|,dashed,blue!70,line width=1pt] (-3,-2.5) -- (-3,-5.5);
\draw[->] (-4.9,-3.8) -- (-4.5,-3.8);
\node at (-4.8,-3.6) {$\rmp$};
\draw[->] (-1.5,-3.8) -- (-1.1,-3.8);
\node at (-1.4,-3.6) {$\rmp$};
\draw[<-] (-3.1,-3.2) -- (-3.5,-3.35);
\node at (-3.2,-3.5) {$\rml_1$};
\draw[<-] (-2.9,-4.8) -- (-2.5,-4.65);
\node at (-2.8,-4.5) {$\rml_2$};
\node at (-3.4,-2.85) {$\IndE$};
\node at (-2.6,-5.15) {$\IndF$};
\node [circle,draw=black!100,fill=black!5,thick,opacity=.95] at (-4,-4) {\footnotesize$\mathcal{F}^{(0)}$\normalsize};
\node [circle,draw=black!100,fill=black!5,thick,opacity=.95] at (-2,-4) {\footnotesize$\mathcal{F}^{(0)}$\normalsize};
\end{tikzpicture}
\caption{Cut of a two-point function obtained by fusing two form factors. The double line indicates an off-shell state.}
\label{2ptfun}\nonumber
\end{center}
\end{figure}
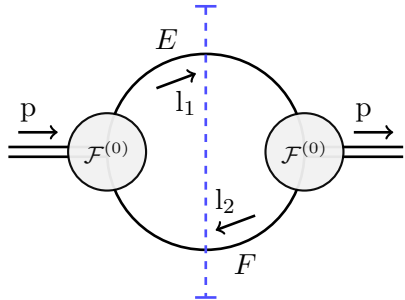

The computation of correlation functions by generalized unitarity was
extensively analyzed in four dimensions in~\cite{Engelund:2012re}, in which it
was shown that the object that needs to be put on either side of the cut is a
form factor as shown in figure
\ref{2ptfun}.
However, let us also note that we will want to expand around the on-shell
condition and hence we ask that the diagram should have a physical cut even
when the external leg is on-shell. This places a restriction on the masses of
the particles involved. In particular they should take the following form;
$m_1$, $m_2$ and $m_1 - m_2$, where we take $m_1>m_2$.

By taking figure \ref{2ptfun} with a mass $m_1-m_2$ external
particle,\footnote{This will be the case we consider for $AdS_3 \times S^3
\times S^3 \times S^1$. One can also consider a mass $m_1$ external particle
and internal particles with masses $m_1-m_2$ and $m_2$ ($m_1>m_2$). In this
case the two loop momenta in figure \ref{2ptfun} should be pointing in the
same direction.\vspace{2pt}} internal particles with masses $m_1$ and $m_2$ corresponding
to momentum $\rml_1$ and $\rml_2$ and returning $\rmp$ off-shell, the explicit
expression for this diagram is given by
\begin{align}
\Sigma^{(1)}(\rmp)|_{cut}&=\int\frac{d^2 \rml_1}{(2\pi)^2} \, i\pi\, \delta^+({\rml_1}^2-m_1)\, i\pi\, \delta^+((\rml_1-\rmp)^2-m_2)\,\\
&\times \mathcal{F}^{(0)}_{\IndE\IndF}(\rmp,\rml_1,\rml_1-\rmp)\,{\mathcal{F}^{(0)}_{\IndE\IndF}}^\dagger(\rmp,\rml_1,\rml_1-\rmp)\ .
\end{align}
Here, as in the unitarity computation of the S-matrix, the cut completely
freezes the internal momenta:
\begin{align}
\rml_1 &= \frac{m_1^2-m_2^2+\rmp^2-\sqrt{\Delta}}{2\, \rmp^2}\, \rmp\equiv \rml_*\ ,\\
\rmp-\rml_1 &= \frac{m_2^2-m_1^2+\rmp^2+\sqrt{\Delta}}{2\, \rmp^2}\, \rmp \equiv \rml'_*\ ,
\end{align}
where $\Delta=\rmp^4+m_1^4+m_2^4-2\, m_1^2 \rmp^2-2\, m_2^2 \rmp^2-2\, m_1^2
m_2^2$. It therefore follows that we can pull the numerators out of the
integrand and uplift the integral as was done for the four-point amplitude.
This gives
\begin{align}\label{ffsquare}
\Sigma^{(1)}(\rmp)=\frac12 \left|\mathcal{F}^{(0)}_{\IndE\IndF}(\rmp,\rml_*,\rml'_*)\right|^2 I(\rmp^2,m_1,m_2)\ ,
\end{align}
with the integral $I(\rmp^2,m_1,m_2)$ defined in \eqref{sbi}. In section
\ref{extlegs} we will apply this formula to a specific example and we will also
point out the limits of its application.

%%%%%%%%%%%%%%%%%%%%%%%%%%%%%%%%%
\subsection{Structure of the result}\label{resstruc}
%%%%%%%%%%%%%%%%%%%%%%%%%%%%%%%%%

To conclude this section let us make some remarks about the features of the
result that are relevant for our discussion. In all the theories of interest for this review the massive excitations can be grouped into particles and
antiparticles transforming with charge $\s = +1$ and $\s = -1$ under a global
$U(1)$ symmetry. Furthermore, not only is the set of incoming momenta preserved
by the scattering process, but so are the $U(1)$ charges associated to the
individual momenta, i.e. $\s_A = \s_B$ and $\s_C = \s_D$. The general structure
of the S-matrix is then
\begin{align}
S_{AB}^{CD}(p,p') &= \exp[i \vp_{\s_A \s_B}(p,p')] \hat{S}_{AB}^{CD}(p,p') \label{structure}\ ,
\end{align}
where $\vp$ are the phases\footnote{In the case of $AdS_5\times S^5$ there is a single dressing factor, but we will see that for $AdS_3\times S^3\times M^4$ there may be more than one.} and the matrix structure $\hat{S}$ is fixed by the
symmetry of the theory. Each of these objects admit a perturbative expansion at
strong coupling:
\begin{align}
S=\mathbf{1}+i\sum_{n=1}^\infty \co^{n}T^{(n-1)} \ , &\qquad~~
\hat{S}=\mathbf{1}+i\sum_{n=1}^\infty \co^{n}\hat{T}^{(n-1)}\ , \nonumber
\\ \vp_{\s_A\s_B}(p,p')&=\sum_{n=1}^\infty \co^{n}\vp^{(n-1)}_{\s_A\s_B}(p,p')\ . \label{phiexp1}
\end{align}
Furthermore, as $\hat{S}$ is fixed by symmetries it should contain no
logarithmic functions of the momenta. Therefore, all the logarithms are
contained in the phases, and to the one-loop order we can separate these off as
follows
\begin{equation}\label{phasestruc}
\vp^{(0)}_{\s_A\s_B}(p,p') = \phi^{(0)}_{\s_A\s_B}(p,p') \ , \qquad
\vp^{(1)}_{\s_A\s_B}(p,p') = \cl_{\s_A\s_B}(p,p')\,\theta + \phi^{(1)}_{\s_A\s_B}(p,p') \ .
\end{equation}
Here $\theta$, defined in eq.~\eqref{thetadef}, is the only possible logarithm
appearing at one loop, and $\phi^{(n)}_{\s_A\s_B}$ are rational functions of
the momenta.

Substituting eqs.~\eqref{phiexp1} and \eqref{phasestruc} into \eqref{structure}
we find
\begin{align}
T^{(0)}&= \phi^{(0)}_{\s_A\s_B}(p,p')\, \mathbf{1}+\hat{T}^{(0)}\ , \label{treestruc}\\
T^{(1)}&= \cl_{\s_A\s_B}(p,p')\, \theta\, \mathbf{1}+\frac i2\left[\phi^{(0)}_{\s_A\s_B}(p,p')\right]^2\mathbf{1}+\phi^{(1)}_{\s_A\s_B}(p,p')\, \mathbf{1}\\
&+i\phi^{(0)}_{\s_A\s_B}(p,p')\,\hat{T}^{(0)}+\hat{T}^{(1)} \ . \label{1Lstruc}
\end{align}
Let us compare the structure of the one-loop result following from
integrability \eqref{1Lstruc} with that following from unitarity methods
\eqref{result}, \eqref{resulttwo}. The comparison between the two expressions
leads to the following identifications (note that by definition the functions
$\cl_{\s_A\s_B}$ and $\phi^{(n)}_{\s_A\s_B}$ are real)
\begin{align}
&\frac{1}{2\p} (T^{(0)} \cc{u} T^{(0)} - T^{(0)} \cc{s} T^{(0)})= \cl_{\s_A\s_B}(p,p')\,\mathbf{1}\ , \label{logsid}
\\
& \frac12 T^{(0)} \cc{s} T^{(0)}=\frac12\left[\phi^{(0)}_{\s_A\s_B}(p,p')\right]^2\mathbf{1}+\phi^{(0)}_{\s_A\s_B}(p,p')\,\hat{T}^{(0)}+\text{Im}(\hat{T}^{(1)}) \nonumber
\\ & \hspace{200pt} \Rightarrow \quad \frac12 \hat T^{(0)} \cc{s} \hat T^{(0)} = \text{Im}(\hat{T}^{(1)}) \ , \label{schannelid}\\
&\tfrac1{16\p} (\tfrac{1}{m^2}\widetilde T^{(0)} \cct{t}{\la} T^{(0)}+\tfrac{1}{m'^2}T^{(0)} \cct{t}{\ra} \widetilde T^{(0)})
+ (\Sigma^{(1)}_1(p) + \Sigma^{(1)}_1(p')) T^{(0)}\\& \hspace{200pt} = \phi^{(1)}_{\s_A\s_B}(p,p')\, \mathbf{1}+ \text{Re}(\hat{T}^{(1)})\ ,\label{tchannelid}
\end{align}
where we have assumed that $T^{(0)}$ is real, which will indeed be the case for
all the models we consider. For the rational terms coming from the s-channel in
\eqref{schannelid}, we have simplified the expression that needs to be checked
by substituting in for $T^{(0)}$ \eqref{treestruc} and using that $\mathbf{1}
\cc{s} \mathbf{1} = \mathbf{1}$, $\hat T^{(0)} \cc{s} \mathbf{1} = \hat T^{(0)}$
and $\mathbf{1} \cc{s} \hat T^{(0)} = \hat T^{(0)}$ are satisfied by definition
(see eq.~\eqref{scont}). In \eqref{tchannelid} we have also included a
possible contribution from external leg corrections to the real rational part
of $T^{(1)}$ (see eq.~\eqref{important}), as discussed in section
\ref{sec:elc}. Eqs.~\eqref{logsid}, \eqref{schannelid} and \eqref{tchannelid}
are therefore the three equations that we need to check to see how much of the
exact S-matrix is recovered from the unitarity construction.

Factoring out an overall phase factor as in \eqref{structure} clearly contains
a degree of arbitrariness. Of course, this choice should not affect the final
result, however, there are certain choices that interplay well with the
unitarity construction. In particular, if there is a scattering process for
which the only possible outgoing two-particle state is the incoming state
($A=C=A_*$, $B=D=B_*$), then the corresponding amplitude must be a phase
factor. In this case we can set
\begin{equation}
\hat S_{A_* B_*}^{A_* B_*} = 1 \ ,
\end{equation}
where $A_*$ and $B_*$ are fixed and there is no sum. This choice is consistent
with \eqref{schannelid} -- both sides are clearly vanishing by construction.
Furthermore, $\phi^{(1)}$ is just given by the t-channel contraction (plus possible
external leg corrections) with indices $A=C=A_*$, $B=D=B_*$.

\section{Worldsheet scattering in \txpf{$AdS_3\times S^3\times M^4$}{AdS3 x S3 x M4}}\label{sec:AdS3}
In this section we apply the methods of section \ref{sec:unitarity} to a class of integrable theories that arise as the light-cone gauge-fixing of the $AdS_3 \times S^3 \times M^4$ string backgrounds described in section \ref{sec:stringAdS3}.
We will focus on the following three cases. The first is the simplest and is
when the compact manifold is $T^4$ with the background supported by RR flux.
The second is when the compact manifold is $S^3 \times S^1$, again supported by
RR flux. For the last we return to $T^4$, but with the background now supported
by a mix of RR and NSNS fluxes. 

In analogy with $AdS_5 \times S^5$ we consider the S-matrix describing the scattering of excitations on the
decompactified string worldsheet in the uniform light-cone gauge. The masses of the asymptotic excitations are given by the
expansion around the BMN string~\cite{Berenstein:2002jq}. For the theories
under consideration we have the following spectra
\renewcommand{\arraystretch}{1.5}
\begin{center}
\begin{tabular}{ll}
\hline\hline Theory & Spectrum
\\ \hline\hline
$AdS_3 \times S^3 \times T^4$ (RR flux) & $(4+4) \times 1~\phantom{-\alpha} \qquad (4+4) \times 0$
\\ \hline
$AdS_3 \times S^3 \times S^3 \times S^1$ (RR flux) \qquad\qquad &
$(2+2) \times 1 ~\phantom{-\alpha} \qquad (2+2) \times \alpha $ \\ & $ (2+2) \times 1-\alpha \qquad (2+2) \times 0$
\\ \hline
$AdS_3 \times S^3 \times T^4$ (mixed flux) & $(4+4) \times \sqrt{1-q^2} \,\,\,\,\, (4+4) \times 0$
\\ \hline\hline \end{tabular}
\end{center}
where $(n+n)$ denotes bosons+fermions. As expected, in each case we have
$(8+8)$ excitations in total and the masses of the bosons match those of the
fermions. All three cases feature massless modes, which need careful treatment
in two dimensions. In the following we will argue that if we restrict to
massive external legs, then we can ignore the massless modes completely in the
one-loop unitarity computation. 

\subsection{Tree-level S-matrices for pure RR flux}
The main input of the one-loop unitarity computation is the tree-level S-matrix
of the theory. Various components of the tree-level S-matrices for the $T^4$
and $S^3 \times S^1$ backgrounds supported by RR flux were computed in~\cite{Sundin:2013ypa,Rughoonauth:2012qd}, and in~\cite{Hoare:2013pma} for the
mixed flux case. These results, along with the symmetries and integrability of
the theory, can be used to completely determine the tree-level S-matrix.

%%%%%%%%%%%%%%%%%%%%%%%%%%%%%%%%%
\subsubsection{Massive sector for \txpf{$AdS_3\times S^3 \times T^4$}{AdS3 x S3 x T4}}\label{sec:t4}
%%%%%%%%%%%%%%%%%%%%%%%%%%%%%%%%%

The quadratic light-cone gauge fixed action for the $AdS_3
\times S^3 \times T^4$ background describes $4+4$ massive and $4+4$ massless
fields. Here we will just consider the scattering of two massive excitations to
two massive excitations. The S-matrix of the theory was fixed up to two phases
in~\cite{Borsato:2013qpa} using symmetries.

Thinking of the particle content of the massive sector as $2+2$ complex degrees
of freedom, we label these fields as $\Phi_{\vf\vf}$, $\Phi_{\psi\psi}$,
$\Phi_{\vf\psi}$ and $\Phi_{\psi\vf}$, and their complex conjugates as
$\Phi_{\bar \vf\bar\vf}$, $\Phi_{\bar \psi \bar \psi}$, $\Phi_{\bar \vf\bar
\psi}$ and $\Phi_{\bar \psi\bar \vf}$, where we understand $\vf$, $\bar \vf$ as
bosonic and $\psi$, $\bar\psi$ as fermionic indices.

As a consequence of the symmetries and integrability of the theory, the
S-matrix factorizes:
\begin{equation}\label{factorize}
\mathbb{S}\ket{\Phi_{A\dot A}(p)\Phi_{B\dot B}(p')} = (-1)^{[\dot A][B]+[\dot C][D]} S_{AB}^{CD}(p,p')S_{\dot A \dot B}^{\dot C\dot D}(p,p')
\ket{\Phi_{C\dot C}(p)\Phi_{D\dot D}(p')} \ ,
\end{equation}
where the indices take the following values:
$\{\vf,\bar{\vf},\psi,\bar{\psi}\}$. One can check that the construction outlined in section \ref{sec:unitarity} gives the same one-loop result whether we consider the factorized or full S-matrix. Therefore, for simplicity we will work with the former. The general structure of the factorized
S-matrix takes the form given in \eqref{structure} with $\s_\vf=\s_\psi=+$ and
$\s_{\bar{\vf}}=\s_{\bar{\psi}}=-$. Charge conjugation symmetry implies that
$\phi_{++}=\phi_{--}$, $\phi_{+-}=\phi_{-+}$, $\cl_{++}=\cl_{--}$ and
$\cl_{+-}=\cl_{-+}$. Therefore, in the following we will focus on the $++$ and
$+-$ sectors. A typical feature of the uniform light-cone gauge is the
dependence of the phase on a gauge-fixing parameter $a$. This dependence has
the following exact form
\begin{equation}\label{gauge}
\exp\big[\frac{i}{2}(a-\tfrac12)(\o'p-\o p')\big] \ ,
\end{equation}
where the all-order energies $\o$ are defined in appendix \ref{notations1}.

As we discussed in section \ref{resstruc} we define the overall phase factors
by setting particular components of $\hat{S}_{AB}^{CD}$ to one
\begin{equation}
\hat{S}_{\vf\vf}^{\vf\vf}(p,p')=1\ , \qquad \hat{S}_{\vf\bar{\psi}}^{\vf\bar{\psi}}(p,p')=1\ . \label{choice1}
\end{equation}
The parametrizing functions of the exact S-matrix are defined as
\begin{align}
{S}_{\vf\vf}^{\vf\vf}(p,p')&=A_{++}(p,p') & {S}_{\vf\bar{\vf}}^{\vf\bar{\vf}}(p,p')&=A_{+-}(p,p')\nonumber\\
{S}_{\vf\psi}^{\vf\psi}(p,p')&=B_{++}(p,p') & {S}_{\vf\bar{\vf}}^{\psi\bar{\psi}}(p,p')&=B_{+-}(p,p')\nonumber\\
{S}_{\vf\psi}^{\psi\vf}(p,p')&=C_{++}(p,p') & {S}_{\vf\bar{\psi}}^{\vf\bar{\psi}}(p,p')&=C_{+-}(p,p')\nonumber\\
{S}_{\psi\vf}^{\psi\vf}(p,p')&=D_{++}(p,p') & {S}_{\psi\bar{\vf}}^{\psi\bar{\vf}}(p,p')&=D_{+-}(p,p')\nonumber\\
{S}_{\psi\vf}^{\vf\psi}(p,p')&=E_{++}(p,p') & {S}_{\psi\bar{\psi}}^{\psi\bar{\psi}}(p,p')&=E_{+-}(p,p')\nonumber\\
{S}_{\psi\psi}^{\psi\psi}(p,p')&=F_{++}(p,p') & {S}_{\psi\bar{\psi}}^{\vf\bar{\vf}}(p,p')&=F_{+-}(p,p')\label{corr}
\end{align}

 The tree-level components computed
directly in~\cite{Sundin:2013ypa,Rughoonauth:2012qd} are consistent with
the near-BMN expansion of the exact result \eqref{funpp}, \eqref{funpm}. The remaining
components of the tree-level S-matrix can then be fixed from the expansion of
the exact result. Here we shall fix $a=\frac12$ as
the dependence on $a$ goes through the unitarity procedure without any
particular subtlety, i.e. it exponentiates as in eq.~\eqref{gauge}. The
tree-level S-matrix reads
\begin{align}
A^{(0)}_{++}(p,p') &= \tfrac{(p+p')^2}{4(e' p-e p')}\ ,
& B^{(0)}_{++}(p,p') &= \tfrac{p'^2-p^2}{4(e' p-e p')}\ , \nonumber \\
C^{(0)}_{++}(p,p') &= p p' \tfrac{\sqrt{(e+p)(e'-p')}+\sqrt{(e-p)(e'+p')}}{2(e' p-e p')}\ ,
& D^{(0)}_{++}(p,p') &= -\tfrac{p'^2-p^2}{4(e' p-e p')}\ , \nonumber \\
E^{(0)}_{++}(p,p') &= p p' \tfrac{\sqrt{(e+p)(e'-p')}+\sqrt{(e-p)(e'+p')}}{2(e' p-e p')}\ ,
& F^{(0)}_{++}(p,p') &= -\tfrac{(p+p')^2}{4(e' p-e p')}\ , \label{treepp}
\end{align}
\begin{align}
A^{(0)}_{+-}(p,p') &= \tfrac{(p-p')^2}{4(e' p-e p')}\ ,
& B^{(0)}_{+-}(p,p') &= p p' \tfrac{\sqrt{(e-p)(e'+p')}-\sqrt{(e+p)(e'-p')}}{2(e' p-e p')}\ , \nonumber \\
C^{(0)}_{+-}(p,p') &= \tfrac{p'^2-p^2}{4(e' p-e p')}\ ,
& D^{(0)}_{+-}(p,p') &= -\tfrac{p'^2-p^2}{4(e' p-e p')}\ , \nonumber \\
E^{(0)}_{+-}(p,p') &= -\tfrac{(p-p')^2}{4(e' p-e p')}\ ,
& F^{(0)}_{+-}(p,p') &= p p' \tfrac{\sqrt{(e-p)(e'+p')}-\sqrt{(e+p)(e'-p')}}{2(e' p-e p')}\ . \label{treepm}
\end{align}

%%%%%%%%%%%%%%%%%%%%%%%%%%%%%%%%%
\subsubsection{Massive sector for \txpf{$AdS_3\times S^3 \times S^3 \times S^1$}{AdS3 x S3 x S3 x S1}}
%%%%%%%%%%%%%%%%%%%%%%%%%%%%%%%%%

The quadratic light-cone gauge fixed action for the $AdS_3 \times S^3
\times S^3 \times S^1$ background describes particles with four different
masses. The field content is summarised in table \ref{masses}. Here we will
focus on the scattering of massive states with masses $\a$ and $\bar\a=1-\a$.
\begin{table}[ht]
\begin{center}
\begin{tabular}[h]{|c|l|}
\hline
Fields& Mass\\
\hline
$\varphi_1,\bar{\varphi}_1,\chi^1,\bar{\chi}^1$&$m_1=1$\\
\hline
$\varphi_2,\bar{\varphi}_2,\chi^2,\bar{\chi}^2$&$m_2=\a$\\
\hline
$\varphi_3,\bar{\varphi}_3,\chi^3,\bar{\chi}^3$&$m_3=\bar\a$\\
\hline
$\varphi_4,\bar{\varphi}_4,\chi^4,\bar{\chi}^4$&$m_4=0$\\
\hline
\end{tabular}
\end{center}
\caption{Field content of the $AdS_3\times S^3 \times S^3 \times S^1$ light-cone gauge fixed string theory.}\label{masses}
\end{table}

Let us first analyze the S-matrix for $AdS_3\times S^3 \times S^3 \times S^1$
describing the scattering of two particles of mass $\a$.\footnote{For particles
of mass $\bar\a$ the corresponding result can be obtained simply by replacing
$\a$ with $\bar\a$.\vspace{2pt}} When we restrict to this sector the S-matrix has the same
structure as the factorized S-matrix for $AdS_3\times S^3 \times T^4$, again
taking the form given in \eqref{structure}. The tree-level S-matrix, however,
is different and this will have non-trivial consequences for the unitarity
calculation. Compared to the $AdS_3\times S^3 \times T^4$ case the dependence
on the gauge-fixing parameter $a$ is modified due to the fact that this is now
the full S-matrix. The new expression reads
\begin{equation}\label{gaugeS1}
\exp\big[i(a-\tfrac12)(\o'p-\o p')\big] \ .
\end{equation}

We again use \eqref{choice1} to choose the overall phase
factors and define the parametrizing functions as in
eq.~\eqref{corr}.\footnote{To be precise we use the definitions \eqref{corr}
with the replacements $\varphi\to \varphi_2$ and $\psi\to\chi^2$ and likewise
for their conjugates.\vspace{2pt}} 

 As in the $AdS_3 \times S^3 \times T^4$ case we shall present the result in the gauge $a=\frac12$. The tree-level S-matrix reads
\begin{align}
A^{(0)}_{++}(p,p') &= \tfrac{\a (p+p')^2}{2(e' p-e p')}\ ,
& B^{(0)}_{++}(p,p') &= \tfrac{\a p' (p+p')}{2(e' p-e p')}\ , \nonumber \\
C^{(0)}_{++}(p,p') &= p p' \tfrac{\sqrt{(e+p)(e'-p')}+\sqrt{(e-p)(e'+p')}}{2(e' p-e p')}\ , &
D^{(0)}_{++}(p,p') &= \tfrac{\a p (p+p')}{2(e' p-e p')}\ , \nonumber \\
E^{(0)}_{++}(p,p') &= p p' \tfrac{\sqrt{(e+p)(e'-p')}+\sqrt{(e-p)(e'+p')}}{2(e' p-e p')}\ , &
F^{(0)}_{++}(p,p') &= 0\ ,\label{tree3}
\end{align}
\begin{align}
A^{(0)}_{+-}(p,p') &= \tfrac{\a(p-p')^2}{2(e' p-e p')}\ , &
B^{(0)}_{+-}(p,p') &= p p' \tfrac{\sqrt{(e-p)(e'+p')}-\sqrt{(e+p)(e'-p')}}{2(e' p-e p')}\ , \nonumber \\
C^{(0)}_{+-}(p,p') &= \tfrac{\a p' (p'-p)}{2(e' p-e p')}\ , &
D^{(0)}_{+-}(p,p') &= \tfrac{\a p (p-p')}{2(e' p-e p')}\ , \nonumber \\
E^{(0)}_{+-}(p,p') &= 0\ , &
F^{(0)}_{+-}(p,p') &= p p' \tfrac{\sqrt{(e-p)(e'+p')}-\sqrt{(e+p)(e'-p')}}{2(e' p-e p')}\ .\label{tree4}
\end{align}

Let us now turn our attention to the scattering between a mode with mass $\a$
and one with mass $\bar\a = 1-\a$. There are no surprises regarding the
gauge-fixing parameter $a$, i.e. eq.~\eqref{gaugeS1} also holds for the two
mass scattering. We again define the parametrizing functions as
\begin{align}
{S}_{\vf_2\vf_3}^{\vf_2\vf_3}(p,p')&=A_{++}(p,p') & {S}_{\vf_2\bar{\vf_3}}^{\vf_2\bar{\vf_3}}(p,p')&=A_{+-}(p,p')\nonumber\\
{S}_{\vf_2\chi^3}^{\vf_2\chi^3}(p,p')&=B_{++}(p,p') & {S}_{\vf_2\bar{\vf_3}}^{\chi^2\bar{\chi}^3}(p,p')&=B_{+-}(p,p')\nonumber\\
{S}_{\vf_2\chi^3}^{\chi^2\vf_3}(p,p')&=C_{++}(p,p') & {S}_{\vf_2\bar{\chi}^3}^{\vf_2\bar{\chi}^3}(p,p')&=C_{+-}(p,p')\nonumber\\
{S}_{\chi^2\vf_3}^{\chi^2\vf_3}(p,p')&=D_{++}(p,p') & {S}_{\chi^2\bar{\vf_3}}^{\chi^2\bar{\vf_3}}(p,p')&=D_{+-}(p,p')\nonumber\\
{S}_{\chi^2\vf_3}^{\vf_2\chi^3}(p,p')&=E_{++}(p,p') & {S}_{\chi^2\bar{\chi}^3}^{\chi^2\bar{\chi}^3}(p,p')&=E_{+-}(p,p')\nonumber\\
{S}_{\chi^2\chi^3}^{\chi^2\chi^3}(p,p')&=F_{++}(p,p') & {S}_{\chi^2\bar{\chi}^3}^{\vf_2\bar{\vf_3}}(p,p')&=F_{+-}(p,p')
\end{align}
and the overall phase factors by setting
\begin{equation}
{\hat S}_{\vf_2\vf_3}^{\vf_2\vf_3}(p,p')=1\ , \qquad {\hat S}_{\vf_2\bar{\chi}^3}^{\vf_2\bar{\chi}^3}(p,p')=1\ .
\end{equation}

As before, the tree-level S-matrix can be extracted from the near-BMN expansion
of the exact result along with those amplitudes computed in~\cite{Sundin:2013ypa,Rughoonauth:2012qd}. For $a=\frac12$ (again the
contribution of the gauge-fixing parameter $a$ to the unitarity computation
goes through without any particular subtlety) it is given by
\begin{align}
{A}^{(0)}_{++}(p,p') &= 0\ , &
{B}^{(0)}_{++}(p,p') &=-\tfrac{ p (\bar\a p+\a p')}{2(e' p-e p')}\ , \nonumber \\
{C}^{(0)}_{++}(p,p') &= p p' \tfrac{\sqrt{(e+p)(e'-p')}+\sqrt{(e-p)(e'+p')}}{2(e' p-e p')}\ , &
{D}^{(0)}_{++}(p,p') &= -\tfrac{ p' (\bar\a p+\a p')}{2(e' p-e p')}\ , \nonumber \\
{E}^{(0)}_{++}(p,p') &= p p' \tfrac{\sqrt{(e+p)(e'-p')}+\sqrt{(e-p)(e'+p')}}{2(e' p-e p')}\ , &
{F}^{(0)}_{++}(p,p') &= -\tfrac{(p+p')(\bar\a p+\a p')}{2(e' p-e p')}\ ,\label{tree5}
\end{align}
\begin{align}
{A}^{(0)}_{+-}(p,p') &= 0\ , &
{B}^{(0)}_{+-}(p,p') &= p p' \tfrac{\sqrt{(e-p)(e'+p')}-\sqrt{(e+p)(e'-p')}}{2(e' p-e p')}\ , \nonumber \\
{C}^{(0)}_{+-}(p,p') &= -\tfrac{ p (\bar\a p-\a p')}{2(e' p-e p')} \ , &
{D}^{(0)}_{+-}(p,p') &= \tfrac{ p' (\bar\a p-\a p')}{2(e' p-e p')}\ , \nonumber \\
{E}^{(0)}_{+-}(p,p') &= -\tfrac{(p-p')(\bar\a p-\a p')}{2(e' p-e p')}\ , &
{F}^{(0)}_{+-}(p,p') &= p p' \tfrac{\sqrt{(e-p)(e'+p')}-\sqrt{(e+p)(e'-p')}}{2(e' p-e p')}\ .\label{tree6}
\end{align}

%%%%%%%%%%%%%%%%%%%%%%%%%%%%%%%%%
\subsubsection{A general tree-level S-matrix for the \txpf{$AdS_3\times S^3 \times M^4$}{AdS3 x S3 x M4} theories}\label{general}
%%%%%%%%%%%%%%%%%%%%%%%%%%%%%%%%%

Comparing the expressions \eqref{treepp}, \eqref{treepm}, \eqref{tree3},
\eqref{tree4}, \eqref{tree5} and \eqref{tree6} we notice their similarity. In
particular, they all differ from one another by a term proportional to the
identity. Therefore in this section we will introduce an additional parameter
$\b$ along with two generic masses $m$ and $m'$, such that, for particular
values of these three parameters the tree-level S-matrices are recovered. The
advantage of this approach is that it demonstrates how some quantities in the
one-loop result are common to all three theories (i.e. $\b$-independent) up to
the right assignment of the masses.

To be concrete the expression for the general tree-level S-matrix is (we use the
notation $\bar \beta=(1-\beta)$)
\begin{align}
{A}^{(0)}_{++}(p,p') &= \beta \tfrac{(p+p')(m' p+m p')}{2(e' p-e p')}\ , &
{B}^{(0)}_{++}(p,p') &=\tfrac{ ( \beta p'-\bar \b p) (m' p+m p')}{2(e' p-e p')}\ , \nonumber \\
{C}^{(0)}_{++}(p,p') &= p p' \tfrac{\sqrt{(e+p)(e'-p')}+\sqrt{(e-p)(e'+p')}}{2(e' p-e p')}\ , &
{D}^{(0)}_{++}(p,p') &= \tfrac{ (\beta p-\bar\beta p') (m' p+m p')}{2(e' p-e p')}\ , \nonumber \\
{E}^{(0)}_{++}(p,p') &= p p' \tfrac{\sqrt{(e+p)(e'-p')}+\sqrt{(e-p)(e'+p')}}{2(e' p-e p')}\ , &
{F}^{(0)}_{++}(p,p') &= -\bar\beta \tfrac{(p+p')(m' p+m p')}{2(e' p-e p')}\ ,\nonumber
\end{align}
\begin{align}
{A}^{(0)}_{+-}(p,p') &= \b \tfrac{(p-p')(m' p-m p')}{2(e' p-e p')} \ , &
{B}^{(0)}_{+-}(p,p') &= p p' \tfrac{\sqrt{(e-p)(e'+p')}-\sqrt{(e+p)(e'-p')}}{2(e' p-e p')}\ , \nonumber \\
{C}^{(0)}_{+-}(p,p') &= \tfrac{ (\bar\b p+\b p') (m p'-m' p)}{2(e' p-e p')} \ , &
{D}^{(0)}_{+-}(p,p') &= \tfrac{ (\bar\b p'+\b p) (m' p-m p')}{2(e' p-e p')}\ , \nonumber \\
{E}^{(0)}_{+-}(p,p') &= -\bar\b \tfrac{(p-p')(m' p-m p')}{2(e' p-e p')}\ , &
{F}^{(0)}_{+-}(p,p') &= p p' \tfrac{\sqrt{(e-p)(e'+p')}-\sqrt{(e+p)(e'-p')}}{2(e' p-e p')}\ . \label{interp}
\end{align}
The explicit assignments that need to be made to recover the various tree-level
S-matrices given in the previous section are shown in table \ref{assign}. For
most of the unitarity computation however, we will keep general values of
$\beta$, $m$ and $m'$ so as to better understand the dependence of the result
on these parameters.
\begin{table}[ht]
\begin{center}
\begin{tabular}[h]{|c|c|}
\hline
$(\b,m,m')$&\textbf{Theory}\\
\hline
$(0,\a,\bar\a)$&$AdS_3\times S^3\times S^3\times S^1$ (two mass scattering)\\
\hline
$(\frac12,1,1)$&$AdS_3\times S^3\times T^4$\\
\hline
$(1,\a,\a)$&$AdS_3\times S^3\times S^3\times S^1$ (one mass scattering)\\
\hline
\end{tabular}
\end{center}
\caption{Assignments of parameters for the various theories of interest.}\label{assign}
\end{table}

%%%%%%%%%%%%%%%%%%%%%%%%%%%%%%%%%
\subsection{Result from unitarity techniques for pure RR flux}
%%%%%%%%%%%%%%%%%%%%%%%%%%%%%%%%%

In this section we compute the one-loop S-matrix from unitarity methods for the
light-cone gauge fixed string theories in the $AdS_3 \times S^3 \times T^4$ and
$AdS_3 \times S^3 \times S^3 \times S^1$ backgrounds supported by RR flux. As
explained in section \ref{resstruc}, we will split the result according to
eqs.~\eqref{logsid}, \eqref{schannelid} and \eqref{tchannelid}, where we recall
that we have chosen $S^{\vf\vf}_{\vf\vf}=A_{++}(p,p')$ and $S^{\vf
\bar\psi}_{\vf \bar\psi}=C_{+-}(p,p')$ as the overall phase factors.

In the general construction described in section \ref{2mass}, we found that
when scattering a particle of mass $m$ with one of mass $m'$, the s- and
u-channel contributions are just
given in terms of the tree-level S-matrices for the same scattering
configuration. Therefore, as the logarithmic terms \eqref{logsid} and the
rational terms \eqref{schannelid} only come from the s-channel
and u-channel contributions, for these we can work with the general
($\b$-dependent) tree-level S-matrix \eqref{interp}. For the t-channel
contribution \eqref{tchannelid} one needs to combine different tree-level
S-matrices, for example the scattering of two particles of mass $m$ with the
scattering of a particle of mass $m$ with one of mass $m'$. Hence for these
terms we will need to restrict to the specific values of $\b$, $m$ and $m'$
given in table \ref{assign}.

%%%%%%%%%%%%%%%%%%%%%%%%%%%%%%%%%
\subsubsection{Coefficients of the logarithms}
%%%%%%%%%%%%%%%%%%%%%%%%%%%%%%%%%

The coefficients of the logarithmic parts were first computed in~\cite{Engelund:2013fja}. As discussed
in section \ref{resstruc} one should always be able to include the logarithmic
terms of the S-matrix in the phases. Therefore at one loop we expect them to
only contribute to the diagonal terms. This is indeed the case and furthermore, the particular
combination governing the logarithmic dependence does not depend on the diagonal
components of the tree-level S-matrix. Therefore, the one-loop logarithmic
terms following from the unitarity construction for the general tree-level
S-matrix \eqref{interp} will be $\b$-independent. Indeed,
\begin{align}
\cl_{++}(p,p')&=-\frac{p^2p'^2}{4 \p (ee'-pp'-mm')}\ , \label{lppS1}\\
\cl_{+-}(p,p')&=-\frac{p^2p'^2}{4 \p (ee'-pp'+mm')}\ , \label{lpmS1}
\end{align}
where the functions $\cl_{\s_M\s_N}$ were introduced in eq.~\eqref{phasestruc}.
Although not transparent from this expression, these functions can be expressed
as
\begin{align}
\cl_{++}(p,p')&=-\frac{1}{2\pi}{C}^{(0)}_{++}(p,p'){E}^{(0)}_{++}(p,p')\ ,\label{logs1}\\
\cl_{+-}(p,p')&=-\frac{1}{2\pi}{B}^{(0)}_{+-}(p,p'){F}^{(0)}_{+-}(p,p')\ .\label{logs2}
\end{align}

%%%%%%%%%%%%%%%%%%%%%%%%%%%%%%%%%
\subsubsection{Rational terms from the s-channel -- The imaginary part.}
%%%%%%%%%%%%%%%%%%%%%%%%%%%%%%%%%

In section \ref{resstruc} we described how the contributions to the rational
part of the S-matrix in the unitarity calculation are split between the
s-channel \eqref{schannelid} and t-channel \eqref{tchannelid}. Let us start by
considering the s-channel, for which we can work with the general
$\b$-dependent tree-level S-matrix \eqref{interp}. From eq.~\eqref{schannelid}
it is clear that we can restrict our attention to $\text{Im}(\hat{T}^{(1)})$,
where we recall that $\hat{T}^{(0)}$ and $\hat{T}^{(1)}$ are the tree-level and
one-loop terms in the expansion of the S-matrix with the overall phase factors,
$S^{\vf\vf}_{\vf\vf}=A_{++}(p,p')$ and $S^{\vf \bar\psi}_{\vf\bar\psi}=C_{+-}(p,p')$,
set to one. The result from the unitarity calculation
is \eqref{schannelid}
\begin{align}
\frac12 \hat{T}^{(0)} \cc{s} \hat{T}^{(0)} \ . \label{subtraction}
\end{align}
Below we give the components of \eqref{subtraction}. These are in perfect
agreement with the one-loop expansion of the exact results \eqref{funpp},
\eqref{funpm}, \eqref{funppS1}, \eqref{funpmS1}, \eqref{funpppS1} and
\eqref{funppmS1} for the appropriate assignments of the masses $m$ and $m'$,
see table \ref{assign}. This is not particularly surprising since the imaginary part of a one-loop S-matrix is completely determined by the optical theorem which was the starting point of our construction. The one-loop expressions are
\begin{align}
{\hat A}^{(1)}_{++}(p,p') &= 0\ , \nonumber \\
{\hat B}^{(1)}_{++}(p,p') &=\frac12\Big[\frac{p(m'p+mp')}{2 (e'p-ep')}\Big]^2+\frac12 \Big[p p' \frac{\sqrt{(e+p)(e'-p')}+\sqrt{(e-p)(e'+p')}}{2(e' p-e p')}\Big]^2\ , \nonumber \\
{\hat{C}}^{(1)}_{++}(p,p') &=-\frac12\Big[\frac{(p+p')(m'p+mp')}{2(e'p-ep')}\Big]\Big[p p' \frac{\sqrt{(e+p)(e'-p')}+\sqrt{(e-p)(e'+p')}}{2(e' p-e p')}\Big]\ , \nonumber \\
{\hat D}^{(1)}_{++}(p,p') &= \frac12\Big[\frac{p'(m'p+mp')}{2 (e'p-ep')}\Big]^2+\frac12 \Big[p p' \frac{\sqrt{(e+p)(e'-p')}+\sqrt{(e-p)(e'+p')}}{2(e' p-e p')}\Big]^2\ , \nonumber \\
{\hat E}^{(1)}_{++}(p,p') &= -\frac12\Big[\frac{(p+p')(m'p+mp')}{2(e'p-ep')}\Big]\Big[p p' \frac{\sqrt{(e+p)(e'-p')}+\sqrt{(e-p)(e'+p')}}{2(e' p-e p')}\Big]\ , \nonumber \\
{\hat F}^{(1)}_{++}(p,p') &= \frac12\Big[\frac{(p+p')(m'p+mp')}{2(e'p-ep')}\Big]^2\ . \label{onelooppp}\\
{\hat A}^{(1)}_{+-}(p,p')&=\frac12\Big[\frac{p(m'p-mp')}{2 (e'p-ep')}\Big]^2+\frac12 \Big[p p' \frac{\sqrt{(e+p)(e'-p')}-\sqrt{(e-p)(e'+p')}}{2(e' p-e p')}\Big]^2\ , \nonumber \\
{\hat B}^{(1)}_{+-}(p,p')&=-\frac12\Big[\frac{(p+p')(m'p-mp')}{2(e'p-ep')}\Big]\Big[p p' \frac{\sqrt{(e+p)(e'-p')}-\sqrt{(e-p)(e'+p')}}{2(e' p-e p')}\Big]\ ,\nonumber \\
{\hat C}^{(1)}_{+-}(p,p')&=0\ ,\nonumber \\
{\hat D}^{(1)}_{+-}(p,p')&=\frac12\Big[\frac{(p+p')(m'p-mp')}{2(e'p-ep')}\Big]^2\ ,\nonumber \\
{\hat E}^{(1)}_{+-}(p,p')&=\frac12\Big[\frac{p'(m'p-mp')}{2 (e'p-ep')}\Big]^2+\frac12 \Big[p p' \frac{\sqrt{(e+p)(e'-p')}-\sqrt{(e-p)(e'+p')}}{2(e' p-e p')}\Big]^2\ ,\nonumber \\
{\hat F}^{(1)}_{+-}(p,p')&=-\frac12\Big[\frac{(p+p')(m'p-mp')}{2(e'p-ep')}\Big]\Big[p p' \frac{\sqrt{(e+p)(e'-p')}-\sqrt{(e-p)(e'+p')}}{2(e' p-e p')}\Big]\ .\label{onelooppm}
\end{align}
Although there are simpler ways to express this result, we have chosen this
form in order to explicitly show the connection with the tree-level functions.
The $\b$-independence of \eqref{onelooppp} and \eqref{onelooppm} is expected
since $\b$ appears only in the phases. As explained earlier in this section
and in section \ref{resstruc}, to check the s-channel rational terms we do not need
to consider the overall phase factors and hence they have been set to one.

Note that expressions for the components of $\frac12 T^{(0)} \cc{s}
T^{(0)}$ in terms of tree-level functions are given in~\cite{Engelund:2013fja}
for $AdS_3\times S^3\times T^4$. These formulae also hold for the general
tree-level S-matrix \eqref{interp}, however, they will depend on $\b$, which
drops out only if we consider $\frac12 \hat{T}^{(0)} \cc{s} \hat{T}^{(0)}$ as
above.  To see explicitly how this works let us consider $F_{++}$.\footnote{For
the remainder of this section the dependence on $p$ and $p'$ is understood.\vspace{2pt}}
From~\cite{Engelund:2013fja} the one-loop expression for $F_{++}$ is simply
given by
\begin{equation}
F^{(1)}_{++}=\frac12 [F^{(0)}_{++}]^2\ ,
\end{equation}
however when we consider \eqref{subtraction} (taking into account that
$\phi^{(0)}_{++}=A^{(0)}_{++}$) we find
\begin{equation}
\hat F^{(1)}_{++}=\frac12 [\hat F^{(0)}_{++}]^2=\frac12 [F^{(0)}_{++}-A_{++}^{(0)}]^2\ .
\end{equation}
Comparing the expressions for $F^{(0)}_{++}$ and $A_{++}^{(0)}$ we can then
observe the cancellation of $\b$. A similar story holds for the other
components
\begin{align}
{\hat A}^{(1)}_{++} &= 0 \ ,&
{\hat B}^{(1)}_{++} &=\frac12 [B^{(0)}_{++}-A_{++}^{(0)}]^2+\frac12 {C}^{(0)}_{++} {E}^{(0)}_{++} \ , \nonumber \\
{\hat{C}}^{(1)}_{++} &=\frac12 [B^{(0)}_{++}+D_{++}^{(0)}-2 A_{++}^{(0)}]\,{C}^{(0)}_{++}\ , &
{\hat D}^{(1)}_{++} &= \frac12 [D^{(0)}_{++}-A_{++}^{(0)}]^2+\frac12 {C}^{(0)}_{++} {E}^{(0)}_{++}\ , \nonumber \\
{\hat E}^{(1)}_{++} &= \frac12 [B^{(0)}_{++}+D_{++}^{(0)}-2 A_{++}^{(0)}]\,{E}^{(0)}_{++}\ , &
{\hat F}^{(1)}_{++} &= \frac12 [F^{(0)}_{++}-A_{++}^{(0)}]^2\ . \label{combTLpp}\\
{\hat A}^{(1)}_{+-}&=\frac12 [A^{(0)}_{+-}-C_{+-}^{(0)}]^2+\frac12 {B}^{(0)}_{+-} {F}^{(0)}_{+-} \ , &
{\hat B}^{(1)}_{+-}&=\frac12 [A^{(0)}_{+-}+E_{+-}^{(0)}-2 C_{+-}^{(0)}]\,{B}^{(0)}_{+-}\ ,\nonumber \\
{\hat C}^{(1)}_{+-}&=0\ ,&
{\hat D}^{(1)}_{+-}&=\frac12[D^{(0)}_{+-}-C_{+-}^{(0)}]^2\ ,\nonumber \\
{\hat E}^{(1)}_{+-}&=\frac12 [E^{(0)}_{+-}-C_{+-}^{(0)}]^2+\frac12 {B}^{(0)}_{+-} {F}^{(0)}_{+-} \ ,&
{\hat F}^{(1)}_{+-}&=\frac12 [A^{(0)}_{+-}+E_{+-}^{(0)}-2 C_{+-}^{(0)}]\,{F}^{(0)}_{+-}\ . \label{combTLpm}
\end{align}
The validity of these relations is rather general and can be applied to any
S-matrix with the same underlying structure. In particular, this allows us to
use them for the mixed flux case in section \ref{sec:phq}.

%%%%%%%%%%%%%%%%%%%%%%%%%%%%%%%%%
\subsubsection{The t-channel contribution and the dressing phases}\label{sec:tchannel}
%%%%%%%%%%%%%%%%%%%%%%%%%%%%%%%%%

As explained in section \ref{sec:unitarity} the t-channel cut requires a non-trivial
generalization of the procedure used for the $AdS_5\times S^5$ case. Furthermore, the t-channel cut for the scattering of
two masses depends on the tree-level S-matrices for the scattering of the same
and different masses. Therefore, in this section it only makes sense to work
with the parameters $\b$, $m$ and $m'$ for the three cases of interest, as
given in table \ref{assign}. Inputting the tree-level S-matrices
\eqref{treepp}, \eqref{treepm}, \eqref{tree3}, \eqref{tree4}, \eqref{tree5} and
\eqref{tree6} into eq~.\eqref{resultsummt} and splitting the result as in
eq.~\eqref{tchannelid} we find for all three scattering processes ($AdS_3
\times S^3 \times T^4$, $AdS_3 \times S^3 \times S^3 \times S^1$ same mass and
$AdS_3 \times S^3 \times S^3 \times S^1$ different mass) the one-loop phases
can be written in the following general form
\begin{align}
\phi^{(1)}_{++}(p,p')&=\frac{p\, p' (m'p+mp')^2}{8 \p \,m m' (e' p-e p') }\ ,\label{eq:phasepp}\\
\phi^{(1)}_{+-}(p,p')&=-\frac{p\, p' (m'p-mp')^2}{8 \p \, m m' (e' p-e p') }\ .\label{eq:phasepm}
\end{align}
The real part of the one-loop cut-constructible S-matrix that is not part of
the overall phase factors is given by
\begin{equation}\label{eq:counterterm}
\text{Re}(\hat{T}^{(1)})|_{\textup{unit.}}=\frac{1}{4\p} \, |1-2\,\b|\left(\frac{{p}^2}{m} +\frac{p'^2}{m'}\right)\, T^{(0)}\ .
\end{equation}
It is important to emphasise that even though we have written them in terms of
$\b$, $m$ and $m'$ the results \eqref{eq:phasepp}, \eqref{eq:phasepm} and
\eqref{eq:counterterm} are only valid for the assignments in table
\ref{assign}.

Two comments are in order here. First, eq.~\eqref{eq:counterterm} is
proportional to $|1-2\b|$. Therefore, this term vanishes for $AdS_3\times S^3
\times T^4$, but does not for $AdS_3\times S^3 \times S^3\times S^1$. However,
we should recall that this is only the contribution to
$\text{Re}(\hat{T}^{(1)})$ coming from unitarity and there are potentially
additional terms arising from external leg corrections \eqref{tchannelid}.
Indeed, one of the main differences between $AdS_3\times S^3 \times T^4$ and
$AdS_3\times S^3 \times S^3\times S^1$ is that the light-cone gauge fixed
Lagrangian of the latter has cubic terms. Furthermore, the tree-level form
factor for one off-shell and two on-shell particles is non-zero and as a
consequence non-trivial external leg corrections are already present at one
loop in the unitarity construction, as described in section \ref{sec:elc}. As
we will see in the following section these precisely cancel
\eqref{eq:counterterm} and re-establish agreement with the exact
result.\footnote{Let us point out that a term like \eqref{eq:counterterm} in
the one-loop S-matrix would prevent the latter from satisfying the Yang-Baxter
equation, conflicting with the integrability of the theory.\vspace{2pt}}$^{,}$\footnote{It
is interesting to note that in the two loop near-flat-space computation of~\cite{Klose:2007rz} for the $AdS_5 \times S^5$ light-cone gauge S-matrix the
external leg corrections also cancelled unwanted terms arising from t-channel
graphs and in the one-loop Feynman diagram computation of~\cite{Roiban:2014cia} external leg corrections were a key ingredient for the cancellation of UV divergences.}

The second comment concerns eqs.~\eqref{lppS1}, \eqref{lpmS1},
\eqref{eq:phasepp} and \eqref{eq:phasepm}, which combined have a natural
interpretation as the one-loop contributions to the phases. It is interesting
to note that they are independent of $\b$, indicating that the phases for all
three scattering processes should be related. This agrees with the
semiclassical computation~\cite{Abbott:2013kka}.\footnote{In~\cite{Abbott:2013ixa} the author states that the one-loop dressing phase of
$AdS_3\times S^3 \times S^3\times S^1$ is half that of $AdS_3\times S^3 \times
T^4$. This is consistent given that we are considering the factorized S-matrix
for $AdS_3\times S^3 \times T^4$.\vspace{2pt}} A natural question is whether this relation
extends to all orders in the coupling. To facilitate comparison with the literature~\cite{Borsato:2013hoa}
we will rewrite the result in terms of the standard strong coupling variables
$x$ and $y$, which we have defined in \eqref{xexp} and \eqref{xofp}
\begin{align}
\vp^{(1)}_{++}(p,p')&=-\tfrac{m m'}{\pi}\tfrac{x^2}{x^2-1}\tfrac{y^2}{y^2-1}\bigg[\tfrac{(x+y)^2 (1-\tfrac{1}{xy})}{(x^2-1)(x-y)(y^2-1)} + \tfrac{2}{(x-y)^2} \log\left(\tfrac{x+1}{x-1}\tfrac{y-1}{y+1}\right)\bigg] \ , \label{1Lphasepp} \\
\vp^{(1)}_{+-}(p,p')&=-\tfrac{m m'}{\pi}\tfrac{x^2}{x^2-1}\tfrac{y^2}{y^2-1}\bigg[\tfrac{(xy+1)^2 (\tfrac{1}{x}-\tfrac{1}{y})}{(x^2-1)(xy-1)(y^2-1)} + \tfrac{2}{(xy-1)^2} \log\left(\tfrac{x+1}{x-1}\tfrac{y-1}{y+1}\right)\bigg] \label{1Lphasepm} \ .
\end{align}
Here $x$ corresponds to momentum $p$ with mass $m$ and $y$ to momentum $p'$
with mass $m'$. Finally, let us stress again that this expression is valid for
all three cases summarized in table \ref{assign}. In particular, for $m=m'=1$
this is consistent with \eqref{AFSstrong}, where the overall sign is
compensated by the fact that $e^{i\vartheta_{\s_M\s_N}(p,p')}\sim
S_{\s_M\s_N}^{11}(p,p')^{-1}$, see eqs.~\eqref{Spp} and \eqref{Spm}.

%%%%%%%%%%%%%%%%%%%%%%%%%%%%%%%%%
\subsubsection{External leg corrections for \txpf{$AdS_3\times S^3 \times S^3 \times S^1$}{AdS3 x S3 x S3 x S1}}\label{extlegs}
%%%%%%%%%%%%%%%%%%%%%%%%%%%%%%%%%

In this section we focus on the $AdS_3\times S^3 \times S^3 \times S^1$
background for which the unwanted term \eqref{eq:counterterm} is present. With
the aim of interpreting this missing term as a contribution cancelled by
external leg corrections let us review the results of~\cite{Sundin:2012gc,
Sundin:2014sfa} for the one-loop two-point functions. The near-BMN expansion of
the light-cone gauge fixed Lagrangian can be schematically written as
\begin{equation}
\mathcal{L}=\mathcal{L}_2+h^{-\frac12}\mathcal{L}_3+h^{-1}\mathcal{L}_4 + \ldots \ .
\end{equation}
The quadratic part is given by\,\footnote{Here we stress that, although the
theory is not Lorentz invariant beyond quadratic order, we are formally
rearranging the fermions into doublets for notational and computational
convenience.\vspace{2pt}}
\begin{align}
\mathcal{L}_2&= \bar{\chi}^a(i\slashed{\pa}-m_a) \chi^a+|\pa \varphi_a|^2-m_a^2 |\varphi_a|^2 \ ,
\end{align}
where our conventions are summarized in appendix \ref{notation} and we have
introduced the index $a=1,\ldots,4$ with the respective masses listed in table
\ref{masses}. The cubic Lagrangian~\cite{Sundin:2012gc,Sundin:2014sfa} is given
by
\begin{align}\label{cublag}
\mathcal{L}_3= \sqrt{\frac{\a \bar\a}{2}}&\Big[
{(\chi^1)}{}^T \gamma^3 (i \slashed{\pa}-\a)\, \varphi_2\, \chi^3 
-i{(\chi^1)}{}^T \gamma^3 (i \slashed{\pa}-\bar\a)\, \varphi_3\, \chi^2 
-2{(\chi^2)}{}^T \gamma^1\pa_1 \varphi_1\, \chi^3 \nonumber \\
& +{\bar{\chi}}^2\gamma^0(i \slashed{\pa}-\a)\, \varphi_2\, \chi^4  \vphantom{\Big]}
+i {\bar{\chi}}^3\gamma^0(i \slashed{\pa}-\bar\a)\, \varphi_3\, \chi^4  \\
& - \left(\bar{\chi}^2 (1-\gamma^3) \chi^2-\bar{\chi}^3(1-\gamma^3) \chi^3
+ 2 \a|\varphi_2|^2-2\bar\a |\varphi_3|^2\right)\pa_0 \varphi_4+\text{h.c.} \Big]\ .\nonumber
\end{align}
Let us start by focusing on the tree-level processes following from the cubic
Lagrangian. The only processes allowed by two-dimensional kinematics involve a
particle of mass 1 decaying into a particle of mass $\a$ and one of mass $\bar\a$
and its reverse.\footnote{Diagrams involving one massless leg are ruled out by
two-dimensional kinematics. In the cubic Lagrangian \eqref{cublag} the massless
modes always couple to massive modes of equal mass. It then follows that
the on-shell condition implies that the massless leg carries vanishing
momentum.\vspace{2pt}} The Feynman rules associated to the relevant vertices are
\begin{align}\nonumber
&\begin{tikzpicture}[scale=1,vertex/.style={circle,fill=black,thick,inner sep=2pt}]
\node (a1) at (-2,0) {$\varphi_1$};
\node (c) at (0,0) [vertex] {};
\node (a2) at (1.5,1) {$\chi^2$};
\node (a3) at (1.5,-1) {$\chi^3$};
\draw[dashed] (a1)--(c);
\begin{scope}[decoration={markings,mark = at position 0.5 with {\arrow[scale=1.4]{latex}}}]
\draw[postaction={decorate}] (a2)--(c);
\draw[postaction={decorate}] (a3)--(c);
\end{scope}
\end{tikzpicture}
&\quad&
\begin{tikzpicture}[scale=1,vertex/.style={circle,fill=black,thick,inner sep=2pt}]
\node (a1) at (-2,0) {$\chi^1$};
\node (c) at (0,0) [vertex] {};
\node (a2) at (1.5,1) {$\varphi_2$};
\node (a3) at (1.5,-1) {$\chi^3$};
\draw[dashed] (a2)--(c);
\begin{scope}[decoration={markings,mark = at position 0.5 with {\arrow[scale=1.4]{latex}}}]
\draw[postaction={decorate}] (a1)--(c);
\draw[postaction={decorate}] (a3)--(c);
\end{scope}
\end{tikzpicture}
&\quad&
\begin{tikzpicture}[scale=1,vertex/.style={circle,fill=black,thick,inner sep=2pt}]
\node (a1) at (-2,0) {$\chi^1$};
\node (c) at (0,0) [vertex] {};
\node (a2) at (1.5,1) {$\chi^2$};
\node (a3) at (1.5,-1) {$\varphi_3$};
\draw[dashed] (a3)--(c);
\begin{scope}[decoration={markings,mark = at position 0.5 with {\arrow[scale=1.4]{latex}}}]
\draw[postaction={decorate}] (a2)--(c);
\draw[postaction={decorate}] (a1)--(c);
\end{scope}
\end{tikzpicture}
\\
&\quad \sqrt{\frac{\a \bar\a}{2}}\, 2\,i\, \g^1 p_1 \ ,
&\quad& \quad -\sqrt{\frac{\a \bar\a}{2}}\, i\, \g^3 (\slashed{\rmp}_2+\a) \ ,
&\quad& \quad -\sqrt{\frac{\a \bar\a}{2}}\, \g^3 (\slashed{\rmp}_3+\bar\a)\ .\label{eq:decayprocess}
\end{align}
To obtain the amplitude one should contract the external legs with the fermion
polarizations and enforce the on-shell condition. The three diagrams share the
same on-shell kinematics, i.e. denoting the incoming momentum of the heavy
particle (with mass $m_1=1$) as $\rmp_1$, the outgoing momenta of the light
particles are given by $\rmp_2=\tfrac{m_2}{m_1} \rmp_1$ and $\rmp_3=\tfrac{m_3}{m_1} \rmp_1$, where
$m_3=m_1-m_2$.\footnote{This is true under the assumption of a relativistic
dispersion relation, which in this case holds just at tree level.\vspace{2pt}} Using the
property that $v(k \rmp_1)=\sqrt{k}v(\rmp_1)$ (see eq.~\eqref{polv}), it is
clear that both the second and the third diagrams vanish as
$(\slashed{\rmp}+1)v(\rmp) = 0$. Furthermore, the first diagram is also
identically zero as a consequence of the identity $v(\rmp)^T \g^1 v(\rmp)=0$.

One may ask how this is compatible with the result of~\cite{Sundin:2012gc}
where the authors find a non-vanishing expression for the one-loop correction
to the propagators coming from the graph formed of two three-point vertices.
Focusing on the one-loop contribution to the self-energy of the heavy boson the
result of~\cite{Sundin:2012gc} reads
\begin{equation}\label{selfenergy}
\Sigma^{(1)}_0(p) = i\braket{\varphi_1 \bar \varphi_1}^{(1)}=\frac{1}{\pi^2}(\a \log\a+\bar\a \log\bar\a)\, p^2\ . %v2
\end{equation}
This result is obtained setting $\rmp^2=1$ (i.e. putting the propagator
on-shell) and its dependence on $p$ is a consequence of the lack of Lorentz
invariance. In a unitarity computation with the setup described in section
\ref{sec:elc} the two tree-level form factors appearing in figure \ref{2ptfun}
would be vanishing in the strict on-shell limit and this contribution would not
be caught. However, as discussed in section \ref{sec:elc}, our treatment
ignored any kind of tadpole diagram contributing to the external leg
corrections. Moreover, as pointed out in~\cite{Sundin:2012gc} the contribution
\eqref{selfenergy} can be understood as the one-loop term in the expansion of
$h(\z)$, an effective coupling featuring all the integrability-based calculations (see section \ref{sec:h} for an extensive discussion of the analogue quantity in \adscp background). This effective coupling enters the dispersion relation of the worldsheet excitations as one easily understands from equation \eqref{selfenergy} and perturbatively it is translated in a shift of the coupling. Combining this observation with the fact that, in a number of other examples we
have considered, ignoring tadpole diagrams gives the S-matrix up to corrections
in $h(\z)$ we may argue that these are coming from tadpole diagrams whose
analysis would require the introduction of a regularization (see also~\cite{Sundin:2012gc}). This is therefore an additional indication that
unitarity techniques, neglecting tadpoles, are blind to shifts in the coupling.

Therefore, we will consider the following alternative question. Are there
external leg corrections that are caught by unitarity and which are relevant
for the one-loop calculation? In the S-matrix computation we consider
scattering processes for which the external legs have masses $\a$ or $\bar\a$.
Therefore, the external leg corrections we compute come from diagrams similar
to the first graph in figure \ref{graphs} with masses $m_1=1$ and $m_2=\a$ or %v2
$m_2=\bar\a$.

We start by considering an external leg of mass $\a$. Using the vertices in
eq.~\eqref{eq:decayprocess} we find the following form factors
{\allowdisplaybreaks
\begin{align}
&\begin{tikzpicture}[scale=1,vertex/.style={circle,fill=black,thick,inner sep=2pt},baseline=-5pt]
\node (a1) at (-2,0) {$\varphi_2$};
\node (c) at (0,0) [vertex] {};
\node (a2) at (1.5,1) {$\chi^1$};
\node (a3) at (1.5,-1) {$\chi^3$};
\draw[dashed] (a1)--(c);
\begin{scope}[decoration={markings,mark = at position 0.5 with {\arrow[scale=1.4]{latex}}}]
\draw[postaction={decorate}] (a2)--(c);
\draw[postaction={decorate}] (a3)--(c);
\end{scope}
\begin{scope}[decoration={markings,mark = at position 1 with {\arrow[scale=1.4]{to}}}]
\draw[postaction={decorate}] (-1.3,-0.25)--(-1,-0.25) node [anchor=north] {$\rmp$}--(-0.7,-0.25);
\draw[postaction={decorate}] (0.2,1/3+0.2)--(0.45,1/2+0.2) node [anchor=south] {$\rml_1$}--(0.7,2/3+0.2);
\draw[postaction={decorate}] (0.7,-2/3-0.2)--(0.45,-1/2-0.2) node[anchor=north] {$\rml_2$}--(0.2,-1/3-0.2);
\end{scope}
\end{tikzpicture} \quad = \quad i\,\sqrt{\frac{\a \bar\a}{2}}\, v(\rml_1)^{T} \gamma^3 (\slashed{\rmp}-\a) u(\rml_2)\ ,\label{diagscalar}
\\ &
\begin{tikzpicture}[scale=1,vertex/.style={circle,fill=black,thick,inner sep=2pt},baseline=-5pt]
\node (a1) at (-2,0) {$\chi^2$};
\node (c) at (0,0) [vertex] {};
\node (a2) at (1.5,1) {$\varphi_1$};
\node (a3) at (1.5,-1) {$\chi^3$};
\draw[dashed] (a2)--(c);
\begin{scope}[decoration={markings,mark = at position 0.5 with {\arrow[scale=1.4]{latex}}}]
\draw[postaction={decorate}] (a1)--(c);
\draw[postaction={decorate}] (a3)--(c);
\end{scope}
\begin{scope}[decoration={markings,mark = at position 1 with {\arrow[scale=1.4]{to}}}]
\draw[postaction={decorate}] (-1.3,-0.25)--(-1,-0.25) node [anchor=north] {$\rmp$}--(-0.7,-0.25);
\draw[postaction={decorate}] (0.2,1/3+0.2)--(0.45,1/2+0.2) node [anchor=south] {$\rml_1$}--(0.7,2/3+0.2);
\draw[postaction={decorate}] (0.7,-2/3-0.2)--(0.45,-1/2-0.2) node[anchor=north] {$\rml_2$}--(0.2,-1/3-0.2);
\end{scope}
\end{tikzpicture} \quad = \quad i\, \sqrt{\frac{\a \bar\a}{2}}\, 2\, u(\rml_2)^{T}\g^1 l_1 u(\rmp)\ ,\label{diagfermion1}
\\ &
\begin{tikzpicture}[scale=1,vertex/.style={circle,fill=black,thick,inner sep=2pt},baseline=-5pt]
\node (a1) at (-2,0) {$\chi^2$};
\node (c) at (0,0) [vertex] {};
\node (a2) at (1.5,1) {$\chi^1$};
\node (a3) at (1.5,-1) {$\varphi_3$};
\draw[dashed] (a3)--(c);
\begin{scope}[decoration={markings,mark = at position 0.5 with {\arrow[scale=1.4]{latex}}}]
\draw[postaction={decorate}] (a2)--(c);
\draw[postaction={decorate}] (a1)--(c);
\end{scope}
\begin{scope}[decoration={markings,mark = at position 1 with {\arrow[scale=1.4]{to}}}]
\draw[postaction={decorate}] (-1.3,-0.25)--(-1,-0.25) node [anchor=north] {$\rmp$}--(-0.7,-0.25);
\draw[postaction={decorate}] (0.2,1/3+0.2)--(0.45,1/2+0.2) node [anchor=south] {$\rml_1$}--(0.7,2/3+0.2);
\draw[postaction={decorate}] (0.7,-2/3-0.2)--(0.45,-1/2-0.2) node[anchor=north] {$\rml_2$}--(0.2,-1/3-0.2);
\end{scope}
\end{tikzpicture} \quad = \quad \sqrt{\frac{\a \bar\a}{2}}\, v(\rml_1)^{T}\g^3(\slashed{\rml}_2-\bar\a)u(\rmp)\ .\label{diagfermion2}
\end{align}}

To apply the construction outlined in section \ref{sec:elc} we need to
compute eq.~\eqref{ffsquare}. In particular, we are interested in expanding the
form factor squared around the on-shell condition. Since we already know that
the tree-level form factor vanishes on-shell, to get the first order in the
expansion there is no need to also expand the integral, i.e. it can be
evaluated strictly on-shell
\begin{align}
I(\a^2,1,\bar\a)=-\frac{i}{4\p\bar\a}\ .
\end{align}
Squaring the form factor \eqref{diagscalar} and expanding around the on-shell
condition we find\,\footnote{A minus sign is included to take account of the
fermion loop.\vspace{2pt}}
\begin{equation}\label{eq:scalarext}
-i\Sigma^{(1)}_{1,\vf_2}(\rmp)=\frac{i}{4\p\a}\, p^2\ .
\end{equation}
Comparing to \eqref{eq:counterterm} this result is promising. However,
\eqref{eq:scalarext} holds only when the external leg is a boson. A non-trivial
check of our procedure is that when the external leg is a fermion the
correction, which comes from two terms associated to the diagrams
\eqref{diagfermion1} and \eqref{diagfermion2}, is exactly the same as for the
boson, i.e.
\begin{equation}\label{eq:fermionext}
-i\Sigma^{(1)}_{1,\chi^2}(\rmp)=\frac{i}{4\p\a}\, p^2\ .
\end{equation}
One might have expected this from worldsheet supersymmetry as discussed in~\cite{Sundin:2014sfa}. Here we have computed the external leg corrections for a
particle of mass $\a$. From the symmetry of the Lagrangian, it is clear
that the result for a particle of mass $\bar\a$ is just given by the
replacement $\a\to\bar\a$.

Once the external leg contributions are computed we can apply
eq.~\eqref{important} to find their contribution to the one-loop S-matrix. To
be general, let us consider the scattering of a particle of mass $m$ with a
particle of mass $m'$. Our result then reads
\begin{equation}
T^{(1)}_{ext}=-\frac{1}{4\pi}\bigg(\frac{p^2}{m}+\frac{p'^2}{m'}\bigg) T^{(0)}\ .
\end{equation}
This contribution exactly cancels \eqref{eq:counterterm} for $\beta=0$ and
$\beta=1$. These are precisely the values associated to the single and mixed
mass scattering processes for $AdS_3\times S^3 \times S^3 \times S^1$, and
hence we have established agreement between the unitarity calculation and the
exact result up to shifts in the coupling.

%%%%%%%%%%%%%%%%%%%%%%%%%%%%%%%%%
\subsection{Tree-level S-matrix for mixed flux}\label{sec:TLmixed}
%%%%%%%%%%%%%%%%%%%%%%%%%%%%%%%%%

The quadratic light-cone gauge fixed action for the $AdS_3 \times S^3 \times
T^4$ background supported by mixed flux again describes $4+4$ massive and $4+4$
massless fields. As usual we restrict ourselves to considering the scattering
of two massive excitations to two massive excitations. Following the RR case
described in section \ref{sec:t4} we group the particle content of the massive
sector into $2+2$ complex degrees of freedom (to recall, $\Phi_{\vf\vf}$,
$\Phi_{\psi\psi}$, $\Phi_{\vf\psi}$, $\Phi_{\psi\vf}$, and their complex
conjugates $\Phi_{\bar \vf\bar\vf}$, $\Phi_{\bar \psi \bar \psi}$, $\Phi_{\bar
\vf\bar \psi}$, $\Phi_{\bar \psi\bar \vf}$). The presence of the NSNS flux
then breaks the charge conjugation invariance, such that the near-BMN
dispersion relations for these complex degrees of freedom are given by
\begin{equation}\label{eq:dr}
e_\pm = \sqrt{(1 -q^2) + (p \pm q)^2} \ .
\end{equation}
where $+$ corresponds to $\Phi_{\vf\vf}$, $\Phi_{\psi\psi}$, $\Phi_{\vf\psi}$,
$\Phi_{\psi\vf}$ and $-$ to their complex conjugates.

As for $q=0$ the S-matrix factorizes as in \eqref{factorize} and the general
structure of the factorized S-matrix takes the form given in \eqref{structure}
with $\s_\vf=\s_\psi=+$ and $\s_{\bar{\vf}}=\s_{\bar{\psi}}=-$. Furthermore,
the construction outlined in section \ref{sec:unitarity} still gives the same one-loop
result whether we consider the factorized or full S-matrix. Therefore, for
simplicity we will again work with the former. Due to the lack of charge
conjugation symmetry all four phases are now different.  However, charge
conjugation along with formally sending $q \to -q$ is a symmetry and hence
$\phi_{++}=\phi_{--}|_{q\to -q}$ and $\phi_{+-}=\phi_{-+}|_{q\to -q}$.
Similarly, for the functions $\cl_{\s_M\s_N}$ we have $\cl_{++}=\cl_{--}|_{q\to
-q}$ and $\cl_{+-}=\cl_{-+}|_{q\to -q}$. Therefore, in the following we will
again focus on the $++$ and $+-$ sectors.
%v2-start
The dependence on the gauge-fixing parameter $a$ is also modified in the
following natural way
%v2-end
\begin{equation}\label{qgauge}
\exp\big[\frac{i}{2}(a-\tfrac12)(\o_{\s_N}'p-\o_{\s_M} p')\big] \ ,
\end{equation}
where the all-order energies $\o_\pm$ are defined in appendix \ref{notations2}.
As discussed in section \ref{resstruc} we choose the overall phase factors by
setting particular components of $\hat{S}_{MN}^{PQ}$ to one
\begin{align}
\hat{S}_{\vf\vf}^{\vf\vf}(p,p')&=1\ , \qquad \hat{S}_{\vf\bar{\psi}}^{\vf\bar{\psi}}(p,p')=1\ . \label{qchoice}
\end{align}

The parametrizing functions of the S-matrix are defined as
\begin{align}
{S}_{\vf\vf}^{\vf\vf}(p,p')&=A_{++}(p,p') & {S}_{\vf\bar{\vf}}^{\vf\bar{\vf}}(p,p')&=A_{+-}(p,p')\nonumber\\
{S}_{\vf\psi}^{\vf\psi}(p,p')&=B_{++}(p,p') & {S}_{\vf\bar{\vf}}^{\psi\bar{\psi}}(p,p')&=B_{+-}(p,p')\nonumber\\
{S}_{\vf\psi}^{\psi\vf}(p,p')&=C_{++}(p,p') & {S}_{\vf\bar{\psi}}^{\vf\bar{\psi}}(p,p')&=C_{+-}(p,p')\nonumber\\
{S}_{\psi\vf}^{\psi\vf}(p,p')&=D_{++}(p,p') & {S}_{\psi\bar{\vf}}^{\psi\bar{\vf}}(p,p')&=D_{+-}(p,p')\nonumber\\
{S}_{\psi\vf}^{\vf\psi}(p,p')&=E_{++}(p,p') & {S}_{\psi\bar{\psi}}^{\psi\bar{\psi}}(p,p')&=E_{+-}(p,p')\nonumber\\
{S}_{\psi\psi}^{\psi\psi}(p,p')&=F_{++}(p,p') & {S}_{\psi\bar{\psi}}^{\vf\bar{\vf}}(p,p')&=F_{+-}(p,p')\label{qcorr}
\end{align}

The input needed for the unitarity construction of section \ref{sec:unitarity} is the
tree-level S-matrix. Various tree-level components were computed directly in~\cite{Hoare:2013pma}. These are in agreement with the near-BMN expansion of the
exact result \eqref{qfunpp}, \eqref{qfunpm}. The remaining
components of the tree-level S-matrix can then be fixed from the expansion of
the exact result. As in the RR case, here we shall present the result in the
gauge $a=\frac12$ -- the dependence on $a$ goes through the unitarity
procedure without any particular subtlety, i.e. it exponentiates as in
eq.~\eqref{qgauge}. The tree-level S-matrix reads \begin{align}
A^{(0)}_{++}(p,p') &= - F^{(0)}_{++}(p,p') = \tfrac{(p+p')(e'_+p+e_+p')}{4\,(p-p')}\ , \qquad \nonumber \\
C^{(0)}_{++}(p,p') &= E^{(0)}_{++}(p,p') = p\, p' \tfrac{\sqrt{(e_++p+q)(e'_++p'+q)}+\sqrt{(e_+-p-q)(e'_+-p'-q)}}{2(p-p')}\ , & \nonumber \\
B^{(0)}_{++}(p,p') &= - D^{(0)}_{++}(p,p') = -\tfrac{e'_+ p-e_+ p'}4\ ,\label{qtrepp}
\end{align}
\begin{align}
A^{(0)}_{+-}(p,p') &= - E^{(0)}_{+-}(p,p') = \tfrac{(p-p')(e'_-p+e_+p')}{4\,(p+p')}\ , \qquad \nonumber \\
B^{(0)}_{+-}(p,p') &= F^{(0)}_{+-}(p,p') = p\, p'\tfrac{\sqrt{(e_+-p-q)(e'_--p'+q)}-\sqrt{(e_++p+q)(e'_-+p'-q)}}{2(p+p')}\ , & \nonumber \\
C^{(0)}_{+-}(p,p') &= - D^{(0)}_{+-}(p,p') = -\tfrac{e'_- p-e_+ p'}4\ ,
\label{qtrepm}
\end{align}
This form of writing the tree-level S-matrix elements is the simplest for the
purposes of introducing the parameter $q$. Agreement with \eqref{treepp} and
\eqref{treepm} for $q=0$ can be checked using the dispersion relation.

%%%%%%%%%%%%%%%%%%%%%%%%%%%%%%%%%
\subsection{Result from unitarity techniques for mixed flux}\label{sec:phq}
%%%%%%%%%%%%%%%%%%%%%%%%%%%%%%%%%

In this section we compute the one-loop S-matrix from unitarity methods for the
light-cone gauge fixed string theory in the $AdS_3 \times S^3 \times T^4$
background supported by a mix of RR and NSNS fluxes. Again, we will split the
result according to eqs.~\eqref{logsid}, \eqref{schannelid} and
\eqref{tchannelid}, where we recall that we have chosen
$S^{\vf\vf}_{\vf\vf}=A_{++}(p,p')$ and $S^{\vf \bar\psi}_{\vf\bar\psi}=C_{+-}(p,p')$
as the overall phase factors.

There is a subtlety regarding the unitarity computation in that the near-BMN
dispersion relations \eqref{eq:dr} are not the standard relativistic ones that
we assumed for the derivation in section \ref{sec:unitarity}. To bypass this problem,
we will first shift the momenta as
\begin{equation} \begin{split}\label{shiftmo}
& p \to p - q \text{ for particles and } p \to p + q \text{ for antiparticles} \ ,
\end{split} \end{equation}
so as to put the near-BMN dispersion relations into the standard form.
At the level of the light-cone gauge fixed Lagrangian this just amounts to a
$\s$-dependent rotation of the complex fields, where $\s$ is the spatial
coordinate on the worldsheet~\cite{Hoare:2013pma}. We can then
straightforwardly use the construction of section \ref{sec:unitarity} for two
particles of mass $\sqrt{1-q^2}$. To construct the one-loop result, we should
then conclude by undoing the shift \eqref{shiftmo}. An analogous approach was used
in~\cite{Engelund:2013fja} to compute the logarithmic terms.

Following this procedure it is apparent that the logarithms appearing in the
one-loop integrals, when written in terms of energy and momentum, are different
for each of the four sectors
\begin{equation}
\theta_{\pm\pm} = \as\big(\frac{e'_\pm(p\pm q)-e_\pm(p'\pm q)}{1-q^2}\big)\ , \qquad
\theta_{\pm\mp} = \as\big(\frac{e'_\mp(p\pm q)-e_\pm(p'\mp q)}{1-q^2}\big)\ .
\end{equation}
The functions $\ell_{\s_M\s_N}$ are then defined as the coefficients of
$\theta_{\s_M\s_N}$ in the one-loop phase, see eq.~\eqref{phasestruc}.

The coefficient of the logarithmic terms were first computed in~\cite{Engelund:2013fja}. Given that
the structure of the S-matrix is not altered by the presence of NSNS flux it
follows from the unitarity computation that the coefficients of the logarithms
written in terms of the tree-level functions, \eqref{qtrepp} and
\eqref{qtrepm}, are still given by \eqref{logs1} and \eqref{logs2}
\begin{align}
\cl_{++}(p,p')&=-\frac{1}{2\pi}{C}^{(0)}_{++}(p,p'){E}^{(0)}_{++}(p,p') =- \frac{p^2p'^2\big(e_+ e_+' + (p+q)(p'+q) +(1-q^2)\big)}{4\pi(p-p')^2} \ ,\label{qlogs1}\\
\cl_{+-}(p,p')&=-\frac{1}{2\pi}{B}^{(0)}_{+-}(p,p'){F}^{(0)}_{+-}(p,p') =- \frac{p^2p'^2\big(e_+ e_-' +(p+q)(p'-q)-(1-q^2)\big)}{4\pi(p+p')^2}\ .\label{qlogs2}
\end{align}
Using the dispersion relation, one can check that these expressions agree with
eqs.~\eqref{lppS1} and \eqref{lpmS1} for $q=0$ and $m=m'=1$.

Furthermore, the rational s-channel terms (with the overall phase factors set
to one) are again given in terms of the tree-level functions as in
eqs.~\eqref{combTLpp} and \eqref{combTLpm}. Plugging in the corresponding
expressions, \eqref{qtrepp} and \eqref{qtrepm}, one can check agreement with
the near-BMN expansion of the exact result \eqref{qfunpp} and \eqref{qfunpm}.

Finally, as for the $AdS_3 \times S^3 \times T^4$ background supported by pure
RR flux, the rational contributions from the t-channel go completely into the
phases. That is $\text{Re}(\hat{T}^{(1)})|_{\textup{unit.}} = 0$. Furthermore,
also as for the case of pure RR flux, the light-cone gauge fixed Lagrangian
contains no cubic terms. Therefore, there are correspondingly no external leg
corrections at one loop in the unitarity computation. It follows from
computing the t-channel cuts that
\begin{align}
\phi^{(1)}_{++}(p,p')&=\frac{p\, p' (p+p')(e_+'p+e_+p')}{8 \p (p-p')}\ ,\label{eq:qphasepp}\\
\phi^{(1)}_{+-}(p,p')&=-\frac{p\, p' (p-p')(e_-'p+e_+p')}{8\p(p+p') }\ .\label{eq:qphasepm}
\end{align}
Using the dispersion relation, one can check that these expressions agree with
eqs.~\eqref{eq:phasepp} and \eqref{eq:phasepm} for $q=0$ and $m=m'=1$.

We conclude this section by giving the generalization of the one-loop dressing
phases \eqref{1Lphasepp} and \eqref{1Lphasepm} in the presence of NSNS flux. As
discussed in appendix \ref{notations2} the standard strong coupling variables
$x$ and $y$ are modified for $q\neq 0$. In particular, we now have a separate
variable for the particle $x_+$, $y_+$ and the antiparticle $x_-$, $y_-$. These
are defined in \eqref{xqexp} and \eqref{xqofp}. Our conjecture for the one-loop
dressing phases is then given by ($x_\pm$ corresponds to $p$ and $y_\pm$ to
$p'$)
\begin{align}
\vp^{(1)}_{++}(p,p')&=-\frac{1}{\pi}\frac{x_+^2}{\qh(x_+^2-1)-2q x_+}\frac{y_+^2}{\qh(y_+^2-1)-2qy_+}\nonumber
\\ & \qquad\quad\bigg[\frac{(x_++y_+)\big(\qh(x_++y_+)(1-\frac{1}{x_+y_+})-4q\big)}{(\qh(x_+^2-1)-2qx_+)(x_+-y_+)(\qh(y_+^2-1)-2qy_+)} \nonumber
\\ & \qquad \quad  + \frac{2}{(x_+-y_+)^2} \log\left(\frac{\qp\, x_++\qm}{\qm\, x_+-\qp}\frac{\qm\, y_+-\qp}{\qp \,y_++\qm}\right)\bigg] \ , \label{1Lphaseppq} \\
\vp^{(1)}_{+-}(p,p')&=-\frac{1}{\pi}\frac{x_+^2}{\qh(x_+^2-1)-2q x_+}\frac{y_-^2}{\qh(y_-^2-1)+2qy_-}\nonumber
\\ & \qquad \quad \bigg[\frac{(x_+y_-+1)\big(\qh(x_+y_-+1)(\frac{1}{x_+}-\frac{1}{y_-}) + 4q\big)}{(\qh(x_+^2-1)-2qx_+)(x_+-y_-)(\qh(y_-^2-1)+2q y_-))} \nonumber
\\ & \qquad \quad  + \frac{2}{(x_+y_--1)^2} \log\left(\frac{\qp\,x_++\qm}{\qm\,x_+-\qp}\frac{\qp\,y_--\qm}{\qm\,y_-+\qp}\right)\bigg] \label{1Lphasepmq} \ .
\end{align}
This result was independently found in~\cite{Bianchi:2014rfa} and~\cite{Babichenko:2014yaa}.

% Chapter Template

\chapter{GKP string and cusp anomalous dimension} % Main chapter title
\label{GKP} % Change X to a consecutive number; for referencing this chapter elsewhere, use \ref{ChapterX}

\lhead{Chapter 4. \emph{GKP string}} % Change X to a consecutive number; this is for the header on each page - perhaps a shortened title

Among the many solitonic classical solutions for string theory in $AdS_5\times S^5$ (see~\cite{Tseytlin:2002ny,Tseytlin:2003ii,Plefka:2005bk,Tseytlin:2010jv} for extensive reviews), one that has deserved a lot of attention is the folded spinning string~\cite{Frolov:2002av,Gubser:2002tv}. This configuration lies in an $AdS_3$ subspace of $AdS_5$ and can be pictured as a closed folded string rotating around its center of mass in $AdS_3$. Parametrizing $AdS_5$ in global coordinates
\begin{equation}\label{eq:metricGKP}
 ds^2_{AdS_5}=-dt^2\cosh^2 \rho+d\rho^2+\sinh^2 \rho\, d\O_3^2\, ,
\end{equation}
we consider the Ansatz
\begin{equation}
 t=\t \qquad \phi=\o \t \qquad \rho=\rho(\s)\, ,
\end{equation}
where $\phi$ is an angle in the $S^3$ parametrized by $d\O_3$ in \eqref{eq:metricGKP}. One can check that the equations of motions are translated in the following 1d sinh-Gordon equation
\begin{equation}
 (\acute\rho)^2=\cosh^2 \rho+\o^2 \sinh^2 \rho\, .
\end{equation}
The general solution (an elliptic sn function) has been studied in full details in~\cite{Frolov:2002av} (see also~\cite{Beccaria:2010ry}). In~\cite{Gubser:2002tv}, Gubser, Klebanov and Polyakov (GKP) observed that, when the folded string stretches all the way to the boundary of $AdS_5$, the relation between the two quantum numbers of the string (energy associated to time translation and spin associated to rotations in the $\phi$ directions) exhibits an intriguing logarithmic behaviour
\begin{equation}\label{eq:logGKP}
 E-S\sim f(\l) \log S\, .
\end{equation}
As discussed in the Introduction, the gauge theory counterpart of equation \eqref{eq:logGKP} is the large spin behaviour of the anomalous dimension for twist-two Wilson operators~\cite{Korchemsky:1988si,Korchemsky:1992xv}. The function $f(\l)$ is then identified with twice the cusp anomalous dimension, governing the UV divergences of a cusped Wilson loop. On the string theory side this identification was clarified in~\cite{Kruczenski:2002fb,Kruczenski:2007cy}, where the authors derived a precise equivalence between the two classical solutions describing the long folded spinning string and a minimal surface ending on a light-like cusp on the boundary. Therefore, the physics of string theory expanded around the null cusp vacuum is equivalent to that of the GKP string. As the former turns out to be more tractable for perturbative computations, hereafter we focus on the study of the quantum fluctuations about the null-cusp vacuum.

The construction of a minimal surface solution is easily achieved in a light-cone gauge with the light-like geodesic lying in $AdS$. In this case the expression of the light-cone gauge fixed action is rather compact and perturbative computations of the free energy up to two loops have been performed in~\cite{Giombi:2009gd,Bianchi:2014ada}. As it should be clear from the previous discussion, the computation of the free energy yields the two-loop expansion of the cusp anomalous dimension and therefore constitutes a highly non-trivial test of the quantum integrability of the model (as mentioned in the Introduction, assuming integrability allows to express the cusp anomalous dimension at finite coupling as the solution of an integral equation). Furthermore, assuming that the ABJM cusp anomalous dimension is related to that of $\mathcal{N}=4$ SYM by a simple replacement $\frac{\sqrt{\l}}{4\pi}\to h(\l)$ (as predicted comparing the asymptotic Bethe Ans\"atze of the two theories~\cite{Gromov:2008qe}) the two-loop result for \adscp 
background~\cite{Bianchi:2014ada} provides additional data on the form of the effective coupling $h(\l)$.

\begin{table}[htbp]
 \begin{center}
  \begin{tabular}[htbp]{|l|cl|cl|}
  \hline
  & Weak coupling&  &Strong coupling& \\
  \hline
   \multirow{4}{*}{$AdS_5\times S^5$} 
   &6 bosons in the $\bf{6}$& $m=1$&5 bosons $y^A$ &$m=0$\\
   &4/4 fermions in the $\bf{4}/\bar{\bf{4}}$&$m=1$ & 8 fermions $\eta_i$, $\theta_i$&$m=1$\\
   &2 bosons in the $\bf{1}$  &$m=1$ & 2 bosons $x$,$\bar x$& $m=\sqrt{2}$\\
   &&&1 boson $\phi$&$m=2$\\
   \hline
    \multirow{5}{*}{\adscp} 
   &4/4 spinons in the $\bf{4}/\bar{\bf{4}}$& $m=\tfrac12$&3 complex bosons $z^a$ &$m=0$\\
   &&& 2 fermions $\eta_4$, $\theta_4$ &$m=0$\\
   &6 fermions in the $\bf{6}$&$m=1$ & 6 fermions $\eta_a$, $\theta_a$ &$m=1$\\
   &1 boson in the $\bf{1}$  &$m=1$ & 1 boson $x$& $m=\sqrt{2}$\\
   &&&1 boson $\varphi$&$m=2$\\
   \hline
  \end{tabular}
 \end{center}
\caption{Summary of the spectra of GKP elementary excitations for $AdS_5\times S^5$ and \adscp backgrounds.}\label{tab:GKPspectra}
\end{table}

Starting from the same light-cone gauge fixed action, one can also estimate the dispersion relations of the worldsheet excitations. For the GKP string the first non-trivial quantum corrections appear at one loop (as predicted in~\cite{Basso:2010in} using integrability). It is therefore interesting to compare the predictions from integrability with those from perturbation theory. In particular, the correspondence between the weak coupling spectrum of elementary excitations and the strong coupling worldsheet modes is not completely straightforward. For $AdS_5\times S^5$ the spectra are resumed in table \ref{tab:GKPspectra} and the relation can be summarized in the following way: 
\begin{itemize}
   \item The mass of the 8 fermionic excitations is protected~\cite{Alday:2007mf} and the mapping between weak and strong coupling is straightforward;
   \item The 2 weak coupling excitations associated to the field strength insertion are mapped to the 2 bosonic mass-$\sqrt{2}$ $AdS_3$ excitations of the string\footnote{This trend is confirmed by the one-loop computation of the dispersion relation of these fields which shows that their mass decreases with the coupling~\cite{Giombi:2010bj}.};
   \item The mass of the 6 scalars, as already clarified in~\cite{Alday:2007mf}, decreases as the coupling gets larger and becomes exponentially small at strong coupling. The semiclassical analysis detects only 5 massless excitations,the Goldstone bosons for rotations in $\rm S^{5}$. Nevertheless the actual spectrum contains 6 massive scalars with mass $m\sim e^{-\sqrt{\l}/4}$, in agreement with the gauge theory expectations~\cite{Alday:2007mf}. This is a phenomenon that has been observed already for the $O(N)$ sigma-model~\cite{Zamolodchikov:1977nu} and is related to the fact that the $AdS$ light-cone gauge fixed sigma-model on $AdS_5\times S^5$ is described in some low-energy limit by the $O(6)$ sigma-model~\cite{Alday:2007mf};
   \item The worldsheet mass-2 boson is not an elementary excitation in the weak coupling description and its role has been object of a long debate in the literature~\cite{Giombi:2010bj,Zarembo:2011ag,Basso:2014koa,Fioravanti:2015dma}. The upshot is that the heavy scalar is most probably a compound state of two mass-1 fermions, whose pole is below the production threshold, but is located in the unphysical strip of the rapidity complex plane~\cite{Basso:2014koa, Zamolodchikov:2013ama}.  
  \end{itemize}
  
A similar comparison can be carried out for \adscp and the outcome of this analysis is:
\begin{itemize}
 \item The six worldsheet massive fermions simply correspond to the twist-one fermions at weak coupling. Their mass is protected and their dispersion relation is the same as in $\mathcal{N}=4$ SYM up to replacing $\sqrt{\l}/4\p\to h(\l)$.
 \item The mass-$\sqrt{2}$ boson is related to the weak coupling insertion of a gauge field and its dispersion relation is once more the same as in $\mathcal{N}=4$ SYM
 \item The massless excitations differ substantially from the previous case. The Lagrangian for the low-energy excitations was written down in~\cite{Bykov:2010tv} and consists of a $\cp^3$ sigma model coupled to a massless Dirac fermion, in agreement with the massless string theory spectrum. The dynamics and S-matrix of this model were then studied in~\cite{Basso:2012bw} using integrability and it turned out that the spectrum is gapped (the excitations acquire a mass which is exponentially suppressed at strong coupling as in $\mathcal{N}=4$ SYM) and spanned by two multiplets of excitations in the $\bf{4}$ and $\bar {\bf{4}}$ of $SU(4)$. They were called spinons and anti-spinons and, interestingly, they are neither fermions nor bosons, but they have a fractional statistics corresponding to spin 1/4.
 \item The story for the mass-$2$ boson is essentially the same as in $\mathcal{N}=4$ SYM.
\end{itemize}

In the following, we will briefly summarize the results of the computations in $AdS_5\times S^5$~\cite{Giombi:2009gd,Giombi:2010bj} and then focus on the case of $AdS_4\times \mathbb{CP}^3$~\cite{Bianchi:2014ada,Bianchi:2015laa}. Despite the difference in the two sigma model actions we will notice a striking similarity in the results. This is not surprising, since the two systems are believed to be described by the same integrable structure up to a non-trivial 
interpolating  function of the 't Hooft coupling $h(\lambda)$. We discuss this important feature in section \ref{sec:h}.

\section{String theory in \txpf{$AdS$}{AdS} light-cone gauge}

The gauge fixing procedure for the $AdS$ light-cone gauge is less involved than for the uniform light-cone gauge and here we describe the general strategy applicable both to the $AdS_5\times S^5$ Lagrangian \eqref{eq:Lkinnew} and to the \adscp one \eqref{eq:lagrangian}.

There are many different equivalent procedures of fixing the light-cone gauge with flat target space~\footnote{See thorough discussion in~\cite{Metsaev:2000yf,Metsaev:2000yu}}. The BDHP formulation~\cite{Brink:1976sc,Polyakov:1981rd}, for instance, consists in fixing the conformal gauge and then the residual conformal diffeomorphism symmetry on the plane by choosing $x^+=\t$. Alternatively, the GGRT~\cite{Goddard:1973qh} approach is based on writing the Nambu action in first-order form and then fixing the diffeomorphisms by the two conditions $x^+=\t$ and $P^+=1$. The first approach does not apply in curved space-time with Killing vectors which are not of the direct product form $R^{1,1}\times M^{d-2}$, and therefore for the case at hand one has to give up the standard conformal gauge. A slight modification of it turns out to be a consistent gauge choice
 \begin{equation}\label{eq:nonconf}
\gamma^{ij} = {\rm diag}\left(-G,G^{-1}\right)\,,
\end{equation}
where $G=e^{2\phi}=|z|^2\equiv z^2$ for $AdS_5\times S^5$ and $G=e^{4\varphi}$ for \adscp. Substituting this worldsheet metric in  \eqref{eq:lccoord} and \eqref{eq:lccoordAdS4} we realize that the resulting action contains $x^-$ only in the kinetic term. Imposing then 
\begin{equation}
 x^+=p^+ \tau 
\end{equation}
completely fixes the two-dimensional diffeomorphism invariance and $x^-$ decouples from the action (it can be determined by the Virasoro constraint where it appears linearly). The final form of the $AdS_5\times S^5$ Lagrangian is
\begin{align}
 \mathcal{L}_{AdS_5}& =  \dot{x}^* \dot x   + 
\big( \dot z^M  +
%- in original non-MTT conventions
 \frac{i   }{ z^{2}} z_N \eta_i  (\rho^{MN}){}^i{}_j \eta^j   \big)^2
+  i  p^+ (\theta^i \dot{\theta}_i +\eta^i\dot{\eta}_i +\theta_i \dot{\theta}^i +\eta_i\dot{\eta}^i )  \nonumber\\[0pt]   
& 
-   \frac{(p^+)^2 }{z^{2}}   (\eta^i\eta_i)^2
-    \frac{1}{z^{4}} ( \acute{x}^*\acute{x}  + \acute{z}_M\acute{z}^M)  \nonumber\\[0pt] 
&-    2 \Big[\ \frac{p^+}{ z^{3}}z_M \eta^i (\rho^M)_{ij}
\big(\acute{\theta}^j - \frac{i }{z} \eta^j  \acute{x}\big)
                    +
%- in original non-MTT conventions
\frac{p^+}{ z^{3}}z^M \eta_i (\rho_M^\dagger)^{ij}
\big(\acute{\theta}_j + \frac{i }{z} \eta_j  \acute{x}^*\big)\Big] \ . 
\label{la}
\end{align}

The \adscp one is more involved and we express it in terms of the functions \eqref{eq:pieces}
\begin{align}
 \mathcal{L}_{AdS_4} &=\frac14
\dot x\dot x -\frac14
e^{-8\varphi}\acute x\acute x+ \dot \varphi
\dot \varphi -e^{-8\varphi}\acute \varphi
\acute \varphi + e^{4\varphi}\O^a_\t \O_{a \t}-e^{-4\varphi}\O^a_\s \O_{a \s}\nonumber\\
 &+p^+ \left(\varpi_\t + h_\t+e^{-4\vf} p^+ B\, +2\, e^{-4\varphi} ~{\omega}_\s -2\, e^{-4\varphi}\ell_\s+2\, e^{-6\varphi} C\,\acute x
~\right)\, .\label{eq:AdS4lcgf}
\end{align}
One important classical solution of this action is the open string solution ending on a null cusp on the boundary. We discuss the expansion around this vacuum in section \ref{sec:nullcusp}. Let us first make a brief summary of the results of~\cite{Giombi:2009gd,Giombi:2010bj} concerning the $AdS_5\times S^5$ background.

\section{Summary of the results for \txpf{$AdS_5\times S^5$}{AdS5 x S5}}
The expansion of the Lagrangian \eqref{la} about the null cusp background yields a two-dimensional quantum field theory which can be studied perturbatively for large values of the string tension. According to the discussion at the beginning of this section, the computation of the free energy would give a prediction for the $\mathcal{N}=4$ SYM cusp anomalous dimension and provide a non-trivial check of the quantum consistency of the non-linear sigma model. The calculation of the free energy up to two loops was performed in~\cite{Giombi:2009gd}. The expansion in term of the inverse string tension ($T=\frac{\sqrt{\l}}{2\p}$) reads
\begin{equation}\label{flambda}
f(\lambda)=\frac{\sqrt{\lambda}}{\pi}\Big[1+\frac{a_1}{\sqrt{\lambda}}+\frac{a_2}{(\sqrt{\lambda})^2}+\cdots\Big]\,,
\end{equation}
with coefficients given by
\begin{equation}
a_1=-3\log 2\, , \qquad a_2=-K\, ,
\end{equation}
where $K$ is the Catalan constant
\begin{equation}\label{eq:catalan}
K \equiv \sum_{n=0}^{\infty} \frac{(-1)^n}{(2n+1)^2} \,.
\end{equation}

The spectrum of worldsheet excitations of the $AdS$ light-cone gauge fixed $AdS_5\times S^5$ superstring expanded around the null cusp vacuum consists of 
\begin{align}
 AdS_3\textup{ transverse mode }(\phi):	\qquad&m_\phi^2=4\\
 AdS_5 \textup{ outside } AdS_3\ (x,\bar x):		\qquad&m_x^2=2\\
 S^5\ (y^A, A=1,...,5):				\qquad&m_y^2=0\\
 \textup{Fermions } (\eta^i,\theta^i,\ i=1,...,4):	\qquad&m_{\eta}^2=m_{\theta}^2=1
\end{align}

Those excitations are non-relativistic and a one-loop estimate of the corrections to the dispersion relations can be obtained by studying their two-point functions. This was done in~\cite{Giombi:2010bj} and we can summarize the result as\footnote{Consistently with chapter \ref{BMN}, here and in the following $\o$ is the energy of the worldsheet excitations and $p$ is the spacial component. We use $\ppp$ to indicate the two-momentum.}
\begin{align}\label{eq:resultAdS5}
 &\o^2(p,\l)=\left[p^2+m^2+\frac{q}{\sqrt{\l}}+\mathcal{O}(\l^{-1})\right]\left[1+\frac{c\, p^2}{\sqrt{\l}}+\mathcal{O}(\l^{-1})\right]\, ,\\
 &\begin{array}{lllll}
 q_\phi=0 \, ,		\quad&q_x=-\pi\, ,  	\quad&q_y=0\, ,  			\quad&q_{\eta}=q_{\theta}=0\, ,  		\\
 c_\phi=-\frac{\pi}{2}\, , 	\quad&c_x=-\pi\, ,	 	\quad&c_y=-\frac73\, ,	\quad&c_{\eta}=c_{\theta}=-2\pi\, .  	
\end{array}
\end{align}
In sections \ref{sec:h}, \ref{comparison} and \ref{sec:comparison} we will describe the same analysis in the setup of \adscp (relevant for the $AdS_4/CFT_3$ case) and compare the two results.

\section{\txpf{$AdS_4/CFT_3$}{AdS4/CFT3} system and \txpf{$h(\l)$}{h(lambda)}}\label{sec:h}

A  powerful attribute that the planar AdS$_4/$CFT$_3$ system~\cite{Aharony:2008ug} shares with its 
higher-dimensional version, is its conjectured 
integrability~\cite{Minahan:2008hf,Gromov:2008qe,Klose:2010ki, Cavaglia:2014exa}.
The explicit realization of the integrable structure is however non-trivial, due to significative 
peculiarites of this case.
 
A first important ingredient, to take into account when comparing string theory calculations with weak coupling results, is the correction
to the effective string tension~\cite{Bergman:2009zh} which must be considered for the
first time at  two loops in sigma-model perturbation theory.
The  original 
``dictionary'' 
proposal~\cite{Aharony:2008ug} for the effective string tension  in terms of the effective 't Hooft 
coupling $\lambda$ of ABJM reads 
\begin{equation}\label{shift}
T= 
\frac{R^2}{2\pi\alpha'}
=2\sqrt{2 \lambda}\,,
\qquad\qquad
\lambda=\frac{N}{k}\,,
\end{equation}
where $R$ is the $\cp^3$ radius. As pointed out in~\cite{Bergman:2009zh}, the geometry 
(and flux, in the ABJ~\cite{Aharony:2008gk} theory)  of the background induces higher order corrections 
to  the radius of curvature in the Type IIA description, which in the planar 
limit %~\footnote{The anomalous 
%radius shift  in the Type 
%IIA description of \adscp\cite{Bergman:2009zh}, due to 
%effects of curvature corrections and discrete torsion, is in general given in 
%terms of units of  torsion (3-form) flux quanta and orbifold parameter $k>1$.} 
of interest here appear in the form of a shift in the square root
\begin{equation}\label{eq:tension}
T = 2\sqrt{2\left(\lambda-\frac{1}{24}\right)}\,.
\end{equation}
We emphasize that the string perturbative expansion is  an expansion in inverse string tension 
whose coefficients are obviously not affected by the correction~\eqref{eq:tension}. 
The radius shift is a (corrected) AdS$_4$/CFT$_3$ dictionary proposal, an assumed, new input which plays a role 
when expressing the result in terms of the 't Hooft coupling. 
 
 Another crucial property of the $AdS_4/CFT_3$ system is the interpolating  function of the 't Hooft coupling $h(\lambda)$, which features all 
the integrability-based calculations in this model\footnote{A possible way to interpret these relations is to consider the triplet $\{T,\l,h\}$ as the three couplings for string theory, quantum field theory and integrability respectively. Whereas in the $AdS_5 \times S^5$ case the relations among them are trivial, here the dictionary is more complicated and one should take this into account when comparing different results. On may argue further that not all the three quantities are physical since the string tension is always defined uo to finite renomarlization and therefore its relation with $h$ is simply a choice of regularization scheme.}. 
 Clearly its knowledge is decisive to grant the conjectured integrability of ABJM theory a full predictive power.
At strong coupling, one way to obtain information on $h(\lambda)$ 
is to evaluate in string theory 
the cusp anomalous dimension
for the ABJM theory $f_{\abjm}(\lambda)$, and then 
compare the result with the asymptotic Bethe Ansatz prediction 
of~\cite{Gromov:2008qe}. The latter is based on the equivalence of the BES~\cite{Beisert:2006ez} equations for the $\mathcal{N}=4$ 
and the ABJM case and reads
%~\footnote{The prediction is based on the equivalence of the BES equations for the $\mathcal{N}=4$ 
%case and the ABJM case, provided the replacement of the coupling constant as in \eqref{cusp_pred}.} 
\begin{equation}\label{cusp_pred}
f_{\abjm}(\lambda)=\left.\frac{1}{2}\, f_{\mathcal{N}=4}(\lambda_\YM)\,\right|_{\frac{\sqrt{\lambda_\YM}}{4\pi}
\rightarrow h(\lambda)}\,,
\end{equation}
which implies
\begin{equation}\label{cusp_pred_2}
f_{\abjm}(\lambda)=2h(\lambda)-\frac{3\log 
2}{2\pi}-\frac{K}{8\pi^2}\frac{1}{h(\lambda)}+\cdots~,\\
\end{equation}
where $f_{\mathcal{N}=4}(\lambda_\YM)$ is the cusp anomaly of $\mathcal{N}=4$ 
SYM  and $K$ is the Catalan constant. The leading strong coupling value for $f(\lambda)$ 
has been given already in~\cite{Aharony:2008ug} and reads 
$f(\lambda\gg1)=\sqrt{2\lambda}$, from which via \eqref{cusp_pred_2} one gets 
$h(\lambda\gg1)=\sqrt{\lambda/2}$. At one loop in sigma-model perturbation theory, 
the scaling function has been evaluated in~\cite{McLoughlin:2008ms,Alday:2008ut,Krishnan:2008zs,McLoughlin:2008he,Gromov:2008fy,
 Astolfi:2008ji,Bandres:2009kw, Abbott:2010yb,Abbott:2011xp,Astolfi:2011ju,Astolfi:2011bg,LopezArcos:2012gb,Forini:2012bb} 
via the energy of closed spinning strings  in the large spin limit or similar means, providing a first subleading
 correction $-\log2/(2\pi)$ to $h(\lambda)$ on which some debate existed~\cite{Shenderovich:2008bs}. 

At two loops the shift \eqref{eq:tension} starts playing a role and the result reads
\begin{equation}\label{cusp_result}
f_{\abjm}(\lambda) = \sqrt{2\lambda } - \frac{5 \log 2}{2 \pi } - \left(\frac{K}{4\pi^2} + 
\frac{1}{24}\right)\frac{1}{\sqrt{2\lambda}} + {\cal O}(\sqrt{\lambda})^{-2}
\,.
\end{equation}
The formula can be rewritten in a more compact way defining the shifted coupling
\begin{equation}
\tilde\lambda \equiv \lambda - \frac{1}{24}\,,
\end{equation}
from which
\begin{equation}
f_{\abjm}\left(\tilde\lambda\right) = \sqrt{2\tilde\lambda } - \frac{5 \log 2}{2 \pi } - 
\frac{K}{4\pi^2\,\sqrt{2\tilde\lambda}} + {\cal O}(\sqrt{\tilde\lambda})^{-2}\,.
\end{equation}
This form of the result makes evident the striking similarity with the  $AdS_5\times S^5$ 
result
\begin{equation}
f_{\YM} (\lambda_\YM) = \frac{\sqrt{\lambda_\YM }}{\pi} - \frac{3 \log 2}{ \pi } 
- \frac{K}{\pi\,\sqrt{\lambda_\YM}} 
+ {\cal O}(\sqrt{\lambda_\YM})^{-2}\,,
\end{equation} 
where  the change in the transcendentality pattern is due to the corresponding difference
in the effective string tensions. %(at variance with the $AdS_5\times S^5$ case, where 
%$T=\sqrt{\lambda}/(2\pi)$, here the tension \eqref{eq:tension} has transcendentality zero).

From \eqref{cusp_result} and via \eqref{cusp_pred} we get then the strong-coupling two-loop correction for the interpolating 
function $h(\lambda)$, that we report here together with the weak coupling results~\cite{Gaiotto:2008cg,Grignani:2008is,Nishioka:2008gz,Minahan:2009aq,Minahan:2009wg,Leoni:2010tb}
\begin{equation}\label{eq:prediction}
\begin{array}{lll}
h^2(\lambda)& = \displaystyle 
\lambda^2 - \frac{2\,\pi^3}{3}\,\lambda^4+ {\cal O}\left(\lambda^6\right) & \quad \lambda \ll 1\quad , \\
h(\lambda)& = \displaystyle\sqrt{\frac{\lambda}{2}} - \frac{\log2}{2\pi} - \frac{1}{48\sqrt{2\lambda}} 
+ {\cal O}(\sqrt{\lambda})^{-2} & \quad \lambda \gg 1 \quad,
\end{array}
\end{equation}
where we emphasize the  a priori non-obvious fact the two-loop coefficient at strong coupling is 
only due to the  anomalous radius shift. 

A conjecture for the exact expression of $h(\lambda)$
has  been recently made~\cite{Gromov:2014eha}, in a spirit quite close to the one followed
in~\cite{Correa:2012at,Correa:2012hh} on the comparison 
between two exact
computations of the same observable.
The authors of~\cite{Gromov:2014eha} elaborated  on the similarity between two all-order 
calculations in ABJM theory: one 
- the ``slope function''~\cite{Basso:2011rs}~-
derived via integrability as exact solution of a quantum spectral curve~\cite{Cavaglia:2014exa} 
and one - a 1/6 BPS Wilson loop~\cite{Kapustin:2009kz,Marino:2009jd,Drukker:2010nc}
 - obtained with supersymmetric localization. 
%and based on previous experience in $\mathcal{N}=4$ with the so-called 
%Bremsstrahlung function~\cite{Drukker:2012de,Correa:2012hh,Gromov:2012eu}. 
As the first of the two exact results is expressed in terms of the 
effective coupling $h(\lambda)$,  an  ``extrapolation'' for the latter has been derived in an 
exact,  implicit, form~\footnote{As noticed in~\cite{Gromov:2014eha}, a more solid derivation of 
$h(\lambda)$ would require comparison between the localization results 
of~\cite{Marino:2009jd,Drukker:2010nc} and  the ABJM 
Bremsstrahlung function~\cite{Lewkowycz:2013laa, Bianchi:2014laa, Correa:2014aga}, similarly to the case of  
the $h(\lambda_\YM)$ of $\mathcal{N}=4$ SYM.}.
It is 
\begin{equation}\label{eq:proposal}
\lambda = \frac{\sinh{2\pi h(\lambda)}}{2\pi}\, _3F_2 \left(\frac12,\frac12,\frac12 ; 1,\frac32 ; 
-\sinh^2{2\pi h(\lambda)} \right) \,,
\end{equation}
with weak and strong coupling expansions 
\begin{align} 
h(\lambda)&= \lambda-\frac{\pi^2}{3}\,\lambda^3+\frac{5\pi^4}{12}\,\lambda^5-
\frac{893\pi^6}{1260}\,\lambda^7+\mathcal{O}(\lambda^9)  & \lambda &\ll 1 \, , 
\\\label{eq:strong}
h(\lambda) &=  \sqrt{\frac12\left( \lambda - \frac{1}{24}\right)} - \frac{\log 2}{2 \pi}  
+ {\cal O}\left( e^{-2\pi\sqrt{2\lambda}} \right)
& \lambda &\gg 1 \, .
\end{align}
We see that  \eqref{eq:strong} above, expanded for large $\lambda$,  agrees with 
\eqref{eq:prediction}.  The aim of the next sections is to provide an explicit string theory computation of the first three terms in \eqref{eq:strong} supporting the conjecture of~\cite{Gromov:2014eha}.

\section{The null-cusp fluctuation in \txpf{$AdS_4 \times \cp^3$}{AdS4 x CP3}}\label{sec:nullcusp}

In this section we consider the Wick-rotated, Euclidean formulation of the Lagrangian \eqref{eq:AdS4lcgf} and compute its fluctuations about the null cusp background.
The equations of motion derived from the (Euclidean) AdS light-cone gauge Lagrangian \eqref{eq:AdS4lcgf} 
admit a classical solution for which the on-shell action is the area of the minimal surface 
ending on a null cusp on the $AdS_4$ boundary.
This configuration is just the $AdS_4$ embedding of the classical string solution found in the $AdS_5$ 
background~\cite{Kruczenski:2002fb, Giombi:2009gd}, and reads
\begin{eqnarray}\label{eq:background}
& w \equiv e^{2\varphi} = \displaystyle\sqrt{\frac{\tau}{\sigma}} \qquad\qquad
x = 0 &\nonumber \\
& x^{+} =\tau \qquad\qquad
x^{-} =-\displaystyle\frac{1}{2\sigma}  \qquad\qquad
z^{M} = 0~. &
\end{eqnarray} 
The requirement that the open string Euclidean world-sheet described by these coordinates ends on a 
cusp at the boundary of $AdS_4$ at $w=0$ is manifestly enforced by the relation $x^+\, x^- = -\frac12 w^2$.
In the AdS/CFT dictionary of~\cite{Maldacena:1998im,Rey:1998ik}, the Wilson loop evaluated on a light-like cusp contour is then given by the superstring partition function
\begin{equation}\label{eq:partition}
\left\langle W_{cusp} \right\rangle = Z_{string} \equiv \int {\cal D}[x,w,z,\eta,\theta]\, e^{-S_E}\,.
\end{equation}
In order to compute it perturbatively, we first construct the Euclidean action $S_E$ for 
fluctuations about the background \eqref{eq:background}. 
Following~\cite{Giombi:2009gd}, we will use a suitable parametrization of fluctuations which, combined with 
a further redefinition of the worldsheet coordinates\footnote{\label{fn:factor2} Compared to~\cite{Bianchi:2014ada} we introduced an additional factor of 2 in the redefinition of the worldsheet coordinates. This effectively doubles the masses of the excitations and does not affect the final result} 
$t =2\, \log \tau$ and $s=2\,\log\sigma$, is such  that  
the coefficients of the fluctuation action become constant, namely 
$(\tau,\sigma)$-independent. It reads~\footnote{The factor 2 in the fluctuation of the field $x$ is
 introduced to normalize the kinetic term of $\tilde x$.}
\begin{eqnarray}\label{eq:fluctuations}
& x = 2\, \displaystyle\sqrt{\frac{\tau}{\sigma}} \tilde x \qquad\qquad w = \displaystyle\sqrt{\frac{\tau}{\sigma}}\, \tilde w \qquad\qquad \tilde w = e^{2\tilde\varphi}  & \nonumber\\
& z^a = \tilde z^a \qquad\qquad \bar z^a = \tilde{\bar z}^a \qquad\qquad a=1,2,3 & \nonumber\\
& \eta = \displaystyle\frac{1}{\sqrt{2\, \sigma}}\, \tilde \eta \qquad\qquad \theta = \displaystyle\frac{1}{\sqrt{2\, \sigma}}\, \tilde \theta\,. & 
\end{eqnarray}
After the Wick rotation $\tau\to -i\,\tau, p^+\to i p^+$ and having set $p^+=1$, 
we end up    with the following action for  fluctuations over the 
null-cusp background \eqref{eq:background}
\begin{equation}\label{eq:Lagrangian_exp}
S_E = \frac{T}{2}\, \int dt\, ds\, {\cal L}\qquad,\qquad
{\cal L} = {\cal L}_B + {\cal L}_F^{(2)} + {\cal L}_F^{(4)}\,,
\end{equation}
where
\begin{align}
 {\cal L}_B& = \left( \partial_t \tilde x +  \tilde x \right)^2 + \frac{1}{\tilde w^4} \left( \partial_s \tilde x - \tilde x \right)^2
+ \tilde w^2\, \left(\partial_t \tilde \varphi \right)^2 + \frac{1}{\tilde w^2}\, \left(\partial_s \tilde \varphi \right)^2 + \frac{1}{4} \left( \tilde w^2 + \frac{1}{\tilde w^2} \right) 
\nonumber\\& 
+ \tilde w^2 \, \tilde g_{MN}\, \partial_t \tilde z^M\, \partial_t \tilde z^N + \frac{1}{\tilde w^{2}}\, \tilde g_{MN}\, \partial_s \tilde z^M\, \partial_s \tilde z^N \, ,\label{boslag}
\\
{\cal L}_F^{(2)} &= i\Big[  \partial_t {\tilde\theta}_a {\tilde\theta}^a - {\tilde\theta}_a\partial_t{\tilde\theta}^a
+ \partial_t{\tilde\theta_4}{\tilde\theta}^4 - {\tilde\theta_4}\partial_t
{\tilde\theta}^4 + \partial_t{\tilde\eta_a}{\tilde\eta}^a - {\tilde\eta_a}\partial_t{\tilde\eta}^a 
+ \partial_t\tilde{\eta_4}\tilde{\eta}^4 - \tilde{\eta_4}\partial_t
\tilde{\eta}^4 \Big] \nonumber\\&
+\frac{2i}{\tilde w^2}\Big[\hat{\tilde\eta}_a \left(\hat{\partial_s}{\tilde\theta}^a - \hat{ \tilde\theta}^a \right)
+ \left(\hat{\partial_s}\tilde\theta_a - {\hat {\tilde\theta}}_a \right) \hat{\tilde{\eta}}^a + \frac12 \left(\partial_s\tilde\theta_4{\tilde\eta}^4 - \partial_s\tilde\eta_4
{\tilde\theta}^4 + \tilde\eta_4\partial_s{\tilde\theta}^4 -\tilde \theta_4
\partial_s{\tilde\eta}^4\right)\Big]\nonumber\\&
+ \partial_t \tilde z^M\, \tilde h_M + \frac{4\,i}{\tilde w^3}\, \tilde C\, \left( \partial_s \tilde x 
- \tilde x \right) - \frac{2i}{\tilde w^2}\, \partial_s \tilde z^M\, \tilde\ell_M  \, ,
\\\label{eq:Lagrangian_exp_fin}
{\cal L}_F^{(4)} &= \frac{1}{\tilde w^2}\, \tilde B \,.
\end{align}
In the expressions above,  with $\tilde B$, $\tilde C$, $\tilde h_M$ and $\tilde \ell_M$ we indicate the
quantities $ B$, $  C$, $  h_M$ and $ \ell_M$ in \eqref{eq:pieces} where a tilde 
over each field appears (namely, the weighting factors for the fluctuations  
in \eqref{eq:fluctuations} have already been made explicit in the derivatives of products). 
%The additional factors $i$ in the last three terms of ${\cal L}_F^{(2)}$ 
%come from Wick-rotating the original action.

\subsection{Feynman rules}

Provided with an explicit Lagrangian for the fluctuations around the cusp background, we can expand it and extract the relevant Feynman rules for performing perturbative computations. Hereafter we drop tildes from fluctuation fields in order not to clutter formulae. All the fields are understood to be fluctuations.

The bosonic propagators are diagonal and read
\begin{equation}
G_{\varphi\varphi}(p) = \frac{1}{T}\, \frac{1}{\ppp^2+4}\, ,\qquad\qquad G_{z_a \bar z^b}(p) = \frac{1}{T}\, \frac{2\, \delta_a^b}{\ppp^2}\, ,\qquad\qquad G_{xx}(p) = \frac{1}{T}\, \frac{1}{\ppp^2+2}\,.
\end{equation}
The fermionic propagators are not diagonal, instead, and take the form
\begin{align}
 G_{\eta_4\eta^4}(p) &= G_{\theta_4\theta^4}(p) = \frac{1}{T}\, \frac{\ppp_0}{\ppp^2} \, ,&
 G_{\eta_4\theta^4}(p) &= G_{\theta_4\eta^4}(-p) = -\frac{1}{T}\, \frac{\ppp_1}{\ppp^2}\, ,\nonumber \\
 G_{\eta_a\eta^b}(p) &= G_{\theta_a\theta^b} (p)= \frac{1}{T}\, \frac{\ppp_0}{\ppp^2+1}\delta_a^b\, , &   G_{\eta_a\theta^b}(p) &=G_{\theta_a\eta^b}(-p)= -\frac{1}{T}\, \frac{\ppp_1+i}{\ppp^2+1}\delta_a^b \,.
\end{align}
The interaction vertices are obtained expanding the Lagrangian \eqref{eq:Lagrangian_exp} in the fluctuation fields. For the one-loop computation only terms with up to four fields are relevant. We spell them out in the appendix \ref{app:lagr_exp}.

\section{Cusp anomaly in \txpf{$AdS_4 \times \cp^3$}{AdS4 x CP3}}\label{sec:cuspAdS4}

Since the Lagrangian has now constant coefficients and is thus translationally invariant, 
the (infinite) world-sheet volume factor $V$ factorizes. The scaling function is then defined 
via the string partition function as~\cite{Giombi:2009gd}
\begin{equation}\label{Wpert}
W=-\ln Z=\frac{1}{2}f(\lambda) \,V=W_0+W_1+W_2+... \,,
\qquad\qquad V\equiv \int dt\, ds\, ,
\end{equation}
where $W_0\equiv S_E$ coincides with the value of the action on the background, 
$W_1, W_2, ...$ are one-, two- and higher loop corrections, and for the volume $V$ 
we use a slightly different convention from~\cite{Giombi:2009gd} due to the different choice of worldsheet coordinates (see footnote \ref{fn:factor2}).
%is due to the coordinate transformations used above%~\footnote{
%The rescaling $t = \log \tau$ and $s=\log\sigma$ of the world-sheet coordinates, which 
%puts the induced worldsheet metric in the conformal gauge 
%$ds^2_{\rm ind}=\frac{1}{4}(dt^2+ds^2)$. The 2-dimensional volume is then $V=\frac{1}{4}\int dt 
%ds$.} 
From \eqref{Wpert} we explicitly define $f(\lambda)$ in terms of the 
effective action $W$
\begin{equation}\label{eq:cuspprescription}
f(\lambda) = \frac{2}{V}\, W\, .
\end{equation}
We are now ready to compute the effective action perturbatively 
in inverse powers of the effective string tension $g \equiv \frac{T}{2}$. From 
this we will extract the corresponding strong coupling perturbative expansion for the scaling function
\begin{equation}\label{eq:expansion}
f(g) = g\, \left[ 1 + \frac{a_1}{g} + \frac{a_2}{g^2} + \dots 
\right]\,,\qquad\qquad g=\frac{T}{2}~.
\end{equation} 
where we have factorized the classical result from $W_0 = S_E$~\cite{Aharony:2008ug} 
and the effective string tension $T$ is defined in \eqref{eq:tension}.
%%%%%%%%%%%%%%%%%%%%%%%%%%%%%%%%%
\subsection{Cusp anomaly at one loop}\label{oneloop}
\label{sec:oneloop}
%%%%%%%%%%%%%%%%%%%%%%%%%%%%%%%%%
We start considering one-loop quantum corrections to the free energy \eqref{eq:partition}, 
which are derived expanding the fluctuation Lagrangian \eqref{eq:Lagrangian_exp} to second order in the fields.\\
For the bosonic part we obtain
\begin{equation}
\mathcal{L}_{B}^{(2)} = \left(\partial_t  x\right)^2 + \left(\partial_s  x\right)^2 + 2\, {x}^2 + \left(\partial_t\varphi\right)^2 + \left(\partial_s\varphi\right)^2 + 4\varphi^2 + \left|\partial_t  z^a\right|^2 + \left|\partial_s  z^a\right|^2\,.
\end{equation}
The bosonic degrees of freedom consist of six real massless scalars (associated to the $\mathbb{CP}^3$ coordinates), one real scalar $ x$ with mass $m^2 = 2$ and one real scalar $\varphi$ with mass $m^2 = 4$.
This is a simple truncation  (one less transverse degree of freedom in the AdS space)  of the bosonic spectrum
found in the $AdS_5\times S^5$~\cite{Giombi:2009gd}.
%Up to the number of scalar fields of each type, this is the same bosonic spectrum as in $AdS^5 \times S^5$.\\
For the fermions one gets an off-diagonal kinetic matrix
\begin{equation}
{\cal L}^{(2)}_F = i\, \Theta\, K_{F}\, \Theta^{T} \quad \text{where} \quad \Theta \equiv \left({\theta}_{a},{\theta}_{4},{{\theta}}^{a},{{\theta}}^{4},{\eta}_{a},{\eta}_{4},{{\eta}}^{a},{{\eta}}^{4}\right)\,,
\end{equation}
which reads{\small
\begin{equation}\label{eq:KF}
K_F=
\begin{pmatrix}
0 & 0 & -\partial_t & 0 & 0 & 0 & -\partial_s-1 & 0 \\
0 & 0 & 0 & -\partial_t  & 0 & 0 & 0 & -\partial_s \\ 
-\partial_t  & 0 & 0 & 0 & \partial_s+1 & 0 & 0 & 0 \\
0 & -\partial_t  & 0 & 0 & 0 & \partial_s & 0 & 0 \\
0 & 0 & \partial_s -1 & 0 & 0 & 0 & -\partial_t  & 0 \\
0 & 0 & 0 & \partial_s & 0 & 0 & 0 & -\partial_t  \\
-\partial_s+1 & 0 & 0 & 0 & -\partial_t  & 0 & 0 & 0 \\
0 & -\partial_s & 0 & 0 & 0 & -\partial_t  & 0 & 0\\
\end{pmatrix}\,.
\end{equation}}
Fermions contribute to the partition function with the determinant 
($\partial_{\mu} = i\, \ppp_\mu\,,\,\mu=0,1 $)
\begin{equation}
\text{det}\ K_F = \left(\ppp^2\right)^2\left(\ppp^2+1\right)^6\,,
\end{equation}
from which we read that the fermionic spectrum is composed of six massive degrees of freedom with mass $m^2=1$ and two massless ones. The latter  are of $\eta_4$ and $\theta_4$ type, namely  those fermionic directions corresponding to the broken supersymmetries.
The presence of massless fermions marks a difference with respect to the ${\cal N}=4$ SYM case,
already noticed in this theory when studying fluctuations over classical  string 
solutions only lying in $AdS_4$~\cite{McLoughlin:2008he,Abbott:2010yb,LopezArcos:2012gb, Forini:2012bb}.

The one-loop effective action is computed as
\begin{equation}
W_1 = -\log Z_1\, ,
\end{equation}
where $Z_1$ is the ratio of fermionic over bosonic determinants. Therefore
\begin{align}
W_1 &= \frac{1}{2}\, V\, \int{\frac{d^2\ppp}{(2\pi)^2}\left\{\log \left(\ppp^2+4\right)+\log \left(\ppp^2+2\right) + 4\log\left(\ppp^2\right) - 6\log\left(\ppp^2+1\right)\right\}} \nonumber \\
&= -\frac{5\log 2}{4\pi}\, V\,.
\end{align}
The one-loop correction to the scaling function reads, according to \eqref{eq:cuspprescription},
\begin{equation}\label{eq:a1}
a_1 = -\frac{5\log 2}{2\pi}\, ,
\end{equation}
and agrees with previous independent results~\cite{McLoughlin:2008he,Abbott:2010yb,LopezArcos:2012gb}.

%%%%%%%%%%%%%%%%%%%%%%%%%%%%%%%%
\subsection{Cusp anomaly at two loops}\label{twoloops}
\label{sec:twoloops}
%%%%%%%%%%%%%%%%%%%%%%%%%%%%%%%%
In this section we provide the details on the computation of the two-loop coefficient of the scaling function. 
The aim is to compute the connected vacuum diagrams of the fluctuation Lagrangian around the null cusp background. Denoting by $W$ the free energy of the theory, $W=-\log Z$, the two-loop contribution is given by
\begin{equation}
 W_2=\braket{S_{int}}-\frac12\braket{S_{int}^2}_c\,,
\end{equation}
where $S_{int}$ is the interacting part of the action at cubic and quadratic order (see appendix \ref{app:lagr_exp}). 
The subscript $c$ indicates that only connected diagrams need to be included. In the following we use $S_{int}=T\int\, dt\, ds\,  \mathcal{L}_{int}$ and we give the expressions of the vertices as they appear in $\mathcal{L}_{int}$. Throughout this section we neglect the string tension $T$ and the volume $V$ in the intermediate steps and reinstate them at the end of the calculation.

\subsubsection{Bosonic sector}
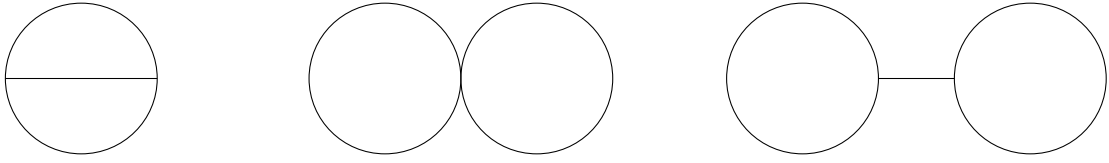
\begin{figure}[h]
\begin{center}
\begin{tikzpicture}[scale=1, vertex/.style={circle,fill=black,thick,inner sep=2pt}]
\draw(-5,0) circle (1 cm);
%\node (c) at (-6,0) [vertex] {};
%\node (c) at (-4,0) [vertex] {};
 \draw[-] (-6,0)--(-4,0);
\draw(-1,0) circle (1 cm);
\draw(1,0) circle (1 cm);
% \node (c) at (0,0) [vertex] {};
 \draw[-] (5.5,0)--(6.5,0);
 \draw(7.5,0) circle (1 cm);
\draw(4.5,0) circle (1 cm);
%\node (c) at (5.5,0) [vertex] {};
%\node (c) at (6.5,0) [vertex] {};
 \end{tikzpicture}
\end{center}
\caption{Sunset, double bubble and double tadpole are the diagrams appearing in the two-loop contribution to the partition function.}\label{fig:vacdiagrams}
\end{figure}

Let us first consider the purely bosonic sector. 
As pointed out in section \ref{oneloop}, the spectrum of the theory contains one real boson of squared mass 4, one real boson of squared mass 2 and three complex massless bosons. The interaction among these excitations involves cubic and quartic vertices which give rise to the diagrams in figure \ref{fig:vacdiagrams}.

When combining vertices and propagators in the sunset diagrams they originate various non-covariant integrals with components of the loop momenta in the numerators. Standard reduction techniques allow to rewrite every integral as a linear combination of the two following scalar ones (explicit reductions for the relevant integrals are spelled out in appendix \ref{app:intred})
\begin{align}
I\left[m^2\right] & \equiv \int \frac{d^2\ppp}{\left(2\pi\right)^2}\, \frac{1}{\ppp^2+m^2}\, , \\
I\left[m_1^2,m_2^2,m_3^2\right] & \equiv \int \frac{d^2\ppp\, d^2\qqq\,d^2\rrr}{\left(2\pi\right)^4}\,  \frac{\delta^{(2)}(\ppp+\qqq+\rrr)}{(\ppp^2+m_1^2)(\qqq^2+m_2^2)(\rrr^2+m_3^2)} \quad. 
\end{align}
The latter integral is finite, provided none of the masses vanishes, and is otherwise IR divergent.
The former is clearly UV logarithmically divergent, and also develops IR singularities in the massless case. 
In our computation we expect all UV divergences to cancel and therefore no divergent integral to appear in the final result.
Nonetheless, performing reduction of potentially divergent tensor integrals to scalar ones still implies the choice of a regularization scheme. In our case we use the one adopted in~\cite{Giombi:2009gd,Roiban:2007jf,Roiban:2007dq}.
This prescription consists of performing all manipulations in the numerators in $d=2$, which has the advantage of simpler tensor integral reductions.
In this process we set to zero power UV divergent massless tadpoles, as in dimensional regularization 
\begin{equation}\label{eq:reg}
\int \frac{d^2\ppp}{(2\pi)^2}\, \left( \ppp^2\right)^n = 0\,, \qquad\qquad n\geq 0.
\end{equation}
All remaining logarithmically divergent integrals happen to cancel out in the computation and there is no need to pick up an explicit regularization scheme to compute them.\\
As an explicit example, we consider   the contribution to the sunset coming from the first vertex in \eqref{bos3ver}
\begin{equation}\label{sunsetxx}
-\frac12\braket{V_{\varphi x x}^2}= -16\int \frac{d^2\ppp\, d^2\qqq\,d^2\rrr}{(2\pi)^4}\,\frac{(1+ \qqq_1^2)\, (1+ \rrr_1^2)\, \delta^{(2)}(\ppp+\qqq+\rrr)}{(\ppp^2+4) (\qqq^2+2) (\rrr^2+2)} =8\, I[4,2,2]\,.
\end{equation}
The integral $I[4,2,2]$ is a particular case of the general class
\begin{equation}\label{intcat}
 I\left[2\,m^2,m^2,m^2\right]=\frac{K}{8\, \p^2\, m^2}\,,
\end{equation}
where $K$ is the Catalan constant \eqref{eq:catalan}.
The contribution of the sunset diagram involving the second vertex in \eqref{bos3ver} is proportional to $I[4]^2$, whereas the contribution of the third vertex vanishes
\begin{equation}
-\frac12\braket{V_{\varphi^3}^2}=8\, I[4]^2 \, , \qquad\qquad -\frac12\braket{V_{\varphi |z|^2}^2}=0\, .
\end{equation}
The final contribution of the bosonic sunset diagrams is
\begin{equation}\label{bossun}
 W_{2,\textup{bos. sunset}}=8\, I[4,2,2]+8\,I[4]^2\,.
\end{equation}
The first two vertices in \eqref{bos3ver} can also be contracted to generate non-1PI graphs, 
namely double tadpoles. However the resulting diagrams turn out to vanish individually.

Despite the lengthy expressions of the vertices (see appendix \ref{app:lagr_exp}), the only non-vanishing double-tadpole comes from $V_{\varphi^4}$ and gives
\begin{equation}
 W_{2,\textup{bos. bubble}}=-8\, I[4]^2\, ,
\end{equation}
and cancels the divergent part of \eqref{bossun}.
As a result, the bosonic sector turns out to be free of divergences without the need of fermonic contributions (as it happens at one loop),
which was already observed in the $AdS_5\times S^5$ case~\cite{Giombi:2009gd}.

%%%%%%%%%%%%%%%%
\subsubsection{Fermionic contributions}\label{fermcontr}
%%%%%%%%%%%%%%%%
We compute the diagrams arising from interactions involving fermions.
The main difference between the spectrum of $AdS_5\times S^5$ and the one introduced in section \ref{oneloop} resides in the fermionic part. Although both theories have eight fermionic degrees of freedom, in $AdS_4\times\mathbb{CP}^3$ they are split into six massive and two massless excitations, which interact non-trivially among themselves.

We start by considering diagrams involving at least one massless fermion.
The quartic interactions are either not suitable for constructing a double tadpole diagram or they produce vanishing integrals. These include vector massless tadpoles, which vanish by parity, and tensor massless tadpoles, which have power UV divergences and are set to zero. For completeness we list them in appendix \ref{app:lagr_exp}. 

Focusing on the Feynman graphs which can be constructed from cubic interactions 
we also note that the only double tadpole diagrams that can be produced using \eqref{fer3verm0} and \eqref{fer3verm1} involve tensor
massless tadpole integrals and therefore vanish.
In the sector with massless fermions we are therefore left with the sunset diagrams, which, thanks to the diagonal structure of the bosonic propagators, turn out to be only five
\begin{equation}
W_{2,\psi_4}=-\tfrac12 \textstyle{ \braket{V_{z\eta_a\eta_4}V_{z \eta_a\eta_4}+V_{z\eta_a\theta_4}V_{z\eta_a\theta_4}+2\,V_{z\eta_a\eta_4}V_{z\eta_a\theta_4}+ V_{\varphi \eta_4 \theta^4} V_{\varphi \eta_4 \theta^4}+ V_{x \psi^4 \psi_4} V_{x \psi^4 \psi_4}}}
\end{equation}
The explicit computation of the individual contributions shows that they are all vanishing. As an example we consider 
\begin{equation}
-\frac12 \braket{V_{\varphi \eta_4 \theta^4} V_{\varphi \eta_4 \theta^4}}=4 \int \frac{d^2\ppp\, d^2\qqq\,d^2\rrr}{(2\p)^4} \frac{(\ppp_1-\qqq_1)^2(\ppp_0\qqq_0-\ppp_1\qqq_1)\, \delta^{(2)}(\ppp+\qqq+\rrr)}{\ppp^2 \qqq^2 (\rrr^2+4)} = 0\, ,
\end{equation}
and similar cancellations happen for the other diagrams. Therefore we conclude that  $W_{2,\psi_4}=0$ and that massless fermions are effectively decoupled at two loops.

We then move to consider massive fermions, starting from their cubic coupling to bosons.
As in the massless case, this generates five possible sunset diagrams. None of them is vanishing. We present the details of a particularly relevant example, i.e. the one involving the vertex  $V_{x \eta \eta}$. This gives
\begin{align}\label{fermsun1}
 -\frac12 \braket{V_{x \eta \eta} V_{x \eta \eta}} &=24 \int \frac{d^2\ppp\, d^2\qqq\,d^2\rrr}{(2 \pi)^4}\, \frac{(\ppp_1^2+1)\,\qqq_0\,\rrr_0\, \delta^{(2)}(\ppp+\qqq+\rrr)}{(\ppp^2+2)(\qqq^2+1)(\rrr^2+1)}\nonumber\\
 &= -6\,  I[2,1,1] + 3\, I[1]^2\,.
\end{align}
We note the appearance of another integral in the class \eqref{intcat}. The coefficient in front of this integral depends on the degrees of freedom of the theory and is thoroughly discussed in section \ref{comparison}. The partial results of the remaining sunset diagrams are
\begin{align}\label{fermsun2}
& -\frac12 \braket{(V_{z\eta\eta}+V_{z\eta\theta})(V_{z\eta\eta}+V_{z\eta\theta})}= 12\, I[1]^2 - 24\, I[1]\, I[0]\, ,\nonumber\\
& -\frac12 \braket{ V_{\varphi\eta \theta}  V_{\varphi\eta \theta} }_{\textup{1PI}}=24\, I[1]\, I[4]+ 3\, I[1]^2\,.
\end{align}
The latter vertices can be contracted also in a non-1PI manner
\begin{equation}
-\frac12 \braket{ V_{\varphi\eta \theta}  V_{\varphi\eta \theta} }_{\textup{non-1PI}}=-\frac12\, G_{\varphi\varphi} (0)\times 2^6 \times 3^2\times \int \frac{d^2\ppp}{(2\p)^2}\, \frac{\ppp_1^2+1}{\ppp^2+1}=-18\, I[1]^2\, ,
\end{equation}
where the factor in front of the integrals comes from the expression of the vertex and from counting the degrees of freedoms that can run in the loops. 
As in~\cite{Giombi:2009gd}, the divergent contribution proportional to $I[1]^2$ cancels exactly those coming from \eqref{fermsun1} and \eqref{fermsun2}.
The total cubic fermionic part reads
\begin{equation}
 W_{2,\textup{ferm. cubic}}=-6\,  I[2,1,1]+24\, I[1]\,I[4] - 24\, I[1]\, I[0]\,.
\end{equation}

Finally, we consider the fermionic double bubble diagrams. These involve the fermionic quartic vertices. However, most of the vertices appearing in the Lagrangian cannot contribute to the partition function, either because the bosonic propagators are diagonal or because they would produce vanishing integrals. The only relevant vertices are
 $V_{\varphi^2 \eta \theta}$ and $ V_{zz\eta\theta}$. Although we can build a diagram with $ V_{\eta^4}$, fermion propagators carry one component of the loop momentum in the numerator and produce vector tadpole integrals, which vanish by parity. We conclude that the contribution from fermionic double bubble graphs is
\begin{equation}
W_{2,\textup{ferm. bubbles}}=-24\, I[1]\,I[4] + 24\, I[1]\, I[0]\,.
\end{equation}

Summing all the partial results and reinstating the dependence on the string tension and the volume, we obtain
\begin{equation}\label{finalres}
 W_2=\frac{V}{T} \left(8\, I[4,2,2]-6 I[2,1,1]\right) =-4 \frac{V}{T} I[4,2,2]=-\frac{K}{4\, \pi^2}\frac{V}{T}\, ,
\end{equation}
where $T$ is defined in \eqref{eq:tension}.
Finally we can plug this expression into equation \eqref{eq:cuspprescription} and read out the second order of the strong coupling expansion \eqref{eq:expansion} of the ABJM cusp anomalous dimension
\begin{equation}\label{eq:a2}
a_2 = -\frac{K}{4\,\pi^2}\,.
\end{equation}
Plugging the result into \eqref{eq:expansion} we find perfect agreement with \eqref{cusp_result}, giving strong support to the conjecture \eqref{eq:strong} formulated in~\cite{Gromov:2014eha}.

%%%%%%%%%%%%%%%%%%%%%%%%%%%%%%%%
\subsection{Comparison with \txpf{$AdS_5\times S^5$}{AdS5 x S5}}\label{comparison}
%%%%%%%%%%%%%%%%%%%%%%%%%%%%%%%%
In this section we point out similarities and differences between the calculation we performed and 
its $AdS_5\times S^5$ analogue~\cite{Giombi:2009gd}. 
The starting points, i.e. the Lagrangians in $AdS$ light-cone gauge, look rather different. Yet the final results of the two-loop computations are strikingly similar.
More precisely, when written in terms of the string tension, the two expressions have exactly the same structure up to the numerical coefficients in front of the integrals.
Indeed the $AdS_5$ computation gives\footnote{We translated the result of~\cite{Giombi:2009gd} to our convention for the worldsheet coordinates.}
\begin{equation}\label{resAdS5}
 W^{(AdS_5)}_2=\frac{V}{T} \left(4\, I[4,2,2]-4\, I[2,1,1]\right)\,,
\end{equation}
which looks very similar in structure to \eqref{finalres}.
Furthermore, using \eqref{intcat}, both combinations sum up to 
\begin{equation}
W_2=-\frac{V}{T}\,4\, I[4,2,2]\, ,
\end{equation}
and only the different relation between the string tension and the 't Hooft couplings distinguishes the final results.
It is easy to trace the origin of the integrals and their coefficients back in the vertices of the Lagrangian and to understand their meaning. 
In particular in both computations only the sunset diagrams involving the interactions $V_{\varphi xx}$ and $V_{x\psi\psi}$ (with massive fermions) seem to effectively contribute. All other terms are also important, but just serve to cancel divergences.
Hence we can now focus on the relevant interactions and point out the differences between the $AdS_5$ and the $AdS_4$ cases. \\
We start from the bosonic sectors.
The two theories differ for the number of scalar degrees of freedom with given masses.
Focussing on massive fluctuations, after gauge fixing we have one scalar with $m^2=4$ associated to the radial coordinate of $AdS_{d+1}$ and $(d-2)$ real scalars with $m^2=2$.
In the metric we chose for the $AdS_4\times \mathbb{CP}^3$ background, the size of the $AdS_4$ part is rescaled by a factor of $r^2=4$. We have compensated this, parametrizing the radial coordinate as $w=e^{r\varphi}$ and introducing a factor $r$ in the fluctuation of $x$, so as to have the same normalization for their kinetic terms as in $AdS_5\times S^5$. This causes some factors $r$ to appear in interaction vertices in our Lagrangian. Apart from this, the relevant interaction vertices are exactly the same.
Then, the number of $x$ fields $(d-2)$ and this factor $r$ determine the coefficient of the integral $I[4,2,2]$ appearing in equations \eqref{finalres} and \eqref{resAdS5}.\\
Turning to fermions, the first striking difference between the $AdS_5$ and $AdS_4$ cases is the presence of massless ones.
As pointed out at the beginning of section \ref{fermcontr} their contribution is effectively vanishing at two loops (though they do contribute at first order).
Focussing on massive fermions, the relevant cubic interactions giving rise to $I[2,1,1]$ look again similar in the $AdS_4$ and $AdS_5$ cases. The difference is given once more by the ratio of the radii $r$ (through the normalization of $\varphi$ and $x$ coordinates) and the number $n_f$ of massive fermions in the spectrum ($n_f=8$ for $AdS_5\times S^5$ and $n_f=6$ for $AdS_4\times \mathbb{CP}^3$).\\ 
The final results \eqref{finalres} and \eqref{resAdS5} can be re-expressed in the general form
\begin{align}\nonumber
 W^{(AdS_{d+1})}_2&=2\frac{V}{T} (d-2)r^2 \left(I[4,2,2]-\frac{n_f}{8}\, I[2,1,1]\right)\\\label{finalresgen}
 &=2\frac{V}{T} (d-2)r^2 \left(1-\frac{n_f}{4}\right)\, I[4,2,2]\,,\qquad \qquad d=3,4\,,
\end{align}
where the cases at hand are $d=4$, $n_f=8$, $r=1$ for ${\cal N}=4$ SYM and $d=3$, $n_f=6$, $r=2$ for ABJM.

\section{Quantum dispersion relations for the \txpf{$AdS_4\times CP^3$}{AdS4 x CP3} GKP string.}

The excitations appearing in \eqref{eq:Lagrangian_exp} are in general non-relativistic beyond the leading order approximation. Moreover, unlike the BMN case the first non-relativistic corrections appear already at one-loop order due to the presence of cubic interactions. Therefore it is an interesting question to study the one-loop correction to the dispersion relation of the worldsheet excitations and compare the result with the integrability predictions of~\cite{Basso:2013pxa}.

\subsection{One-loop dispersion relations}\label{sec:dispersion}

\begin{figure}[h]
\begin{center}
\begin{tikzpicture}[scale=1, vertex/.style={circle,fill=black,thick,inner sep=2pt}]
\draw(-5,0) circle (1 cm);
\draw[-] (-7,0)--(-6,0);
\draw[-] (-4,0)--(-3,0);
%\node (c) at (-6,0) [vertex] {};
%\node (c) at (-4,0) [vertex] {};
\draw(0,0) circle (1 cm);
\draw[-] (-2,-1)--(2,-1);
% \node (c) at (0,0) [vertex] {};
 \draw[-] (5,0)--(5,-1);
 \draw[-] (4,-1)--(6,-1);
 \draw(5,0.5) circle (0.5 cm);
%\node (c) at (5.5,0) [vertex] {};
%\node (c) at (6.5,0) [vertex] {};
 \end{tikzpicture}
\end{center}
\caption{Diagram topologies for the two-point function one-loop corrections.}
\label{fig:diagrams}
\end{figure}
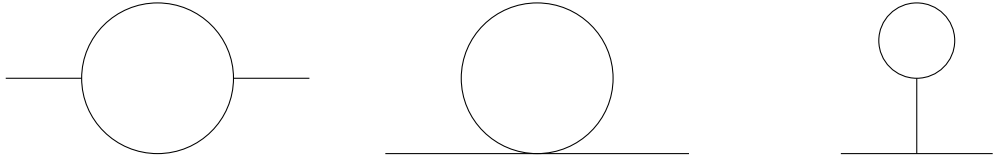

We consider the one-loop corrections to the two-point functions of the elementary fields of the action \eqref{eq:Lagrangian_exp}.
One-loop self-energy diagrams come in three different topologies: bubble, tadpole and non-1PI contributions, which are depicted in Figure \ref{fig:diagrams}.
The latter are allowed since the heavy scalar $\varphi$ has a non-trivial expectation value~\cite{Bianchi:2014ada}.
Indeed the only one-loop contribution comes from a fermionic loop giving
\begin{equation}
\langle \varphi \rangle = 3\, {\rm I}[1]\,,
\end{equation}
with the tadpole integral ${\rm I}[m^2]$ defined below in \eqref{eq:masterint}.
Bubble and tadpole diagrams give rise to integrals with several powers of loop momentum (up to six) in the numerator.
These are reduced to scalar integrals via Passarino-Veltman reduction. We use the same regularization prescription adopted in section \ref{sec:cuspAdS4} (see comments around \eqref{eq:reg}).

After tensor reduction one is left with two kinds of integral: tadpoles and bubbles\footnote{The bubble integral is the same appearing in \eqref{sbi}, however here we do not indicate explicitly the dependence on the external momentum since there is only one external legf in the problem and this will not generate any confusion. Moreover we use the representation \eqref{eq:Ires} of the result which is more convenient than \eqref{intu} in this context.}
\begin{align}
{\rm I}[m^2] &\equiv \int \frac{d^2\qqq}{(2\pi)^2}\, \frac{1}{\qqq^2+m^2}\, ,
\nonumber\\
{\rm I}[m_1^2,m_2^2] &\equiv \int \frac{d^2\qqq}{(2\pi)^2}\, \frac{1}{\left[\qqq^2+m_1^2 \right]\left[(\qqq+\ppp)^2+m_2^2 \right]}\label{eq:masterint}\,.
\end{align}
The latter are ultraviolet convergent and IR finite if both propagators are massive and evaluate to 
\begin{equation}\label{eq:Ires}
{\rm I}[m_1^2,m_2^2] = \frac{\log \frac{\ppp^2+m_1^2+m_2^2+\sqrt{(\ppp^2+m_1^2+m_2^2)^2-4\, m_1^2\, m_2^2}}{\ppp^2+m_1^2+m_2^2-\sqrt{(\ppp^2+m_1^2+m_2^2)^2-4\, m_1^2\, m_2^2}}}{4\,\pi\, \sqrt{(\ppp^2+m_1^2+m_2^2)^2-4\, m_1^2\, m_2^2}}\,.
\end{equation}
Whenever one of the masses vanishes the bubble suffers from infrared singularities which can be isolated in terms of tadpole integrals using~\cite{Giombi:2010bj}
\begin{equation}\label{eq:bubble0}
{\rm I}[0,m^2] = \frac{1}{\ppp^2+m^2} \left( \frac{1}{2\pi} \log \frac{\ppp^2+m^2}{m^2} - {\rm I}[m^2] + {\rm I}[0] \right)\,.
\end{equation}
Tadpoles are UV divergent. We verify that in dispersion relations they always drop out because they are multiplied by factors going to zero on-shell.
Nevertheless, they are present in the off-shell corrections to the two-point functions. In some cases they appear in finite combinations, but in other they do produce ultraviolet singularities, indicating that the corresponding fields undergo a non-trivial wave function renormalization. 

We collect the tree level structure of propagators according to
\begin{equation}
\langle \bullet(\ppp)\star(-\ppp) \rangle^{(1)} = \frac{1}{T}\, \frac{G_{\bullet\star}(\ppp)}{\ppp^2+m_{\bullet}^2}\, F_{\bullet\star}^{(1)}\,,
\end{equation}
for generic fields $\bullet$ and $\star$. When performing the usual one-loop resummation of non-1PI contributions the on-shell ($\ppp_0=\sqrt{-m^2-\ppp_1^2}$) value of the function $F_{\bullet\star}^{(1)}$ shifts the pole of the propagator. From this shift one can read off the corrections to the dispersion relations in \eqref{eq:result}. In particular, evaluating the shift at $\ppp_1=0$ one computes the mass shift $q$ in equation \eqref{eq:result} and subsequently the coefficients $c$ and $d$ by subtraction.
We now spell out the details of the results for the perturbative one-loop corrections to the dispersion relations and masses of each particle in the fluctuation Lagrangian \eqref{eq:Lagrangian_exp}.

\subsubsection{Light scalar}

The $x$ scalar self-energy one-loop correction reads
\begin{align}
F_{xx}^{(1)} & = \frac{\left(\ppp_1^2+1\right) \left(-12\, \ppp^2 {\rm I}[1,1] \left(\ppp^4+4\, \ppp_1^2\right)-16 \left(\ppp^4+8\, \ppp^2+4\right) {\rm I}[2,4] \left(\ppp^2-2\, \ppp_1^2\right)\right)}{\ppp^4}+
\nonumber\\&
+ \frac{16 \left({\rm I}[2]-{\rm I}[4]\right) \left(\ppp^2+2\right) \left(\ppp_1^2+1\right) \left(\ppp^2-2 \ppp_1^2\right)}{\ppp^4}\,,
\end{align}
where the difference of UV divergent tadpoles gives a finite remainder ${\rm I}[2]-{\rm I}[4] = \log 2$ and hence $x$ does not need any renormalization.
The self-energy evaluated on-shell reads
\begin{equation}
F_{xx}^{(1)} \Big|_{\ppp^2=-2} = \left(\ppp_1^2+1\right)^2\,.
\end{equation}
The one-loop corrected dispersion relation then becomes
\begin{equation}
\ppp^2+2 = \frac{1}{2\sqrt{2\lambda}}\, F_{xx}^{(1)} \Big|_{\ppp^2=-2} + {\cal O}(\lambda^{-1})\,,
\end{equation}
that is, in Lorentzian signature $(\ppp_0,\ppp_1)\rightarrow (-i \o, p)$,
\begin{equation}\label{eq:dispx}
\o^2 = p^2 + 2 - \frac{1}{4\, h(\lambda)}\, \left(p^2+1\right)^2 + {\cal O}(\lambda^{-1})\,.
\end{equation}
At $p=0$ one can read off the one-loop correction to the mass
\begin{equation}
m_{x}^2 = 2 - \frac{1}{4\, h(\lambda)} + {\cal O}(\lambda^{-1}) < 2\,.
\end{equation}
The fact that the first perturbative correction to the mass at strong coupling is decreasing its value is in general agreement with the trend put forward in~\cite{Basso:2013pxa}, according to which the masses of all elementary excitations should tend to 1 at weak coupling.

\subsubsection{Heavy scalar}

We now turn to the heavy scalar mode  $\varphi$, whose one-loop correction to the self-energy is found to be
\begin{align}\label{eq:phi2pt}
F_{\varphi\varphi}^{(1)} & = 4 \left(3 {\rm I}[1]-{\rm I}[2] - 2 {\rm I}[4] \right) \left(\ppp^2+4\right) -\frac{12 \left(\ppp^2+4\right) \ppp_1^2 {\rm I}[1,1] \left(\ppp^4+4\, \ppp_1^2\right)}{\ppp^4}
\nonumber\\&
+\frac{8 \left(\ppp^2+4\right)^2 {\rm I}[4,4] \left(\ppp^2-2\, \ppp_1^2\right)^2}{\ppp^4}+2 {\rm I}[2,2] \left(\frac{64\, \ppp_1^4}{\ppp^4}-\frac{64\, \ppp_1^2}{\ppp^2}+\left(\ppp^2+4\right)^2\right)\,.
\end{align}
Again, the difference of UV divergent tadpoles leave a finite remainder $3 {\rm I}[1]-{\rm I}[2] - 2 {\rm I}[4] = 5\log 2$. Therefore the field $\varphi$ does not renormalize, to one loop order.
Evaluating the self-energy on-shell we obtain
\begin{equation}
F_{\varphi\varphi}^{(1)} \Big|_{\ppp^2=-4} =\frac12 \ppp_1^2 \left(\ppp_1^2+4\right)\,.
\end{equation}
In going on-shell the integral ${\rm I}[1,1]$ is singular, which is explained as coinciding with the threshold energy for production of a pair of fermions. This integral is multiplied by a power of $(\ppp^2+4)$, enforcing the limit to vanish.
Then the one-loop corrected dispersion relation reads
\begin{equation}
\ppp^2+4 = \frac{1}{2\sqrt{2\lambda}}\, F_{\varphi\varphi}^{(1)} \Big|_{\ppp^2=-4} + {\cal O}(\lambda^{-1})\,.
\end{equation}
Switching to Lorentzian signature it becomes
\begin{equation}\label{eq:dispphi}
\o^2 = p^2 + 4 - \frac{1}{8\, h(\lambda)}\, p^2 \left(p^2+4\right) + {\cal O}(\lambda^{-1})\,.
\end{equation}
The one-loop correction to the mass is clearly seen to vanish. This agrees with the analysis of~\cite{Alday:2007mf}, according to which the mass of this mode is protected.
In section \ref{heavy} we discuss more deeply the analytic structure of the one-loop correction \eqref{eq:phi2pt} and its implications for the role of the heavy scalar in the asymptotic states of the model.

\subsubsection{Massless scalars}

The one-loop contribution to the two-point function of the massless scalars suffers from both IR and UV divergences, which can be expressed in terms of tadpoles using the identity \eqref{eq:bubble0}.
The $z$ scalar self-energy one-loop correction reads
\begin{align}
F_{zz}^{(1)} & = \frac{1}{2 \pi  \ppp^4} \left[8 \pi  \ppp^2\, {\rm I}[1,1] (\ppp^2-\ppp_1^2) \left(\ppp^4+4\, \ppp_1^2\right)+2 \left(\ppp^2+4\right) \left(\ppp^4-8 \ppp^2 \ppp_1^2+8 \ppp_1^4\right) \log (\tfrac{\ppp^2+4}{4}) 
\right.\nonumber\\&\left. 
+ \left(\ppp^6-\ppp^4 \left(2 \ppp_1^2+1\right)+8 \ppp^2 \ppp_1^2-8 \ppp_1^4\right) \log \left(\ppp^2+1\right) \right] + \frac43 \left({\rm I}[0]-3{\rm I}[1]\right) \ppp^2\,.
\end{align}
Then one can see that ${\rm I}[4]$ tadpoles cancel and the rest is proportional to ${\rm I}[0]-3\, {\rm I}[1]$ which is UV (and IR) divergent, but it is multiplied by $\ppp^2$ and vanishes on-shell. 
The on-shell self-energy evaluates
\begin{equation}
F_{zz}^{(1)} \Big|_{\ppp^2=0} = \frac{11}{3 \pi}\, \ppp_1^4\,,
\end{equation}
where the residual UV and IR divergences disappear.
Hence the one-loop corrected dispersion relation reads
\begin{equation}\label{eq:dispz}
\o^2 = p^2 - \frac{1}{h(\lambda)}\, \frac{11}{12\, \pi}\, p^4 + {\cal O}(\lambda^{-1})\,.
\end{equation}
At $p=0$ one can read off the one-loop correction to the mass, which is seen to vanish.

\subsubsection{Massive fermions}

The kinetic terms of the fermion Lagrangian mix the fermion fields.
Hence we have to consider separately the corrections to the two-point functions $\braket{\eta_a \eta^a}$, $\braket{\theta_a \theta^a}$ and $\braket{\eta_a \theta^a}$.
Their computation involves several contributions and the final forms are not particularly illuminating; we spell them out in appendix \ref{app:fermse}.
We point out that the off-shell one-loop corrections to $\langle\eta_a\eta^a\rangle$ and $\langle\theta_a\theta^a\rangle$ are finite, whereas that for $\langle\eta_a\theta^a\rangle$ is UV divergent, although the divergent term cancels on-shell. This implies that the massive fermions, like the massless scalars, undergo wave-function renormalization. The correction to the $\braket{\eta_a \theta^a}$ two-point function is also IR divergent off-shell. Once more the divergent term vanishes on-shell. We will comment on the role of IR divergences in section \ref{sec:comparison}. 

The different two-point functions all coincide on-shell, corroborating the hypothesis that all the massive fermions have the same dispersion relation
\begin{align}
F_{\eta_a\eta^a}^{(1)} \Big|_{\ppp^2=-1} = 
F_{\theta_a\theta^a}^{(1)} \Big|_{\ppp^2=-1} = 
F_{\eta_a\theta^a}^{(1)} \Big|_{\ppp^2=-1} = 2\, \ppp_1^2 \left( \ppp_1^2 + 1 \right)\,.
\end{align}
Thus, the one-loop corrected dispersion relation takes the form
\begin{equation}\label{eq:dispferm1}
\o^2 = p^2 +1 - \frac{1}{2\, h(\lambda)}\, p^2 \left( p^2 + 1\right)\,,
\end{equation}
from which one sees that the mass does not receive corrections.
Again, this conclusion is in agreement with the integrability prediction that the massive fermion mass is protected from strong to weak coupling.

\subsubsection{Massless fermions}

The two-point functions for massless fermions are different, depending on the fields, but coincide on-shell, where they are all finite
\begin{equation}
F_{\eta_4\eta^4}^{(1)} \Big|_{\ppp^2=0} = 
F_{\theta_4\theta^4}^{(1)} \Big|_{\ppp^2=0} = 
F_{\eta_4\theta^4}^{(1)} \Big|_{\ppp^2=0} = \frac{\ppp_1^2 \left(7 \ppp_1^2-4\right)}{\pi }\,.
\end{equation}
Hence the one-loop correction to the dispersion relation reads
\begin{equation}\label{eq:dispferm2}
\o^2 = p^2 - \frac{1}{4\, \pi\, h(\lambda)}\, p^2 \left( 7 p^2 - 4 \right)\,,
\end{equation}
from which the mass is not corrected.

\subsubsection{Summary of the results}

We collect here the results of our perturbative computation for the excitations of the ABJM GKP string in a compact form. The string theory spectrum at $\l\to \infty$ consists of
\begin{align}
 AdS_3\textup{ transverse mode }(\varphi):	\qquad&m_\varphi^2=4\\
 AdS_4 \textup{ outside } AdS_3\ (x):		\qquad&m_x^2=2\\
 \mathbb{CP}^3\ (\{z^a,\bar z_a\}, a=1,2,3):	\qquad&m_z^2=0\\
 \textup{Massive fermions } (\eta^a,\theta^a):	\qquad&m_{\eta^a}^2=m_{\theta^a}^2=1\\
 \textup{Massless fermions } (\eta^4,\theta^4):	\qquad&m_{\eta^4}^2=m_{\theta^4}^2=0
\end{align}

We find the following quantum corrections to the dispersion relations and masses of those excitations, which can be compared to the results \eqref{eq:resultAdS5} by replacing $h(\l)\to \frac{\sqrt{\l}}{4\,\pi}$
\begin{align}\label{eq:result}
 &\o^2(p,\l)=\left[p^2+m^2+\frac{q}{h(\l)}+\mathcal{O}(\l^{-1})\right]\left[1+\frac{c\, p^2+d }{h(\l)}+\mathcal{O}(\l^{-1})\right]\, ,\\
 &\begin{array}{lllll}
 q_\varphi=0\, , 	\quad&q_x=-\frac14  \, , \quad&q_z=0 \, , 		\quad&q_{\eta^a}=q_{\theta^a}=0 \, , 		\quad&q_{\eta^4}=q_{\theta^4}=0\, ,\\
 c_\varphi=-\frac18\, , \quad&c_x=-\frac14 \, , \quad&c_z=-\frac{11}{12\,\p}\, ,\quad&c_{\eta^a}=c_{\theta^a}=-\frac12 \, ,  \quad&c_{\eta^4}=c_{\theta^4}=-\frac{7}{4\,\p}\, ,\\
 d_\varphi=0 \, ,	\quad&d_x=0\, ,  	\quad&d_z=0\, ,  	\quad&d_{\eta^a}=d_{\theta^a}=0\, ,  	\quad&d_{\eta^4}=d_{\theta^4}=\frac{1}{\p}\, .
\end{array}
\end{align}

\subsection{Comparison and comments}\label{sec:comparison}

Provided with the result \eqref{eq:result} we can compare it to the higher-dimensional case and discuss some interesting implications of it.

\subsubsection{Comparison with \txpf{$AdS_5\times S^5$}{AdS5 x S5} and integrability predictions}

The physics of the excitations on top of the GKP vacuum for the ABJM model has been extensively analysed using integrability in~\cite{Basso:2013pxa}. In particular the dispersion relations of its modes were computed exactly. The Bethe Ansatz analysis reveals a remarkable similarity to the $AdS_5\times S^5$ spinning string setting.
Therefore we start commenting on the results of the previous section by comparing them with the corresponding findings of ${\cal N}=4$ SYM \eqref{eq:resultAdS5}.
We observe that all the dispersion relations for massive modes are related to those of the corresponding fields in the $AdS_5\times S^5$ sigma model by
\begin{equation}
\o(p)^{(1)}_{AdS_5\times S^5} = \left.\o(p)^{(1)}_{AdS_4 \times \mathbb{CP}^3}\right|_{h(\l)\to\frac{\sqrt{\l}}{4\pi}} \,.
\end{equation}
For massless modes such a comparison is not possible, since it is not even clear what to compare: in $AdS_5\times S^5$ there are only massless scalars, whereas for $AdS_4\times \mathbb{CP}^3$ these are coupled to a massless fermion. Also the two low-energy models and their fundamental excitations are rather different: in $AdS_5\times S^5$ the relevant model in the Alday-Maldacena limit is the $O(6)$ sigma model, whose fundamental excitations are six massive scalars in the $\bf{6}$ of $SO(6)$~\cite{Alday:2007mf}; on the other hand for \adscp the fundamental excitations of the Bykov model~\cite{Bykov:2010tv} turned out to be 4 spinons and 4 antispinons transforming in the $\bf{4}$ and $\bar{\bf{4}}$ of $SU(4)$~\cite{Basso:2012bw}. For both models the strong coupling perturbative interpretation is far from obvious due to the exponentially small mass of the excitations in such regime.

Turning to the comparison with integrability, it turns out that, as in the $AdS_5/CFT_4$ case, it hides some subtleties.
We start commenting on massive modes.
In the asymptotic Bethe Ansatz approach the dispersion relation of the massive modes of ${\cal N}=4$ SYM is predicted to be the same as that of the corresponding massive excitations of ABJM.

For the bosons, the quantum correction to the dispersion relation of the light massive scalar agrees with the integrability result.
The heavy scalar, as in ${\cal N}=4$ SYM, is absent in the Bethe Ansatz description. Therefore its role in the sigma model should be analysed carefully and we postpone a thorough discussion of this issue to section \ref{heavy}. Here we stress that at one loop order the heavy scalar has the same dispersion relation as the corresponding heavy field in ${\cal N}=4$ SYM, despite the fact that there is no direct integrability based argument explaining that (although one may argue that the similarity of the two Bethe Ans\"atze would make the predictions for $\mathcal{N}=4$ SYM valid also in the present case). 

For the fermions, the one-loop corrected dispersion relation for massive modes is in full agreement with the integrability prediction.

Turning to the massless modes, only the fact that the mass does not receive perturbative corrections is compatible with the integrability predictions.
Indeed, the Bethe equations analysis reveals that the model has a gap and such modes acquire non perturbatively an exponentially small mass. This parallels what occurs to the scalars of the $O(6)$ sigma model emerging in $AdS_5\times S^5$ in the Alday-Maldacena limit~\cite{Alday:2007mf,Basso:2008tx}.
Apart from that, there is no direct identification between the dispersion relations of massless fields of the superstring description and the non-perturbative modes of integrability.
As pointed out in~\cite{Zarembo:2011ag}, the presence of perturbatively massless fields induces IR divergences in loop computations, which appear as logarithms of the infrared scale of the theory.
Indeed the explicit computation of some one-loop two-point functions already shows the presence of IR divergences, though they always drop out from the dispersion relations.
The infrared cutoff of the theory is set by the non-perturbative mass of the particles which, roughly, scales exponentially with the coupling $\sqrt{\lambda}$. This implies that logarithms of this scale behave like powers of the coupling, effectively lowering the perturbative order to which these terms contribute. In practice this means that an IR divergence appearing at $l$ loops contributes to the $(l-1)$-loop result, invalidating the perturbation theory predictivity at that order.
Therefore it is likely that the one-loop dispersion relations for massless modes \eqref{eq:dispz} and \eqref{eq:dispferm2} are not trustworthy due to two-loop IR divergences, despite being IR finite at one loop.
This argument could actually spoil the computation of the one-loop dispersion relations for massive fields, where IR divergences could also appear at two loops. However the theorems in~\cite{Elitzur:1978ww,David:1980gi} suggest that $O(6)$ invariant quantities should be IR finite, and since $\varphi$ and $x$ are singlets under $O(6)$ we expect their correlation function to be reliable in perturbation theory.
It would be interesting to ascertain this explicitly via a two-loop computation of the two-point functions. 

Let us also mention an additional striking feature of the comparison with integrability. The scalar excitations over the GKP vacuum in the integrability analysis of~\cite{Basso:2013pxa} transform in the $\mathbf{4}$ and $\mathbf{\bar 4}$ of $SU(4)$, whereas the superstring elementary excitations transform only in the fundamental representation of the $SU(3)$ symmetry which survives in the Goldstone vacuum. This is similar to what happens in $\mathcal{N}=4$ SYM where the scalar excitations in the string picture are organized in vectors of $SO(5)$, the explicit symmetry of the $O(6)$ sigma model expanded around the Goldstone vacuum. In this context the analysis of~\cite{Elitzur:1978ww} gives a recipe for computing $O(N)$ invariant correlation functions in the $O(N)$ sigma model and in~\cite{David:1980gi} it was proven that they are free of IR divergences. It is an interesting question whether the same technique can be applied to the Bykov model or even to the full non-linear string sigma model in $AdS_5\times 
S^5$ or in $AdS_4 \times \mathbb{CP}^3$.

\subsubsection{Comments on the heaviest scalar}\label{heavy}

As is the case for ${\cal N}=4$ SYM, the heaviest scalar mode $\varphi$, which is present in the Lagrangian \eqref{eq:action}, does not correspond to an elementary excitation in the Bethe ansatz description, based on the conjectured integrability of the model.
The r\^ole of this field was deeply analysed in the literature for $AdS_5\times S^5$ \cite{Giombi:2010bj,Zarembo:2011ag,Basso:2014koa,Fioravanti:2015dma}.
A possible explanation that was put forward to explain this mismatch is that the $\varphi$ field is not an asymptotic state of the quantum theory, along the lines of the arguments of \cite{Zarembo:2009au}.
This latter hypothesis and its consequences can be studied perturbatively.
In particular the analytic structure of the two-point function should tell whether it exists as an asymptotic state and whether it is stable or it can decay into lighter particles, such as a pair of massive fermions. This kind of analysis was performed at one loop in \cite{Giombi:2010bj} and \cite{Zarembo:2011ag}.
The punchline is that up to one-loop order the scalar $\varphi$ is a stable threshold composite state of two fermions. Its would be pole in the two-point function coincides with the branching point of the two-fermion continuum square root and hence the scalar cannot be interpreted as a genuine asymptotic bound state.
However, depending on the next order corrections, this conclusion can vary according to how the $\varphi$ and the fermion dispersion relations get modified.

In \cite{Basso:2014koa} the contribution of the heavy scalar appears naturally as a $SU(4)$-singlet compound state of two fermions which perfectly reproduces one of the two-particle contributions to the excited flux-tube. The energy and the momentum of this two-particle state at finite coupling are simply related to the energy and momentum of the fermionic excitations. In particular analysing this relation at strong coupling one finds that
\begin{equation}\label{eq:phifermion}
E_{\varphi}(\ppp)-2\, E_{\psi}\left(\frac{\ppp}{2}\right)=-\frac{\p^2 \ppp^4(\ppp^2+4)^{\frac32}}{8\, \l}+\mathcal{O}(\l^{-\frac32})\,,
\end{equation}
where $\lambda$ is the ${\cal N}=4$ SYM 't Hooft coupling.
The minus sign in the r.h.s of this equation predicts that at two-loops the pole of the heavy scalar two-point function actually moves below the threshold. The results of \cite{Basso:2014koa} show that this property holds also at finite coupling preventing $\varphi$ from decaying into two fermions. Although the pole of the heavy scalar two-point function is shifted below the threshold, the analysis of the singlet channel in the scattering phase of two fermions shows that the unwanted pole is located in the unphysical strip of the rapidity complex plane \cite{Basso:2014koa, Zamolodchikov:2013ama}. This in turn means that $\varphi$ cannot be a true asymptotic state of the theory.

The same arguments should also apply to the heavy scalar in the $AdS_4\times \mathbb{CP}^3$ model.
However they go beyond the one-loop computation carried out in this paper.
What our analysis can test is the integrability prediction that up to one-loop the $\varphi$ scalar should appear as a stable threshold bound state of two fermions.
This expectation can be verified along the lines of \cite{Giombi:2010bj} and \cite{Zarembo:2011ag} as follows.
The one-loop contribution to the denominator of the resummed two-point function has the form
\begin{equation}
F_{\varphi\varphi}^{(1)}(p) = a_0 + a_{1/2} (p^2+4)^{\tfrac12} + \dots\,,
\end{equation}
where all other terms vanish more rapidly in the vicinity of the tree-level mass condition.
In particular we note the presence of the square root $\sqrt{p^2+4}$. Although it is not immediate to see the emergence of this term from \eqref{eq:phi2pt}, it arises from the denominator of ${\rm I}[1,1]$, appearing in the fermion loop diagram.
Close to the threshold, the inverse corrected two-point function
\begin{equation}
G_{\varphi}^{-1}(p) = p^2 + 4 - \frac{1}{2\sqrt{2\lambda}}\, F_{\varphi\varphi}^{(1)}(p) + {\cal O}(\lambda^{-1})
\end{equation}
vanishes at
\begin{equation}
p^2 = -4 + \frac{1}{2\sqrt{2\lambda}}\, a_0 + {\cal O}(\lambda^{-1})\,,
\end{equation}
where here $a_0 = \tfrac12 p_1^2(p_1^2+4)$.
This location lies below the branch cut threshold induced by the square root, meaning that it corresponds to a genuine pole.
From this one would conclude that the $\varphi$ scalar does represent an asymptotic state of the theory.
However this does not take into account that the physical threshold for fermion production is also shifted by quantum corrections.
One can imagine the structure of the resummed two-point function to all orders to have the form (in Lorentz signature)
\begin{equation}\label{eq:inverseprop}
G_{\varphi}^{-1}(\ppp) = -E^2 + 4 E_{\psi_i}^2\left(\frac{\ppp}{2}\right) - \frac{a_{1/2}}{2\sqrt{2\lambda}} \left(-E^2 + 4 E_{\psi_i}^2\left(\frac{\ppp}{2}\right)\right)^{\tfrac12} + \dots\,,
\end{equation}
where $ 4E_{\psi_i}(\ppp/2)=4-\frac{a_0}{2\sqrt{2\lambda}}+\mathcal{O}(\l^{-1})$ is the quantum corrected dispersion relations of the massive fermions.
Its expansion to first order in $\lambda^{-\tfrac12}$ would be in agreement with the perturbative computation \eqref{eq:phi2pt}, although the latter does not guarantee nor hint that \eqref{eq:inverseprop} should hold at higher order.
Assuming this is the case, the would be pole at $E^2 = 4 E_{\psi_i}^2\left(\frac{\ppp}{2}\right)$ coincides with the branching point of the square root. Moreover if the coefficient of the square root $a_{1/2}$ is positive (as the one-loop computation shows it is the case) no other physical poles are present in the two-point function, but only a pole on the second, unphysical, sheet of the square root, located at
\begin{equation}
E^2 = 4\, E_{\psi_i}^2\left(\frac{\ppp}{2}\right) - \frac{a_{1/2}^2}{8\,\lambda} + {\cal O}(\l^{-\frac32})
\end{equation}
where $a_{1/2}$ can be extracted expanding \eqref{eq:phi2pt} near the threshold and reads
\begin{equation}
a_{1/2} = \frac{3\, \ppp^2 (\ppp^2+4)}{4}
\end{equation}
As a result $\varphi$ does not represent a real asymptotic state of the theory.
Insisting on this logic, we can derive a conjectural analogue of \eqref{eq:phifermion}, for the $AdS_4\times \mathbb{CP}^3$ case 
\begin{equation}
E_{\varphi}(\ppp)-2\, E_{\psi_i}\left(\frac{\ppp}{2}\right)=-\frac{9\, \ppp^4(\ppp^2+4)^{\frac32}}{256\, \l}+\mathcal{O}(\l^{-\frac32})\,,
\end{equation}
which would be interesting to check against an integrability based prediction and a full two-loop perturbative computation.

\section{Bound states for the \txpf{$AdS_4\times \mathbb{CP}^3$}{AdS4 x CP3} GKP string.}\label{sec:boundstates}

The Bethe equation analysis of the GKP excitations shows that the light scalars $x$ can form bound states, whose energy can be computed.
Although they are not immediately detectable in a superstring approach, following~\cite{Zarembo:2011ag} we can attempt to estimate their energy to leading order.
This is done treating the $x$ fields as non-relativistic and computing the scattering amplitude of a pair of them.
From the amplitude one can extract the effective (attractive) potential experienced by the two particles.
In particular, this is done by computing their $2\to 2$ scattering amplitude and comparing it with the Born approximation in quantum mechanics
\begin{equation}
{\cal M}(k) = - 2\, (2m)^2 \int dx\, e^{-ikx}\, V(x)\,,
\end{equation}
where $k$ is the momentum transfer of the scattering process. This means that the effective potential $V(x)$ is basically the Fourier transform of the amplitude up to numerical constants due to different normalization of the wave-function and Bose statistics.
To lowest order in a momentum expansion, the scattering amplitudes become constants and their Fourier transform is proportional to a $\delta$-function.
The problem then reduces to a many-body system of particles interacting pairwise with a $\delta$-function potential $V_{ij}(x)= - g\, \delta(x_{i}-x_{j})$.
Such a model admits a two-particle bound state with one energy level $\o=-\tfrac{\mu\,g^2}{2}$, where $\mu$ is the reduced mass of the system ($\mu=\tfrac{1}{\sqrt{2}}$ for the $x$ scalars).
More generally, the binding energies for bound states of $\ell$ particles of mass $m$ are~\cite{McGuire:1964zt}
\begin{equation}\label{eq:multidelta}
\o_{\ell} = - \frac{m\, g^2}{24}\, \ell (\ell^2-1)\,.
\end{equation}
This energy can be compared to the static limit of the lowest order expansion for $\lambda\gg 1$ of the binding energy derived from integrability. This is given by
\begin{equation}
\o_{binding,\ell}(p) = \o_{\ell}(p) - \ell\, \o_1(\tfrac{p}{\ell})\,,
\end{equation}
where $\o_{\ell}(p)$ is the dispersion relation for the relevant twist-$\ell$ excitation.

\begin{figure}
\centering
\includegraphics[width=0.7\textwidth]{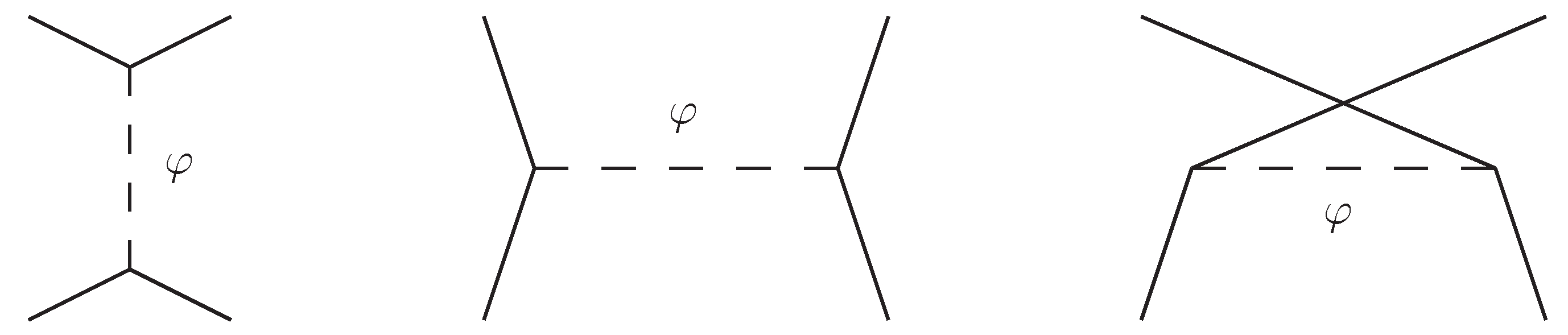}
\caption{Tree level scattering $xx\to xx$.}
\label{fig:scatx}
\end{figure}

In ${\cal N}=4$ SYM such a program was successfully carried out for the gauge excitation, showing agreement with the integrability prediction at $p=0$.
In this section we perform a similar computation for the mass $\sqrt{2}$ mode of the $AdS_4\times \mathbb{CP}^3$ superstring.
At tree level the amplitude for $xx\to xx$ scattering receives contributions from all $s$, $t$ and $u$ channels, as in Figure \ref{fig:scatx}.
In the zero-momenta limit, the contributions from the $t$ and $u$ channels are equal and give
\begin{equation}
{\cal M}_{xx\to xx,\, t} = {\cal M}_{xx\to xx,\, u} = 2^5\, \sqrt{2\lambda} + {\cal O}(k)\,,
\end{equation}
whereas the $s$-channel contributes with an opposite result, corresponding to a repulsive interaction.
Altogether the amplitude gives
\begin{equation}
{\cal M}_{xx\to xx} = 2^5\, \sqrt{2\lambda} + {\cal O}(k)\,,
\end{equation}
from which we find the effective potential (after properly rescaling fields by a $T^{-1/2}$ factor and introducing $h(\lambda)$ \eqref{eq:prediction})
\begin{equation}
V_{xx}(x) = -\frac{1}{4\, h(\lambda)}\, \delta(x)\,.
\end{equation}
Plugging this into \eqref{eq:multidelta} we give an estimate for the binding energy of the twist $\ell$ gauge bound state
\begin{equation}
\o_{binding,\ell}(0) = - \frac{\sqrt{2}\, \ell (\ell^2-1)}{384\, h(\lambda)^2} + {\cal O}(\lambda^{-2})\,,
\end{equation}
which is equivalent to the corresponding one for $AdS_5\times S^5$, once the replacement $h(\lambda) \to \tfrac{\sqrt{\lambda}}{4\pi}$ is performed. Thus it agrees with the integrability prediction of~\cite{Basso:2010in} at first order at strong coupling.

According to the parallel analysis of~\cite{Basso:2014koa} in $AdS_5\times S^5$, multi-fermion states are also present in the theory.
These appear as bound states of the two-fermion composites which we have identified as the mass 2 excitations $\varphi$ of the sigma model. These composite states of $2n$ fermions are expected to have mass $2n$, and consequently the bound states of $\varphi$ to have zero binding energy at vanishing momentum\footnote{We would like to thank B. Basso for explaining this to us.}.
We therefore repeat the same analysis as above for the scalars $\varphi$, in order to check whether the binding energy is vanishing at leading order in the static limit.
\begin{figure}
\centering
\includegraphics[width=0.9\textwidth]{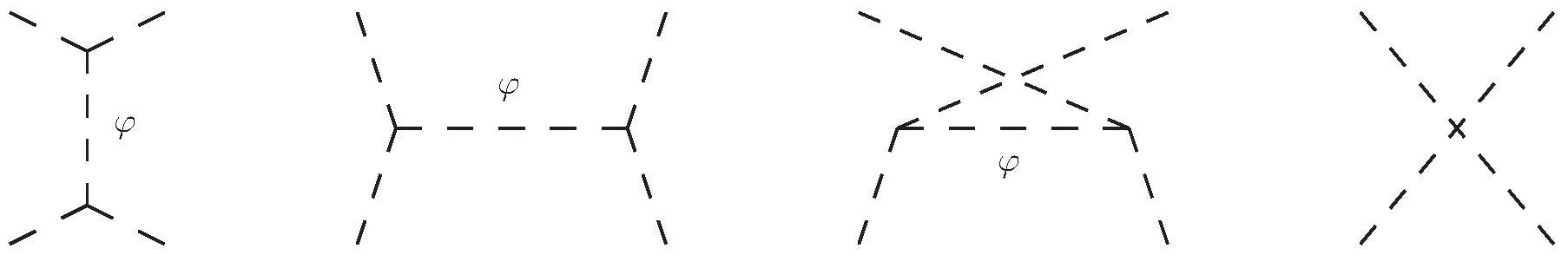}
\caption{Tree level scattering $\varphi\varphi \to \varphi\varphi$.}
\label{fig:scatphi}
\end{figure}
The lowest order scattering amplitude for $\varphi\varphi\to \varphi\varphi$ is given by the sum of the diagrams in Figure \ref{fig:scatphi}. Once again the $t-$ and the $u-$channel give two identical contributions in the static limit
\begin{equation}
 {\cal M}_{\varphi\varphi\to \varphi\varphi,\, t} = {\cal M}_{\varphi\varphi\to \varphi\varphi,\, u} = 2^7\, \sqrt{2\lambda} + {\cal O}(k)\,.
\end{equation}
In this case also the four point vertex gives an attractive contribution, which is once more equal to
\begin{equation}
 {\cal M}_{\varphi\varphi\to \varphi\varphi,\, 4}= 2^7\, \sqrt{2\lambda} + {\cal O}(k)\,.
\end{equation}
The s-channel contribution, as in the previous case, contributes with a repulsive interaction which compensates exactly the other terms
\begin{equation}
{\cal M}_{\varphi\varphi\to \varphi\varphi,\, s} =-3\times  2^7\, \sqrt{2\lambda} + {\cal O}(k)\,.
\end{equation}
In conclusion
\begin{equation}
{\cal M}_{\varphi\varphi\to \varphi\varphi} =  {\cal O}(k)\,,
\end{equation}
which implies that the bound state of $\varphi$ has vanishing binding energy in the static limit, in agreement with the integrability prediction.
As a further check we performed the same computation in $AdS_5\times S^5$, where the vertices are modified by relative factors and we found that the mechanism is exactly the same. Therefore, as expected, the binding energy vanishes also in that case. 

% Chapter Template

\chapter{Conclusions and outlook} % Main chapter title

\label{conclusion} % Change X to a consecutive number; for referencing this chapter elsewhere, use \ref{ChapterX}

\lhead{Chapter 5. \emph{Conclusion}} % Change X to a consecutive number; this is for the header on each page - perhaps a shortened title

In this thesis we have reviewed the construction of superstring theory for various $AdS$ backgrounds and we have shown several examples of perturbative computations in the strong coupling regime of the AdS/CFT correspondence. The main purpose of these calculations is to provide perturbative checks of the quantum integrability and quantum consistency of the string sigma models. We probed these features in perturbation theory for a number of interesting observables, finding strong support for their validity.

In particular, in the context of the near-BMN expansion of the $AdS_5\times S^5$ and $AdS_3\times S^3 \times M^4$ superstring actions, we have shown how the introduction of new powerful techniques allows to overcome the obstacles to the computation of the one-loop correction to the worldsheet S-matrix. Those obstacles had mainly to do with the several complicated interaction vertices appearing in the action and with the subtleties related to different possible regularization procedures. The unitarity methods provide an efficient way to bypass these difficulties~\cite{Bianchi:2013nra,Engelund:2013fja}. Indeed, the only ingredient for the computation of a one-loop amplitude via unitarity is the tree-level amplitude and, since the result is expressed only in terms of bubble integrals, it is inherently finite and any regularization issue is avoided.
Driven by this line of thought, we explicitly reproduced all the steps leading to a compact and rather general formula expressing the one-loop S-matrix of any two-dimensional massive field theory in terms of the tree-level one. There is an important caveat, though. The result obtained by unitarity is guaranteed to work as far as the logarithmic (and imaginary) part of the one-loop amplitude is concerned (the so-called cut constructible part). The result is therefore determined, in general, up to a rational function of the kinematical variables. 

Nonetheless, the various examples we collected allow us to postulate that for integrable theories the cut constructible part should coincide with the full result, up to something proportional to the tree-level S-matrix which can be interpreted as a shift in the coupling. Under this assumption, following~\cite{Bianchi:2014rfa} we provided one-loop predictions for the dressing phases of $AdS_3\times S^3 \times M^4$ backgrounds. For $AdS_3\times S^3 \times S^3 \times S^1$ and $AdS_3\times S^3 \times T^4$ supported by pure RR flux we found complete agreement with the results available in the literature~\cite{Abbott:2012dd,Abbott:2013ixa}, while for $AdS_3\times S^3 \times T^4$ supported by a mix of RR and NSNS flux the unitarity result allowed to predict the previously unknown form of the one-loop dressing factor.

In chapter \ref{GKP} we analyzed the quantum fluctuations about the null-cusp background for the $AdS$ light-cone gauge fixed superstring in \adscp. The study of the free energy of this model entails information about the cusp anomalous dimension of ABJM and, indirectly, about the interpolating function $h(\l)$ (see section \ref{sec:h}). We reproduced in full details the calculation of~\cite{Bianchi:2014ada} for the one- and two-loop correction to the string free energy, which allowed to extract the order $\l^{-1/2}$ (NNLO) contribution to $h(\l)$~\cite{Bianchi:2014ada}. While the one-loop result was already available~\cite{McLoughlin:2008he,Abbott:2010yb,LopezArcos:2012gb}, the two-loop result was first computed in~\cite{Bianchi:2014ada} and provided strong support for a recent conjecture on the exact form of $h(\l)$~\cite{Gromov:2014eha}.

Elaborating further on the same  $AdS$ light-cone gauge fixed action in \adscp, we reviewed the evaluation of the one-loop dispersion relation of the fundamental excitations on top of the GKP vacuum~\cite{Bianchi:2015laa}. This was done by studying the two-point functions of the fundamental fields and comparing the result with the predictions from integrability~\cite{Basso:2013pxa}. This comparison however is not completely straightforward. Indeed, while for massive modes we have ascertained that the dispersion relation coincides with that predicted by the asymptotic Bethe Ansatz, for massless modes it is hard to match the string elementary excitations with the spinons of the integrability description, and consequently there is no clear identification of their dispersion relations. Furthermore these quantities are probably plagued by IR divergences at higher loops and this fact, combined with an IR cut-off (the mass of the low-energy excitations) that is exponentially small at strong coupling, would 
completely 
invalidate the reliability of perturbation theory at this order (see section \ref{sec:comparison}).
Another intriguing issue in the comparison with the integrability picture is the fate of the heaviest scalar in the spectrum. Such a mode is absent as an elementary state in the integrability approach and it has been interpreted as a two-fermion virtual state, i.e. a state whose pole in the fermion S-matrix lies in the unphysical strip of the rapidity plane~\cite{Basso:2014koa}. This pole is expected to appear at two-loop order, since up to one loop it merges with the branching point of the two-fermion continuum. The explicit computation of~\cite{Bianchi:2015laa}, reviewed in section \ref{sec:dispersion}, confirmed this fact.

We finally comment (see section \ref{sec:boundstates}) on the possible bound states that the GKP excitations can form. Integrability predicts the binding energy of possible bound states of several mass-$\sqrt{2}$ or mass-$1$ scalars at finite coupling. Using the non-relativistic limit of their scattering amplitudes we have estimated those binding energies, finding consistency with the integrability predictions in the static approximation.

\section{Future directions}

An immediate follow-up of the work described in this thesis is the study of the worldsheet S-matrix for the GKP excitations in $AdS_5\times S^5$. Recently, this object has been extensively studied using integrability~\cite{Basso:2013pxa,Fioravanti:2013eia,Fioravanti:2015dma}, especially due to its primary role in the OPE approach to light-like polygonal Wilson loops and scattering amplitudes~\cite{Alday:2010ku,Gaiotto:2011dt,Basso:2013vsa,Basso:2013aha,Basso:2014koa,Basso:2014jfa,
Basso:2014nra,Basso:2014hfa}. The string perturbative analysis of this object presents various subtleties and interesting aspects (see also the recent works~\cite{Bianchi:2015iza,Bianchi:2015vgw}). First of all, since the perturbative action is not $O(6)$ invariant it is important to understand which is the mechanism that enhances the symmetry non-perturbatively and leads to a finite coupling $O(6)$-invariant S-matrix. This question, pretty well understood in the case of the six massless bosons, has not been analyzed yet for the case of fermions.

Pushing the computation at one-loop order would allow to consider the application of unitarity techniques for this model. This is particularly interesting since the diversified mass spectrum of the GKP excitations would provide a very non-trivial setting where to test the conjecture of a connection between integrability and cut-constructibility~\cite{Bianchi:2013nra}. Furthermore, it would be interesting to analyze the implications of this result for the computation of the pentagon transition, i.e. the building block of the OPE construction of~\cite{Basso:2013vsa,Basso:2013aha,Basso:2014koa,Basso:2014jfa,
Basso:2014nra,Basso:2014hfa}, and consequently of the gluon scattering amplitudes beyond the leading order approximation at strong coupling.

One of the complications that one may encounter during this analysis is the presence of massless modes. The perturbative interpretation of massless modes in two dimensions is rather tricky since there is only one spatial dimension and all the left (or right) moving particles have the same speed (the speed of light). However, in the context of the GKP excitations, the spectrum is gapped, and there is no issue in writing down an exact S-matrix for the sextuplet of massive excitations at finite coupling (the S-matrix is actually a non-relativistic generalization of the $O(6)$ sigma model S-matrix). At strong coupling, though, the mass of those excitations is exponentially suppressed, and in a perturbative setting they are effectively massless. It would be interesting to understand how these two pictures can be combined.

A rather different context is represented by the massless modes appearing in the near-BMN $AdS_3\times S^3 \times T^4$ spectrum. In that case, indeed, the excitations are massless at any value of the coupling, and the S-matrix for those excitations may look like a completely meaningless object. Nevertheless, an exact expression for it can be extracted by symmetry considerations~\cite{Sax:2012jv,Lloyd:2013wza,Borsato:2014exa,Lloyd:2014bsa} and the authors of~\cite{Sax:2012jv} argued that the scattering is actually well-defined due to the non-relativistic nature of the excitations. How this is translated to the perturbative picture is still an open and interesting question.

Whenever dealing with perturbation theory, a natural possible development is the exploration of higher and higher orders in the perturbative expansion. Let us mention some of the problematics that one may encounter in the extension of the results presented in this thesis.
The recipe to compute the two-loop logarithmic part of the worldsheet S-matrix via unitarity has been given already in~\cite{Engelund:2013fja} and then perfected in~\cite{Engelund:2014pla}. The recovery of the correct rational term, already at one-loop, is tied to the interpretation of the singular t-channel cut. While at one loop the prescription described in section \ref{sec:unitarity} looks pretty natural, the higher number of diagrams and cuts involved in the two-loop computation seems to obstruct the proposal of a correct prescription. Still, the evaluation of the complete two-loop worldsheet S-matrix is an interesting and  challenging problem which deserves further analysis.

The kind of obstacles one has to face in the extension of our results for the free-energy of the GKP string and the dispersion relations of the GKP excitations are slightly different. The regularization employed for the two-loop calculation in section \ref{sec:twoloops} (or equivalently the one used in section \ref{sec:dispersion} for the one-loop dispersion relation) does not admit an immediate higher loop extension, and one would have to explore possible generalizations of that procedure. Furthermore, the presence of cubic interactions and of a diversified mass spectrum dramatically increases the difficulty in the computation of the Feynman integrals, compared to the BMN picture. However, the recent developments in the computation of Feynman integrals~\cite{Henn:2013pwa,Henn:2013nsa} may offer a valuable tool to overcome this obstruction.

Let us conclude mentioning a possible future direction which lies outside the context of perturbation theory, but it is closely related to the subjects discussed here. Following~\cite{McKeown:2013vpa}, it would be interesting to discretize the $AdS$-light-cone gauge action \eqref{la} on the lattice and analyze numerically various features of the GKP string at finite coupling~\footnote{Preliminary results in this direction appeared in \cite{Forini:2016sot} }, providing a formidable test of quantum integrability.

%\input{Chapters/Chapter6} 
%\input{Chapters/Chapter7} 

%----------------------------------------------------------------------------------------
%	THESIS CONTENT - APPENDICES
%----------------------------------------------------------------------------------------

\addtocontents{toc}{\vspace{2em}} % Add a gap in the Contents, for aesthetics

\appendix % Cue to tell LaTeX that the following 'chapters' are Appendices

% Include the appendices of the thesis as separate files from the Appendices folder
% Uncomment the lines as you write the Appendices

% Appendix A

\chapter{Notations and conventions} % Main appendix title

\label{notation} % For referencing this appendix elsewhere, use \ref{AppendixA}

\lhead{Appendix A. \emph{Notations and conventions}} % This is for the header on each page - perhaps a shortened title
\section{\txpf{$SO(5)$}{SO(5)} gamma matrices}
Throughout the text we use the following representation of the $SO(5)$ gamma matrices.
\begin{align}
 \g^1&={\scriptsize\begin{pmatrix} 
 0	& 0 	& 0 	& -1\\
 0	& 0 	& 1 	& 0 \\
 0	& 1 	& 0 	& 0 \\
 -1	& 0 	& 0 	& 0 \\
 \end{pmatrix} }\, ,
 & \g^2&={\scriptsize\begin{pmatrix} 
 0	& 0 	& 0 	& i\\
 0	& 0 	& i 	& 0 \\
 0	& -i 	& 0 	& 0 \\
 -i	& 0 	& 0 	& 0 \\
 \end{pmatrix}}\, ,
 & \g^3&={\scriptsize\begin{pmatrix} 
 0	& 0 	& 1 	& 0\\
 0	& 0 	& 0 	& 1 \\
 1	& 0 	& 0 	& 0 \\
 0	& 1 	& 0 	& 0 \\
 \end{pmatrix}}\, , \nonumber \\
\g^4&={\scriptsize\begin{pmatrix} 
 0	& 0 	& -i 	& 0\\
 0	& 0 	& 0 	& i \\
 i	& 0 	& 0 	& 0 \\
 0	& -i 	& 0 	& 0 \\
 \end{pmatrix} }\, ,
 & \g^5&={\scriptsize\begin{pmatrix} 
 1	& 0 	& 0	& 0\\
 0	& 1 	& 0 	& 0 \\
 0	& 0 	& -1 	& 0 \\
 0	& 0 	& 0 	& -1 \\
 \end{pmatrix}}\, .\label{eq:gammamatrices}
\end{align}

\section{\txpf{$\rho$}{rho} matrices}

Our convention on the $\rho$ matrices appearing in section \ref{sec:4+6} and then in chapter \ref{GKP} is the following
\begin{align}
\rho^1&={\scriptsize\begin{pmatrix}0&0&i&0\\0&0&0&i\\i&0&0&0\\0&i&0&0
\end{pmatrix}}\,,&
\rho^2&={\scriptsize\begin{pmatrix}0&0&-1&0\\0&0&0&1\\-1&0&0&0\\0&1&0&0
\end{pmatrix}}\,,&
\rho^3&={\scriptsize\begin{pmatrix}0&0&0&i\\0&0&-i&0\\0&-i&0&0\\i&0&0&0
\end{pmatrix}}\,,\nonumber \\
\rho^4=&-{\scriptsize\begin{pmatrix}0&0&0&1\\0&0&1&0\\0&1&0&0\\ 1&0&0&0
\end{pmatrix}}\,,&
\rho^5&={\scriptsize\begin{pmatrix}0&i&0&0\\-i&0&0&0\\0&0&0&i\\0&0&-i&0
\end{pmatrix}}\,,&
\rho^6&={\scriptsize\begin{pmatrix}0&1&0&0\\-1&0&0&0\\0&0&0&-1\\0&0&1&0
\end{pmatrix}}\, .\label{eq:rhomat}
\end{align}

The definition of  the matrices $\rho^{MN}$ is given in \eqref{eq:rhomndef} and they enjoy the following properties
and $\rho^{MN}=-\rho^{NM}$, $\rho^{MN\, \dagger}= - \rho^{MN}$, $(\rho^T)^{MN}= - (\rho^*)^{MN}$.

The following identities hold
\begin{equation}
 \rho^M_{ij}=-\rho^M_{ji}\, , \qquad (\rho^M)^{ij}=-(\rho^M_{ij})^*\, , \qquad (\rho^M)^{il} \rho^N_{lj}+(\rho^N)^{il} \rho^M_{lj}=2\, \d^{MN} \d^i_j\, .
\end{equation}

\section{Uniform light-cone gauge generators}
In section \ref{sec:unilccoset} we introduced a parametrization of $\mathfrak{psu}(2,2|4)$ particularly suitable for fixing a light-cone gauge involving one big angle coordinate on $S^5$. In that contest it emerged that the bosonic subalgebra $\mathfrak{su}(2,2)\oplus \mathfrak{su}(4)$ is conveniently represented in terms of the set of generators $\{\G^0,\G^i,\G^{i0},\G^{ij}\}\oplus\{\tilde\G^{A},\tilde\G^{AB}\}$ with $i,j=1,...,4$ and $A,B=1,...,5$ satisfying the following commutation relations
\begin{align}
 [\G^i,\G^j]&=\G^{ij}\, , & [\G^i,\G^0]&=\G^{i0}\, , & [\tilde\G^A,\tilde\G^B]&=-\tilde\G^{AB}\, , \\
 [\G^0,\G^{i0}]&=\G^i\, , & [\G^i,\G^{j0}]&=\d_{ij} \G^0\, , & [\G^i,\G^{jk}]&=\d^{i[j}\G^{k]}\, ,\\
 [\G^{i0},\G^{j0}]&=\G^{ij}\, , & [\G^{i0},\G^{jk}]&=\d^{i[j}\G^{k]0}\, , & [\G^{ij},\G^{kl}]&=\d^{l[i}\G^{j]k}+\d^{k[j}\G^{i]l}\, ,
 \end{align}\vspace{-1.5cm}
 \begin{align}
 [\tilde\G^A,\tilde\G^{BC}]&=\d^{A[B}\tilde\G^{C]}\, , & [\tilde\G^{AB},\tilde\G^{CD}]&=\d^{D[A}\tilde\G^{B]C}+\d^{C[B}\tilde\G^{A]D}\, . \label{eq:ulccommrel}
\end{align}
It is worthwhile noting that the generators $\{\tilde\G^{A},\tilde\G^{AB}\}$ are organized in such a way that the subset $\{\tilde \G^{AB}\}$ alone generates the $\mathfrak{so}(5)$ algebra which appears in the denominator of the coset. As a consequence the generators $\{\tilde\G^{A}\}$ are associated to the coset $\frac{SO(6)}{SO(5)}=S^5$, and indeed they generate translations in the directions $\{y^i,\phi\}$ introduced in \eqref{eq:Lambda} and \eqref{eq:gyz}. In particular $\tilde \G^5$ generates a translation in the direction of $\phi$, and finding the centralizer of the $\mathfrak{u}(1)$ isometry associated to shifts of $\phi$ coincides with finding the maximal subset of $SO(6)$ generators commuting with $\tilde\G^5$. This is clearly given by $\{\tilde \G^{ij}\}, i,j=1,...,4$, which generates a $\mathfrak{so}(4)\subset \mathfrak{so}(5)$. 
A similar reasoning applies to the subset $\{\G^0,\G^i,\G^{i0},\G^{ij}\}$, with the only difference that now the index $0$ is special due to the different signature. As before, the generators $\{\G^{i0},\G^{ij}\}$ alone generate $\mathfrak{so}(1,4)$ and the remaining ones $\{\G^0,\G^i\}$ are associated to translations in the $AdS_5$ coordinates $\{t,z_i\}$. Once more finding the centralizer of the $\mathfrak{u}(1)$ isometry associated to shifts of $t$ is equivalent to find the subset of generators commuting with $\G^0$. This is simply $\{\G^{ij}\}$, which provides the second $\mathfrak{so}(4)$ algebra appearing in \eqref{eq:foursu2}.

An explicit representation in terms of supermatrices is given by 
\begin{align}
 \G^0&=\begin{pmatrix}\frac{i}{2}\g^5 &0\\ 0&0\end{pmatrix}\, , & \G^i&=\begin{pmatrix}\frac{1}{2}\g^i &0\\ 0&0\end{pmatrix}\, ,\nonumber\\
 \G^{i0}&=\begin{pmatrix}\frac{i}{4}[\g^i,\g^5] &0\\ 0&0\end{pmatrix}\, , & \G^{ij}&=\begin{pmatrix}\frac{1}{4}[\g^i,\g^j] &0\\ 0&0\end{pmatrix}\, ,\nonumber\\
 \tilde\G^{A}&=\begin{pmatrix} 0&0\\ 0&\frac{i}{2}\g^A \end{pmatrix}\, , & \tilde\G^{AB}&=\begin{pmatrix} 0&0\\ 0&\frac{1}{4}[\g^A,\g^B]\end{pmatrix}\, ,\label{eq:Gsupermat}
\end{align}
with the gamma matrices given in \eqref{eq:gammamatrices}.

The fermionic degrees of freedom are best dealt with using the supermatrix representation of $\mathfrak{su}(2,2|4)$. A generic element is represented by
\begin{equation}
 \chi=\begin{pmatrix}
             0 & \Theta\\
             -\Theta^\dagger \S & 0
            \end{pmatrix}\, ,
\qquad
\Theta =\begin{pmatrix}
         \theta_{11} & \theta_{12} & \theta_{13} & \theta_{14}\\
         \theta_{21} & \theta_{22} & \theta_{23} & \theta_{24}\\
         \theta_{31} & \theta_{32} & \theta_{33} & \theta_{34}\\
         \theta_{41} & \theta_{42} & \theta_{43} & \theta_{44}\\
         \end{pmatrix}\, . \label{eq:ulcfermions}
\end{equation}
As we have shown explicitly in section \ref{sec:unilccoset}, the constraint coming from $\k$-symmetry reduce the 16 complex degrees of freedom of matrix to \eqref{eq:ulcfermions} by a factor one half, leaving a matrix like \eqref{eq:chifix}.

\section{\txpf{$AdS$}{AdS} light-cone gauge basis for \txpf{$AdS_5\times S^5$}{AdS5 x S5}}
Here we describe the $AdS$ light-cone basis for the generators of $\mathfrak{psu}(2,2|4)$. We spell out the expressions of all the non-vanishing commutators and provide a representation in terms of $8\times 8$ supermatrices as in \eqref{eq:supermatrix}. As we mentioned in footnote \ref{fn:u1}, there is no explicit supermatrix representation of the whole $\mathfrak{psu}(2,2|4)$ superalgebra, therefore we need to include the identity among the list of generators. Of course the identity commutes with all the other generators, but it appears on the right-hand-side of the anticommutator of two supercharges and is necessary for the closure of the algebra of $\mathfrak{su}(2,2|4)$.

The bosonic subalgebra consists of the direct sum $\mathfrak{su}(2,2)\oplus \mathfrak{su}(4)$. As mentioned above \eqref{eq:lcbosgen}, we interpret $\mathfrak{su}(2,2)$ as the conformal group in 4 dimension whose commutation relations are
\begin{align}
 [P^{\mu},J^{\n\r}]&=\eta^{\mu\nu} P^{\r}-\eta^{\mu\r} P^{\n}\, , & [K^{\mu},J^{\n\r}]&=\eta^{\mu\nu} K^{\r}-\eta^{\mu\r} K^{\n}\, , \label{eq:confcomm1}\\
 [P^\m,K^\n]&=-2\, \eta^{\m\n} D +2\,   J^{\m\n}\, , & [J^{\m\n},J^{\r\s}]&=\eta^{\m[\r} J^{\s]\n}-\eta^{\n[\r} J^{\s]\m}\, , \label{eq:confcomm2}\\
 [D,P^\m]&=-P^{\m}\, ,  & [D,K^\m]&=K^{\m} \, .\label{eq:confcomm3}
\end{align}
In the light-cone coordinates \eqref{eq:lccoord} one can introduce the generators \eqref{eq:lcbosgen} which are given by
\begin{align}
 P^{\pm}&=\frac{P^3\pm P^0}{\sqrt{2}} \, ,& P&=\frac{-P^2+i\, P^1}{\sqrt{2}}\, , & \bar P&=\frac{-P^2-i\, P^1}{\sqrt{2}}\, ,\\
  K^{\pm}&=\frac{K^3\pm K^0}{\sqrt{2}}\, , & K&=\frac{-K^2+i\, K^1}{\sqrt{2}}\, , & \bar K&=\frac{-K^2-i\, K^1}{\sqrt{2}}\, .
  \end{align}\vspace{-0.8 cm}
  \begin{align}
  J^{+-}&=J^{03} , & J^{+x}&=\frac{-J^{02}-J^{32}+i\, J^{01}+i\, J^{31}}{2} , & J^{+\bar x}&=\frac{-J^{02}-J^{32}-i\, J^{01}-i\, J^{31}}{2} ,\\
  J^{x \bar x}&=-i\, J^{12}\, ,& J^{-x}&=\frac{J^{02}-J^{32}-i\, J^{01}+i\, J^{31}}{2}\, , & J^{-\bar x}&=\frac{J^{02}-J^{32}+i\, J^{01}-i\, J^{31}}{2}\, .
\end{align}
The commutation relations of the new generators are given by \eqref{eq:confcomm1}, \eqref{eq:confcomm2} and \eqref{eq:confcomm3} provided that $\eta^{+-}=\eta^{-+}=\eta^{x\bar x}=\eta^{\bar x x}=1$.
The $\mathfrak{su}(4)$ commutation relations read
\begin{align}
 [{J^i}_j,{J^k}_l]=\d^i_l {J^k}_j-\d^k_j {J^i}_l\, .
\end{align}

The 32 supercharges of $\mathfrak{psu}(2,2|4)$ are chosen to be diagonal under the action of $D$, $J^{+-}$ and $J^{x\bar x}$, i.e.
\begin{align}
 [D,Q^{\pm\, i}]&=-\frac12 Q^{\pm\, i} & [D,Q^{\pm}_i]&=-\frac12 Q^{\pm}_i & [D,S^{\pm\, i}]&=\frac12 S^{\pm\, i} & [D,S^{\pm}_i]&=\frac12 S^{\pm}_i \label{eq:dilQS}\\
 [J^{\scriptscriptstyle{+-}},Q^{\pm\, i}]&=\pm \frac12 Q^{\pm\, i} & [J^{\scriptscriptstyle{+-}},Q^{\pm}_i]&=\pm\frac12 Q^{\pm}_i & [J^{\scriptscriptstyle{+-}},S^{\pm\, i}]&=\pm \frac12 S^{\pm\, i} & [J^{\scriptscriptstyle{+-}},S^{\pm}_i]&=\pm \frac12 S^{\pm}_i \label{eq:JpmQS}\\
  [J^{x\bar x},Q^{\pm\, i}]&=\pm \frac12 Q^{\pm\, i} & [J^{x\bar x},Q^{\pm}_i]&=\mp\frac12 Q^{\pm}_i & [J^{x\bar x},S^{\pm\, i}]&=\mp \frac12 S^{\pm\, i} & [J^{x\bar x},S^{\pm}_i]&=\pm \frac12 S^{\pm}_i. \label{eq:JxxQS}
\end{align}
They carry an $SU(4)$ index and they rotate under the action of $\mathfrak{su}(4)$ generators
\begin{align}\label{eq:QSSU4}
 [Q^{\pm}_i,{J^j}_k]&=-\d_i^j Q^{\pm}_k+\frac14 \d^j_k Q^{\pm}_i, &   [Q^{\pm\, i},{J^j}_k]&=\d^i_k Q^{\pm\, j}-\frac14 \d^j_k Q^{\pm\, i} ,
\end{align}
and similarly for the $S$ supercharges. The action of translations and conformal boosts are given by
\begin{align}
 [S^{\pm}_i,P^\mp]&=\pm i\, \sqrt2\, Q^\mp_i, & [S^{+}_i,\bar P]&=i\,\sqrt2\,Q^+_i, & [S^{-}_i, P]&=i\, \sqrt2\, Q^-_i,\\
 [Q^{\pm\, i},K^\mp]&=\mp i\,\sqrt2\, S^{\mp\, i}, & [Q^{+\, i},\bar K]&=i\, \sqrt2\, S^{+\, i}, & [Q^{-\, i}, K]&=i\, \sqrt2\, S^{-\, i},
\end{align}
whereas Lorentz transformations act as
\begin{align}
 [Q^{-\, i}, J^{+ x}]&=Q^{+\, i}, & [Q^{+\, i}, J^{- \bar x}]&=-Q^{-\, i}, & [S^{-\, i}, J^{+ \bar x}]&=-S^{+\, i}, & [S^{+\, i},J^{- x}]=S^{-\, i}.
\end{align}
Finally, the anticommutation relations of two supercharges are given by
\begin{align}
 \{Q^{\pm\, i}, Q^{\pm}_j\}&=\mp i\, P^\pm \d^i_j, &  \{Q^{+\, i}, Q^{-}_j\}&=i\, P \d^i_j, & \{Q^{+\, i}, S^{+}_j\}&=\sqrt2\, J^{+ x} \d^i_j, \\
 \{S^{\pm\, i}, S^{\pm}_j\}&=\mp i K^\pm \d^i_j, &  \{S^{-\, i}, S^{+}_j\}&= -i\, K \d^i_j, & \{Q^{-\, i}, S^{-}_j\}&=-\sqrt2\, J^{- \bar x} \d^i_j,
\end{align}
\begin{equation}
 \{Q^{\pm\, i}, S^{\mp}_j\}=\sqrt2\left(\mp \frac12(J^{+-}+J^{x\bar x}\mp D)\d^i_j- J^i_j+\frac14 \mathbb{1} \d^i_j\right).
\end{equation}
%The remaining non-trivial commutation relations can be simply found using the conjugation rules
%\begin{align}
% (P^\pm)^\dagger&=K^\mp\, , & P^\dagger&=\bar K\, , & \bar P^\dagger&= K\, , & D^\dagger&=D\, , & ({J^i}_j)^\dagger&={J^j}_i\, , \nonumber  \\  
% (J^{+-})^\dagger&=J^{+-}\, , & (J^{x\bar x})^\dagger&=J^{x \bar x}\, , & (J^{\pm x})^\dagger&=-J^{\mp \bar x}  & (Q^{\pm\, i})^\dagger&=S^{\mp}_i & (Q^\pm_i)^\dagger&=S^{\mp\, i} \label{eq:conrules}
%\end{align}

The supermatrix representation which we employ in the text and which reproduces these commutation relations can be represented as follows. The bosonic generators of $\mathfrak{su}(2,2)$ are $4\times 4$ matrices in the upper left corner. The translation generators are given by
\begin{align}
 P^0&={\scriptsize\begin{pmatrix}
      0& 0& 0& 0\\
      0& 0& 0& 0\\ 
      i& 0& 0& 0\\
      0& i& 0& 0\\
     \end{pmatrix} }\, ,
& 
P^1&={\scriptsize\begin{pmatrix}
      0& 0& 0& 0\\
      0& 0& 0& 0\\ 
      0& 1& 0& 0\\
      -1& 0& 0& 0\\
     \end{pmatrix}}\, ,
\\
P^2&={\scriptsize\begin{pmatrix}
      0& 0& 0& 0\\
      0& 0& 0& 0\\ 
      0& i& 0& 0\\
      i& 0& 0& 0\\
     \end{pmatrix}}\, ,
&
P^3&={\scriptsize\begin{pmatrix}
      0& 0& 0& 0\\
      0& 0& 0& 0\\ 
      i& 0& 0& 0\\
      0& -i& 0& 0\\
     \end{pmatrix}}\, .
\end{align}
The diagonal generators are
{\small
\begin{equation}
 D=\frac12 {\scriptsize\begin{pmatrix}
      1& 0& 0& 0\\
      0& 1& 0& 0\\ 
      0& 0& -1& 0\\
      0& 0& 0& -1\\
     \end{pmatrix} }\, ,
\quad 
J^{+-}=\frac12 {\scriptsize\begin{pmatrix}
      -1& 0& 0& 0\\
      0& 1& 0& 0\\ 
      0& 0& 1& 0\\
      0& 0& 0& -1\\
     \end{pmatrix}}\, ,
\quad
J^{x\bar x}=\frac12{\scriptsize\begin{pmatrix}
      -1& 0& 0& 0\\
      0& 1& 0& 0\\ 
      0& 0& -1& 0\\
      0& 0& 0& 1\\
     \end{pmatrix}}\, .
\end{equation}}
The remaining Lorentz generators read
\begin{equation}
 J^{+x}={\scriptsize\begin{pmatrix}
      0& 0& 0& 0\\
      1& 0& 0& 0\\ 
      0& 0& 0& 0\\
      0& 0& 0& 0\\
     \end{pmatrix} }\, ,
\quad 
J^{-x}={\scriptsize \begin{pmatrix}
      0& 0& 0& 0\\
      0& 0& 0& 0\\ 
      0& 0& 0& 0\\
      0& 0& 1& 0\\
     \end{pmatrix}}\, .
\end{equation}

The $\mathfrak{su}(4)$ generators occupy the lower right corner of the supermatrix and we use the following convention for the entries of the matrices ${J^i}_j$
\begin{equation}
 {({J^i}_j)^k}_l=-\d^{ik}\d_{jl}+\frac14 \d^i_j \d^k_l\, .
\end{equation}
Finally, the supercharges are represented by non-vanishing entries in the odd part of the supermatrix. We provide here the representation of a generic odd element of the algebra $\mathfrak{su}(2,2|4)$, from which it is easy to extract supermatrix representations for the single supercharges
\begin{multline}
 \theta^-_i Q^{+\, i}+\theta^{-\, i}Q^+_i+\eta^-_i S^{+\, i}+\eta^{-\, i}S^+_i+  \theta^+_i Q^{-\, i}+\theta^{+\, i}Q^-_i+\eta^+_i S^{-\, i}+\eta^{+\, i}S^-_i=\\
 2^{\frac14}
{\scriptsize \left(\begin{array}{cccc|cccc}
        0 & 0 & 0 & 0 & \eta^{+\,1} & \eta^{+\,2} & \eta^{+\,3} & \eta^{+\,4}\\
        0 & 0 & 0 & 0 & \eta^{-\,1} & \eta^{-\,2} & \eta^{-\,3} & \eta^{-\,4}\\
        0 & 0 & 0 & 0 & \theta^{-\,1} & \theta^{-\,2} & \theta^{-\,3} & \theta^{-\,4}\\
        0 & 0 & 0 & 0 & \theta^{+\,1} & \theta^{+\,2} & \theta^{+\,3} & \theta^{+\,4}\\ \hline
        \theta^-_1 & \theta^+_1 & \eta^+_1 & \eta^-_1 & 0 & 0 & 0 & 0\\
        \theta^-_2 & \theta^+_2 & \eta^+_2 & \eta^-_2 & 0 & 0 & 0 & 0\\
        \theta^-_3 & \theta^+_3 & \eta^+_3 & \eta^-_3 & 0 & 0 & 0 & 0\\
        \theta^-_4 & \theta^+_4 & \eta^+_4 & \eta^-_4 & 0 & 0 & 0 & 0\\
       \end{array}\right) }\, .\label{eq:lcfermions}
\end{multline}

\section{\txpf{$AdS$}{AdS} light-cone gauge basis for \txpf{$AdS_4\times S^7$}{AdS4 x S7}}
In equation \eqref{eq:cartanformAdS4first} we represented a generic element of $\mathfrak{osp}(4|8)$ using the set of generators: $\{M_{\m\n},M_{\m},V_{IJ},V_{8I},Q_{A'}\}$. Here we give a detailed description of the procedure to change from this basis to the one that we used for the construction of the Lagrangian. Let us start from the $AdS_4$ bosonic part. We define the 3d conformal generators as 
\begin{equation}\label{3dconf}
P_m=-M_{3m}+\frac12M_{m},\qquad K_m=M_{3m}+\frac12M_{m},\qquad D=-M_{3}, \qquad J_{mn}=M_{mn},
\end{equation}
with the standard commutation relations 
\begin{align}
 [P_m,J_{nr}]&=\eta_{mn} P_{r}-\eta_{mr} P_{n}, & [K_{m},J_{nr}]&=\eta_{mn} K_{r}-\eta_{mr} K_{n},\\
 [P_m,K_n]&= \eta_{mn} D +2\,  J_{mn}, & [J_{mn},J_{rs}]&=\eta_{m[r} J_{s]n}-\eta_{n[r} J_{s]m}, \\
 [D,P_m]&=2\, P_{m} , & [D,K_m]&=-2\, K_{m} .
\end{align}
Using the representation \eqref{eq:cartanformAdS4first} of the Cartan form one can relate the coefficients of \eqref{eq:cartanformAdS4first} with the ones in \eqref{eq:AAds4}. This yields
\begin{equation}
\omega^m=E^{m}-\o^{3m},\qquad c^m=E^{m}+\o^{3m},\qquad \Delta=-E^{3}.
\end{equation}

The light-cone basis is simply introduced by the change of variables $P^{\pm}=P^2\pm P^0$ and similarly for $K$.

The $SO(8)$ generators in \eqref{eq:cartanformAdS4first} are $\{V_{IJ},V_{I8}\}$ with the commutation relations
\begin{align}
 [V_{I8},V_{J8}]&=-V_{IJ},\qquad
[V_{IJ},V_{K8}]=\delta_{JK}V_{I8}-\delta_{IK}V_{J8},\\
\left[V_{IJ},V_{KL}\right]&=\delta_{IL}V_{JK}-\delta_{IK}V_{JL}+\delta_{JK}V_{IL}-\delta_{JL}V_{IK}.
\end{align}
These generators can be split further to the set $\{V_{MN},V_{78},V_{8M},V_{7M}\}$ appearing in equation \eqref{eq:cfso8}. The $\mathfrak{so}(6)$ generators $V_{MN}$ are projected to an $SU(4)$ basis via
\begin{equation}
{V^i}_j=\frac{i}{4}(\rho^{MN})^i{}_jV_{MN}
\end{equation}
and then further reduced to the $SU(3)$ irreducible parts
\begin{equation}\label{newso8basis}
\{V^{78},\ V^7{}_{4a},\ V^8{}_{4a},\ V_a{}^4,\
V_a{}^b-\frac13\delta_a^bV_c{}^c,\ V_a{}^a\}\, ,
\end{equation}
and their complex conjugates using 
\begin{align}
 {V^i}_j&=\begin{pmatrix}
V_a{}^b & V_a{}^4\\
V_4{}^b & V_4{}^4\\
\end{pmatrix}
,& V_4{}^4&=-V_a{}^a ,&
V^{7(8)}{}_{4a}&=V_{7(8)M}(\rho^M)_{4a}\, .
\end{align}
To provide a representation which make the Hopf fibration explicit we need to further redefine the generators \eqref{newso8basis}. The $U(1)$ fiber generator is
\begin{equation}\label{deft2}
H=2V_a{}^a-V^{78},
\end{equation}
and the $U(3)$ subrgoup of $SU(4)$ is generated by 
 \begin{equation}\label{defu3mod}
\tilde V_a{}^b=V_a{}^b-\frac12\delta_a^bV_c{}^c-\frac14\delta_a^bV^{78},
\end{equation}
with the trace $\tilde V_a{}^a=-\frac12V_a{}^a-\frac34V^{78}$ identifying with the $U(1)$ subgroup 
of $U(3)$. The form of $T^a$ and $T_a$ in \eqref{cfso6so2} is dictated by the generation of the $\mathfrak{su}(4)$ algebra commutation relations and commutativity with $H$
 \begin{equation}\label{newcp3gen}
T_a=\frac12(V^7{}_{4a}-iV^8{}_{4a}),\quad T^a=-\frac12(V^{74a}+iV^{84a}).
\end{equation}
Finally the remaining generators are
 \begin{equation}\label{so8su4}
\tilde T_a=-\frac12(V^7{}_{4a}+iV^8{}_{4a}),\quad\tilde T^a=\frac12(V^{74a}-iV^{84a}),\quad V_a{}^4,\quad V_4{}^a.
\end{equation}

Using all these relations one can easily work out the relations between the coefficients in \eqref{eq:cartanformAdS4first} and those in \eqref{cfso6so2} first using
\begin{align}
\Omega_i{}^j&=\frac{i}{2}\Omega^{MN}(\rho_{MN})_i{}^j=\begin{pmatrix}
\tilde \Omega_a{}^b &\Omega_a{}^4\\
\Omega_4{}^b &\tilde \Omega_4{}^4\, ,
\end{pmatrix}, &
\tilde \Omega_4{}^4&=-\tilde \Omega_a{}^a\, , \\
\Omega^{7}{}_{4a}&=\Omega^{7I}(\rho_I)_{4a}\, , & \Omega^{8}{}_{4a}&=\Omega^{8I}(\rho_I)_{4a}\, ,
\end{align}
and then
\begin{align}
\Omega_a&=\Omega^7{}_{4a}-\frac{i}{2}\Omega^8{}_{4a},& \Omega^a&=-\Omega^{74a}-\frac{i}{2}\Omega^{84a},\\
\tilde\Omega_a&=-\Omega^7{}_{4a}-\frac{i}{2}\Omega^8{}_{4a},&\tilde\Omega^a&=\Omega^{74a}-\frac{i}{2}\Omega^{84a},\\
 \Omega_a{}^b&=\tilde\Omega_a{}^b-\delta^b_a\tilde\Omega_c{}^c+\delta^b_ah, & \Omega^{78}&=-\tilde\Omega_a{}^a-h.
\end{align}

\section{Conventions for the exact S-matrices}
\subsection{\txpf{$AdS_5\times S^5$}{AdS5 x S5}}
The $AdS_5 \times S^5$ S-matrix reported in \ref{eq:exactSAdS5} is expressed in terms of the Zhukovsky variables $x^\pm$ defined by the following relations
\begin{align}
\frac{x^+}{x^-}&=e^{i p}\ , & x^+ - \frac{1}{x^+}-x^- + \frac{1}{x^-}&=\frac{2\,i\,\o}{h} \ , & x^+ + \frac{1}{x^+}-x^- - \frac{1}{x^-}&=\frac{2\,i}{h} \ .
\end{align}
Solving for energy and momentum we get
\begin{align}\label{eq:xpm}
x^\pm&=\frac{e^{\pm i \frac{p}{2}} (1+\o(p))}{2h \sin \frac{p}{2}}\, ,
&             \o(p)&=\sqrt{1+4h^2 \sin^2 \frac{p}{2}}\, ,
&                                \gamma&=|x^- -x^+|^{1/2}\, .
\end{align}
Here $h$ is in general a different coupling from the string tension $T$ and one could define a non-trivial interpolating function $h(T)$ which builds a connection between the result from integrability and the one from perturbation theory. However for $AdS_5\times S^5$ large evidence has been provided, both at weak and strong couplingm for the equality $h=T$.

When expanding the exact result in the near-BMN limit, we should understand how the
spin chain momenta are related to the worldsheet momenta. As part of
the gauge fixing of the worldsheet theory we chose the density of the
light-cone momentum to be a constant, which in turn
fixed the string length to be $\frac{P_+}{2T}$. Then, we took
$\mathcal{P_+}$ to be infinite, which
allowed for a sensible definition of the $S$-matrix, and
expanded in powers of $\co$ which acts as a
loop-counting parameter. This should be contrasted with the spin chain picture, where the spin chain length
$L$ is identified with the momentum $J$ plus an
additional term that depends on the number of
excitations%
: $L=J+M$. Going from
the spin chain to the string worldsheet involves the rescaling by a
factor of $\co$, which affects all dimensional
quantities and in particular all momenta, which
should be rescaled as
\[
 p\,\,\longrightarrow\,\, \co p\, ,
~~~~~~~~
p_{\rm chain}=\co p_{\rm string}\;.
\]
 Therefore, the strong-coupling expansion is
equivalent to the low-momentum expansion of the spin chain
S-matrix. For the kinematical variables
\eqref{eq:xpm} the rescaling of momenta yields:
\begin{equation}\label{}
 x^\pm=\frac{1+\o }{p}\left(1\pm\frac{i\co p}
 {2}+\mathcal{O}(\co^2)\right)~~.
\end{equation}
Note that in the limit we are considering here all information about
bound states appears at higher orders in the $\co$-
expansion.
%%%%%%%%%%%%%%%%%%%%%%%%%%%%%%%%%
\subsection{\txpf{$AdS_3\times S^3\times M^4$}{AdS3 x S3 x M4} supported by RR flux}\label{notations1}
%%%%%%%%%%%%%%%%%%%%%%%%%%%%%%%%%

In appendix \ref{app:exactSmat} the exact S-matrices are written as functions of the
Zhukovsky variables $x^{\pm}$ and $y^{\pm}$. These are defined in terms of the
energy and momentum as follows
\begin{align}
\frac{x^+}{x^-}&=e^{i p}\ , & x^+ - \frac{1}{x^+}-x^- + \frac{1}{x^-}&=\frac{2\,i\,\o}{h} \ , & x^+ + \frac{1}{x^+}-x^- - \frac{1}{x^-}&=\frac{2\,i\,m}{h} \ , \\
\frac{y^+}{y^-}&=e^{i p}\ , & y^+ - \frac{1}{y^+}-y^- + \frac{1}{y^-}&=\frac{2\,i\, \o}{h}\ ,& y^+ + \frac{1}{y^+}-y^- - \frac{1}{y^-}&=\frac{2\,i\,m'}{h}\ ,
\end{align}
where $h$ is the integrable coupling that is (potentially non-trivially)
related to the string tension $T$. The third equation of each line is a
constraint that is interpreted as the dispersion relation. In particular, $m$
and $m'$ are the masses of the respective particles. The variables $x'^{\pm}$
and $y'^{\pm}$ are simply given by sending $p\to p'$ and $\o \to \o'$.
Solving for $x^{\pm}$ and $y^{\pm}$ in terms of $p$ we find
\begin{align}
x^{\pm}&=\frac{e^{\pm i \frac{p}{2}}(m+\o)}{2\, h\, \sin\frac{p}{2}}\ , & \o&=\sqrt{m^2+4\,h^2\sin^2\frac{p}{2}}\ , \label{xpm}\\
y^{\pm}&=\frac{e^{\pm i \frac{p}{2}}(m'+\o)}{2\, h\, \sin\frac{p}{2}}\ , & \o&=\sqrt{m'^2+4\,h^2\sin^2\frac{p}{2}}\ . \label{ypm}
\end{align}
When expanding in near-BMN regime, the spatial momenta should first be rescaled
as $p\to \co\, p$ where $\co$ is the inverse of the string tension. The integrable coupling
$h$, in principle, is related to $\co$ in a non-trivial way, however, its
strong coupling (small $\co$) expansion starts with $h(\co)= \co^{-1} + \mathcal{O}(\co^0)$.
Therefore, at leading order in the near-BMN expansion the dispersion relation
is given by its relativistic counterpart. The two additional functions
that we use to write the expressions for the exact S-matrices are
\begin{equation}\label{etanu}
\eta=\sqrt{i(x^- - x^+)} \ , \qquad \n=\sqrt{\frac{x^+}{x^-}}\ ,
\end{equation}
and similarly for $y^\pm$ when referring to a particle of mass $m'$.

In section \ref{sec:tchannel} we are interested in expanding the functions
$x^\pm$ and $y^\pm$ at strong coupling. To do so it is convenient to introduce
a new variable $x$ such that
\begin{equation}\label{xexp}
x^\pm=x\pm\frac{im}{h} \frac{x^2}{x^2-1}+\mathcal{O}(h^{-3})\ .
\end{equation}
Expressing $x$ in terms of $p$ in the near-BMN expansion (i.e. first rescaling
$p$) one finds
\begin{equation}\label{xofp}
x(p)=\frac{m+\sqrt{m^2+p^2}}{p}+\mathcal{O}(\co^2)\ .
\end{equation}
Using the new variable one can easily expand the dressing phase at strong
coupling as shown in appendix \ref{app:exactSmat}.

In the discussion of the $AdS_3 \times S^3 \times S^3 \times S^1$ background, we
need to use the cubic terms in the expansion of the light-cone gauge-fixed
Lagrangian. We use a worldsheet metric with signature $(+,-)$. Light-cone
coordinates are defined for a generic two-dimensional vector $v^\mu$ as
$v^{\pm}=\frac12(v^0\pm v^1)$ and for a covector $v_\mu$ as $v_{\pm}=v_0\pm
v_1$. The non-vanishing elements of the metric in light-cone coordinates are
$\eta_{+-}=\eta_{-+}=2$. Correspondingly $\eta^{+-}=\eta^{-+}=\frac12$. The
Levi-Civita tensor is defined as $\e^{01}=1=-\e_{01}$.

As usual, gamma matrices are defined by the anti-commutation relation
\begin{equation}
\{\c^\mu,\c^\nu\}=2\,\eta^{\mu\nu}\ .
\end{equation}
An explicit representation is given by
\begin{equation}
\c^0=\begin{pmatrix}
0&1\\
1&0
\end{pmatrix}, \qquad
\c^1=\begin{pmatrix}
0&1\\
-1&0
\end{pmatrix}, \qquad
\c^3=-\c^0\c^1=\begin{pmatrix}
1&0\\
0&-1
\end{pmatrix}\ .
\end{equation}
A generic spinor is represented as
\begin{equation}
\chi=\begin{pmatrix}
\chi_+\\
\chi_-
\end{pmatrix} \ ,
\end{equation}
where $\chi_{\pm}$ are the chiral projections of $\chi$ by the projectors
$P_{\pm}=\frac12(1\pm\g^3)$. The conjugation is defined in the usual way
$\bar{\chi}=\chi^\dagger\g^0$ and to make contact with \cite{Sundin:2012gc,
Sundin:2014sfa} we define $\bar\chi_\pm\equiv\chi_\pm^\dagger$. The
polarization vectors can be chosen to be purely real and given by
\begin{align}
&\label{polu}\begin{tikzpicture}[baseline=-3pt]
\begin{scope}[decoration={markings,mark = at position 0.5 with {\arrow[scale=1.4]{latex}}}]
\draw[postaction={decorate}] (-1,0)--(0,0);
\end{scope}
\draw [pattern=crosshatch dots] (0.5,0) circle (0.5);
\end{tikzpicture}
\qquad u(\rmp) = \begin{pmatrix}\sqrt{p_+}\\\sqrt{p_-}\end{pmatrix} \ ,
\\\nonumber \\
&\label{polv}\begin{tikzpicture}[baseline=-3pt]
\draw [pattern=crosshatch dots] (0.5,0) circle (0.5);
\begin{scope}[decoration={markings,mark = at position 0.5 with {\arrow[scale=1.4]{latex}}}]
\draw[postaction={decorate}] (2,0)--(1,0);
\end{scope}
\end{tikzpicture}
\qquad v(\rmp) = \begin{pmatrix}\sqrt{p_+}\\-\sqrt{p_-}\end{pmatrix}\ .
\end{align}

%%%%%%%%%%%%%%%%%%%%%%%%%%%%%%%%%
\subsection{\txpf{$AdS_3\times S^3\times T^4$}{AdS3 x S3 x T4} supported by mixed flux}\label{notations2}
%%%%%%%%%%%%%%%%%%%%%%%%%%%%%%%%%

In the mixed flux case discussed in appendix \ref{app:exactSmat}, the S-matrix is again
written in terms of Zhukovsky-type variables. However, the dispersion relation
is modified and is different for particles ($x^\pm_+$) and antiparticles
($x^\pm_-$). The Zhukovsky variables are defined in terms of the energy and
momentum as follows
\begin{align}
\frac{x^+_\spm}{x^-_\spm}&=e^{i p}\ , & x^+_\spm - \frac{1}{x^+_\spm}-x^-_\spm + \frac{1}{x^-_\spm}&=\frac{2\,i\,\e_\pm}{h\sqrt{1-q^2}}\ .
\end{align}
However, the dispersion relation \cite{Hoare:2013lja} is now given by
\begin{equation}
\sqrt{1-q^2}\big( x_\spm^+ + \frac{1}{x^+_\spm}-x^-_\spm- \frac{1}{x^-_\spm}\big) \mp 2\,q \log \frac{x^+_\spm}{x^-_\spm}=\frac{2i}{h} \ , \\
\end{equation}
The variables $x'^{\pm}_+$ and $x'^{\pm}_-$ are simply given by sending $p\to
p'$ and $\e_\pm \to \e_\pm'$. Solving for $x^{\pm}_+$ and $x^{\pm}_-$ in terms
of $p$ we find
\begin{equation}\begin{split}
x^{\pm}_+&\,=\frac{e^{\pm i \frac{p}{2}}(1+ q\,p + \e_+(p))}{2\, h \sqrt{1-q^2}\, \sin\frac{p}{2}} \ , \qquad \quad
x^{\pm}_-=\frac{e^{\pm i \frac{p}{2}}(1- q\,p + \e_-(p))}{2\, h \sqrt{1-q^2}\, \sin\frac{p}{2}} \ ,
\\ & \qquad \qquad \e_\pm =\sqrt{(1\pm q\,h\, p)^2+4\,h^2(1-q^2)\sin^2\frac{p}{2}}\ . \label{xqpm}
\end{split}\end{equation}
As expected, at leading order in the near-BMN expansion the dispersion relation
is given by $e_\pm$ as defined in \eqref{eq:dr}. The functions $\eta_\pm$ and
$\nu_\pm$ are generalized in the obvious way from \eqref{etanu}.

In section \ref{sec:phq} we are interested in expanding the functions $x^\pm_+$
and $x^\pm_-$ at strong coupling. To do so it is convenient to introduce new
variables $x_\pm$ such that
\begin{equation}\begin{split}\label{xqexp}
x^\pm_+= & \, x_+ \pm \frac{i}{h} \frac{x_+^2}{\sqrt{1-q^2}(x_+^2-1)- 2\, q\, h\, x_+}+\mathcal{O}(h^{-3})\ , \qquad
\\ x^\pm_- = & \, x_- \pm \frac{i}{h} \frac{x_-^2}{\sqrt{1-q^2}(x_-^2-1)+ 2\, q\, h\, x_-}+\mathcal{O}(h^{-3})\ .
\end{split}\end{equation}
Expressing $x_\spm$ in terms of $p$ in the near-BMN expansion (i.e. first
rescaling $p$) one finds
\begin{equation}\label{xqofp}
x_\spm(p)=\frac{1 \pm q \,p+\sqrt{(1\pm q\,p)^2+(1-q^2)p^2}}{\sqrt{1-q^2}\,p}+\mathcal{O}(\z^2)\ .
\end{equation}

% Appendix Template

\chapter{Exact S-matrices} % Main appendix title

\label{app:exactSmat} % Change X to a consecutive letter; for referencing this appendix elsewhere, use \ref{AppendixX}

\lhead{Appendix B. \emph{Exact S-matrices and BMN expansion}} % Change X to a consecutive letter; this is for the header on each page - perhaps a shortened title
\section{\txpf{$AdS_5\times S^5$}{AdS5 x S5}}
\def\bes{{\bf B}}
The exact $SU(2|2)$ S-matrix was first evaluated in \cite{Beisert:2005tm}. The parametrizing functions used there are slightly different from the one used here \eqref{eq:comps1}-\eqref{eq:comps4}. The precise relation is \cite{Klose:2007wq}
\begin{equation}\begin{split}
A = & \frac1{2\sqrt{A^\bes}}(A^\bes - B^\bes) \ , \qquad B = \frac1{2\sqrt{A^\bes}}(A^\bes + B^\bes) \ , \qquad C = \frac1{2\sqrt{A^\bes}}C^\bes \ ,
\\
D = & \frac1{2\sqrt{A^\bes}}(-D^\bes + E^\bes) \ , \quad \, E = \frac1{2\sqrt{A^\bes}}(-D^\bes - E^\bes) \ , \quad  \,F = - \frac1{2\sqrt{A^\bes}}F^\bes \ ,
\\
H = & \frac1{\sqrt{A^\bes}}H^\bes \ , \qquad
K = \frac1{\sqrt{A^\bes}}K^\bes \ , \qquad
G = \frac1{\sqrt{A^\bes}}G^\bes \ , \qquad
L = \frac1{\sqrt{A^\bes}}L^\bes \ ,
\end{split}\end{equation}
where the label $\bes$ refers to the functions used in \cite{Beisert:2005tm}. Translating the result of \cite{Beisert:2005tm} to our language we find
\begin{align}
A&=S_0 
\frac{x'^- - x^-}{x'^- - x^+} \frac{1 -\frac{1}{ x'^- x^+}}{1 -\frac{1}{ x^+ x'^+}}\, ,
&
B&=S_0\left( \frac{x'^+ - x^-}{x'^- - x^+} +\frac{ 1 -\frac{1}{x^+ x'^-}}{ 1- \frac{1}{x^+ x'^+}}
\frac{x'^- - x^-}{x'^- - x^+}\right)\, , \nonumber
\\
C&=-S_0 \frac{x'^- - x^-}{x'^- - x^+} \frac{\gamma \gamma'}{x^+ x'^+} \frac{1}{1 -\frac{1}{x^+ x'^+}}\, 
&
D&=S_0 \frac{x'^- - x^-}{x'^- - x^+}  \frac{1 - \frac{1}{x^- x'^+}}{1 - \frac{1}{x^+ x'^+}}\, ,\nonumber
\\
E&=S_0\left(1-\frac{1-\frac{1}{x'^+ x^-}}{1-\frac{1}{x^- x'^-}} \frac{x'^+ - x^+}{x'^- - x^+}\right)\, ,
&
F&=S_0 \frac{x'^+ - x^+}{x'^- - x^+} \frac{\gamma \gamma'}{x^- x'^-}  \frac{1}{1 - \frac{1}{x^- x'^-}}\, ,\nonumber
\\
G&=S_0 \frac{x'^+-x^+}{ x'^- - x^+}\, ,
&
H&=S_0 \frac{x'^+-x'^- }{ x'^- - x^+ }\frac{\gamma}{\gamma'}\, ,\nonumber
\\
L&=S_0 \frac{x'^--x^-}{ x'^- - x^+}\, ,
&
K&=S_0 \frac{x^+-x^- }{ x'^- - x^+}\frac{\gamma'}{\gamma}\, .\label{eq:exactSAdS5}
\end{align}
The definitions of the variables $x^\pm$ entering these expressions are given in \eqref{eq:xpm} and the overall factor $S_0$ is related to the BES \cite{Beisert:2006ez} dressing phase 
\begin{align}
S_0^2=\frac{x'^--x^+}{x'^+-x^-} \frac{1-\frac{1}{x'^+ x^-}}{1-\frac{1}{x'^- x^+}}e^{i\vartheta_{BES}(x^\pm, x'^{\pm})}\, ,
\end{align}
with $\vartheta_{BES}(x^\pm, x'^{\pm})$ expressible in the following way
\begin{eqnarray}\label{afsphase}
 \vartheta_{BES}(p,p')&=&\frac1{\co}\sum_{r,s=\pm}^{}rs\,\chi_{BES} (x_{p}^r,x_{p'}^s).
 \end{eqnarray}
The function $\chi_{BES}$ can be represented compactly as a contour integral 
 \begin{align}
\chi^{\BES}(x,y)&= i \ointc \frac{dw}{2 \pi i} \ointc \frac{dw'}{2 \pi i} \, \frac{1}{x-w}\frac{1}{y-w'} \log{\frac{\Gamma[1+i \hh(w+1/w-w'-1/w')]}{\Gamma[1-i \hh(w+1/w-w'-1/w')]}}\ .
\label{eq:besdhmrep1}
\end{align}
The first few orders in the near-BMN expansion read
\begin{align}
 \chi_{BES}(x,y)=&\sum_{l=0}^\infty \co^{l-1} (\chi^{(l)}(x,y)-\chi^{(l)}(y,x))\, ,\\
 \chi^{(0)}
 (x,y)=&-\frac{1}{y}+\left(\frac{1}{y}-x\right)\log\left(1-\frac{1}{xy}\right) \; ,\\
 \chi^{(1)}(x,y)=&-\frac{1}{2\pi} \Li\frac{\sqrt{x}-\frac{1}{\sqrt{y}}}{\sqrt{x}-\sqrt{y}}-\frac{1}{2\pi} \Li\frac{\sqrt{x}+\frac{1}{\sqrt{y}}}{\sqrt{x}+\sqrt{y}}\nonumber\\
  &+\frac{1}{2\pi} \Li\frac{\sqrt{x}+\frac{1}{\sqrt{y}}}{\sqrt{x}-\sqrt{y}}+\frac{1}{2\pi} \Li\frac{\sqrt{x}-\frac{1}{\sqrt{y}}}{\sqrt{x}+\sqrt{y}}\, ,\\
  \chi^{(2)}(x,y)=&-\frac{y}{24(xy-1)(y^2-1)}\, ,\\
  \chi^{(3)}(x,y)=&0\, ,\\
  \chi^{(4)}(x,y)=&-\frac{y^3+4y^5-9xy^6+y^7+3x^2 y^7-3xy^8+3x^2y^9}{720 (xy-1)^3(y^2-1)^5}\, ,\\
\chi^{(5)}(x,y)=&0\, .
  \end{align}
  Let us stress that the whole logarithmic dependence has a one-loop origin and this is an important constraint for the unitarity-cut computation.

 \section{Massive sector for \txpf{$AdS_3\times S^3 \times T^4$}{AdS3 x S3 x T4}}

The exact S-matrix for the massive sector of light-cone gauge fixed sigma model on $AdS_3\times S^3 \times T^4$ supported by pure RR flux was first computed in \cite{Borsato:2013qpa}.  In string frame it reads
\begin{align}
A_{++}(p,p') &= S^{11}_{++}(p,p') \ , &
B_{++}(p,p') &= S^{11}_{++}(p,p') \frac{x'^+ - x^+}{x'^+ - x^-} \frac{1}{\n}\ , \nonumber \\
C_{++}(p,p') &= S^{11}_{++}(p,p') \frac{x'^+ - x'^-}{x'^+ - x^-} \frac{\eta}{\eta'} \sqrt{\frac{\n'}{\n}}\ , &
D_{++}(p,p') &= S^{11}_{++}(p,p') \frac{x'^- - x^-}{x'^+ - x^-} \n'\ , \nonumber \\
E_{++}(p,p') &= S^{11}_{++}(p,p') \frac{x^+ - x^-}{x'^+ - x^-} \frac{\eta'}{\eta} \sqrt{\frac{\n'}{\n}}\ , &
F_{++}(p,p') &= S^{11}_{++}(p,p') \frac{x'^- - x^+}{x'^+ - x^-}\frac{\n'}{\n}\ , \label{funpp}
\end{align}
\begin{align}
A_{+-}(p,p') &= S^{11}_{+-}(p,p') \frac{1-\frac{1}{x^+\, x'^-}}{1-\frac{1}{x^-\, x'^-}} \n\ , &
B_{+-}(p,p') &= -S^{11}_{+-}(p,p') \frac{i\, \eta\eta'}{x^- x'^-}\frac{(\n\n')^{-\frac12}}{1-\frac{1}{x^-\, x'^-}} \ , \nonumber \\
C_{+-}(p,p') &= S^{11}_{+-}(p,p')\ , &
D_{+-}(p,p') &= S^{11}_{+-}(p,p') \frac{1-\frac{1}{x^+\, x'^+}}{1-\frac{1}{x^-\, x'^-}}\n\n'\ , \nonumber \\
E_{+-}(p,p') &= S^{11}_{+-}(p,p') \frac{1-\frac{1}{x^-\, x'^+}}{1-\frac{1}{x^-\, x'^-}} \n'\ , &
F_{+-}(p,p') &= -S^{11}_{+-}(p,p') \frac{i\, \eta\eta'}{x^+ x'^+}\frac{(\n\n')^\frac32}{1-\frac{1}{x^-\, x'^-}}\ . \label{funpm}
\end{align}
The definitions of the variables $x^\pm$ entering these expressions are given
for general mass in appendix \ref{notations1}. Here the masses should be set to
one. The functions $S^{11}_{++}(p,p')$ and $S^{11}_{+-}(p,p')$ are two overall
phase factors, i.e. in the notation of eq.~\eqref{phasestruc}
$S^{11}_{\s_M\s_N}(p,p')=e^{i\vp^{11}_{\s_M\s_N}(p,p')}$. The superscripts
refer to the masses of the two particles being scattered. These phase factors
are not fixed by symmetry. They are, however, constrained by crossing symmetry
and a conjecture for their exact expressions was given in
\cite{Borsato:2013hoa}, supported by semiclassical one-loop computations
\cite{Abbott:2012dd,Beccaria:2012kb}. %v2+
The proposal reads
\begin{align}
S^{11}_{++}(p,p')^{-1}&=e^{-\frac{i}{2} a (\e' p-\e p')}\sqrt{\frac{x'^- - x^+}{x'^+ - x^-}\, \frac{1-\frac{1}{x^+ x'^-}}{1-\frac{1}{x^- x'^+}}} \frac{\n'}{\n} \, e^{i\, \vartheta^{11}_{++}(x^\pm,x'^\pm)}\ ,\label{Spp}\\
S^{11}_{+-}(p,p')^{-1}&=e^{-\frac{i}{2} a (\e' p-\e p')}\sqrt{\frac{1-\frac{1}{x^+ x'^+}}{1-\frac{1}{x^- x'^-}} \frac{1-\frac{1}{x^+ x'^-}}{1-\frac{1}{x^- x'^+}}}\, \n' \, e^{i\, \vartheta^{11}_{+-}(x^\pm,x'^\pm)}\ .\label{Spm}
\end{align}
The functions $\vartheta^{11}_{++}(p,p')$ and $\vartheta^{11}_{+-}(p,p')$ can
be expressed in terms of an auxiliary function $\chi$
\begin{align}
\vartheta^{11}_{++}(x^\pm,x'^\pm)&=\chi(x^+,x'^+)+\chi(x^-,x'^-)-\chi(x^+,x'^-)-\chi(x^-,x'^+)\ ,\\
\vartheta^{11}_{+-}(x^\pm,x'^\pm)&=\tilde\chi(x^+,x'^+)+\tilde\chi(x^-,x'^-)-\tilde\chi(x^+,x'^-)-\tilde\chi(x^-,x'^+)\ ,
\end{align}
and the explicit all-order expressions for $\chi$ and $\tilde \chi$ are
\begin{align}
\chi(x,y)&=\chi^{\BES}(x,y)+\frac12\big(-\chi^{\HL}(x,y)+\chi^-(x,y)\big)\ ,\\
\tilde\chi(x,y)&=\chi^{\BES}(x,y)+\frac12\big(-\chi^{\HL}(x,y)-\chi^-(x,y)\big)\ .
\end{align}
Here the function $\chi^{\BES}$ is the same which appears in the
$AdS_5\times S^5$ dressing factor \cite{Beisert:2006ez}, $\chi^{\HL}$ is the
Hernandez Lopez phase \cite{Hernandez:2006tk} and is given by the one-loop term
in the strong coupling expansion of $\chi^{\BES}$, while the function $\chi^-$
does not appear in the $AdS_5\times S^5$ light-cone gauge S-matrix. The three
functions can be expressed compactly as contour integrals
\begin{align}
\chi^{\BES}(x,y)&= i \ointc \frac{dw}{2 \pi i} \ointc \frac{dw'}{2 \pi i} \, \frac{1}{x-w}\frac{1}{y-w'} \log{\frac{\Gamma[1+i \hh(w+1/w-w'-1/w')]}{\Gamma[1-i \hh(w+1/w-w'-1/w')]}}\ ,
\label{eq:besdhmrep}\\
\chi^{\HL}(x,y)&= \frac{\pi}{2} \ointc \frac{dw}{2 \pi i} \ointc \frac{dw'}{2 \pi i} \, \frac{1}{x-w}\frac{1}{y-w'} \, \text{sign}(w'+1/w'-w-1/w)\ , \\
\chi^-(x,y) &=\ointc \, \frac{dw}{8\pi} \frac{1}{x-w} \log{\left[ (y-w)\left(1-\tfrac{1}{yw}\right)\right]} \, \text{sign}((w-1/w)/i) \ - x \leftrightarrow y \ .
\end{align}

We are interested in the near-BMN expansion of these expressions. Therefore,
let us quote the first two orders of $\vartheta^{11}_{++}(x^\pm,x'^\pm)$ and
$\vartheta^{11}_{+-}(x^\pm,x'^\pm)$
\begin{align}\label{strong}
\vartheta^{11}_{++}(x^\pm,x'^\pm)&=\frac{1}{h}\vartheta^{\AFS}(x,x')+\frac{1}{h^2}\vartheta_{++}^{(1)}(x,x')+\mathcal{O}(h^{-3})\ ,\\
\vartheta^{11}_{+-}(x^\pm,x'^\pm)&=\frac{1}{h}\vartheta^{\AFS}(x,x')+\frac{1}{h^2}\vartheta_{+-}^{(1)}(x,x')+\mathcal{O}(h^{-3})\ . \label{strong2}
\end{align}
The functions appearing in \eqref{strong} and \eqref{strong2} are given by
\begin{align}
\vartheta^{\AFS}(x,y)&=\tfrac{2 (x-y)}{(x^2-1)(xy-1)(y^2-1)}+\mathcal{O}(h^{-2})\ , \label{AFSstrong}\\
\vartheta_{++}^{(1)}(x,y)&=\tfrac{1}{\pi}\tfrac{x^2}{x^2-1}\tfrac{y^2}{y^2-1}\left[\tfrac{(x+y)^2 \left(1-\tfrac{1}{xy}\right)}{(x^2-1)(x-y)(y^2-1)} + \tfrac{2}{(x-y)^2} \log\left(\tfrac{x+1}{x-1}\tfrac{y-1}{y+1}\right)\right]+\mathcal{O}(h^{-1})\ ,\nonumber \\
\vartheta_{+-}^{(1)}(x,y)&=\tfrac{1}{\pi}\tfrac{x^2}{x^2-1}\tfrac{y^2}{y^2-1}\left[\tfrac{(xy+1)^2 \left(\tfrac{1}{x}-\tfrac{1}{y}\right)}{(x^2-1)(xy-1)(y^2-1)} + \tfrac{2}{(xy-1)^2} \log\left(\tfrac{x+1}{x-1}\tfrac{y-1}{y+1}\right)\right]+\mathcal{O}(h^{-1})\ . \nonumber
\end{align}
It is important to point out that the pre-factors appearing in \eqref{Spp} and
\eqref{Spm} can be written as a phase factor whose exponent has a vanishing
one-loop ($\mathcal{O}(h^{-2})$) term. This property, together with
\eqref{AFSstrong}, allows us to compare $\vartheta^{(1)}_{++}$ and
$\vartheta^{(1)}_{+-}$ directly with our perturbative result following from
unitarity-cut methods.

%%%%%%%%%%%%%%%%%%%%%%%%%%%%%%%%%
\section{Massive sector for \txpf{$AdS_3\times S^3 \times S^3 \times S^1$}{AdS3 x S3 x S3 x S1}}
%%%%%%%%%%%%%%%%%%%%%%%%%%%%%%%%%
The exact S-matrix for the massive sector of the light-cone gauge fixed sigma model on $AdS_3\times S^3 \times S^3\times S^1$ supported by pure RR flux was first computed in \cite{Borsato:2012ud,Borsato:2012ss}.  In string frame it reads
\begin{align}
A_{++}(p,p') &= S^{\a\a}_{++}(p,p') \ , &
B_{++}(p,p') &= S^{\a\a}_{++}(p,p') \frac{x'^+ - x^+}{x'^+ - x^-} \frac{1}{\n}\ , \nonumber \\
C_{++}(p,p') &= S^{\a\a}_{++}(p,p') \frac{x'^+ - x'^-}{x'^+ - x^-} \frac{\eta}{\eta'} \sqrt{\frac{\n'}{\n}}\ , &
D_{++}(p,p') &= S^{\a\a}_{++}(p,p') \frac{x'^- - x^-}{x'^+ - x^-} \n'\ , \nonumber \\
E_{++}(p,p') &= S^{\a\a}_{++}(p,p') \frac{x^+ - x^-}{x'^+ - x^-} \frac{\eta'}{\eta} \sqrt{\frac{\n'}{\n}}\ , &
F_{++}(p,p') &= S^{\a\a}_{++}(p,p') \frac{x'^- - x^+}{x'^+ - x^-}\frac{\n'}{\n}\ , \label{funppS1}
\end{align}
\begin{align}
A_{+-}(p,p') &= S^{\a\a}_{+-}(p,p') \frac{1-\frac{1}{x^+\, x'^-}}{1-\frac{1}{x^-\, x'^-}} \n\ , &
B_{+-}(p,p') &= -S^{\a\a}_{+-}(p,p') \frac{i\, \eta\eta'}{x^- x'^-}\frac{(\n\n')^{-\frac12}}{1-\frac{1}{x^-\, x'^-}} \ , \nonumber \\
C_{+-}(p,p') &= S^{\a\a}_{+-}(p,p')\ , &
D_{+-}(p,p') &= S^{\a\a}_{+-}(p,p') \frac{1-\frac{1}{x^+\, x'^+}}{1-\frac{1}{x^-\, x'^-}}\n\n'\ , \nonumber \\
E_{+-}(p,p') &= S^{\a\a}_{+-}(p,p') \frac{1-\frac{1}{x^-\, x'^+}}{1-\frac{1}{x^-\, x'^-}} \n'\ , &
F_{+-}(p,p') &= -S^{\a\a}_{+-}(p,p') \frac{i\, \eta\eta'}{x^+ x'^+}\frac{(\n\n')^\frac32}{1-\frac{1}{x^-\, x'^-}}\ . \label{funpmS1}
\end{align}
The structure of the S-matrix is identical to \eqref{funpp} and \eqref{funpm},
the only differences being the overall phase factors, $S^{\a\a}_{++}(p,p')$ and
$S^{\a\a}_{+-}(p,p')$, and that in the definition of the variables $x^\pm$
given in appendix \ref{notations1} the mass should be set to $\a$. The phase
factors $S^{\a\a}_{\pm\pm}$ and $S^{\a\a}_{\pm\mp}$ have been computed
semiclassically in \cite{Abbott:2013ixa}.

For the scattering of a mass $\a$ with a mass $\bar \a$ the functions in string frame are given by \cite{Borsato:2012ud,Borsato:2012ss}
\begin{align}
A_{++}(p,p') &= S^{\a\bar \a}_{++}(p,p') \ , &
B_{++}(p,p') &= S^{\a\bar \a}_{++}(p,p') \frac{y'^+ - x^+}{y'^+ - x^-} \frac{1}{\n}\ , \nonumber \\
C_{++}(p,p') &= S^{\a\bar \a}_{++}(p,p') \frac{y'^+ - y'^-}{y'^+ - x^-} \frac{\eta}{\eta'}\sqrt{\frac{\n'}{\n}}\ , &
D_{++}(p,p') &= S^{\a\bar \a}_{++}(p,p') \frac{y'^- - x^-}{y'^+ - x^-} \n'\ , \nonumber \\
E_{++}(p,p') &= S^{\a\bar \a}_{++}(p,p') \frac{x^+ - x^-}{y'^+ - x^-} \frac{\eta'}{\eta} \sqrt{\frac{\n'}{\n}}\ , &
F_{++}(p,p') &= S^{\a\bar \a}_{++}(p,p') \frac{y'^- - x^+}{y'^+ - x^-} \frac{\n'}{\n}\ , \label{funpppS1}
\end{align}
\begin{align}
A_{+-}(p,p') &= S^{\a\bar \a}_{+-}(p,p') \frac{1-\frac{1}{x^+\, y'^-}}{1-\frac{1}{x^-\, y'^-}} \n\ , &
B_{+-}(p,p') &= -S^{\a\bar \a}_{+-}(p,p') \frac{i\, \eta\eta'}{x^-y'^-}\frac{(\n\n')^{-\frac12}}{1-\frac{1}{x^-\, y'^-}} \ , \nonumber \\
C_{+-}(p,p') &= S^{\a\bar \a}_{+-}(p,p') \ , &
D_{+-}(p,p') &= S^{\a\bar \a}_{+-}(p,p') \frac{1-\frac{1}{x^+\, y'^+}}{1-\frac{1}{x^-\, y'^-}}\n\n'\ , \nonumber \\
E_{+-}(p,p') &= S^{\a\bar \a}_{+-}(p,p') \frac{1-\frac{1}{x^-\, y'^+}}{1-\frac{1}{x^-\, y'^-}} \n'\ , &
F_{+-}(p,p') &= -S^{\a\bar \a}_{+-}(p,p') \frac{i\, \eta\eta'}{x^+ y'^+}\frac{(\n\n')^\frac32}{1-\frac{1}{x^-\, y'^-}}\ . \label{funppmS1}
\end{align}

Here we have defined the overall phase factors by setting
\begin{equation}
{\hat S}_{\vf_2\vf_3}^{\vf_2\vf_3}(p,p')=1\ , \qquad {\hat S}_{\vf_2\bar{\chi}^3}^{\vf_2\bar{\chi}^3}(p,p')=1\ .
\end{equation} 
and we can express them as
\begin{align}
S^{\a\a}_{++}(p,p')^{-1}&=e^{-i a (\e' p-\e p')} \frac{1-\frac{1}{x^+ x'^-}}{1-\frac{1}{x^- x'^+}}\frac{x'^- - x^+}{x'^+ -x^-} \left(\frac{\n'}{\n}\right)^2 \, e^{ i\, \vartheta^{\a\a}_{++}(x^\pm,x'^\pm)}\ ,\label{SppS1} \\
S^{\a\a}_{+-}(p,p')^{-1}&=e^{-i a (\e' p-\e p')}\sqrt{\frac{1-\frac{1}{x^+ x'^+}}{1-\frac{1}{x^- x'^-}} \frac{1-\frac{1}{x^+ x'^-}}{1-\frac{1}{x^- x'^+}} }\, \n' \, e^{ i\, \vartheta^{\a\a}_{+-}(x^\pm,x'^\pm)}\ , \label{SpmS1}
\end{align}
and
\begin{align}
S^{\a\bar\a}_{++}(p,p')^{-1}&=e^{-i a (\e' p-\e p')}\ \frac{1-\frac{1}{x^+ y'^-}}{1-\frac{1}{x^- y'^+}} \frac{\n'}{\n} \, e^{ i\, \vartheta^{\a\bar\a}_{++}(x^\pm,x'^\pm)}\ ,\label{SpppS1} \\
S^{\a\bar\a}_{+-}(p,p')^{-1}&=e^{-i a (\e' p-\e p')}\ \sqrt{\frac{1-\frac{1}{x^+ y'^+}}{1-\frac{1}{x^- y'^-}}} \left(\frac{1-\frac{1}{x^+ y'^-}}{1-\frac{1}{x^- y'^+}} \right)^{\frac32}\, \n' \, e^{ i\, \vartheta^{\a\bar\a}_{+-}(x^\pm,x'^\pm)}\ . \label{SppmS1}
\end{align}
Unlike the $AdS_3\times S^3 \times T^4$ case, all-order expressions for
$\vartheta^{\a\a}_{\s_M\s_N}$ and $\vartheta^{\a\bar\a}_{\s_M\s_N}$ are not
known. The one-loop near-BMN expansions for these phases have been computed semiclassically at one loop in \cite{Abbott:2013ixa} and we displayed them in
eqs.~\eqref{1Lphasepp}, \eqref{1Lphasepm}. They are essentially the same as
\eqref{AFSstrong} up to an overall scaling depending on the masses.

\section{Massive sector for \txpf{$AdS_3\times S^3 \times T^4$}{AdS3 x S3 x T4} supported by mixed flux}
The functions in string frame given by \cite{Hoare:2013ida}
\begin{align}
A_{++}(p,p') &= S_{++}(p,p') \ , &
B_{++}(p,p') &= S_{++}(p,p') \frac{x'^+_+ - x^+_+}{x'^+_+ - x^-_+} \frac{1}{\n_+}\ , \nonumber \\
C_{++}(p,p') &= S_{++}(p,p') \frac{x'^+_+ - x'^-_+}{x'^+_+ - x^-_+} \frac{\eta_+}{\eta'_+} \sqrt{\frac{\n'_+}{\n_+}}\ , &
D_{++}(p,p') &= S_{++}(p,p') \frac{x'^-_+ - x^-_+}{x'^+_+ - x^-_+} \n'_+\ , \nonumber \\
E_{++}(p,p') &= S_{++}(p,p') \frac{x^+_+ - x^-_+}{x'^+_+ - x^-_+} \frac{\eta'_+}{\eta_+} \sqrt{\frac{\n'_+}{\n_+}}\ , &
F_{++}(p,p') &= S_{++}(p,p') \frac{x'^-_+ - x^+_+}{x'^+_+ - x^-_+}\frac{\n'_+}{\n_+}\ , \label{qfunpp}
\end{align}
and
\begin{align}
A_{+-}(p,p') &= S_{+-}(p,p') \frac{1-\frac{1}{x^+_+\, x'^-_-}}{1-\frac{1}{x^-_+\, x'^-_-}} \n_+\ , &
B_{+-}(p,p') &= -S_{+-}(p,p') \frac{i\, \eta_+\eta'_-}{{x}^-_+ x'^-_-}\frac{(\n_+\n'_-)^{-\frac12}}{1-\frac{1}{x^-_+\, x'^-_-}} \ , \nonumber \\
C_{+-}(p,p') &= S_{+-}(p,p') \ , &
D_{+-}(p,p') &= S_{+-}(p,p') \frac{1-\frac{1}{x^+_+\, x'^+_-}}{1-\frac{1}{x^-_+\, x'^-_-}}\n_+\n'_-\ , \nonumber \\
E_{+-}(p,p') &= S_{+-}(p,p') \frac{1-\frac{1}{x^-_+\, x'^+_-}}{1-\frac{1}{x^-_+\, x'^-_-}} \n'_-\ , &
F_{+-}(p,p') &= -S_{+-}(p,p') \frac{i\, \eta_+\eta'_-}{x^+_+ x'^+_-}\frac{(\n_+\n'_-)^\frac32}{1-\frac{1}{x^-_+\, x'^-_-}}\ . \label{qfunpm}
\end{align}
The definitions of the variables entering these expressions are given in
appendix \ref{notations2}. The functions $S_{++}(p,p')$ and $S_{+-}(p,p')$ are
two of the overall phase factors. These phase factors are not fixed by global
symmetry, but constrained by crossing symmetry, however, they are
currently unknown.

% Appendix Template

\chapter{Details on the expanded Lagrangian for null cusp fluctuations in \txpf{$AdS_4\times \mathbb{CP}^3$}{AdS4 x CP3}} % Main appendix title

\label{app:lagr_exp} % Change X to a consecutive letter; for referencing this appendix elsewhere, use \ref{AppendixX}

\lhead{Appendix C. \emph{Details on the expanded Lagrangian}} % Change X to a consecutive letter; this is for the header on each page - perhaps a shortened title

In this appendix we provide the expanded fluctuation Lagrangian \eqref{eq:Lagrangian_exp} up to quartic order in the fields. 
The vertices come with a factor $\frac12$, with respect to the original Lagrangian, from the prefactor $\tfrac{T}{2}$ in the action.
In order not to clutter the expressions we drop the tildes and the coupling $T$, which is understood to appear in each vertex insertion in Feynman diagrams. We also introduce the notation $\nabla_s=\pa_s-1$.
The cubic interactions read
\begin{equation}\label{bos3ver}
V_{\varphi x x}=-4\varphi \left[\nabla_s\, x\right]^2, \quad  V_{\varphi^3}=2\varphi\left[(\pa_t \varphi)^2-(\pa_s\varphi)^2\right],\quad
V_{\varphi |z|^2}=2\varphi\left[|\pa_t z|^2-|\pa_s z|^2\right],
\end{equation}\vspace{-30pt}
\begin{align}
V_{z\eta\eta}&=-\e^{abc} \pa_t \bar z_a \eta_b \eta_c + h.c. ,&
 V_{z\eta\theta}&=-2\, \e^{abc} \bar z_a \eta_b \nabla_s\theta_c-h.c., \\
 V_{\varphi\eta \theta}&=-4\, i\, \varphi\, \eta_a \nabla_s\theta^a -h.c. ,&
 V_{x \eta \eta}&=-4\, i\,  \eta^a \eta_a \nabla_s x, \\
 V_{z\eta_a\eta_4}&=\ -2\, \pa_t z^a \eta_a \eta_4 + h.c., &
 V_{z\eta_a\theta_4}&=2\, \pa_s z^a \eta_a \theta_4-h.c., \label{fer3verm0} \\
 V_{\varphi \eta_4 \theta^4}&= -2\,i\, \varphi\, (\theta^4 \pa_s \eta_4-\pa_s\theta^4 \eta_4)-h.c., &
 V_{x \psi^4 \psi_4}&= -2\, i\, (\eta^4 \eta_4+\theta^4 \theta_4) \nabla_s x, \label{fer3verm1} 
\end{align}
whereas the quartic vertices are
\begin{align}
V_{z^4}&=\frac16 \left[ \left(\bar z_a \pa_t z^a\right)^2 + \left(\bar z_a \pa_s z^a\right)^2 + \left(z^a \pa_t \bar z_a\right)^2 + \left(z^a \pa_s \bar z_a\right)^2
\right. & &\nonumber\\& \hspace{1cm} \left.
- |z|^2\left(|\pa_t z|^2+|\pa_s z|^2\right) - \left|\bar z_a \pa_t z^a\right|^2 - \left|\bar z_a \pa_s z^a\right|^2\right],
\end{align}\vspace{-40pt}
\begin{align}
&V_{\varphi^2 x x} =16\,\varphi^2\, \left[\nabla_s\, x\right]^2, &
&V_{\varphi^4} =4\, \varphi^2\left[(\pa_t \varphi)^2+(\pa_s\varphi)^2+\frac23\varphi^2\right],\nonumber\\
&V_{\varphi^2 |z|^2} =4\, \varphi^2 \left[|\pa_t z|^2+|\pa_s z|^2\right], &
&V_{\dot z\bar z \psi^4 \psi_4} = -2\, i\, (\eta^4 \eta_4+\theta^4 \theta_4)\bar z _b\pa_t z^b+h.c.,  \nonumber\\
&V_{\eta^2 \eta^4 \eta_4} = 8\,   \eta^4 \eta_4  \eta^a \eta_a , &
& V_{ z'\bar z \psi^4 \psi_4} = -2\, i\, (\eta^4 \theta_4-\theta^4 \eta_4)\bar z _b\pa_s z^b-h.c., \nonumber\\
&V_{\eta^4} =4 (\eta^a \eta_a)^2, &
& V_{\varphi^2 \eta_4 \theta^4} = 4\,i\, \varphi^2\, (\theta^4 \pa_s \eta_4-\pa_s\theta^4 \eta_4)-h.c. ,\nonumber\\
&V_{\eta_4 \eta^4\theta_4 \theta^4} =-8\,  \eta^4 \eta_4  \theta^4 \theta_4, &
& V_{\varphi\, x \psi^4 \psi_4} =12\,i\, \varphi\, (\eta^4 \eta_4+\theta^4 \theta_4) \nabla_s x,\nonumber\\
&V_{\eta^3 \eta_4} =4\,  \e^{abc}\eta_a\eta_b\eta_c\eta_4+h.c., &
&V_{zz\eta^a\eta_4} =-2\, i\, \e_{abc} \pa_t z^a z^b  \eta^c \eta_4+ h.c.,\nonumber\\
&V_{\varphi\, z\eta_a\theta_4} =-8\, \varphi\,  \pa_s z^a \eta_a \theta_4-h.c., &
& V_{\varphi\, z\eta\theta} =8\, \varphi \e^{abc} \bar z_a \eta_b \nabla_s\theta_c-h.c.,\nonumber \\
&V_{zz\eta^a\theta_4} =2\, i\, \e_{abc} \pa_s z^a z^b  \eta^c \theta_4- h.c. ,&
&  V_{zz\eta\eta} =-2\, i\, (\bar z_a \pa_t z^a \eta^b \eta_b-\bar z_b \pa_t z^a  \eta^b \eta_a) +h.c.,\nonumber \\
&V_{\varphi\, x \eta \eta} =24\, i\,\varphi\,  \eta^a \eta_a \nabla_s x, &
&  V_{zz\eta\theta} =-2\, i\, [|z|^2 \eta_a \nabla_s\theta^a - \bar z_b z^a\eta_a \nabla_s\theta^b] -h.c., \nonumber\\
&V_{\varphi^2 \eta \theta} =8\, i\, \varphi^2\, \eta_a \nabla_s\theta^a -h.c. ,&
& V_{x z\eta\eta} =-4\, \nabla_sx \e^{abc} \bar z_a \eta_b \eta_c - h.c.. &
\end{align} 

% Appendix Template

\chapter{Integral reductions for vacuum diagrams} % Main appendix title

\label{app:intred} % Change X to a consecutive letter; for referencing this appendix elsewhere, use \ref{AppendixX}

\lhead{Appendix C. \emph{Integral reductions}} % Change X to a consecutive letter; this is for the header on each page - perhaps a shortened title

In this appendix we provide the relevant tensor integral reductions in two dimensions that 
we used in the computation of the two-loop correction to the partition function in section \ref{twoloops}.
We define the two basic scalar integrals
\begin{align}
I\left[m^2\right] & \equiv \int \frac{d^2\ppp}{\left(2\pi\right)^2}\, \frac{1}{\ppp^2+m^2}, \\
I\left[m_1^2,m_2^2,m_3^2\right] & \equiv \int \frac{d^2\ppp\, d^2\qqq\,d^2\rrr}{\left(2\pi\right)^4}\,  \frac{\delta^{(2)}(\ppp+\qqq+\rrr)}{(\ppp^2+m_1^2)(\qqq^2+m_2^2)(\rrr^2+m_3^2)} \,. 
\end{align}
Then we have (the factors $(2\pi)^4$ in the denominator of the integrands are understood)
{\allowdisplaybreaks
\begin{align}
& \int \frac{d^2\ppp\, d^2\qqq\,d^2\rrr\, \ppp^{\mu} \qqq^{\nu}\, \delta^{(2)}(\ppp+\qqq+\rrr)}{(\ppp^2+m_1^2)(\qqq^2+m_2^2)(\rrr^2+m_3^2)} = \\& 
\quad= \frac{\delta^{\mu\nu}}{4} \left[ I(m_1^2)I(m_2^2) - I(m_1^2)I(m_3^2) - I(m_2^2)I(m_3^2) + (m_1^2+m_2^2-m_3^2) I(m_1^2,m_2^2;m_3^2) \right],
\nonumber\\&
I^{\mu}_{\mu}(m_1^2,m_2^2;m_3^2)=\int \frac{d^2\ppp\, d^2\qqq\,d^2\rrr\, \left(\ppp\cdot \qqq\right)\, \delta^{(2)}(\ppp+\qqq+\rrr)}{(\ppp^2+m_1^2)(\qqq^2+m_2^2)(\rrr^2+m_3^2)} = \\& 
\quad= \frac{1}{2} \left[ I(m_1^2)I(m_2^2) - I(m_1^2)I(m_3^2) - I(m_2^2)I(m_3^2) + (m_1^2+m_2^2-m_3^2) I(m_1^2,m_2^2;m_3^2) \right],\nonumber
\\&
\int \frac{d^2\ppp\, d^2\qqq\,d^2\rrr\, \ppp^{\mu}\, \ppp^{\nu}\, \delta^{(2)}(\ppp+\qqq+\rrr)}{(\ppp^2+m_1^2)(\qqq^2+m_2^2)(\rrr^2+m_3^2)} 
= \frac{\delta^{\mu\nu}}{2} \left[ I(m_2^2)I(m_3^2) - m_1^2\, I(m_1^2,m_2^2;m_3^2) \right],
\\&
J\equiv \int \frac{d^2\ppp\, d^2\qqq\,d^2\rrr\, \ppp^2 \qqq^2\, \delta^{(2)}(\ppp+\qqq+\rrr)}{(\ppp^2+m_1^2)(\qqq^2+m_2^2)(\rrr^2+m_3^2)} 
\\&\quad= m_1^2 m_2^2\, I(m_1^2,m_2^2;m_3^2) - m_1^2\, I(m_1^2)I(m_3^2) - m_2^2\, I(m_2^2)I(m_3^2) ,
\nonumber\\&
K\equiv \int \frac{d^2\ppp\, d^2\qqq\,d^2\rrr\, (\ppp\cdot \qqq)^2\, \delta^{(2)}(\ppp+\qqq+\rrr)}{(\ppp^2+m_1^2)(\qqq^2+m_2^2)(\rrr^2+m_3^2)} 
\\&\quad= \frac{1}{2} \left[ - m_2^2\, I(m_2^2)I(m_3^2) - m_1^2\, I(m_1^2)I(m_3^2)  + (m_1^2+m_2^2-m_3^2) I^{\mu}_{\mu}(m_1^2,m_2^2;m_3^2) \right],\nonumber
\\&
\int \frac{d^2\ppp\, d^2\qqq\,d^2\rrr\, \ppp^{\mu}\, \ppp^{\nu}\, \qqq^{\rho}\, \qqq^{\s}\, \delta^{(2)}(\ppp+\qqq+\rrr)}{(\ppp^2+m_1^2)(\qqq^2+m_2^2)(\rrr^2+m_3^2)} 
\\&\quad= \left(\frac38 J - \frac14 K \right) \delta^{\mu\nu}\delta^{\rho\sigma} + 
\left(\frac14 K - \frac18 J \right) \left( \delta^{\mu\rho}\delta^{\nu\sigma} + \delta^{\mu\sigma}\delta^{\nu\rho} \right),
\nonumber\\&
\int \frac{d^2\ppp\, d^2\qqq\,d^2\rrr\, \ppp^{\mu}\, \ppp^{\nu}\, \ppp^{\rho}\, \qqq^{\s}\, \delta^{(2)}(\ppp+\qqq+\rrr)}{(\ppp^2+m_1^2)(\qqq^2+m_2^2)(\rrr^2+m_3^2)} 
\\&\quad= \frac18 \left( \delta^{\mu\nu}\delta^{\rho\sigma} + \delta^{\mu\rho}\delta^{\nu\sigma} + \delta^{\mu\sigma}\delta^{\nu\rho} \right)
\left[ m_2^2\, I(m_2^2)I(m_3^2) - m_1^2\, I^{\mu}_{\mu}(m_1^2,m_2^2;m_3^2) \right],
\nonumber\\&
L\equiv \int \frac{d^2\ppp\, d^2\qqq\,d^2\rrr\, \ppp^2\, \left(\qqq\cdot \rrr\right)\, \delta^{(2)}(\ppp+\qqq+\rrr)}{(\ppp^2+m_1^2)(\qqq^2+m_2^2)(\rrr^2+m_3^2)} 
= -m_1^2\, I^{\mu}_{\mu}(m_3^2,m_2^2;m_1^2),
\\&
M\equiv \int \frac{d^2\ppp\, d^2\qqq\,d^2\rrr\, (\ppp\cdot \qqq) (\ppp\cdot \rrr)\, \delta^{(2)}(\ppp+\qqq+\rrr)}{(\ppp^2+m_1^2)(\qqq^2+m_2^2)(\rrr^2+m_3^2)} 
\\&\quad= \frac{1}{2} \left[ (m_1^2+m_3^2-m_2^2) I^{\mu}_{\mu}(m_1^2,m_2^2;m_3^2) +  m_1^2\, I(m_1^2)I(m_3^2) - m_2^2\, I(m_2^2)I(m_3^2)  \right],
\nonumber\\&
\int \frac{d^2\ppp\, d^2\qqq\,d^2\rrr\, \ppp^{\mu}\, \ppp^{\nu}\, \qqq^{\rho}\, \rrr^{\sigma} \delta^{(2)}(\ppp+\qqq+\rrr)}{(\ppp^2+m_1^2)(\qqq^2+m_2^2)(\rrr^2+m_3^2)} 
\\&\quad= \left(\frac38 L - \frac14 M \right) \delta^{\mu\nu}\delta^{\rho\sigma} + 
\left(\frac14 M - \frac18 L \right) \left( \delta^{\mu\rho}\delta^{\nu\sigma} + \delta^{\mu\sigma}\delta^{\nu\rho} \right).\nonumber
\end{align}}

% Appendix Template

\chapter{Self-energies of fermions in the null cusp background for \txpf{$AdS_4\times \mathbb{CP}^3$}{AdS4 x CP3}} % Main appendix title

\label{app:fermse} % Change X to a consecutive letter; for referencing this appendix elsewhere, use \ref{AppendixX}

\lhead{Appendix E. \emph{Self-energies of fermions}} % Change X to a consecutive letter; this is for the header on each page - perhaps a shortened title

In this appendix we collect the off-shell fermion self-energies entering the computation of the one-loop dispersion relations in section \ref{sec:dispersion}.
\small
\begin{align}
F_{\eta_a\eta^a}^{(1)} &= \frac{2}{\ppp^6} \bigg[
\left(\ppp^2+1\right) \Big(\left(-2 {\rm I}[1]-{\rm I}[2]-{\rm I}[4
]\right) \ppp^6 
\nonumber\\&
+\ppp^4 \left(\left(6 {\rm I}[1]+{\rm I}[2]-7 {\rm I}[4]\right) \ppp_1^2-2 {\rm I}[1]+{\rm I}[2]+{\rm I}[4]\right)
\nonumber\\&
+ \ppp^2 \left( \left(26 {\rm I}[1]-5 {\rm I}[2]-21 {\rm I}[4]\right) \ppp_1^2+ \left(16 {\rm I}[4]-16 {\rm I}[1]\right)\ppp_1^4\right)\nonumber\\
&-4 \left(10 {\rm I}[1]-{\rm I}[2]-9 {\rm I}[4]\right) \ppp_1^4\Big) 
\nonumber\\&
- \left(\ppp^2+1\right) \Big(
\frac{2 \ppp_1^2\left(\ppp^4-\ppp^2+4\, \ppp_1^2\right) \log \left(\ppp^2+1\right)}{\pi } 
\nonumber\\&
-\left(3 \ppp^4 + 4 \ppp^6 + \ppp^8 - 63 \ppp^2 \ppp_1^2 - 56 \ppp^4 \ppp_1^2 - 9 \ppp^6 \ppp_1^2 + 
 108 \ppp_1^4 + 108 \ppp^2 \ppp_1^4 + 16 \ppp^4 \ppp_1^4\right) {\rm I}[1,4] \Big) 
\nonumber\\&
+\left(\ppp^2-\ppp_1^2\right) \left(\ppp^2 \left(\ppp^2+1\right)^3+4 \left(\ppp^4-4\, \ppp^2-1\right) \ppp_1^2\right) {\rm I}[1,2]
\bigg],
\end{align}
\begin{align}
F_{\theta\bar\theta}^{(1)} &= 2 \frac{\ppp_1^2+1}{\ppp^6} \bigg[
\left(\ppp^2+1\right) \Big( 
\ppp^4 \left(6 {\rm I}[1]+{\rm I}[2]-7 {\rm I}[4]\right) 
\nonumber\\&
+ \ppp^2 \left( 10 {\rm I}[1]- {\rm I}[2]-9 {\rm I}[4] + \left(16 {\rm I}[4]-16 {\rm I}[1]\right)\ppp_1^2\right)-4 \left(10 {\rm I}[1]-{\rm I}[2]-9 {\rm I}[4]\right) \ppp_1^2\Big) 
\nonumber\\&
- \left(\ppp^2+1\right) \Big(
\frac{2 \left(\ppp^4-\ppp^2+4\, \ppp_1^2\right) \log \left(\ppp^2+1\right)}{\pi }  
\nonumber\\&
-\left(27 \ppp^2 + 36 \ppp^4 + 9 \ppp^6 - 108 \ppp_1^2 - 108 \ppp^2 \ppp_1^2 - 16 \ppp^4 \ppp_1^2\right) {\rm I}[1,4]\Big)
\nonumber\\&
- \left(\ppp^2 \left(\ppp^2+1\right)^3+4 \left(\ppp^4-4\, \ppp^2-1\right) \ppp_1^2\right) {\rm I}[1,2]
\bigg],
\end{align}
\begin{align}
F_{\eta\bar\theta}^{(1)} &= \frac{2}{\ppp^6} \bigg[
\left(\ppp^2+1\right) \Big(
\left(-4 {\rm I}[0]+2 {\rm I}[1]-{\rm I}[2]+{\rm I}[4]\right) \ppp^6
\nonumber\\&
+\ppp^4 \left(\left(14 {\rm I}[1]+ {\rm I}[2]-15 {\rm I}[4]\right) \ppp_1^2-4 {\rm I}[1]+{\rm I}[2]+3 {\rm I}[4]\right)
\nonumber\\&
+\ppp^2 \ppp_1^2 \left(\left(16 {\rm I}[4]-16 {\rm I}[1]\right) \ppp_1^4+38 {\rm I}[1]-5 {\rm I}[2]-33 {\rm I}[4]\right)+4 \left({\rm I}[2]+9 {\rm I}[4]-10 {\rm I}[1]\right) \ppp_1^4 \Big)
\nonumber\\&
- \left(\ppp^2+1\right) \Big(
\frac{2 \left(\ppp^4-3\ppp^2+4\, \ppp_1^2\right) \log \left(\ppp^2+1\right)}{\pi }
\nonumber\\&
+\left(9 \ppp^4 + 12 \ppp^6 + 3 \ppp^8 - 99 \ppp^2 \ppp_1^2 - 100 \ppp^4 \ppp_1^2 - 17 \ppp^6 \ppp_1^2 + 
 108 \ppp_1^4 + 108 \ppp^2 \ppp_1^4 + 16 \ppp^4 \ppp_1^4\right) {\rm I}[1,4]\Big) 
\nonumber\\&
+ \left(\ppp^2-\ppp_1^2\right) \left(\ppp^2(\ppp^2+1)^3+4 \left(\ppp^4-4\ppp^2-1\right) \ppp_1^2\right) {\rm I}[1,2]
\bigg],
\end{align}
\begin{align}
F_{\eta_4\bar\eta^4}^{(1)} &= F_{\theta_4\bar\theta^4}^{(1)} = \frac{1}{4 \pi  \ppp^6}\bigg(6 \left(1 + \ppp^2\right) \left(\ppp^6 + 12 \ppp^2 \ppp_1^2 - 16 \ppp_1^4 - \ppp^4 \left(1 + 4 \ppp_1^2\right)\right) \log \left(\ppp^2+1\right)
\nonumber\\&
+\left(\ppp^4 \left(4 + \ppp^2\right)^2 - 32 \ppp^2 \left(6 + 5 \ppp^2 + \ppp^4\right) \ppp_1^2 +
  64 \left(2 + \ppp^2\right)^2 \ppp_1^4\right) \log \left(\tfrac{\ppp^2}{4}+1\right)
\nonumber\\&
+\left(4 \ppp^4 + \ppp^8 - 48 \ppp^2 \ppp_1^2 + 64 \ppp_1^4\right) \log \left(\tfrac{\ppp^2}{2}+1\right)\bigg)-\left(6 {\rm I}[1]+{\rm I}[2]+{\rm I}[4]\right) \ppp^2\, ,
\end{align}
\begin{align}
F_{\eta_4\bar\theta^4}^{(1)} &= \frac{1}{4 \pi  \ppp^6}\bigg(6 \left(1 + \ppp^2\right) \left(3\ppp^6 + 20 \ppp^2 \ppp_1^2 - 16 \ppp_1^4 - \ppp^4 \left(5 + 4 \ppp_1^2\right)\right) \log \left(\ppp^2+1\right)
\nonumber\\&
+\left(\ppp^4 \left(4 + \ppp^2\right)(20+9\ppp^2) - 32 \ppp^2\left(2 + \ppp^2\right) \left(5 +2 \ppp^2\right) \ppp_1^2 + 
  64 \left(2 + \ppp^2\right)^2 \ppp_1^4\right) \log \left(\tfrac{\ppp^2}{4}+1\right)
\nonumber\\&
+\left(20 \ppp^4 + \ppp^8 - 80 \ppp^2 \ppp_1^2 + 64 \ppp_1^4\right) \log \left(\tfrac{\ppp^2}{2}+1\right)\bigg)-\left(6 {\rm I}[1]+{\rm I}[2]+{\rm I}[4]\right) \ppp^2\, .
\end{align}

\addtocontents{toc}{\vspace{2em}} % Add a gap in the Contents, for aesthetics

\backmatter

%----------------------------------------------------------------------------------------
%	BIBLIOGRAPHY
%----------------------------------------------------------------------------------------

\label{Bibliography}

\lhead{\emph{Bibliography}} % Change the page header to say "Bibliography"

\bibliographystyle{nb} % Use the "unsrtnat" BibTeX style for formatting the Bibliography

\bibliography{biblio} % The references (bibliography) information are stored in the file named "Bibliography.bib"

\end{document}